\documentclass[12pt]{report}

\def\proof{\noindent{\sl Proof:}\kern0.6em}

\def\frac#1#2{\hbox{$#1\over#2$}}
\def\dual{\mathstrut^*\kern-0.1em}

\def\lvec#1{\setbox0=\hbox{$#1$}
    \setbox1=\hbox{$\scriptstyle\leftarrow$}
    #1\kern-\wd0\smash{
    \raise\ht0\hbox{$\raise1pt\hbox{$\scriptstyle\leftarrow$}$}}
    \kern-\wd1\kern\wd0}
\def\rvec#1{\setbox0=\hbox{$#1$}
    \setbox1=\hbox{$\scriptstyle\rightarrow$}
    #1\kern-\wd0\smash{
    \raise\ht0\hbox{$\raise1pt\hbox{$\scriptstyle\rightarrow$}$}}
    \kern-\wd1\kern\wd0}

\def\nabstar#1{\nabla\kern-0.5pt\smash{\raise 4.5pt\hbox{$\ast$}}
               \kern-4.5pt_{#1}}

\def\drvstar#1{\partial\kern-0.5pt\smash{\raise 4.5pt\hbox{$\ast$}}
               \kern-5.0pt_{#1}}

\def\momp#1#2{
    \setbox0=\hbox{${#1}$}\setbox1=\hbox{${#1}_{#2}$}
    {#1}_{#2}\kern-\wd1\kern\wd0
    \smash{\raise4.5pt\hbox{$\scriptscriptstyle +$}}}
\def\momm#1#2{
    \setbox0=\hbox{${#1}$}\setbox1=\hbox{${#1}_{#2}$}
    {#1}_{#2}\kern-\wd1\kern\wd0
    \smash{\raise4.5pt\hbox{$\scriptscriptstyle -$}}}
\def\mompm#1#2{
    \setbox0=\hbox{${#1}$}\setbox1=\hbox{${#1}_{#2}$}
    {#1}_{#2}\kern-\wd1\kern\wd0
    \smash{\raise4.5pt\hbox{$\scriptscriptstyle \pm$}}}
\def\smomp#1#2{
    \setbox0=\hbox{${#1}$}\setbox1=\hbox{${#1}_{#2}$}
    {#1}_{#2}\kern-\wd1\kern\wd0
    \smash{\raise3pt\hbox{$\scriptscriptstyle +$}}}
\def\smomm#1#2{
    \setbox0=\hbox{${#1}$}\setbox1=\hbox{${#1}_{#2}$}
    {#1}_{#2}\kern-\wd1\kern\wd0
    \smash{\raise3pt\hbox{$\scriptscriptstyle -$}}}
\def\smompm#1#2{
    \setbox0=\hbox{${#1}$}\setbox1=\hbox{${#1}_{#2}$}
    {#1}_{#2}\kern-\wd1\kern\wd0
    \smash{\raise3pt\hbox{$\scriptscriptstyle \pm$}}}

\def\rhoprime{\rho\kern1pt'}
\def\rhobar{\bar{\rho}}
\def\rhobarprime{\rhobar\kern1pt'}
\def\rhobartilde{\kern2pt\tilde{\kern-2pt\rhobar}}
\def\rhobartildeprime{\kern2pt\tilde{\kern-2pt\rhobar}\kern1pt'}

\def\zetabar{\bar{\zeta}}
\def\zetaprime{\zeta\kern1pt'}
\def\zetabarprime{\zetabar\kern1pt'}
\def\zetar{\zeta_{\raise-1pt\hbox{\rm R}}}
\def\zetabarr{\zetabar_{\raise-1pt\hbox{\rm R}}}

\def\phiimpr{\phi_{\kern0.5pt{\rm I}}}

\def\diracstar#1#2{
    \setbox0=\hbox{$\gamma$}\setbox1=\hbox{$\gamma_{#1}$}
    \gamma_{#1}\kern-\wd1\kern\wd0
    \smash{\raise4.5pt\hbox{$\scriptstyle#2$}}}

\def\faprime{\setbox0=\hbox{$f$}\setbox1=\hbox{$f_{\rm A}$}
    f_{\rm A}\kern-\wd1\kern\wd0
    \smash{\raise5.5pt\hbox{\kern0.5pt$\scriptstyle\prime$}}\kern1pt}
\def\fpprime{\setbox0=\hbox{$f$}\setbox1=\hbox{$f_{\rm P}$}
    f_{\rm P}\kern-\wd1\kern\wd0
    \smash{\raise5.5pt\hbox{\kern0.5pt$\scriptstyle\prime$}}\kern1pt}
\def\fxprime{\setbox0=\hbox{$f$}\setbox1=\hbox{$f_{\rm X}$}
    f_{\rm X}\kern-\wd1\kern\wd0
    \smash{\raise5.5pt\hbox{\kern0.5pt$\scriptstyle\prime$}}\kern1pt}

\def\opprime#1{\setbox0=\hbox{${\cal O}$}\setbox1=\hbox{${\cal O}_{\rm #1}$}
    {\cal O}_{\rm #1}\kern-\wd1\kern\wd0
    \smash{\raise4.5pt\hbox{\kern1pt$\scriptstyle\prime$}}\kern1pt}

\def\ophatprime#1{\setbox0=\hbox{$\widehat{\cal O}$}
    \setbox1=\hbox{$\widehat{\cal O}_{\rm #1}$}
    \widehat{\cal O}_{\rm #1}\kern-\wd1\kern\wd0
    \smash{\raise4.5pt\hbox{\kern1pt$\scriptstyle\prime$}}\kern1pt}

\def\bopprime#1{\setbox0=\hbox{${\cal O}$}\setbox1=\hbox{${\cal O}_{\rm #1}$}
    {\cal L}_{\rm #1}\kern-\wd1\kern\wd0
    \smash{\raise4.5pt\hbox{\kern1pt$\scriptstyle\prime$}}\kern1pt}

\def\blagprime#1{\setbox0=\hbox{${\cal B}$}\setbox1=\hbox{${\cal B}_{#1}$}
    {\cal B}_{#1}\kern-\wd1\kern\wd0
    \smash{\raise5.2pt\hbox{\kern1pt$\scriptstyle\prime$}}\kern1pt}

\def\s{s}

\def\mbar{\kern1pt\overline{\kern-1pt m\kern-1pt}\kern1pt}

\def\msbar{{\rm \overline{MS\kern-0.05em}\kern0.05em}}

\def\blackboardrrm{\mathchoice
{\rm I\kern-0.21 em{R}}{\rm I\kern-0.21 em{R}}
{\rm I\kern-0.19 em{R}}{\rm I\kern-0.19 em{R}}}
 
\def\blackboardzrm{\mathchoice
{\rm Z\kern-0.32 em{Z}}{\rm Z\kern-0.32 em{Z}}
{\rm Z\kern-0.28 em{Z}}{\rm Z\kern-0.28 em{Z}}}
 
\def\blackboardh{\mathchoice
{\rm I\kern-0.14 em{H}}{\rm I\kern-0.14 em{H}}
{\rm I\kern-0.11 em{H}}{\rm I\kern-0.11 em{H}}}
 
\def\blackboardp{\mathchoice
{\rm I\kern-0.14 em{P}}{\rm I\kern-0.14 em{P}}
{\rm I\kern-0.11 em{P}}{\rm I\kern-0.11 em{P}}}
 
\def\blackboardt{\mathchoice
{\rm T\kern-0.52 em{T}}{\rm T\kern-0.52 em{T}}
{\rm T\kern-0.40 em{T}}{\rm T\kern-0.40 em{T}}}
  
\def\deltaoneprime{\Delta\kern-1.0pt
    \smash{\raise 4.5pt\hbox{$\scriptstyle\prime$}}
    \kern-1.5pt_{1}}

\def\zetaprime#1{\zeta'(0|#1)}
 
\def\Tr{{\rm Tr}}
 
\def\delstar#1{\Delta\kern-1.0pt\smash{\raise 4.5pt\hbox{$\ast$}}
               \kern-4.0pt_{#1}}

\def\nabstar#1{\nabla\kern-0.5pt\smash{\raise 4.5pt\hbox{$\ast$}}
               \kern-4.5pt_{#1}}
\def\cdev#1{D\kern-0.2pt\smash{\raise 4.2pt
            \hbox{$\scriptstyle\phantom{\ast}$}}
            \kern-4.8pt_{#1}}
\def\cdevstar#1{D\kern-0.2pt\smash{\raise 4.2pt
                \hbox{$\scriptstyle\ast$}}
                \kern-4.8pt_{#1}}

\newcommand{\bes}{\begin{eqnarray}}
\newcommand{\ees}{\end{eqnarray}}

\newcommand{\eq}[1]{(\ref{#1})}

 \usepackage[utf8]{inputenc}
\usepackage{dsfont}

\usepackage{tikz}
\usepackage[compat=1.1.0]{tikz-feynman}
\usetikzlibrary{shapes,positioning,automata,backgrounds,calc,er,patterns}
    \usepackage{pgfplots}
\usetikzlibrary{positioning,arrows}
\usetikzlibrary{decorations.pathmorphing}
\usepackage[english]{babel}
\usepackage{color}
\usepackage{graphicx}
\usepackage{latexsym}
\usepackage{mathtools}
\usepackage{amssymb, amsmath, amsfonts,bbold}
\usepackage{indentfirst}
\usepackage{float,times}
\usepackage{bbm}
\usepackage{subfigure}
\usepackage{slashed}
\usepackage{ifthen}
\usepackage{mathrsfs}

\usepackage{array}    
\usepackage{setspace}
 \usepackage{pdfpages}
\usepackage{pxfonts}
\usepackage{blindtext}
\usepackage{youngtab}
\usepackage{ytableau} 
\usepackage{cases}
\usepackage{comment}
\usepackage{placeins}
\usepackage{xspace}
\usepackage{cancel} 
\usepackage{soul}
  
\usepackage{subfigure}
 \usepackage{ifthen}
 \usepackage{array}    
\usepackage{setspace}
\usepackage{pdfpages}
\usepackage{multirow}
\usepackage{youngtab}
\newcommand{\FC}{\mathrm{FC}}
\newcommand{\SM}{\mathrm{SM}}
\newcommand{\SU}{\mathrm{SU}}
\newcommand{\U}{\mathrm{U}}
\newcommand{\Sp}{\mathrm{Sp}}
\newcommand{\SO}{\mathrm{SO}}
\newcommand{\tcf}{\mathcal{F}}

 \graphicspath{{Figs/}{LatticeFigs/}}
\usepackage[
bookmarks]{hyperref}

\hypersetup{colorlinks, bookmarksopen, bookmarksnumbered,
citecolor=verdeCP3, linkcolor=bluCP3, pdfstartview=FitH, urlcolor=rossoCP3
}

\def\s#1{\setbox0=\hbox{$#1$}%
\rlap{\ifdim\wd0>.7em\kern.22\wd0\else\kern.1\wd0\fi /}#1}
\def\circa#1{\,\raise.3ex\hbox{$#1$\kern-.75em\lower1ex\hbox{$\sim$}}\,}

\newskip\humongous \humongous=0pt plus 1000pt minus 1000pt

\newif\ifdtup

\def\oldreffmt#1{\rlap{[#1]} \hbox to 2\parindent{}}

\def\figfmt#1{\rlap{Figure {#1}} \hbox to 1in{}}

\def\Tr{\mathop{\rm Tr}}

\def\abs#1{\left| #1\right|}

\def\beq{\begin{eqnarray}}
\def\eeq{\end{eqnarray}}

 \newcommand{\ba}{\begin{array}}
\newcommand{\ea}{\end{array}}
\newcommand{\bi}{\begin{itemize}}
\newcommand{\ei}{\end{itemize}}
\newcommand{\bn}{\begin{enumerate}}
\newcommand{\en}{\end{enumerate}}
\newcommand{\bc}{\begin{center}}
\newcommand{\ec}{\end{center}}

 \newcommand{\gsim}{\lower.7ex\hbox{$\;\stackrel{\textstyle>}{\sim}\;$}}
\newcommand{\lsim}{\lower.7ex\hbox{$\;\stackrel{\textstyle<}{\sim}\;$}}

\newcommand{\SP}{\,{\rm Sp}}

\newcommand{\mF}{\mathcal F}

\newcommand{\spur}[2]{\psi^{#1}\phantom{}_{#2} }
\newcommand{\spurbar}[2]{\bar{\psi}^{#1 #2}} 
\def\reffig#1{Fig.~\ref{#1}}

\def\be{\begin{equation}} 
\def\ee{\end{equation}} 
\def\bea{\begin{eqnarray}}
\def\eea{\end{eqnarray}}
\def\ba{\begin{array}}
\def\ea{\end{array}}
\def\kslash{\raise.15ex\hbox{/}\kern-.57em k}
\newcommand{\bear}{\begin{eqnarray}}
\newcommand{\eear}{\end{eqnarray}}

   \newcommand{\cO}{\mbox{\cal O}}
\def\bal{\begin{align}}
\def\eal{\end{align}}

\def\Tr{\textrm{Tr}}

\def\drawbox#1#2{\hrule height#2pt
        \hbox{\vrule width#2pt height#1pt \kern#1pt
              \vrule width#2pt}
              \hrule height#2pt}

\def\Asym#1#2{\vcenter{\vbox{\drawbox{#1}{#2}
              \kern-#2pt 
              \drawbox{#1}{#2}}}}

\definecolor{rossoCP3}{cmyk}{0,.88,.77,.40}
\definecolor{verdeCP3}{rgb}{0.09765625, 0.57421875, 0.1015625}
\definecolor{bluCP3}{rgb}{0, 0.23, 0.67}

\begin{document}

\setlength{\unitlength}{1mm}


\begin{titlepage}
\begin{center}
{\LARGE {\color{rossoCP3}
\bf Fundamental Composite Dynamics: A Review  }} 
\end{center}
\par \vskip .2cm \noindent
\begin{center}
{  Giacomo Cacciapaglia$^{\color{rossoCP3} {\spadesuit}}$, Claudio Pica$^{\color{rossoCP3} {\vardiamondsuit}}$ and Francesco Sannino$^{\color{rossoCP3} {\varheartsuit}}$ 
}
\end{center}
\begin{center}
 \noindent
  { 
$^{\color{rossoCP3} {\spadesuit}}$ \mbox{\small \it Universit\'e de Lyon, F-69622 Lyon, France: Universit\'e Lyon 1, Villeurbanne} \\ \mbox{\it CNRS/IN2P3, UMR5822, Institut de Physique des deux Infinis (IP2I) de Lyon.}}\\ 

\vskip .2cm
  { \it
$^{\color{rossoCP3} {\vardiamondsuit}}${ CP}$^{ \bf 3}${-Origins}, Institute of Mathematics and Computer Science,\\ University of Southern Denmark, Odense, Denmark 
}
 
\vskip .2cm
  { \it
$^{\color{rossoCP3} {\varheartsuit}}${ CP}$^{ \bf 3}${-Origins} and Danish-IAS, University of Southern Denmark, Odense, Denmark \\ \& \\ 
Dept. of Physics "Ettore Pancini", Univ. di Napoli "Federico II", Napoli, Italy} 
 
\end{center}     
\vskip 1cm
 
 We introduce fundamental gauge theories that can be employed to construct informed composite bright and dark extensions of the Standard Model, within and beyond the standard paradigms. The gap between theory and experiments is bridged by providing  predictions and ways to test them, for example, at the Fermi scale and via precision flavor experiments. We will review time-honoured paradigms from (walking) technicolor to composite Goldstone Higgs and discuss their features and differences. Standard model fermion mass generation in composite models will also be discussed along with the challenges and opportunities that it offers. To be concrete and pedagogical we will concentrate on minimal constructions featuring strongly coupled gauge theories supporting the global symmetry breaking pattern SU(4)/Sp(4). The most minimal underlying fundamental description consists of an SU(2) gauge theory with two Dirac fermions transforming according to the fundamental representation of the gauge group. This minimal choice enables us to use first principle lattice results to predict the massive spectrum for models of composite (Goldstone) Higgs dynamics and strongly interacting  dark matter, of immediate impact for current and future experimental searches. 
   Because composite dynamics embraces a rich spectrum of theories with dynamics ranging from  QCD-like behaviour to (near) conformal one, we also report here the state-of-the-art of numerical and analytic properties of  several strongly coupled theories including their spectrum, phase diagrams and, when applicable, their (near) conformal data. 
  

\end{titlepage}

         \newpage

\def\baselinestretch{1.0}
\tiny
\normalsize

\tableofcontents

\newpage

\chapter{Composite extensions of the Standard Model, why?}

 The Standard Model (SM) of particle interactions is an excellent description of physical phenomena down to the astonishing scale of $10^{-17}~{\rm m}$, which corresponds to a high energy scale of the order of a fraction of a TeV. However, little is known about the fundamental laws of nature at even shorter distances and how they will interface with quantum gravity when reaching distances of the order of $10^{-35}~{\rm m}$. The latter is the distance where  the quantum nature of gravity cannot be ignored. 
Additionally, despite its unquestionable success, the SM falls short to explaining the matter-antimatter asymmetry in the universe, i.e. why there is something rather than nothing. Furthermore, it doesn't account for a particle interpretation of dark matter and/or dark energy. In other words, the SM cannot account for 95\% of the universe and therefore must be extended.

The discovery of the Higgs particle at the Large Hadron Collider (LHC) has added robustness to the SM. It also mutated the Higgs particle and its sector from being missing ingredients to being tools to explore and test extensions of the SM. 
 
But, how to choose on which extension of the SM to work on? It is a risky business to spend time working on extensions of the SM  that might not be selected by nature and some play (mostly) safe by working on bottom-up approaches rooted into the effective field theory approach. The obvious advantage is the non-commitment to a specific underlying realization. This is, however, also a limiting factor: beyond the leading order in whatever expansion parameter is used, the effective approach generally introduces a large number of operators with unknown coefficients, thus reducing the {\it effectiveness} and prediction power of the  approach. Marrying effective approaches and true underlying realizations has the beneficial effect of remaining sufficiently general at low energies while guaranteeing that a deeper fundamental realization exists and can be tested experimentally. This marriage is by no means always straightforward rending the entire project demanding and therefore highly relevant and interesting. This is part of what we will show in this review, i.e. how to match a fundamental field theory dynamics to an effective approach. So far,  however, we have not yet answered the question on which extensions to work on because there is obviously no \emph{right} answer. We believe that one  should work on the extensions that you feel most comfortable about and can learn most from. Extra theoretical guidance can further influence the choice, however we believe that this should not be a limiting factor. 

We chose to review here fundamental composite extensions of the SM because they are phenomenologically relevant, theoretically appealing and technically challenging. In fact,  these extensions require to determine the non-perturbative dynamics of unexplored gauge-Yukawa theories. The results will therefore remain in the literature ready to be employed in other realms of particle physics and cosmology.  Another important aspect is the highly interdisciplinary nature of these explorations since they touch upon collider physics, model building, effective field theories, conformal methods, and they further require state-of-the-art first principle lattice field theory studies. 

\section{From Technicolor to Composite Higgs}

A time-honored avenue to render the SM Higgs sector natural is to replace it with a fundamental gauge dynamics featuring fermionic matter fields. Naturalness is solved because  the chiral symmetry of the new fermions forbids power law sensitivity of physical quantities to a higher energy cutoff.  The oldest incarnation of this idea goes under the name of Technicolor (TC) \cite{Weinberg:1975gm,Susskind:1978ms}. In these models the Higgs-sector, and therefore the Higgs boson itself, are made of new fundamental dynamics. Variations on the theme appeared later in the literature \cite{Kaplan:1983fs,Kaplan:1983sm}. 

In the traditional Technicolor setup, the electroweak (EW) symmetry breaks thanks to the gauge dynamics of the new underlying gauge theory. The Technicolor Higgs is then identified with the lightest scalar excitation of the fermion condensate responsible for the electroweak symmetry breaking (EWSB). However, the Technicolor theory per se is not able to provide mass to the SM fermions and therefore a new sector must be introduced \cite{Eichten:1979ah}. This extended Technicolor (ETC) sector cannot be neglected since it  modifies the overall dynamics of the theory \cite{Fukano:2010yv} as proven in \cite{Rantaharju:2017eej,Rantaharju:2019nmh}. In fact, even the physical mass of the Technicolor Higgs is typically reduced due to the ETC interactions \cite{Foadi:2012bb}. 

Additionally the full, i.e. TC + ETC, dynamics can be of near-conformal type. This  dynamics can emerge near the number-of-flavour-driven quantum phase transition from an IR fixed point to a non-conformal phase where chiral symmetry is broken~\cite{Miransky:1996pd}.  Several scenarios have been envisioned for this type of phase transition ranging from  a Berezinskii--Kosterlitz--Thouless (BKT)-like phase transition~\cite{Kosterlitz:1974sm}, used in four dimensions in~\cite{Miransky:1984ef,Miransky:1996pd,Holdom:1988gs,Holdom:1988gr,
Cohen:1988sq,Appelquist:1996dq,Gies:2005as}, to a jumping (non-continuous) phase transition~\cite{Sannino:2012wy}. The discovery that higher-dimensional representations could be (near) conformal~\cite{Sannino:2004qp} for a small number of flavours led to the well-known conformal window phase diagram of~\cite{Dietrich:2006cm} that  guides lattice investigations~\cite{Pica:2017gcb}.   
A smooth quantum phase transition for four-dimensional gauge-fermion theories is also known as \emph{walking}~\cite{Holdom:1988gs,Holdom:1988gr}. It is expected, as we shall review, to enhance the effect of bilinear fermion operators. This is a relevant property for the generation of SM fermion masses \cite{Holdom:1988gs,Holdom:1988gr}.  However, for gauge-fermion theories, it is impossible to approach the transition in a continuous manner (since the number of flavors is an integer), limiting studies of the nature of the transition. Fortunately, the ETC interactions come to the rescue \cite{Fukano:2010yv}. They allow to consider, instead, the quantum phase phase transition  in the gauged Nambu--Jona-Lasinio (NJL) model varying the four-fermion coupling continuously, starting at zero value, within a gauge-fermion theory that is IR conformal. At a large enough NJL coupling, the four fermion interaction is expected to break chiral symmetry. If the transition is continuous, it will lead to what is known as {\it ideal walking}  \cite{Fukano:2010yv,Yamawaki:1996vr}. That such a possibility occurs has been recently demonstrated in \cite{Rantaharju:2017eej,Rantaharju:2019nmh}.

Another possibility is that the gauge dynamics underlying the Higgs sector does not break the EW symmetry but breaks a global symmetry of the new fermions: a Higgs-like state is therefore identified with one of the Goldstone Bosons (GB) of the global symmetry breaking \cite{Kaplan:1983fs}. In this case the challenges are not only to provide masses to the SM fermions but also to break the EW symmetry in the first place via another sector that, in turn, should also contribute to give mass to the would--be pseudo-Nambu--Goldstone boson (pNGB) Higgs. In any event, a true improvement with respect to the SM Higgs sector shortcomings would arise only if a more fundamental description exists.

The exciting physics above triggered, in recent years, much dedicated analytic and numerical work aimed at identifying a fundamental description of models of composite dynamics for the EW sector of the SM including mass generation for the SM fermions. 

In the following we will clarify the differences, similarities, interplay, and shortcomings of the various composite approaches. We will also provide specific underlying realizations in terms of fundamental strongly coupled gauge theories. This will permit us to use recent first principle lattice results to make predictions for the massive spectrum of the composite theories, which is of the utmost relevance to guide searches of new physics at the LHC and future colliders. Furthermore, we will review that, for a generic vacuum alignment, the observed Higgs is neither a purely pNGB state nor the TC--Higgs, but rather a mixed state. This fact impacts on its physical properties and associated phenomenology. Of particular relevance, for a walking transition, is the interplay between a dilaton-like state and a pNGB Higgs. The existence of a dilaton-like Higgs in the walking regime could be enforced by approximate conformal invariance~\cite{Leung:1985sn,Bardeen:1985sm,Yamawaki:1985zg,Sannino:1999qe,Hong:2004td,Dietrich:2005jn,Appelquist:2010gy}.  
This subject has  received renewed interest~\cite{Hong:2004td,Dietrich:2005jn,Goldberger:2008zz,Appelquist:2010gy,Hashimoto:2010nw,Matsuzaki:2013eva,Golterman:2016lsd,Hansen:2016fri,Golterman:2018mfm} due to recent lattice studies ~\cite{Appelquist:2016viq,Appelquist:2018yqe,Aoki:2014oha,Aoki:2016wnc,Fodor:2012ty,Fodor:2017nlp,Fodor:2019vmw} that reported evidence of the presence of a light singlet scalar particle in the spectrum.

As possible underlying gauge theories we will consider, at first, those featuring fermionic matter. The pattern of chiral symmetry breaking depends on the underlying gauge dynamics \cite{Peskin:1980gc,Preskill:1980mz,Kosower:1984aw}. Of course, the initial hypothesis that the global symmetry breaks dynamically should be verified. In fact we know that, depending on the number of matter fields, the choice of the underlying gauge group and the dimension of the gauge group (e.g. the number of underlying fermionic matter), the symmetry might not break at all because the theory can develop large distance conformality, as discussed in \cite{Sannino:2004qp} for fermions in two-index representations, in \cite{Dietrich:2006cm} for a universal classification of $\SU(N)$ gauge theories and their applications to Technicolor and composite dark matter models, in \cite{Sannino:2009aw} for orthogonal and symplectic groups, and in \cite{Mojaza:2012zd} for exceptional underlying gauge groups. Furthermore, even assuming global chiral symmetry breaking, it remains to be seen if the breaking is to the maximal diagonal subgroup. We shall see that for certain phenomenologically relevant gauge theories, using first principle lattice simulations, there has been substantial progress to answering precisely these questions.
We note that the classification of relevant underlying gauge theories for Technicolor models appeared in \cite{Dietrich:2006cm}, while a classification of the symmetry breaking patterns relevant for composite models of the Higgs as a pGB can be found in~\cite{Mrazek:2011iu,Bellazzini:2014yua}.

\section{Patterns of Chiral Symmetry breaking}

The first relevant observation is that the largest non-abelian global quantum flavor symmetry, for any underlying gauge theory with one Dirac species of fermions, is bound to be $\SU(2N_f)$ or $\SU(N_f)\times\SU(N_f)$, depending on whether the underlying fermion representation is (pseudo-)real or complex. 

An $\SU(2 N_f)$ flavor symmetry can be achieved only if the new fermions belong to a real (like the adjoint) or pseudo-real (like the fundamental of $\SU(2) = \Sp(2)$) representation of the underlying gauge group.
In either case, both left-handed fermions and charge-conjugated right-handed anti-fermions can be recast, via a similarity transformation, to transform according to the same representation of the underlying fundamental gauge group. One can organize the two fermion components as a $2 N_f$ column with complex Weyl fermions indicated by the vector $\psi^f_c$ with $f = 1,..,2N_f$ and $c$ the new color index. 
 At this point, the simplest gauge invariant fermion bilinear we can construct  is $\psi^f_c \psi^{f'}_{c^{\prime}}$ with the color contracted either via a delta function for real representations or an antisymmetric epsilon term for pseudo-real ones. As fermions anticommute, a non vanishing condensate can be formed only if the full wave-function is completely antisymmetric with respect to spin, new color and flavor. Since Lorentz symmetry is conserved, the spin indices are contracted via an antisymmmetric tensor and therefore, according to the reality or pseudo--reality of the representation, we can have the following two patterns of chiral symmetry breaking \cite{Kosower:1984aw}:

\begin{itemize}
\item[Real:]  In this case we expect $\SU(2 N_f)\to \SO(2 N_f)$. The point is that the invariant product is symmetric (for instance $3 \times 3$ for the adjoint of $\SU(2)$), and the flavor contraction must also be symmetric implying that the condensate belongs to the symmetric 2-index representation of SU($2 N_f$).
\item[Pseudo:] Here we expect $\SU(2 N_f) \to \Sp(2 N_f)$. In this case, as explained before, the invariant gauge singlet tensor is antisymmetric, therefore the condensate transforms as the antisymmetric 2-index representation of $\SU(2 N_f)$.
\end{itemize}

Comparing these possible symmetry breaking patterns, coming from a fundamental theory with fermions, with the list of composite pNGB Higgs possibilities, we can conclude that:
\begin{itemize}

\item[-] the minimal composite Goldstone Higgs scenario~\cite{Agashe:2004rs} cannot be realized in a simple minimal way~\cite{Caracciolo:2012je}: in fact, it is based on $\SO(5)\to\SO(4)$. However this chiral symmetry pattern cannot occur naturally because the minimal flavor symmetry SO(5) cannot be realized by an underlying fundamental fermionic matter theory.

\item[-] the next to minimal scenario is based on an enlarged $\SO(6) \sim \SU(4)$ global symmetry. The breaking $\SU(4)\to\SO(4)$ is possible via a condensate belonging to the symmetric $10$ dimensional representation: however, such a breaking will generate 9 GBs belonging to a $(3,3)$ of $\SO(4)$, therefore no GB Higgs boson can be generated in the coset~\footnote{The $\SU(4)\to\SO(4)$ breaking with two Higgs doublets used in~\cite{Mrazek:2011iu} is generated by an adjoint $15$ of $\SU(4)$, which however cannot be a condensate of fermions.}. Nevertheless this pattern of chiral symmetry is extremely interesting for modern versions of minimal walking Technicolor models \cite{Sannino:2004qp,Dietrich:2006cm,Foadi:2007ue,Frandsen:2009mi} and their lattice studies \cite{Hietanen:2012sz,Hietanen:2013gva,Rantaharju:2017eej,Rantaharju:2019nmh}. 

\item[-] the symmetry breaking $\SO(6)\sim\SU(4)\to\Sp(4)\sim\SO(5)$ is also an interesting possibility both for  (ultra) minimal Technicolor models \cite{Appelquist:1999dq,Duan:2000dy,Ryttov:2008xe} and the composite Goldstone Higgs example \cite{Katz:2005au,Gripaios:2009pe,Galloway:2010bp,Barnard:2013zea,Ferretti:2013kya}. Here the breaking is generated by an antisymmetric, with respect to the global flavor symmetry, $6$-dimensional representation, and the coset contains 5 GBs. In terms of the $\SO(4)$ subgroup of $\SO(5)$, the GBs decompose into a $(2,2) + (1,1)$, thus also allowing, as we shall see,  a GB Higgs. In the following we will pursue this chiral symmetry breaking pattern, which at the fundamental level is also being studied on the lattice \cite{Lewis:2011zb,Hietanen:2013fya}. 

\item[-] the next interesting chiral symmetry breaking pattern, from the composite GB Higgs boson point of view, is $\SU(5)\to\SO(5)$. Here we have 14 GBs, decomposing as $(3,3) + (2,2) + (1,1)$ of $\SO(4)$~\cite{Katz:2005au,Ferretti:2013kya}.

\item[-] for $\SU(6) \to \Sp(6)$, also featuring 14 GBs, we have two composite Higgs doublets and 6 singlets~\cite{Katz:2005au,Cai:2018tet}.

\item[-] finally, complex representations allow for $\SU(4)\times\SU(4)\to\SU(4)$ \cite{Ma:2015gra} with 15 GBs\footnote{The case with $N_f = 3$ also contains a Higgs candidate, however it does not feature a custodial symmetry SO(4), which is crucial in obtaining a realistic model.}, organised as two Higgs doublets, one triplet and 4 singlets. Interestingly, this coset also allows for a pNGB Dark Matter candidate \cite{Ma:2017vzm}.

\end{itemize}

From this list it is clear that, from the point of view of a fundamental theory with fermionic matter, the minimal scenario to investigate is $\SU(4)\to\Sp(4)$, for both a minimal Technicolor as well as minimal composite GB Higgs scenario, as well as $\SU(4)\to\SO(4)$ for minimal walking Technicolor.  The difference lies in the way one embeds the electroweak theory within the global flavor symmetry. 

In this review, we will concentrate on the  $\SU(4)\to\Sp(4)$ scenario to provide the reader with a concrete template showing how one can bridge phenomenology, underlying dynamics, effective field theories and first principle lattice simulations in a coherent way. We will introduce the tools and provide sufficient details on how to adoperate them so that one will be able to employ them when going beyond the presented template, or even for other applications of composite dynamics.  
To avoid confusion with previous literature, we will dub the new strong gauge interactions as \emph{Fundamental Color}, or Fun. Color (FC) for short.


 \section{What comes next}
 We organized the rest of the review so that each chapter is fairly independent from other chapters, while they naturally integrate with each other, so that the reader will have quick access to the needed information. 
 
 In the next chapter \ref{ch:su2} we will introduce the minimal composite template \cite{Appelquist:1999dq,Cacciapaglia:2014uja} which, at the underlying theory level, is constituted by an $\SU(2)$ gauge theory with two Dirac fermions in the fundamental representation of the gauge group. Because, as listed above, the fermions belong to the pseudo-real representation of the gauge group, the global symmetry is $\SU(4)$ broken spontaneously to $\Sp(4)$, as also observed on the lattice \cite{Lewis:2011zb,Hietanen:2013fya,Hietanen:2014xca,Arthur:2016dir}. 
 
 We will introduce the low energy effective field theory for the Goldstone excitations and link it to the observed lattice spectrum. We will then gauge some of the global symmetries to introduce weak interactions in a way that accommodates the composite (Goldstone) Higgs framework. We shall go beyond the leading approximation and determine how the vacuum aligns, the composite Higgs boson emerges and interacts with SM fields. Experimental constraints and theoretical considerations will also be presented. 
 
 In the same chapter we will also provide selected relevant examples of composite dark matter models that can emerge from the same  underlying dynamics.  The latter will be taken either to still be connected to EWSB or to be an independent sector entirely\footnote{By independent, here we mean that it will couple to the SM fields in a way not to affect the EW breaking sector.}. This part will show the richness of the composite paradigm beyond the dynamical EWSB scenario. 

We will then address the SM fermion mass generation problem in chapter~\ref{SMFMgenration}. It is notoriously difficult to give mass to all SM fermions within models of composite dynamics powered by underlying gauge-fermion theories. We will elucidate the various ways one can circumvent some of the issues via walking dynamics and partial compositeness. We will also show that, when theories of composite dynamics are allowed to include composite scalars, one can build viable and testable composite alternatives to the SM Higgs sector. We will briefly comment also on extra-dimensional approaches to the same problem. 

In chapter~\ref{Lattice} we will review the state of the art of the numerical simulations for models of composite dynamics as function of the number of colors, number of matter fields and their representation under the gauge group. We will also discuss which model is expected to be (near) conformal at large distances. 

We will conclude in chapter~\ref{conclusions} by providing our final comments and directions for the future.


\newcommand{\cL}{\mathcal{L}}

\chapter{$\SU(2)$ Minimal Template}
\label{ch:su2}

In this chapter we introduce the most minimal and simplest theory of strong dynamics that can generate a composite Goldstone Higgs boson. 
Here by simplest, we mean that it is based on the smallest asymptotically free gauge group with the smallest number of fermions needed to obtain a custodial--preserving model.  
The model~\cite{Cacciapaglia:2014uja,Ryttov:2008xe,Galloway:2010bp} relies on a gauged $\SU(2)_{\FC}$ theory with 2 Dirac flavors in the fundamental representation.
As $\SU(2)$ can be viewed as the first of the symplectic groups \cite{Sannino:2009aw}  the 
phenomenological analysis, and model building, can be generalized to $\Sp(2N)$~\cite{Galloway:2010bp}. 

We will use this minimal model as a template to describe the main features of a composite Higgs boson, which can arise as a pNGB or as a light resonance: these two situation are two different limits in the same theory, based on the misalignment of the condensate with respect to the EW gauge symmetry~\cite{Cacciapaglia:2014uja}. This same model can be used to describe many incarnations of a composite Dark Matter candidate: arising as a ``baryon'' protected by a $\U(1)$ global symmetry~\cite{Ryttov:2008xe}, as a stable pNGB protected by a discrete or continuous accidental symmetry, or, when the strong sector does not couple to the SM, as a strongly interacting state~\cite{Hochberg:2014dra}.

\section{Introducing the theory and its symmetries}

The minimal template we consider here is based on a Fun.-Color gauged group $\mathcal{G}_\FC = \SU(2)_\FC$ with 2 Dirac (4 Weyl) FC-fermions in the fundamental representation.
The underlying Lagrangian of the strong sector is:
\begin{equation}
 \mathcal{L}=-\frac{1}{4}F^{a}_{\mu\nu}F^{a\mu\nu}+\bar{\tcf}_j (i\sigma^{\mu}D_{\mu})\tcf_j - M^{ij}_\tcf \tcf_i \tcf_j + h.c.
\end{equation}
where $F^{a}_{\mu\nu}$ is the field strength of the FC group, and $M_\tcf$ is a general mass matrix for the FC-fermions $\tcf^i$.
The covariant derivative $D_\mu$ contains the FC interactions, plus, eventually, SM gauge bosons, depending on the quantum number assignment of the FC-fermions.
First principle numerical simulations~\cite{Lewis:2011zb,Hietanen:2013fya,Hietanen:2014xca} have demonstrated that this $\SU(2)_\FC$ model does lead to a fermion condensate in the chiral limit, breaking the global symmetry $\SU(4)\to \Sp(4)$. The group-theoretical properties of the condensate are as follows:
\begin{equation}
\langle \tcf^i \tcf^j \rangle = {\bf 6}_{\SU(4)} \to {\bf 5}_{\Sp(4)} \oplus {\bf 1}_{\Sp(4)}\,,
\end{equation}
transforming as a 2-index anti-symmetric representation of $\SU(4)$.
The coset space $\SU(4)/\Sp(4)$ is parametrized by 5 pNGBs, transforming as a $\bf 5$ of $\Sp(4)$~\cite{Katz:2005au}.

The most general gauge-invariant mass term can be written as:
\begin{equation} \label{eq:MQ}
M_\tcf = \begin{pmatrix}
\mu_1\; i \sigma^2 & 0 \\
0 & \mu_2\; i \sigma^2
\end{pmatrix}\,,
\end{equation}
where $\sigma^2$ is the second Pauli matrix, and its appearance is a sign that the FC-fermions are doublets of the gauge symmetry. The phases of the Weyl fields can be used to make the two parameters $\mu_{1,2}$ real. The simple form of the mass allows to clearly identified two Dirac spinors
\begin{equation}
\psi_1 = \begin{pmatrix} \tcf^1 \\ i \sigma^2\ \tcf^{2,c} \end{pmatrix}\,, \qquad \psi_2 = \begin{pmatrix} \tcf^3 \\ i \sigma^2\ \tcf^{4,c} \end{pmatrix}\,,
\end{equation}
where the super-script ``$c$'' indicates the charge conjugate, with mass $\mu_i$ respectively.
The mass term explicitly breaks $\SU(4)$ to $\SO(4) \sim \SU(2)\times\SU(2)$, while in the symmetric case $|\mu_1| = |\mu_2|$ an enhanced $\Sp(4)$ remains unbroken. 
We can thus use the form of the mass term in the symmetric case to define the vacuum of the theory: this is actually not a choice, as it is the FC-fermion mass term that fixes the alignment of the vacuum in the $\SU(4)$ space.
In the following, we will choose $\mu_2 = - \mu_1$ in order to use the same alignment of the vacuum as in~\cite{Cacciapaglia:2014uja}:
\begin{equation} \label{eq:Sigma0}
\Sigma_0 = \left( \begin{array}{cc}
i \sigma^2 & 0 \\
0 & - i \sigma^2
\end{array} \right)\,.
\end{equation}
Note, however, that this sign choice is completely arbitrary, and it can always be reversed by a change in the phase of the constituent quarks.
All the physical results, therefore, are independent on the phases appearing in the mass matrix and in the condensate, provided we do not include the topological term  \cite{DiVecchia:2013swa}. 
The 5 broken generators $X^i$ in the vacuum of \eq{eq:Sigma0} are:
\bea \label{eq:Xgen}
& X^1 = \frac{1}{2 \sqrt{2}} \left( \begin{array}{cc}
0 & \sigma^3 \\
\sigma^3 & 0
\end{array} \right)\,, \quad  X^2 = \frac{1}{2 \sqrt{2}} \left( \begin{array}{cc}
0 & i \\
-i & 0
\end{array} \right)\,, & \nonumber \\  
& X^3 = \frac{1}{2 \sqrt{2}} \left( \begin{array}{cc}
0 & \sigma^1 \\
\sigma^1 & 0
\end{array} \right)\,, \quad  X^4 = \frac{1}{2 \sqrt{2}} \left( \begin{array}{cc}
0 & \sigma^2 \\
\sigma^2 & 0
\end{array} \right)\,, & \\ 
& X^5 = \frac{1}{2 \sqrt{2}} \left( \begin{array}{cc}
1 & 0 \\
 0 & -1
\end{array} \right)\,. & \nonumber
\eea
We remark that the $X^5$ generator corresponds to the phase redefinition of the FC-fermions: a transformation $e^{i \alpha X^5}$ generates a relative phase between the mass terms $\mu_1$ and $\mu_2$ of the two FC-fermion doublets. Our choice to have real masses already fixed $\alpha = 0$ and introducing a phase $\alpha$ in the vacuum alignment will therefore not add any new physical effects in the theory.

Based on the above symmetry considerations, one can describe the low-energy physics of the 5 pNGBs via the CCWZ formalism~\cite{Coleman:1969sm,Callan:1969sn}: here we will use a linearly transforming matrix defined as
\begin{equation}
\Sigma = e^{2 \sqrt{2} i \sum_{j=1}^5 X^j \pi_j/f} \cdot \Sigma_0\,,
\end{equation}
where $\pi_j$ are the pNGB fields.
For our purposes, this formalism is completely equivalent to the one based on 1-forms.
The lowest order chiral Lagrangian is therefore given by:
\bea \label{eq:LchiLO}
\mathcal{L}_{(p^2)} = \frac{f^2}{8}\ \Tr [(D_\mu \Sigma)^\dagger D^\mu \Sigma] + f^2\ \Tr [\chi \Sigma^\dagger + \Sigma \chi^\dagger]\,,
\eeq
where $f$ is the decay constant of the pNGBs\footnote{Note that the normalization of $f$ adopted here is different from the one used in \cite{Cacciapaglia:2014uja} by a factor $2\sqrt{2}$.}, $D_\mu$ contains the axial/vector sources and $\chi$ the scalar ones.
The covariant derivative is defined as
\bea \label{eq:avsources}
D_\mu \Sigma = \partial_\mu \Sigma - i\ J_\mu \Sigma - i\ \Sigma J_\mu^T\,, \quad J_\mu = v_\mu^a S^a + a_\mu^i X^i
\eeq
where $v_\mu$ and $a_\mu$ are vector and axial-vector sources respectively, and $S^a$ are the unbroken generators, given by the following matrices:
\beq \label{eq:S1}
S^{1,2,3} = \frac{1}{2} \left(  \begin{array}{cc}
\sigma_i & 0 \\
0 & 0
\end{array} \right)\,, \qquad S^{4,5,6} = \frac{1}{2} \left(  \begin{array}{cc}
0 & 0 \\
0 & - \sigma_i^T 
\end{array} \right)\,,
\eeq
which form an $\SO(4)$ subgroup of $\Sp(4)$, and
\beq \label{eq:S2}
S^{7,8,9} =  \frac{1}{2\sqrt{2}} \left(  \begin{array}{cc}
0 &i  \sigma_i \\
-i \sigma_i & 0
\end{array} \right)\,, \qquad S^{10} = \frac{1}{2\sqrt{2}} \left(  \begin{array}{cc}
0 &  1\\
1 & 0 
\end{array} \right)\,.
\eeq
Note that the couplings of the FC-fermions to gauge interactions external to the strong dynamics can be easily introduced by replacing the sources with the external gauge bosons. The scalar source $\chi$, on the other hand, can be used to parametrize the effect of the FC-fermion mass term, by replacing
\beq
\chi^\dagger \to 2 B_0 M_\tcf\,,
\eeq
where the coefficient $B_0$ can be measured on the lattice.

\subsection{Extensions of the Chiral Lagrangian} \label{sec:Lext}

The Chiral Lagrangian introduced earlier can be extended in several ways.
As a first step, it's important to consider operators that arise at the next order in the expansion, i.e. $\mathcal{O} (p^4)$. These terms may be very important phenomenologically, in particular in models where the mass of the pNGBs is sizeable with respect to the decay constants. This is the case in QCD, where NLO terms are indeed well studied.
For the symmetry breaking pattern of interest,
the NLO chiral Lagrangian at order ${\cal O} (p^4)$ is given by \cite{Bijnens:2009qm}:
\bea
&    {\cal L}_{(p^4)}  = L_0 \mathrm{Tr} [D_\mu \Sigma (D_\nu \Sigma)^\dagger D^\mu \Sigma (D^\nu \Sigma)^\dagger] 
    +L_1 \mathrm{Tr} [D_\mu \Sigma(D^\mu \Sigma)^\dagger]^2   & \nonumber \\
 &   + L_2 \mathrm{Tr} [D_\mu \Sigma (D_\nu \Sigma)^\dagger] \mathrm{Tr}[D^\mu \Sigma (D^\nu \Sigma)^\dagger]
    + L_3 \mathrm{Tr} [D_\mu \Sigma (D^\mu \Sigma)^\dagger D_\nu \Sigma (D^\nu \Sigma)^\dagger]  & \nonumber \\
&    +L_4 \mathrm{Tr}[(D_\mu \Sigma)(D^\mu \Sigma)^\dagger] \mathrm{Tr}[\chi \Sigma^\dagger+\Sigma \chi^\dagger]
    +L_5 \mathrm{Tr}[(D_\mu \Sigma)(D^\mu \Sigma)^\dagger (\chi \Sigma^\dagger+\Sigma \chi^\dagger)]  & \nonumber \\
&    +L_6 \mathrm{Tr}[\chi \Sigma^\dagger + \Sigma \chi^\dagger]^2
    +L_7 \mathrm{Tr}[\chi \Sigma^\dagger - \Sigma \chi^\dagger]^2 
    + L_8 \mathrm{Tr}[\chi \Sigma^\dagger \chi \Sigma^\dagger + \Sigma \chi^\dagger \Sigma \chi^\dagger] & \nonumber \\
&    -i L_9 \mathrm{Tr} [j_{\mu \nu} D^\mu \Sigma (D^\nu \Sigma)^\dagger-j_{\mu\nu}^T (D^\mu \Sigma)^\dagger D^\nu \Sigma] 
    + L_{10} \mathrm{Tr} [\Sigma j_{\mu\nu}^T \Sigma^\dagger j^{\mu\nu}] & \nonumber \\
&    +2 H_1 \mathrm{Tr} [j_{\mu\nu} j^{\mu\nu}] + H_2 \mathrm{Tr} [\chi \chi^\dagger], &
    \label{Gasser-Leutwyler-Lagrangian}
\eeq
where $J_{\mu \nu}$ is the energy-stress tensor associated to the current $J_\mu$ and the coefficients $L_i$ and $H_i$ are low-energy constants (LECs), which only depend on the strong dynamics and can be computed on the lattice once the details of the underlying theory are specified.
The above Lagrangian is expressed in a particular basis where we remove the redundant operators\footnote{When the number of flavors, $N_\tcf$, is small, the Caley--Hamilton relations 
may be used to remove additional redundant operators.
The equation of motions have also been used to remove two other operators.} in complete analogy with the Gasser and Leutwyler~\cite{Gasser:1983yg} Lagrangian for the complex case. Note that the above list of operators also applies to a symmetric vacuum.

Another natural extension is the inclusion of other massive resonances generated by the strong dynamics. One example is provided by vector and axial-vector resonances, which may be light relative to the cut-off of the effective theory. Phenomenologically, the lightest spin-1 resonances may play an important role in determining the feasibility of a model. In fact, they are often relatively light, as expected in large $N_c$ expansions~\cite{Manohar:1998xv} or assumed in some composite Higgs models~\cite{Contino:2011np,Contino:2015mha}. We also recall that they may be important in determining other low energy constants, following the so-called vector meson dominance~\cite{Sakurai:1960ju}. Finally, via the mixing to the EW gauge bosons, when relevant, they may be produced at colliders via Drell-Yan and thus give detectable signals at colliders.
For the template we consider in this section, the inclusion of spin-1 resonances has been detailed in~\cite{Franzosi:2016aoo}, following the technique of Hidden Local Symmetry~\cite{Bando:1987br}. This technique consists on doubling the global symmetry of the chiral theory, and it introduces the heavy vectors by gauging the copy symmetry group. 
We refer the reader to \cite{Franzosi:2016aoo} for more details on this implementation. Here, we simply recall that these resonances transform under the global symmetry as the axial/vector sources introduced in Eq.\eqref{eq:avsources}, i.e.:
\beq
v_\mu = {\bf 10}_{\Sp(4)}\,, \quad a_\mu = {\bf 5}_{\Sp(4)}\,.
\eeq
What is mostly notable about this structure is that it is quite far from the minimal case often considered in the literature, where the spin-1 sector consists of one vector triplet and one axial-vector one~\cite{Pappadopulo:2014qza}.

Another resonance that is worthy considering is a scalar singlet $\sigma$. The reason for this is that this scalar is often the lightest state in composite models, like for instance in QCD~\cite{Sannino:1995ik}. Furthermore, it may be sizeably lighter than the other resonances if the theory has a conformal symmetry above the cut-off of the chiral theory. As long as the mass of the scalar $\sigma$ is light compared to the pNGB decay constant, $m_\sigma \lesssim f$, we can threat it effectively as a light degree of freedom in the chiral Lagrangian, at par with the pNGBs~\cite{Soto:2011ap}.
The most general chiral Lagrangian is therefore given by:
\begin{multline}
\mathcal{L}_{\sigma} = \kappa_G (\sigma)\ \frac{f^2}{8} \mbox{Tr} [(D_\mu \Sigma)^\dagger D^\mu \Sigma] + \frac{1}{2} \partial_\mu \sigma \partial^\mu \sigma - \frac{1}{2} M^2 \kappa_M (\sigma) \sigma^2  + \\
 f^2 \kappa_m (\sigma)\  \Tr[\chi \Sigma^\dagger + \sigma \chi^\dagger]\,. \label{eq:sigma}
\end{multline}
The couplings of the $\sigma$ are effectively included in the functions $\kappa_X (\sigma)$, which are not constrained by the symmetries at low energy.
The extension at NLO has been studied in~\cite{Hansen:2016fri}.

Finally, a phenomenologically relevant interaction is provided by the topological term~\cite{Wess:1971yu,Witten:1983tw,Witten:1983tx}, known as the Wess-Zumino-Witten (WZW) term, which contains interactions with an odd number of pNGBs. 
It can be written in a compact local form as a five-dimensional integral 
\begin{equation} \label{eq:SWZW}
 S_{WZW} = \frac{N_c}{240\pi^2}\int_0^1 d\alpha\int d^4x~\epsilon^{abcde}\Tr [u_au_bu_cu_du_e],
\end{equation}
were $\alpha$ is the fifth spacetime coordinate. We furthermore redefine the two quantities
\begin{align}
 u &= \exp\left(\sqrt{2} \frac{i\alpha}{f}X^a\pi^a\right), \\
 u_a &= i(u^\dagger\partial_a u - u\partial_a u^\dagger),
\end{align}
with $\partial_a$ being a five-dimensional derivative. Ordinary Minkowski space is now defined as the surface of the five-dimensional space where $\alpha=1$. Furthermore the pre-factor of the WZW term contains direct information about the underlying gauge dynamics. The gauged version of this term and its generalization to include left and right vector sources can be found in \cite{Duan:2000dy}.

\subsection{Spectrum on the Lattice} \label{sec:spectrum}

\begin{figure}[tbh]
     \center
     \includegraphics[width=.49\textwidth]{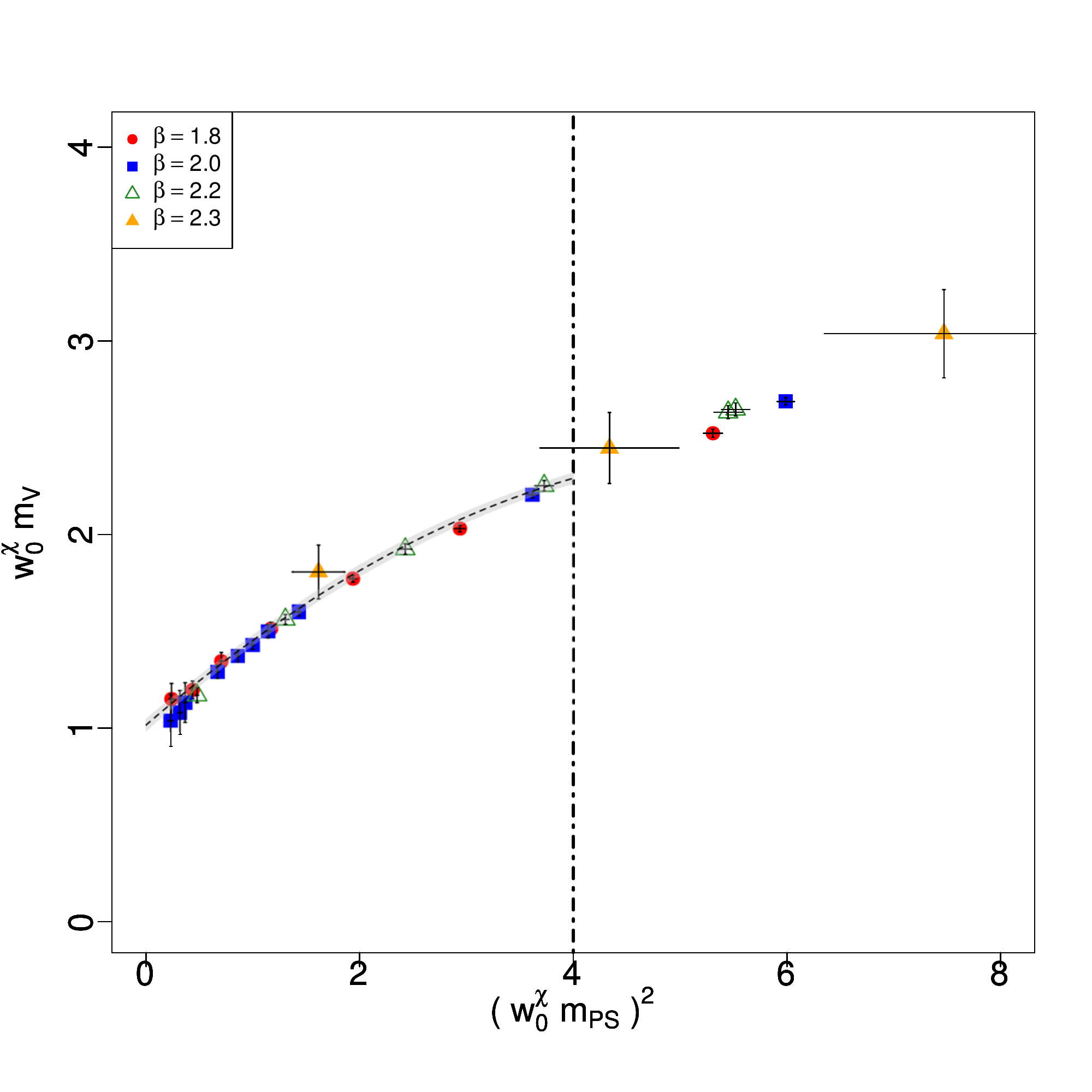}\hfill
     \includegraphics[width=.49\textwidth]{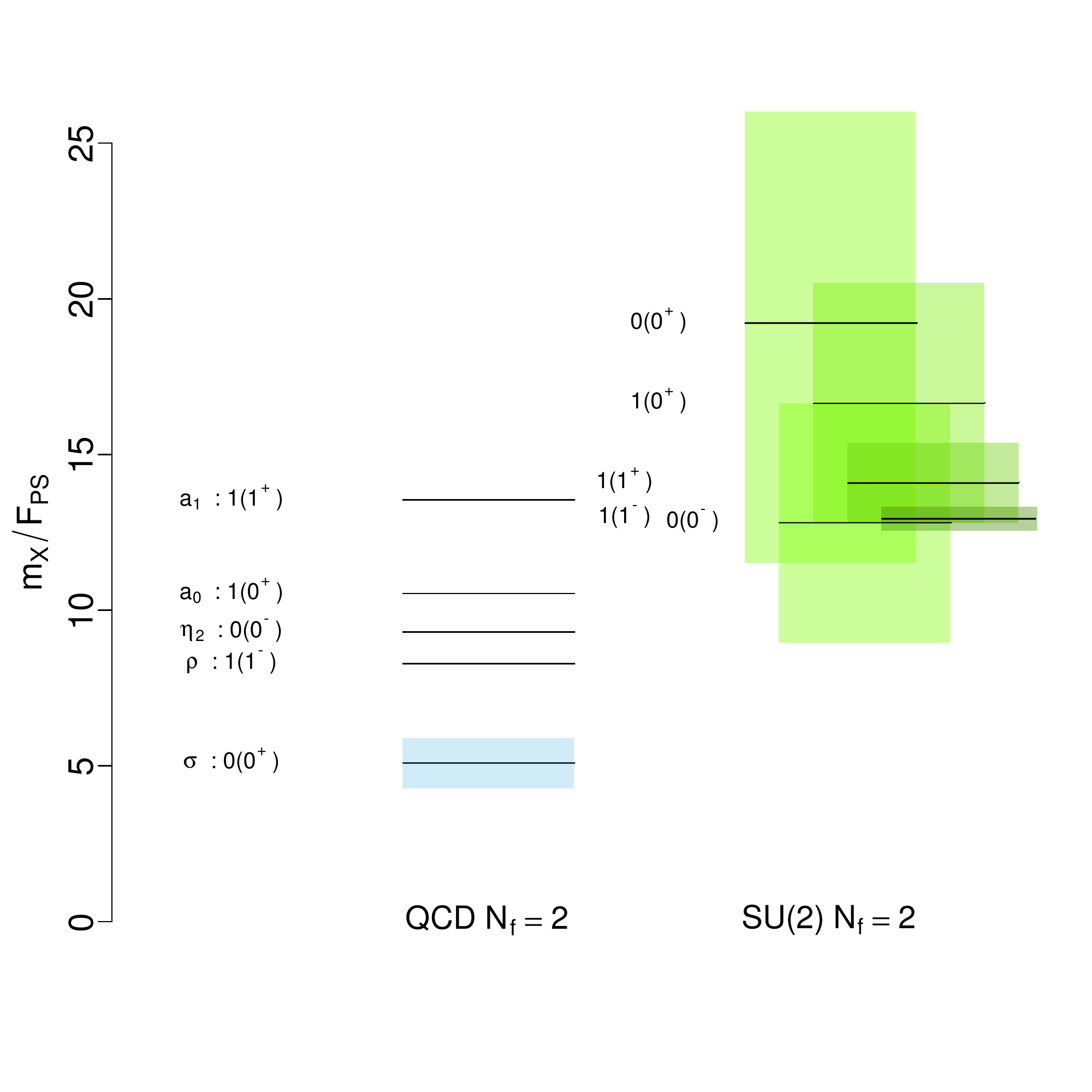}
     \caption{Left panel: Example of combined chiral and continuum extrapolation for the $\rho$ meson. Right panel: Summary plot comparing the spectrum of the $\SU(2)$ model to $\SU(3)$ (QCD). Note that $F_{\rm PS} \equiv f$. Figures from \cite{Arthur:2016dir, Arthur:2016ozw}. }
     \label{fig:drach}
     \end{figure}

Lattice simulations of the $\SU(2)_\FC$ template model are straightforward, requiring only two colors, two Dirac fermions and no high dimensional representations. The challenge is then to perform all the necessary extrapolations, i.e infinite volume, zero FC-fermion mass and continuum limit, for the spectrum of the model, as this has been done for any other BSM model studied so far.
Results for the spectrum of the model were presented in \cite{Lewis:2011zb,Hietanen:2013fya,Hietanen:2014xca,Arthur:2014zda,Arthur:2014lma,Arthur:2016dir,Arthur:2016ozw,Pica:2016zst,Drach:2017jsh}. We summarize the present status here.

Lattice simulations were performed at four different lattice spacings, for a number of FC-fermion masses at each lattice spacing, while keeping large enough volumes to reduce finite-volume effects as much as possible.
The scale was set by using the $w_0$ observable \cite{Borsanyi:2012zs} and the RI-MOM scheme \cite{Martinelli:1994ty} was used to measure the required non-perturbative renomalization constants.
Continuum extrapolated results were obtained for the decay constant $f$, the fermion condensate, and the lightest spin one and zero resonances, analogue to the QCD $\rho$, $a_1$, $\sigma$, $\eta'$, $a_0$ resonances.
A combined chiral and continuum extrapolation was used to extract the physically interested quantities. The left panel of Fig.~\ref{fig:drach} shows an example of such extrapolation for the $\rho$ meson.

The final spectrum for the model is shown in the right panel of Fig.~\ref{fig:drach}, in units of $f \equiv F_{\rm PS}$ and compared to the corresponding QCD spectrum ($\SU(3)$ with 2 Dirac flavors).
Taken at face value, these results indicate a spectrum which is quite different from the QCD one. For Technicolor-like theories, where $f = v = 246$~GeV, the heavier resonances seems beyond the present LHC constraints, while for Goldstone Higgs models, where $f \approx 1$~TeV is preferred, they are well beyond being produced directly at the LHC.
These results are still affected by large systematic errors, as shown in Fig.~\ref{fig:drach}, mainly due to the chiral and continuum extrapolations required to obtain phenomenological predictions. The accuracy of these results will be increased in the future.

Simulations of a model based on a gauged $\Sp(4)$ group with 2 Dirac flavors, which share the same properties of the template we consider here, are underway~\cite{Bennett:2017tum,Bennett:2017ttu,Bennett:2017kbp,Bennett:2017kga}. The first results for the meson spectrum can be found in~\cite{Lee:2018ztv}.

\section{Composite electroweak Higgs sector}
\label{sec:compoHiggs}

\begin{table}[tb] \begin{center}
\begin{tabular}{c|ccc|c|}
   &  $\SU(2)_{\rm FC}$  &  $\SU(2)_L$ & $\U(1)_Y$ & $\SU(2)_R$\\
$\tcf^1$  & \multirow{2}{*}{${\tiny{\yng(1)}}$} & \multirow{2}{*}{{\bf 2}} & \multirow{2}{*}{0} & \multirow{2}{*}{\bf 1} \\
$\tcf^2$   &   &   &  &   \\
$\tcf^3$ & \multirow{2}{*}{${\tiny{\yng(1)}}$} &  \multirow{2}{*}{{\bf 1}} & -1/2 & \multirow{2}{*}{$\bf \bar{2}$}  \\
$\tcf^4$ & &  &1/2 & \\
\end{tabular}
\caption{Quantum numbers under the FC group and the electroweak gauge interactions of the fundamental constituent fermion. The last column indicates the custodial global $SU(2)_R$.} \label{tab:tab1} 
\end{center} \end{table}

As a first application, we can use the theory detailed above to characterize the phenomenology of a Higgs boson arising as a pNGB.
We need thus to specify the embedding of the electroweak symmetry in the model: the simplest choice is to assign the first two $\tcf^i$ to a doublet of SU(2)$_L$, and the second two to an anti-doublet of SU(2)$_R$ (the diagonal generator of SU(2)$_R$ being the generator of hypercharge), as listed in Table~\ref{tab:tab1}. In this way, all gauge anomalies vanish, and we can keep track explicitly of the custodial symmetry built in the model.
Another point is that, with this embedding, we can choose an alignment of the condensate in SU(4) that does not break the EW symmetry: this direction is in fact determined by the mass matrix $M_\tcf$. As the mass term in Eq.\eqref{eq:MQ} are explicitly gauge invariant, the $\SU(2)\times \SU(2)$ group left unbroken coincides with the partly-gauged $\SU(2)_L\times \SU(2)_R$ group of the SM. The vacuum like for instance $\Sigma_0$ in Eq.\eqref{eq:Sigma0} leaves invariant a subgroup $\Sp(4) \sim \SO(5)$ which contains the custodial $\SO(4)$.

In this basis, the $\SU(2)_L$ generators are $S^{1,2,3}$, while the $\SU(2)_R$ ones are $S^{4,5,6}$ (thus, the hypercharge is associated with $S^6$), with explicit expressions given in Eqs~\eqref{eq:S1} and \eqref{eq:S2}.
The alignment of the condensate can be changed by applying an $\SU(4)$ transformation generated by the 5 broken generators: as $X^{1,2,3,4}$ (see Eq.~\eqref{eq:Xgen}) form an $\SU(2)_L$ doublet, one can use gauge transformations to align the vacuum along the Higgs direction ($X^4$ in our notation) without loss of generality.

\subsection{The choice of the correct vacuum} \label{sec:vacuumchoice}

It is most common in the composite pNGB Higgs literature, starting from the seminal work of Georgi and Kaplan~\cite{Kaplan:1983fs}, to define the theory around the EW preserving vacuum. In a second step, a potential for the pNGBs is constructed, which induces a vacuum expectation value (VEV) for the composite Higgs doublet. This approach has a serious drawback in the fact that the Higgs VEV will appear in the lowest order chiral Lagrangian, c.f. Eq.\eqref{eq:LchiLO}, as a parameter that breaks the shift symmetry for the pNGBs. As a consequence, derivative couplings that violate the Higgs shift invariance will appear at LO~\cite{Gripaios:2009pe,Frigerio:2012uc}. 
Here we will follow a more conservative and theoretically clean approach: a VEV for a pNGB can be used to define an $\SU(4)$ rotation, generated by the generator associated to the pNGB, which misaligns the vacuum with respect to the starting one. 
In this approach, the parameter breaking the EW symmetry will not violate the shift symmetry of the pNGBs, thus explicitly revealing that it is intrinsically associated to the compositeness scale. The only drawback of this formalism is that the component of composite states, which are classifies as multiplets of the unbroken $\Sp(4)$ symmetry, cannot be decomposed into multiplets of the EW symmetry, which is broken by the vacuum, but only approximately for $v/f \ll 1$. Nevertheless, the physical results in the two formalisms are the same, as the two parametrizations of the vacuum and pNGBs are related by a field redefinition.

Even the choice of the EW preserving vacuum is not as simple as it may look, as there are generators, which correspond to singlets of the EW symmetry, that can be used to change it. In the case under study here, it is the generator $X^5$ that corresponds to a singlet, plus an overall phase shift. The most general EW preserving vacuum is therefore:
\begin{equation}
\Sigma' = e^{i \gamma} e^{i \alpha 2 \sqrt{2} X^5} \Sigma_0 = \left( \begin{array}{cc}
e^{i (\gamma+\alpha)} (i \sigma^2) & 0 \\
0 & - e^{i (\gamma-\alpha)} (i \sigma^2) 
\end{array} \right)\,.
\end{equation}
We already commented that the two phases correspond to phases of the spinor fields of the FC-fermions $\tcf$, so they must not have any physical relevance. In fact, once a mass term for $\tcf$ is defined, as in Eq.\eqref{eq:MQ}, a potential is generated for them that fixes the proper values. The second term in Eq.\eqref{eq:LchiLO} gives
\beq
\Tr[\chi {\Sigma'}^\dagger + \Sigma' \chi^\dagger ] = B_0  \left[ - (\mu_1-\mu_2) \cos \alpha \cos \gamma + (\mu_1 + \mu_2) \sin \alpha \sin \gamma \right]\,.
\eeq
The above potential has minima for $(\alpha, \gamma) = (0,0)$ or $(\pi, \pi)$ for $|\mu_1-\mu_2| > |\mu_1+\mu_2|$, which corresponds to $\Sigma_0$, and minima at $(\alpha, \gamma) = (\pi/2, 3 \pi/2)$ and $(3\pi/2, \pi/2)$ for $|\mu_1+\mu_2| > |\mu_1-\mu_2|$ corresponding to the other real vacuum~\cite{Galloway:2010bp}
\begin{equation}
\Sigma_0' = \left( \begin{array}{cc}
 i \sigma^2 & 0 \\
0 &  i \sigma^2 
\end{array} \right)\,.
\end{equation}
This shows that the choice of the EW preserving vacuum depends on the choice of phases in the spurions that break explicitly the global symmetries in the composite sector, the mass of $\tcf$ in this case. A non-zero value of the phase $\alpha$ can only be associated to a physical phase appearing in the spurions (C.f.~\cite{Alanne:2018wtp}).
 The phase $\gamma$, generated by the anomalous U(1) symmetry,  generates CP violation in the chiral Lagrangian via the Pfaffian of the pion matrix~\cite{Galloway:2010bp} and can be associated to CP violation in the strong sector. In the following, for simplicity, we will limit ourselves to a CP-invariant model, thus setting $\gamma = \alpha = 0$ and choosing the EW preserving vacuum $\Sigma_0$.
 
The vacuum will then be misaligned along the directions of the composite Higgs doublet  once the EW symmetry is broken. Hence, without loss of generality, we can choose the direction to be $X^4$.
The most general EW breaking vacuum, therefore, can be written as:
\begin{equation}
\Sigma_\theta = e^{i \theta X^4} \cdot \Sigma_0 =  \left( \begin{array}{cc}
\cos \theta\ (i \sigma^2) & \sin \theta  \\
- \sin \theta & -\cos \theta\ (i \sigma^2)
\end{array} \right)\,.
\end{equation}
The only free parameter $\theta$ aligns the condensate to a direction that does break the EW symmetry, and its value will be determined once quantum corrections are added.

Based on the above symmetry considerations, one can describe the physics of the 5 GBs via the CCWZ formalism~\cite{Coleman:1969sm,Callan:1969sn}: here we will use a linearly transforming matrix defined as
\begin{equation}
\Sigma = e^{2 \sqrt{2} i \sum_{j=1}^5 Y^j \chi_j/f} \cdot \Sigma_\theta\,,
\end{equation}
where $\chi_j$ are the pNGB fields around the vacuum $\Sigma_\theta$, and $Y^j = e^{i \theta/2 X^4} \cdot X^j \cdot e^{-i \theta/2 X^4}$ are the broken generators in that vacuum.
For our purposes, this formalism is completely equivalent to the one based on 1-forms.
The most general lowest order chiral Lagrangian is the one in Eq.\eqref{eq:LchiLO}
where $D_\mu$ contains the EW gauge bosons, and the couplings of the Sp(4) singlet $\sigma$ are parametrized in terms of unknown functions $\kappa_x$ (normalized to $\kappa_x (0) = 1$). 
This is complemented by interactions terms that are equivalent to the Yukawa couplings in the SM:
\begin{multline} \label{eq:effYuk1}
\mathcal{L}_Y = - f \left( \kappa_t (\sigma)\, {y'}^{ij}_u (Q_{L,i} u_{R,j}^c)_\alpha^\dagger +  \kappa_b (\sigma)\, {y'}^{ij}_d (Q_{L,i} d_{R,j}^c)_\alpha + \right.  \\
\left.  \kappa_l (\sigma)\, {y'}^{ij}_l (L_i l_j^c)_\alpha^\dagger\right) \mbox{Tr} [P^\alpha \Sigma]  + h.c.
\end{multline}
The matrices $P^\alpha$~\cite{Cacciapaglia:2014uja} are spurions that project the pion matrix on its components transforming as a doublet of $\SU(2)_L$ (explicit formulas are given in Eq.~\eqref{eq:projtop} below).
As we shall see later, the $\sigma$ can also play the role of the Higgs boson, even though naively its mass is expected to be large and its couplings to the SM particles arbitrary.
The effective Yukawa couplings are necessary in order to give mass to the fermions in a similar way as Yukawa couplings do in the SM.
A possible origin of such terms can be traced back to four-Fermi interactions in the form (for the up-sector):
\begin{equation} \label{eq:EFCD}
\mathcal{L}_{\rm EFCD} = - \frac{y_u^{ij}}{\Lambda_u^2}\, (\tcf \tcf)^\alpha (Q_{L,i} u_{R,j}^c)_\alpha^\dagger + h.c.
\end{equation}
As all the Yukawa terms have the same 4-Fermi origin, one may expect $\kappa_t = \kappa_b = \kappa_l$.
The origin of such effective interactions is unspecified here, and their presence alone does not generate any problem with flavor observables. Addressing flavor in this class of models can only go via a detailed description of the UV completion responsible for the generation of the effective interactions: here, we will not attempt to address this issue. The issue of flavor is, in fact, the most challenging aspect in models based on a fundamental composite dynamics. The choice of adding effective Yukawa couplings as in Eq.~\eqref{eq:effYuk1} can be seen as the analog of minimal flavor violation~\cite{DAmbrosio:2002vsn}.
For a bookkeeping description of the possible origin of the couplings for the top alone, we refer the interested reader to~\cite{Cacciapaglia:2015yra}.

\subsection{Higgs as a pNGB}

Having analyzed the general properties and the chiral symmetry breaking pattern relevant for the electroweak symmetry, we move to investigate the various phenomenological aspects of the model. 
We will focus first on the Chiral Lagrangian for the pNGBs (or pions) of the model, which include a Higgs-like state.
Several of the results in this section were first obtained in~\cite{Katz:2005au,Galloway:2010bp} were more details can be found.\footnote{Note that we adopt here a different normalization for the decay constant, defined such that $f=v$ in the TC limit.} Here, we will recall the main features, and stress the connection between the Technicolor and Goldstone Higgs vacua.
The angle $\theta$ that characterixes the vacuum $\Sigma_\theta$ is, at this stage, a free parameter, which interpolates between a purely Technicolor model when $\theta=\pi/2$ to an unbroken phase for $\theta=0$, passing through a model of Goldstone Higgs for small $\theta \ll 1$.
With the above condensate, the 5 broken generators are explicitly given by
\beq
& Y^1  = c_\theta\, X^1 - s_\theta\, \frac{S^1-S^4}{\sqrt{2}}\,, \quad  Y^2 = c_\theta\, X^2 + s_\theta\, \frac{S^2-S^5}{\sqrt{2}}\,, & \nonumber \\
&  Y^3 = c_\theta\, X^3 + s_\theta\, \frac{S^3-S^6}{\sqrt{2}}\,, \quad Y^4 = X^4\,, & \nonumber \\
&  Y^5 = c_\theta\, X^5 - s_\theta\, S^{8}\,; &
\eeq
where $c_\theta = \cos \theta$ and $s_\theta = \sin \theta$.
The exact Goldstone bosons, that become the longitudinal components of  the $W$ and $Z$ gauge bosons, are associated with the $Y^{1,2,3}$ generators. Working in the Unitary gauge, we use explicitly only the fields associated to $Y^{4,5}$ and write:
\beq \label{eq:SigmapNGB}
\Sigma = e^{\frac{2 \sqrt{2} i}{f} (h Y^4 + \eta Y^5)}\cdot \Sigma_\theta = \left[ \cos \frac{x}{f}\; 1 + \frac{2 \sqrt{2} i}{x} \sin \frac{x}{f}\; \left( h Y^4 + \eta Y^5 \right) \right] \cdot \Sigma_\theta\ ,
\eeq
where $x = \sqrt{h^2 + \eta^2}$.
The kinetic term of $\Sigma$, upgraded to include the interactions with the gauge bosons via minimal coupling, yields:
\beq
& \frac{f^2}{8} \,\, \mbox{Tr} (D_\mu \Sigma)^\dagger D^\mu \Sigma = \frac{1}{2} (\partial_\mu h)^2  + \frac{1}{2} (\partial_\mu \eta)^2  - \frac{1}{6 f^2} \left[ (h \partial_\mu \eta - \eta \partial_\mu h)^2\right] + \mathcal{O} (f^{-3})  &  \nonumber \\
&  +\left(  \frac{g^2}{4} W^+_\mu W^{- \mu} + \frac{(g^2 + {g'}^2)}{8} Z_\mu Z^\mu \right) \left[ f^2 s_\theta^2 + s_{2 \theta} f\ h \left(1-\frac{2}{3 f^2} (h^2 + \eta^2) \right)  \right. & \nonumber \\
& \left. +  (c_{2\theta} h^2 - s_\theta^2 \eta^2) \left( 1 -\frac{1}{{3} f^2} (h^2 + \eta^2) \right) + \mathcal{O} (f^{-3}) \right] \ . &
\eeq
From the above expansion, we can pull out the values of the $W$ and $Z$ masses
\beq
m_W^2 = \frac{g^2}{4} f^2 s_\theta^2\,, \qquad m_Z^2 = \frac{(g^2 + {g'}^2)}{4} f^2 s_\theta^2 = m_W^2/c_W^2\,,
\eeq
thus identifying $v = f s_\theta$.
Furthermore, only the scalar $h$ couples singly to the massive gauge bosons, therefore it is the candidate to play the role of the Higgs boson. Its couplings are:
\beq
g_{h W W} &=& \frac{g^2}{2} f s_\theta c_\theta = g m_W\ c_\theta = g_{hWW}^{SM} \ c_\theta\,, \\ 
g_{h Z Z} &=& \frac{(g^2 + {g'}^2)}{2} f s_\theta c_\theta =  \sqrt{g^2 + {g'}^2} m_Z \ c_\theta = g_{hZZ}^{SM}\ c_\theta\,, \\ 
g_{h h WW} &=& \frac{g^2 c_{2\theta}}{4}  = g_{hhWW}^{SM}\ c_{2 \theta}\,, \\
g_{h h ZZ} &=&  g_{h h WW}/c_W^2\,.
\eeq
The second scalar $\eta$ has couplings
\beq
g_{\eta \eta WW} &=& - \frac{1}{ 4} g^2 s^2_{\theta} = - g_{hhWW}^{SM}\ s_{\theta}^2\,, \\
g_{\eta \eta ZZ} &=&  g_{\eta \eta WW}/c_W^2\,.
\eeq
The relations between the couplings to $WW$ and $ZZ$ are guaranteed by custodial invariance.

It is noteworthy that the kinetic term of $\Sigma$ is invariant under the parity transformation $\eta \to - \eta$, therefore $\eta$ is protected and will be stable. This property derives form the fact that non-topological Lagrangians for Goldstone bosons respect a parity operation according to which the only possible terms must be even in the number of Goldstone bosons.  This property has been used in~\cite{Frigerio:2012uc} to study $\eta$ as a composite dark matter candidate. However this apparent discrete symmetry is not a symmetry of the underlying theory. The breaking of this symmetry manifests itself at the effective Lagrangian level via topological-induced terms, like the Wess-Zumino-Witten (WZW) anomaly terms~\cite{Wess:1971yu,Witten:1983tw}. These have been constructed explicitly for the chiral symmetry breaking pattern envisioned here in \cite{Duan:2000dy}. The singlet $\eta$ can therefore decay as $\eta \to Z \gamma,\ ZZ,\ W^+ W^-$.

\subsection{Loop induced Higgs potential}

While the dynamics does not have any preference to where the condensate is aligned whitin the SU(4) space, gauge interactions do because they only involve a subgroup of the flavor symmetry.
The same is true, as we will see, for the top Yukawa, or generically the mechanism that generates a mass for the top.
The breaking of the flavor symmetry will then be communicated to the pNGBs via loops, which will therefore induce a potential determining the value of the angle $\theta$.
The loop-induced potential for this model has been computed in~\cite{Katz:2005au,Galloway:2010bp}, here we will simply recap the origin of the main components and discuss their physical interpretation.

\subsubsection*{Gauge contributions}

The contribution to the one-loop potential of the gauge boson loops can be estimated by constructing the lowest order operator invariant under the flavor symmetry $\SU(4)$.
The gauge generators of $\SU(2)_L$ are  $S^{1,2,3}$ while the one for $\U(1)_Y$ is $S^6$.  Under a  vacuum rotation preserving the unbroken subgroup, the gauged generators transfrom as $S^i \to U S^i U^\dagger$, and the associated relevant term in the effective potential reads \cite{Peskin:1980gc,Preskill:1980mz}:
\beq
& V_{\rm gauge} = - C_g f^4 \sum_{i} g_i^2 \mbox{Tr} \left(S^i \cdot \Sigma \cdot (S^i \cdot \Sigma)^\ast \right) & \nonumber \\
 & \sim  C_g \frac{3 g^2 + {g'}^2}{2} \left( -  f^4 c_\theta^2 +  f^3 s_{2 \theta} h +  f^2 (c_{2\theta} h^2 - s_\theta^2 \eta^2) + \dots \right)\,;&
\eeq
where $C_g$ is an unknown LEC, and we have explicitly shown an expansion in powers of $f$ up to quadratic terms in the fields.
To find the value of $\theta$ it is enough to minimize the field-independent term: $\partial V (\theta)/\partial \theta = 0$.
The constant $C_g$ encodes the loop factor, and it is expected to be positive: in this case, this part of the potential has a minimum for $\theta=0$, which therefore does not break the electroweak symmetry. Note also that the term with linear coupling of the ``Higgs'' $h$ is always proportional to the derivative of the potential, thus it is bound to vanish at the minimum.

\subsubsection*{Top contribution}

To calculate the effects on the vacuum alignment induced by the top corrections we will follow the procedure established in~\cite{Peskin:1980gc,Preskill:1980mz}. Not having at our disposal a complete theory of flavor we assume that the top mass is generated via the following 4-fermion operator
\beq
\frac{y_t}{\Lambda_t^{{2}}} (Q_L t_R^c)^\dagger_\alpha \ \tcf^T P^\alpha \tcf
\eeq
where $\alpha$ is an $\SU(2)_L$ index and the projectors $P^\alpha$ select the components of $\tcf^T \tcf$ that transform as a doublet of $\SU(2)_L$ (i.e. the linearly transforming Higgs boson doublet properties). $\Lambda_t $ is some new dynamical scale.  
The projectors can be written as~\cite{Galloway:2010bp}
\beq \label{eq:projtop}
P^1 = \frac{1}{2} \left( \begin{array}{cccc}
0 & 0 & 1 & 0 \\
0 & 0 & 0 & 0 \\
-1 & 0 & 0 & 0 \\
0 & 0 & 0 & 0
\end{array} \right)\,, \qquad P^2 = \frac{1}{2} \left( \begin{array}{cccc}
0 & 0 & 0 & 0 \\
0 & 0 & 1 & 0 \\
0 & -1 & 0 & 0 \\
0 & 0 & 0 & 0
\end{array} \right)\,.
\eeq
When the FC-fermions condense, this term generates the following operator:
\beq \label{eq:topyuk}
y'_t  f (Q t^c)^\dagger_\alpha \mbox{Tr} (P^\alpha \Sigma) \sim - y'_t  \left( f s_\theta +  c_\theta h  - \frac{1}{2 f} s_\theta (h^2 + \eta^2) + \dots \right) t_R t_L^c
\eeq
Here $y'_t$ is proportional to $y_t (4\pi f)^2/\Lambda^2_t$. We have not assumed the underlying fermionic dynamics to be near conformal. If this were the case the relation changes as it is the case for walking Technicolor. 
The first term in the expansion generates a top mass $m_{top} = y'_t f s_\theta$ when $\theta \neq 0$, and the coupling of the Higgs to the top is
\beq
y_{h\bar{t}t} = y'_t c_\theta = \frac{m_{top}}{v} c_\theta\,.
\eeq
From the form of the operator above, the contribution of the top loop can be estimated as
\beq
V_{top} &=& - C_t {y'_t}^2 f^4 \sum_{\alpha=1}^2 \left[  \mbox{Tr} (P^\alpha \Sigma) \right]^2 \nonumber \\
 & \sim & - C_t  {y'_t}^2 \left[f^4 s_\theta^2 +  f^3  s_{2 \theta} h + f^2 (c_{2\theta} h^2 - s_\theta^2 \eta^2) + \dots \right]
 \eeq
where again we expect the coefficient $C_t$ to be positive. 
In this case, the minimum is located at $\theta = \pi/2$, which would break the electroweak symmetry at the condensate scale. The vacuum preferred by the top corrections corresponds, therefore, to the standard Technicolor-like limit \cite{Dietrich:2006cm,Appelquist:1999dq,Ryttov:2008xe}.

This Technicolor vacuum limit is quite interesting: in fact, one would have that all the linear couplings of $h$ to gauge bosons and to the top vanish, the reason being that in this limit an extra $\U(1)$ symmetry remains unbroken upon gauging the electroweak symmetry. This global $\U(1)$ symmetry is reminiscent of the QCD-like underlying FC-baryon symmetry linking $h$ and $\eta$ into a complex di-FC-fermion pNGB. Intriguingly a similar decoupling property for the would-be composite Higgs pNGB was observed in the Hosotani model \cite{Hosotani:2009jk}. Here too, in the decoupling limit, the near decoupled state becomes a dark matter candidate. 
This property of the Technicolor vacuum has been used extensively for dark matter model building \cite{Gudnason:2006ug,Gudnason:2006yj,Nardi:2008ix,Ryttov:2008xe,Foadi:2008qv,Frandsen:2009mi,DelNobile:2011je} and it is supported also by recent pioneering lattice investigations \cite{Lewis:2011zb,Hietanen:2012sz,Hietanen:2013fya}. We will discuss this class of models further in Sec.\ref{sec:TIMP}. In the Technicolor limit, the pNGB $h$ ceases to be a Higgs-like particle: the physical Higgs-like state now becomes the lightest $\tcf$-flavor (and SM) singlet composite scalar state associated with the fluctuations of the condensate orthogonal to the pNGB directions. The coupling of this state to the gauge bosons and fermions does not vanish for $\theta=\pi/2$: we will investigate the properties of this state in Section~\ref{sec:technihiggs}.

In this vacuum, as explained before, $\eta$ and $h$ are degenerate and acquire the following loop-induced mass term:
\beq
{ m^2_{DM}} = m_h^2 = m_\eta^2 = 2 f^2 \left( C_t {y'_t}^2 - C_g \frac{3 g^2 + {g'}^2}{2}  \right)\,.
\eeq
Interestingly the weak interactions like to misalign the Technicolor vacuum while the top corrections tend to re-align the vacuum in the Technicolor direction. 

\subsubsection*{Explicit breaking of SU(4)}
\label{ESBSU}

Another source for the Higgs potential is given by eventual terms that break explicitly $\SU(4)$. Sources that do not upset the $\theta=\pi/2$ vacuum were constructed in  \cite{Gudnason:2006yj,Foadi:2007ue,Ryttov:2008xe} assuming natural breaking of the $\SU(4)$ symmetry via four-fermion interactions. However in~\cite{Katz:2005au}, for minimal models of composite (Goldstone) Higgs,  such a term is added ad-hoc to give mass to $\eta$; in~\cite{Galloway:2010bp} they are generated by gauge invariant masses of the FC-fermions.
In both cases, the results are the same: here we will follow the idea that such a term is generated by the explicit $\SU(4)$ violating masses of the FC-fermions.
As we want such masses to be invariant under the gauged symmetry, we can assume that the mass term is aligned with the condensate $\Sigma_0$, so that $M_\tcf = \mu \Sigma_0$, as in Eq.\eqref{eq:MQ} for $\mu_1 = - \mu_2 = \mu$.
In this case, the contribution to the potential can be written as~\cite{Galloway:2010bp}:
\beq
V_{m} &=& C_m f^4 \mbox{Tr} (\Sigma_0 \cdot \Sigma)\nonumber \\
& \sim & 4 C_m \left( - f^4 c_\theta + f^3 s_\theta h + \frac{1}{2} f^2 c_\theta (h^2 + \eta^2) + \dots \right)
\eeq
Note that contrary to the gauge and top loops, the coefficient $C_m$ is not expected to be positive and can have both signs.\footnote{In the limit of vanishing top and gauge couplings, the positivity of the pNGB mass squared would suggest $C_m > 0$.}
This potential term contributes to off-set the ground state from $\theta = \pi/2$.

\subsection{Vacuum alignment, fine tuning and Higgs mass}

The total potential, used in~\cite{Cacciapaglia:2014uja,Galloway:2010bp}, consists of the above 3 contributions:
\begin{equation}
V_{\rm scalars} = V_{\rm gauge} + V_{\rm top} + V_m \,.
\end{equation}
 To further the analysis we neglect the gauge boson contribution. This is justified by the fact that it is smaller than the top one.
First, we can compute the potential for $\theta$:
\begin{equation}
V(\theta) = {y'_t}^2 C_t\, \cos^2 \theta - 4 C_m\, \cos \theta + \mbox{constant}
\end{equation}
where $C_{t,m}$ are $\mathcal{O}(1)$ coefficients determined by the dynamics ($C_t$ is expected to be positive to match the sign of a fermion loop).
Notably, the contribution of the gauge loops have the same form and $\theta$ dependence as that of the top.
The minimum of the potential is given by
\begin{equation}
\cos \theta_{\rm min} = \frac{2 C_m}{{y'_t}^2 C_t}\,, \qquad \mbox{for}\;\; {y'_t}^2 C_t > 2 |C_m|\,.
\end{equation}
Note that a small $\theta$ can only be achieved for $2 C_m \to {y'_t}^2 C_t$: in order to reach the pNGB Higgs limit, one needs therefore to fine-tune two contributions in the potential which are of very different origins.
This is the only severe fine-tuning required in the model, if a small $\theta$ needs to be achieved.
Note also that, in the limit of a small mass for the FC-fermions, $C_m \ll C_t$, the vacuum moves towards the TC limit $\theta = \pi/2$. Remarkably, a non fine-tuned (in $\theta$) realization of a pNGB Higgs may occur if its nature is elementary \cite{Alanne:2014kea}, the reason being that the corrections to the potential, once the quadratic divergences are properly subtracted, derive from the dependence of the potential on the fourth power of the couplings (corresponding to logarithmically divergent corrections to the quartic coupling) rather than on the quadratic power (corresponding to the quadratically divergent contribution to the mass). It is also noteworthy that here we used an explicit mass term for the FC-fermions to stabilize the potential, while in other models of composite Goldstone Higgs in the market the stabilization is due to quartic terms: as we just discussed, quartic terms are the dominant ones in elementary realizations of the pNGB Higgs, however they are subleading in composite realizations.

This potential also determines the masses of the pNGBs:
\begin{eqnarray}
m_{\chi_{1,2,3}}^2 &=& 2 f^2 \left(2 C_m  -  {y'_t}^2 C_t\cos \theta\right) \cos \theta = 0\,,\\
m_{h}^2 &=& 2 f^2 \left( 2 C_m\cos\theta -  {y'_t}^2 C_t \cos(2 \theta)\right) = 2 {y'_t}^2 C_t f^2 \sin^2 \theta\,,\\
m_{\eta}^2 &=&  2 f^2 \left( 2 C_m \cos \theta + {y'_t}^2 C_t \sin^2 \theta \right)  =2 {y'_t}^2 C_t f^2 \,,
\end{eqnarray}
where we have used the minimum condition to remove the dependence on $C_m$.
We notice here that, as expected, the new fundamental elementary fermion mass term gives the same mass (of order $f$) to all pions.
On the other hand, the top loop gives a mass of order $f$ to the pNGB Higgs, and a mass of order $f \sin \theta$ to the EW singlet.
This can be easily understood: the top couples via 4-fermi interactions to the FC-fermion doublet that transforms as a doublet of $\SU(2)$, thus the top loop will generate the usual divergent contribution to its mass that, following naive dimensional analysis, can be approximated as
\begin{equation} \label{eq:mh}
\Delta m_h^2 (\mbox{top}) \approx C \frac{{y'_t}^2}{16 \pi^2} \Lambda^2 \approx C {y'_t}^2 f^2\,.
\end{equation}
This large contribution, however, is cancelled by the contribution of the explicit mass at the minimum, so that the final value of the pNGB Higgs mass is
\begin{equation}
m_h^2 = 2 {y'_t}^2 C_t f^2 \sin^2 \theta = m_\eta^2 \sin^2 \theta = 2 C_t m_{\rm top}^2\,.
\end{equation}
Note that it would be enough to have $C_t \sim 1/4$ to generate the correct value for the Higgs mass. The value of $C_t$ is not a free parameter, but it can be determined by the dynamics.
At present, no calculation of such coefficients is available.
Nevertheless, no additional fine-tuning is in principle necessary for the Higgs mass, once the fine-tuning in the alignment is paid off.
The relation between the masses of $h$ and $\eta$ also survives after the gauge corrections are included, however it can easily be spoiled by other corrections, like for instance the mixing between $h$ and $\sigma$.
Finally, the pions eaten by the $W$ and $Z$ are massless on the correct vacuum, as expected for exact GBs.

The parametric smallness of the pNGB Higgs mass can also be understood in terms of symmetries. The 3 GBs eaten by $W$ and $Z$ 
are always massless, for any value of $\theta$. Therefore, if we go continuously to the limit $\theta\to 0$, where the EW symmetry is restored, the 
mass of the pNGB Higgs must also vanish in order to reconstruct a complete massless $\SU(2)$ doublet.
The same argument cannot be applied to $\eta$, which is a singlet unrelated to EW symmetry breaking.

\subsection{Variations on the theme}

We can now explore some variations of the above scenario. For example we can gauge an extra $\U(1)$. The only possibility is to gauge the symmetry generated by $X^5$ which commutes with both $\SU(2)_L$ and $\U(1)_Y$.\footnote{Note that this $\U(1)$ is responsible for the stability of the $h$--$\eta$ system in the Technicolor limit.}
The new gauge boson $X^\mu$ will contribute to the effective potential as follows
\beq
& V_{X} = - C_g g_X^2 f^4  \mbox{Tr} \left(X^5 \cdot \Sigma \cdot (X^5 \cdot \Sigma)^\ast \right) & \nonumber \\
 & \sim   C_g g_X^2 \left( \frac{1}{2} f^4 (2 c_\theta^2-1) -  f^3  s_{2\theta} h -  f^2 (c_{2\theta} h^2 - s_\theta^2 \eta^2) + \dots \right)\,. & 
\eeq
Although this contribution has the familiar form of the contribution coming from the electroweak gauge terms, the preferred minimum is at $c_\theta = 0$.  This happens because the Technicolor ground state does not break the $\U(1)_X$ symmetry thus leaving $X_\mu$ massless. In fact:
\beq
m_X^2 =  \frac{g_X^2}{2} f^2 c_\theta^2\,.
\eeq
The net effect of this contribution would therefore be to add to the top loop.

Another variant is to generate a mass for the top via a heavy mediator $\Psi$.
The idea is to complement the theory with a new fermion belonging to a complete representation of $\SU(4)$, and couple it to $\Sigma$ in an $\SU(4)$ invariant way; the mass is then communicated to the top sector by an $\SU(4)$ violating mixing term of the form~\cite{Katz:2005au}
\beq
\lambda_1 f \bar{\Psi} \Sigma \Psi + \lambda_2 f Q_\psi Q_L + \lambda_3 f T_\psi t_R^c
\eeq
where $Q_\psi$ and $T_\psi$ are components of $\Psi$ with the same quantum numbers as the quark doublet and the right-handed top singlet.
However, it is not possible to find a representation of $\SU(4)$ that contains a doublet and a singlet with the correct hypercharge, following the embedding of the hypercharge generator discussed above.
One may think of embedding the $\U(1)_Y$ as a superposition of the two possible $\U(1)$'s, i.e. $Y = c_\alpha S^6 + s_\alpha X^5$.
However, there is no vacuum that can preserve both $S^6$ and $X^5$ and as a consequence there would be no QED unbroken $\U(1)_{\rm em}$.
This is interesting as it allows us to discard the construction used in~\cite{Katz:2005au} to generate the top mass.
In \cite{Gripaios:2009pe}, the global flavor symmetry is extended to $\SU(4)\times\U(1)$ and the hypercharge is identified with a linear combination of $S^6$ and the external $\U(1)$: this is a trick commonly used in models of composite Goldstone Higgs.
However, it is very unlikely that the condensate in any realistic fundamental theory, featuring fundamental vector-like fermionic matter, can break $\SU(4)\times\U(1)\to\SO(5)$.
In fact, one could imagine to generate the extra $\U(1)$ by adding a single fermion in the adjoint representation, which carries hypercharge.
However, the two sectors of fermions do not talk to each other, and the dynamics will generate, at first, two condensates: one breaking $\SU(4)$ and the other breaking $\U(1)$ independently. Gauge dynamics for vector-like theories with several fermionic matter fields were investigated in~\cite{Ryttov:2009yw,Chen:2010er}. To achieve the desired symmetry breaking pattern, for these theories, one would need to introduce fields transforming simultaneously under the two flavor symmetries. Fundamental scalar fields can easily accommodate this patterns at the price of introducing unnatural scenarios~\cite{Sannino:2016sfx}. On the other hand, chiral gauge theories (where a mass term for the fermions is prohibited) can break simultaneously different global symmetries as well as the underlying gauge dynamics and would constitute and interesting avenue to explore~\cite{Appelquist:2000qg}.

\subsection{Experimental constraints} \label{sec:expconstr}

The main constraints in this model come from modifications of the Higgs couplings to other SM states, as discussed in the previous sections. We recap here that, for the couplings to the massive EW gauge bosons, we have
\beq
\kappa_V = \frac{g_{hVV}}{g_{hVV}^{\rm SM}} = c_\theta\,, \quad \frac{g_{hhVV}}{g_{hhVV}^{\rm SM}} = c_{2 \theta}\,,
\eeq
with the normalized modifications being the same for $W$ and $Z$ due to the custodial invariance of the strong dynamics. It's interesting to note that such modifications are universal, in the sense that they do not depend on the specific details of the coset~\cite{Low:2018acv,Liu:2018vel}. For the Yukawa coupling of the top, we find
\beq
\kappa_t = \frac{g_{ht\bar{t}}}{g_{ht\bar{t}}^{\rm SM}} = c_\theta\,,
\eeq
based on the bilinear coupling of Eq.~\eqref{eq:topyuk}. This is not universal, as it depends crucially on the mechanism giving origin to the fermion masses.

Direct probes of this model would include the detection of the singlet pNGB $\eta$: however, due to its feeble couplings to the SM states, suppressed by $s_\theta$, the production rates are too small to be detectable at the LHC, even after the high-luminosity phase~\cite{Galloway:2010bp,Arbey:2015exa}.

\subsubsection{Higgs coupling measurements}

The couplings of the Higgs boson have been measured by the LHC Collaborations ATLAS and CMS in a series of papers that can be used to extract the constraints from the Higgs couplings. In the following we use the combined results given by the two collaborations in a joint effort \cite{Khachatryan:2016vau} for the Run-I data, and a collection of the most up-to-date results from Run-II from both collaborations in the $ZZ^\ast$~\cite{Aaboud:2017vzb,Sirunyan:2017exp}, $\gamma \gamma$~\cite{Aaboud:2018xdt,CMS:2017rli}, $WW^\ast$~\cite{CMS:2017pzi} and $\tau^+ \tau^-$~\cite{CMS:2017wyg} channels. Note that these data are updated with respect to~\cite{Arbey:2015exa} where this analysis was first published, and the new results appeared in the PhD thesis in~\cite{LeCorre:2018}.
The results of the experimental analyzes are provided as exclusion contours in terms of signal strengths, and treated in the way described in \cite{Cacciapaglia:2012wb}. These experimental plots represent regions allowed at $68\%$ confidence level (C.L.) by the analyzes, in the plane of cross sections rescaling factors for the  main Higgs decay channels $H \to \gamma \gamma, WW^*, Z Z^* , \bar \tau \tau, \bar b b$, under the assumption that $W$ and $Z$-strahlung (VH) and vector boson fusion (VBF) modes are rescaled by the same factor, as well as  the gluon fusion (ggH) and $t\bar{t}H$. 
We fitted these lines as ellipses, therefore extrapolating the $\chi_i^2$ for each channel as a paraboloid, i.e. approximating the likelihood functions with a gaussian. The 2--dimensional likelihood for each channel,  neglecting  the $t \bar t H$ contribution,  is given by:
\beq
 \chi^2_i=\binom{\mu_{ggH}^i-\hat{\mu}_{ggH}^i}{\mu_{VBF/VH}^i-\hat{\mu}_{VBF/VH}^i}^T\, \cdot \, M_{i}^{-1} \,
\cdot\, \binom{\mu_{ggH}^i-\hat{\mu}_{ggH}^i}{\mu_{VBF/VH}^i-\hat{\mu}_{VBF/VH}^i} \,,
\eeq 
where $(\hat{\mu}_{ggH}^i, \hat{\mu}_{VBF/VH}^i)$ is the center of  each ellipse and  $M_{i}$ is a symmetric matrix encoding information about its axes. In our model,  we  define  the signal strengths to be:
\beq 
\mu_{ggH}^i =  \frac{ \sigma_{ggH} (\kappa_t^2, \kappa_b^2) }{ \sigma_{ggH, SM}}
\times \frac{\kappa_i^2 }{\sum_m \kappa_m^2 Br_m^{SM}} \,, \quad 
\mu_{VBF/VH}^i = \kappa_V^2
\times \frac{\kappa_i^2 }{\sum_m \kappa_m^2 Br_m^{SM}}\,,
\eeq
where those $\kappa_i^2$'s  are rescaling factors for  Higgs couplings with gauge bosons and fermions.

Through the $\chi^2$ function  we will  determine the best fit point  and   we can then use the reconstructed  quantity  
$\Delta \chi^2  = \chi^2 - \chi^2_{min}$  to draw the exclusion limits. This method has been validated to reproduce the 
experimental results~\cite{Flament:2015wra}.

\subsubsection{Oblique parameters}
\label{sec:EWPT}

Another source of important constraints is given by EW precision tests (EWPTs), which derive from very accurate measurements at the per-mille level in the EW sector of the SM. Such effects can be effectively described in terms of the oblique parameters S and T~\cite{Peskin:1990zt,Peskin:1991sw} (see also \cite{Barbieri:2004qk}).
The precise determination of the oblique corrections is a delicate issue in composite extensions of the SM. A well defined procedure must be employed that allows to clearly disentangle the intrinsic contribution stemming from strong dynamics from the one coming from the genuine SM contribution \cite{Foadi:2012ga}.   Once such procedure is established, an estimate from the strongly coupled sector is needed. First principle lattice simulations are the primary method to determine this contribution. However, as a very rough estimate, one can use the one-loop contribution from the fundamental fermions with heavy constituent mass terms.  In this case for one SU(2)$_L$ doublet we have $\Delta S = 1/(6\pi)$~\cite{Peskin:1991sw}. In the fundamental model under consideration, this translates into the following contribution
\begin{equation} \label{eq:naiveS}
\Delta S_{UV} = \frac{\sin^2 \theta}{6\pi}
\end{equation}
for each fundamental doublet \footnote{We note that this estimate is modified when the underlying dynamics is near conformal  because of the violation of the second Weinberg's sum rule \cite{Appelquist:1998xf}. There is a limit, however, when this estimate turns into a precise result. This occurs close to the upper limit of the conformal window \cite{Sannino:2010ca}  provided the correct kinematical limits are chosen. The two-loop contributions have been computed in \cite{DiChiara:2010xb} where it is clearly shown how the S parameter increases when moving deeper into the non-perturbative region.  In the non-perturbative regimes recent comprehensive holographic estimates have appeared \cite{Jarvinen:2015ofa}. }.
The reason for the presence of the $\sin^2 \theta$ term can be understood in terms of symmetries: in the radial composite Higgs limit $\theta \to \pi/2$, the fundamental fermions pick up a dynamical mass from the condensate which is aligned with the EW breaking direction, thus the calculation satisfies some of the assumptions in~\cite{Peskin:1991sw}; on the other hand, in the limit $\theta\to 0$, the EW symmetry is recovered and the $S$ parameter must vanish.
The power is understood in terms of masses: in fact, it is expected to be proportional to the square of the ratio of the dynamical mass aligned to the EW breaking direction, $v^2 \sim f^2 \sin^2 \theta$, and the total dynamical mass of the fermions, $\sim f^2$.
This expectation is also confirmed by an operator analysis of this contribution, as shown in~\cite{Galloway:2010bp}.
The strongly interacting contribution to the $T$ parameter vanishes because the dynamics respects the $\SU(2)_V$ custodial symmetry.

The underlying strong dynamics contribution must then be matched with the  important one coming from the quantum corrections in the effective Lagrangian for the lightest states considered here. 
We will use a more naive way to estimate the total correction: we explicitly include the contribution of the loops of the lightest composite states~\footnote{The $\eta$ does not contribute: in fact, its couplings can only generate corrections to the masses and, because of the custodial symmetry, such corrections do not enter the $T$ parameter.}, i.e. the $125$~GeV Higgs $h$, which contributes due to the modified couplings to gauge bosons.
Then, we will assume that the contribution of the heavier resonances can be approximated by the FC-fermion loop in Eq.~\eqref{eq:naiveS}, as one would expect if the contribution were dominated by the lightest vector and axial resonances. 
The net effect can be estimated starting from the contribution of the Higgs loops and we can summarize the results in the following:
\begin{eqnarray}
\Delta S &=& \frac{1}{6 \pi} \left[ (1-\kappa_V^2) \ln \frac{\Lambda}{m_{h}} + N_D \sin^2 \theta \right]\,, 
\nonumber \\
\Delta T &=& - \frac{3}{8 \pi \cos^2 \theta_W} \left[ (1-\kappa_V^2) \ln \frac{\Lambda}{m_{h}} \right]\,, \label{eq:ST1}
\end{eqnarray}
where  $N_D$ is the number of FC-fermion doublets ($N_D = 2$ for $\SU(2)_{\rm FC}$, and $2N$ for $\Sp(2N)_{\rm FC}$). In this analysis we assumed the presence of physical cutoff $\Lambda$ to be identified with the next massive state.  The dependence on the cutoff emerges because the scalar loop contributions are divergent, as a sign of the effective nature of the Lagrangian. 
The divergence is corrected once the proper matching to the underlying UV dynamics is taken into account. In our phenomenological estimates, we will use
\begin{equation}
\Lambda  = \Lambda_{\rm FC} \approx 2 \pi f = \frac{2 \pi v}{\sin \theta}\,,
\end{equation}
which is very close to the mass of the spin-1 resonances as shown by first-principle lattice simulations \cite{Lewis:2011zb,Hietanen:2013fya,Hietanen:2014xca,Arthur:2014lma}. We also added to $\Delta S$ the naive strongly coupled contribution that should partially take into account the heavier states. 
This estimate is clearly naive but should capture at least the correct order of magnitude of the corrections. 
It should be stressed that a more appropriate calculation should be employed if one wanted to use Lattice calculations of the contribution of the strong dynamics to $S$, as thoroughly discussed in \cite{Foadi:2012ga}, where one finds also the discussion of the needed counter-terms in the effective Lagrangian. 

The bound from the EWPTs is taken from \cite{Baak:2014ora}: $S_{U=0} =0.06 \pm 0.09 $ and $T_{U=0} = 0.10 \pm 0.07$ with correlation $0.91$.

\subsubsection*{Results}

Both the Higgs couplings and EWPTs depend only on  $\theta$, thus allowing us to extract an upper bound on the value of this angle. The limits at 3$\sigma$ are summarized in the following table~\cite{LeCorre:2018}:
\begin{center}
\begin{tabular}{l|c|c|c|c|}
  & Higgs (Run-I) & Higgs (Run-II)& EWPTs ($\SU(2)_{\FC}$) & $\Sp(4)_{\FC}$ \\ 
$\theta <$ & $0.46$  & $0.39$ & $0.25$ & $0.24$\\
\end{tabular}
\end{center}
The numbers show that the bound from EWPTs (indicated by the symmetry group name) is much stronger than the bounds from the Higgs couplings
(indicated as ``Higgs"), and points to values $\sin \theta \lesssim 0.2$. This value is consistent with bounds obtained in other models of composite Goldstone 
Higgs~\cite{Barbieri:2007bh}. This result may seem trivial, however the methods employed to estimate the contribution of the strong dynamics are very 
different. While we use a simple FC-fermion loop, most results in the literature are based on the calculation of loops of spin-1 
resonances~\cite{Contino:2015mha,Ghosh:2015wiz} and often include the effects of loops of top partners~\cite{Grojean:2013qca}. From the values in 
the table, we also see that there is a mild dependence on the number of doublets in the dynamical model, thus signalling that the bound is dominated by 
the contribution of the Higgs boson.

\section{Introducing the Techni-Higgs (the $\sigma$)} \label{sec:technihiggs}

As already mentioned in Section~\ref{sec:Lext}, the dynamics also contains a would-be Higgs boson, besides the pNGBs of the flavor symmetry breaking, which behaves like the $\sigma$ particle in QCD and is a singlet under the flavor symmetry.
This state is expected to be heavy (see Sec.~\ref{sec:spectrum}), however it may become lighter if the theory is extended to have a conformal window above the condensation scale~\cite{Ryttov:2008xe,Galloway:2010bp}. Its presence in the spectrum is important for EW symmetry breaking models, as it may play the role of the Higgs boson in Technicolor-like theories~\cite{Belyaev:2013ida} or help easing the EW precision bounds if the SM-like Higgs is dominantly of Goldstone nature~\cite{Arbey:2015exa,BuarqueFranzosi:2018eaj}.

To study its effect when its mass is light, of the order of the pNGB decay constant $f$, we can add it to the chiral Lagrangian as in Eq.\eqref{eq:sigma}, with the addition of the coupling to tops:
\begin{multline}
\mathcal{L} = \frac{1}{2} (\partial_\mu \sigma)^2 - \frac{1}{2} \kappa_M (\sigma) M^2 \sigma^2 + \kappa_G (\sigma)\; \frac{f^2}{8} \,\, \mbox{Tr} (D_\mu \Sigma)^\dagger D^\mu \Sigma  \\ + \kappa_m (\sigma)\Tr[\chi \Sigma^\dagger + \Sigma \chi^\dagger] + \left( \kappa_t (\sigma)\; y'_t  f (Q t^c)^\dagger_\alpha \mbox{Tr} (P^\alpha \Sigma) + \mbox{h.c.}\right)\ , \label{eq:Lsigma}
\end{multline}
where we have introduced the field $\sigma$ with a potential term, together with the modified pNGB kinetic term and the top Yukawa operator as in Eq.~\eqref{eq:topyuk}.
The dynamics generates the couplings of $\sigma$ to the above operators, encoded into the functions $\kappa_{M,G,t,m}$.
The Lagrangian is well defined when the space-time fluctuations of the field $\sigma$ are small compared to the scale of the dynamics $\Lambda_{\rm FC} \approx 4\pi f$. De facto $\sigma$ becomes a background field. In this case one can indeed safely neglect terms with higher derivatives acting on $\sigma$, and naturally assume that the functions $\kappa$ can be expanded for small values of $\sigma$ as
\beq
\kappa_{M,G,t,m} (\sigma) \sim 1 + k^{(1)}_{M,G,t,m} \frac{\sigma}{4 \pi f} + \frac{1}{2} k^{(2)}_{M,G,t,m} \frac{\sigma^2}{(4 \pi f)^2} + \dots \ ,
\eeq
with the coefficients $k^{(n)}$ of order unity. We can therefore use the Lagrangian in Eq.~\eqref{eq:Lsigma} to determine the loop-induced potential for the pNGBs:
\beq \label{eq:Vpotsigma}
V_{\rm 1-loop} = \kappa_G(\sigma) \; V_{\rm gauge}  + \kappa_t (\sigma)^2\; V_{top}\ .
\eeq
The potential above, augmented by the intrinsic $\sigma$ potential term (with its squared mass term factored out) and the FC-fermion mass operator, needs to be minimized with respect to the angle $\theta$ parameterising a given vacuum choice, as well as $\sigma$. It is convenient to split $\sigma$ into a vacuum expectation value and the slowly fluctuating field $\varphi$ as follows: $\sigma = \sigma_0 + \varphi$, where $\varphi$ is identified with the massive physical degree of freedom, i.e. the {\it ``techni-Higgs''}.
For the angle $\theta$, the minimization condition is similar to the one performed in the previous sections, up to appropriate factors of $\kappa_X (\sigma_0)$:
\beq
\cos \theta = \frac{2 \kappa_m(\sigma_0)\, C_m}{{y'_t}^2 \kappa_t^2 (\sigma_0)\, C_t - \frac{3 g^2 + {g'}^2}{2} \kappa_G (\sigma_0)\, C_g}\,.
\eeq
To determine $\sigma_0$, on the other hand, we need to solve:
\beq
\frac{M^2}{f^4} \left( \kappa_M  \sigma_0 + \frac{1}{2} \kappa'_M  \sigma_0^2 \right) - 2 C_t {y'_t}^2 s^2_\theta \kappa_t \kappa'_t - C_g \frac{3 g^2 + {g'}^2}{2} c^2_\theta  \kappa'_G - 4 C_m c_\theta \kappa'_m = 0\,,
\eeq
where all the $\kappa$ functions and their derivatives with respect to $\sigma$ are evaluated at $\sigma_0$.
Consistency requires $\sigma_0\ll 4 \pi f$ for the Taylor expansion of the $\kappa$ functions to be valid.

Furthermore, from \eqref{eq:Lsigma} we see that the $\kappa$ functions also encode the couplings of the techni-Higgs to gauge bosons and the top. In fact, expanding around $\sigma_0$ we find:
\beq
& m_W^2 = \kappa_G (\sigma_0) \cdot \frac{g^2}{4} f^2 s^2_\theta\,, \qquad \mbox{and} \quad g_{WW\varphi} = \frac{\kappa'_G (\sigma_0)}{\kappa_G (\sigma_0)} m_W^2\ , & \\
& m_t = \kappa_t (\sigma_0) \cdot y'_t f s_\theta\,, \qquad \mbox{and} \quad g_{t\bar{t} \varphi} = \frac{\kappa'_t (\sigma_0)}{\kappa_t (\sigma_0)} m_t\,. &
\eeq
These relations become relevant when identifying the techni-Higgs as the SM Higgs. We want to recall here that the couplings of the $\sigma$ can, in principle, be computed on the Lattice and are therefore an intrinsic property of the underlying dynamics.

We will now study this scenario in some limiting cases, starting from the Technicolor limit, reached for $\theta = \pi/2$.

\subsection{The Technicolor vacuum} \label{sec:techniHiggs_TCvac}

Let us now focus on the Technicolor vacuum, i.e. $\theta = \pi/2$, which is obtained in the limit $C_m = 0$.
In this case the SM Higgs can only be played by the techni-Higgs scalar $\varphi$~\cite{Belyaev:2013ida}, while the two pNGBs $h$ and $\eta$ become degenerate. The associated complex state is stable and can play the role a complex dark matter state~\cite{Ryttov:2008xe} (see also Sec.~\ref{sec:TIMP}). Phenomenology requires the couplings of $\varphi$ to the gauge bosons to be close to the SM ones, yielding the following constraints on the function $\kappa_G$:
\beq
\frac{\kappa'_{G} (\sigma_0)}{\kappa_G (\sigma_0)} \sim \frac{k^{(1)}_G}{4 \pi f} \approx \frac{2}{v} = \frac{2}{f} \Rightarrow k^{(1)}_G \sim 8 \pi\,,
\eeq
where we are neglecting higher order terms in $\sigma_0/(4 \pi f)$.
The same analysis for the top implies
\beq
\frac{\kappa'_t (\sigma_0)}{\kappa_t (\sigma_0)} \sim \frac{k^{(1)}_t}{4 \pi f} \approx \frac{1}{v} = \frac{1}{f} \Rightarrow k^{(1)}_t \sim 4 \pi\,. \label{eq:tH2top}
\eeq

We now consider the various contributions to the physical mass for the would-be Higgs $\varphi$, and  to the one of the complex dark matter particle formed by the two pNGBs.
Truncating the Taylor expansion of the $\kappa$ function to the first relevant orders we find:
\beq
m_\varphi^2 &=& M^2 \left( 1 + 3 k_M^{(1)} \frac{\sigma_0}{4 \pi f} \right) - \frac{\left((k_t^{(1)})^2 +k_t^{(2)} \right)}{2 \pi^2} 2 C_t {y'_t}^2 f^2\,, \\
m_{DM}^2 &=&  2 C_t {y'_t}^2 f^2 \left[ 1 - \frac{3 g^2 + {g'}^2}{2 {y'_t}^2} \frac{C_g}{C_t} + \left( 2 k_t^{(1)} - \frac{k_G^{(1)} C_g}{C_t} \frac{3 g^2 + {g'}^2}{2 {y'_t}^2} \right) \frac{\sigma0}{4 \pi f} \right]  \label{eq:massTIMP1}
\,.
\eeq
In our calculation, $\sigma_0 \ll 4 \pi f$, therefore, at leading order the dark matter mass is the same as we obtained in the previous section
\beq
m_{DM}^2 \sim 2 C_t {y'_t}^2 f^2  \simeq 2 C_t m_t^2\,,
\eeq
where $m_t = y'_t  f$ for $\theta = \pi/2$ and $\kappa_t (\sigma_0) \sim 1$. 
Analogously, neglecting terms suppressed by $\sigma_0/(4 \pi f)$, the techni-Higgs mass is given by
\beq
m_{\varphi}^2 \simeq M^2 - \frac{\left((k_t^{(1)})^2 +k_t^{(2)} \right)}{2 \pi^2}  m_{DM}^2 \sim M^2 - C_t\frac{\left((k_t^{(1)})^2 +k_t^{(2)} \right)}{\pi^2} m_t^2 \ .
\eeq
The above formula provides a nice correlation between the would-be Higgs mass, the mass of the dark matter candidate and the top mass under the assumption that no other explicit $\SU(4)$ breaking terms are present in the theory. 
In particular, this formula shows the possibility that a potentially large value of $M$ generated by the strong dynamics, may be cancelled by the mass of the dark matter candidate, which is generated by the top loop as suggested in \cite{Foadi:2012bb}.

Finally, we need to check the consistency of our calculation by checking the value of $\sigma_0$, and making sure that it is not too big to invalidate the expansion. An approximate solution for $\sigma_0$ reads:
\beq
\sigma_0 \sim \frac{4 \pi f^3 k_t^{(1)} C_t {y'_t}^2}{8 \pi^2 M^2 - \left((k_t^{(1)})^2+ k_t^{(2)} \right) C_t {y'_t}^2 f^2} \sim  f \cdot \frac{k_t^{(1)}}{2 \pi} \frac{m_{DM}^2}{m_\varphi^2}\,.
\eeq
From this equation we see that a small correction $\sigma_0 \ll 4 \pi f$ would require either $m_{DM} < m_\varphi$, or small $k_t^{(1)} < 4 \pi$ thus implying that the techni-Higgs has a coupling to the top smaller that the SM expectation (see Eq.\eqref{eq:tH2top}). Nevertheless, this analysis does not exclude the possibility to achieve a light techni-Higgs within a consistent dynamical framework. Furthermore the resulting value of the techni-Higgs mass can be further lowered because the intrinsic value of $M$ itself can be small with respect to $4\pi f$. The most discussed example in the literature is the one according to which $M$ is reduced because the underlying dynamics is near conformal. In this case the techni-Higgs is identified with the techni-dilaton~\cite{Yamawaki:1985zg,Dietrich:2005jn,Appelquist:2010gy}. Several model computations have been used to estimate $M$ ranging from the use of the truncated Schwinger-Dyson equations~\cite{Gusynin:1989jj,Holdom:1986ub,Holdom:1987yu,Harada:2003dc,Kurachi:2006ej,Doff:2008xx,Doff:2009nk,Doff:2009kq,Doff:2009na}  to computations making use of orientifold field theories~\cite{Sannino:2003xe,Hong:2004td}. Perturbative examples have proven useful to demonstrate the occurrence of a calculable dilaton   state parametrically lighter than the other states in the theory \cite{Grinstein:2011dq,Antipin:2011aa,Antipin:2012sm}. Last but not the least recent first principle lattice simulations \cite{Fodor:2014pqa,Fodor:2012ni} support this possibility for certain fundamental gauge theories  put forward in \cite{Hong:2004td,Dietrich:2005jn,Dietrich:2006cm}.

\subsubsection*{Experimental constraints}

\begin{figure}[tb]
\center
\includegraphics[scale=0.38]{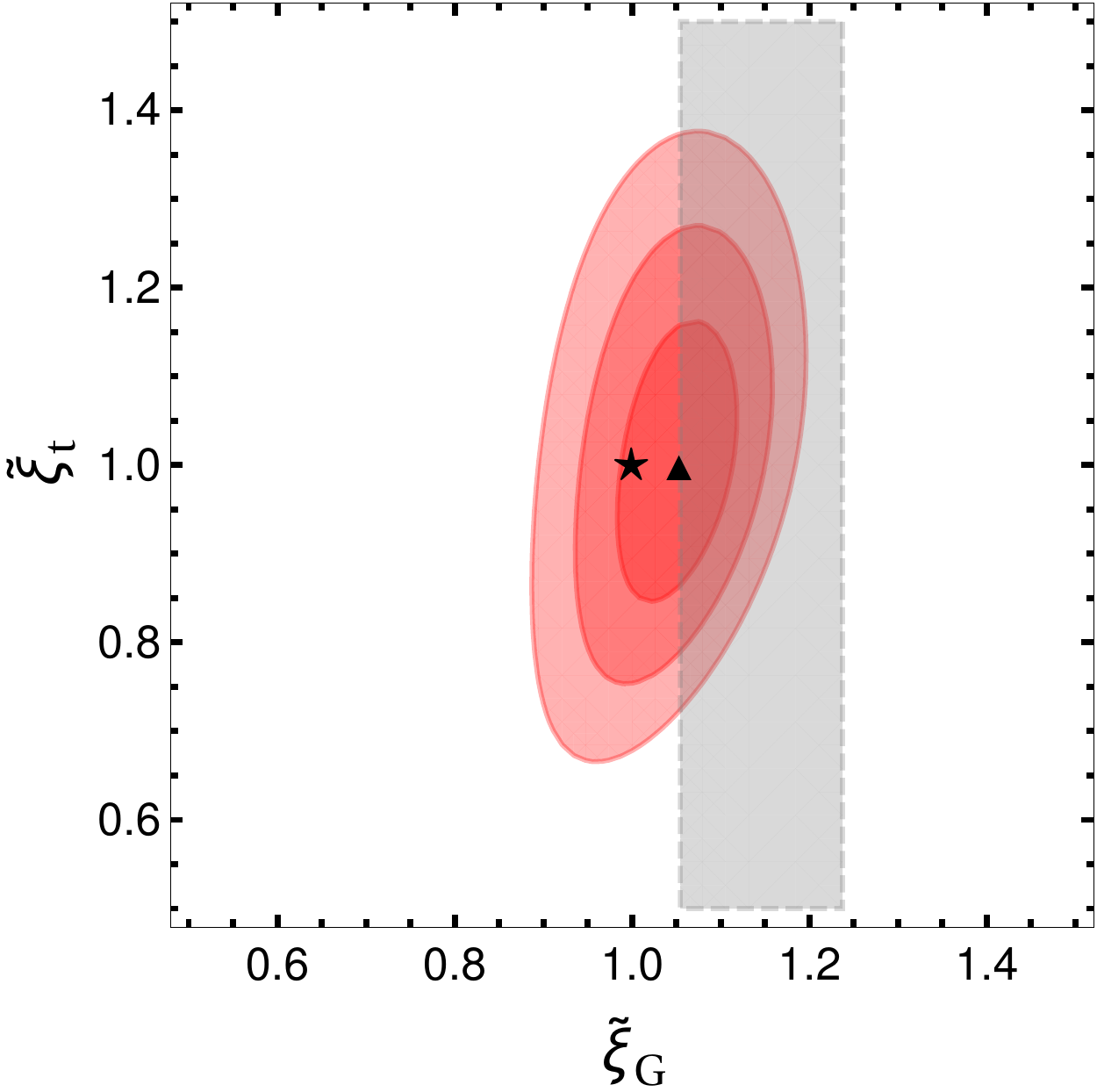}
\includegraphics[scale=0.38]{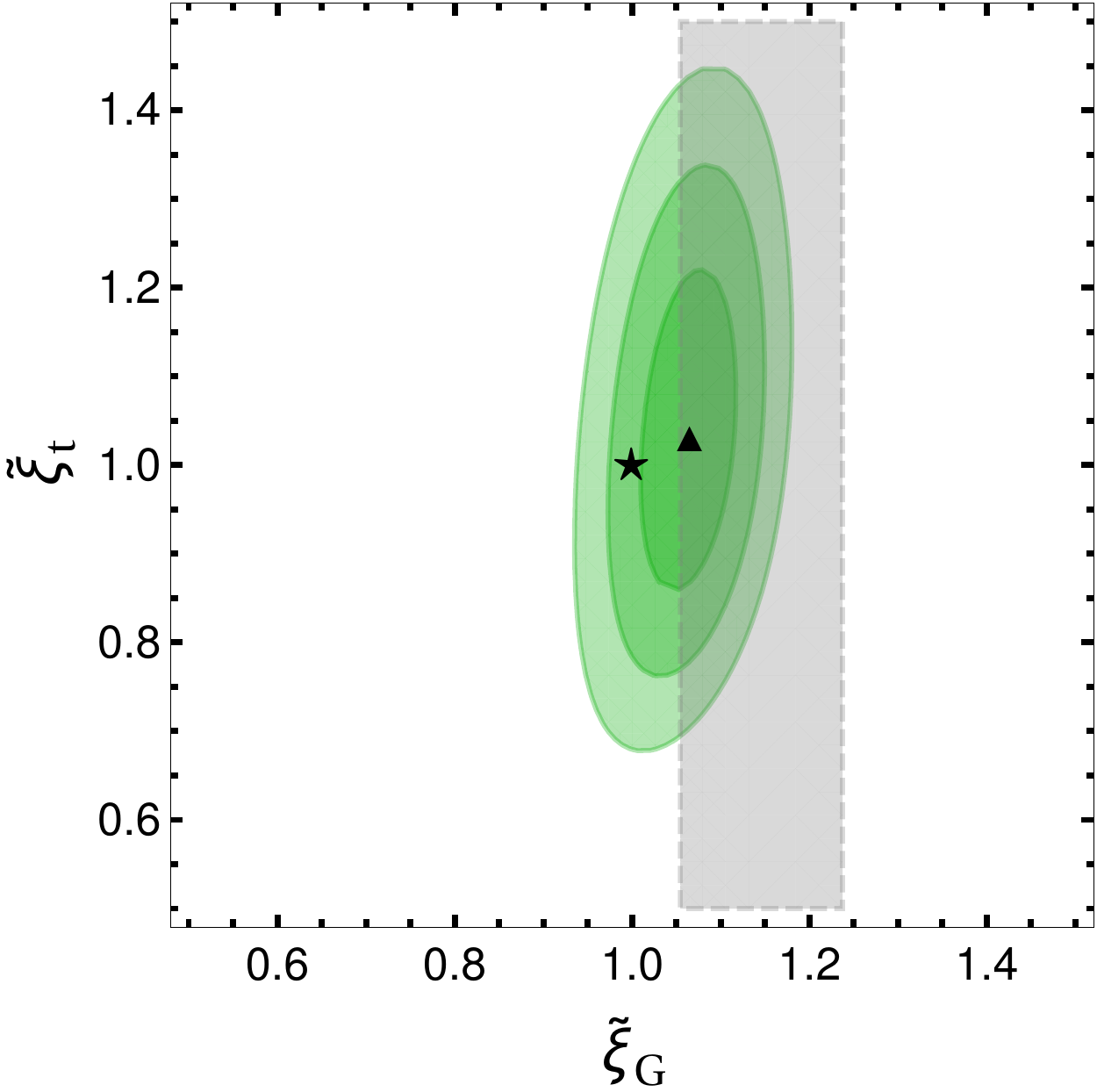}
\caption{Region allowed by the Higgs couplings in the TC limit $\theta = \pi/2$, with $1\sigma$, $2\sigma$ and $3\sigma$ contours from Run-I (left) and 
Run-II (right) data. The gray vertical region is allowed by EWPTs at $3\sigma$ for $\SU(2)_{\FC}$. Figs from~\cite{LeCorre:2018}.}
\label{fig:EWPTTC}
\end{figure}

The couplings of the 125 GeV Higgs depend on the details of the underlying dynamics and are associated to the derivatives of the $\kappa_X$ functions. 
For convenience, we define the following ratios:
\begin{equation} \label{eq:deftildexi}
\tilde{\xi}_G = \frac{\kappa^{(1)}_G}{8 \pi}\,, \quad \tilde{\xi}_f = \frac{\kappa^{(1)}_f}{4 \pi}\,, \quad \tilde{\xi}_G^{(2)} = \frac{\kappa_G^{(2)}}{32 \pi^2}\,, \quad \tilde{\xi}_f^{(2)} = \frac{\kappa_f^{(2)}}{16 \pi^2}\,.
\end{equation}
The normalizations are chosen such that $\tilde{\xi}_G = \tilde{\xi}_t = 1$ would correspond to a SM-like Techni-Higgs.
Assuming the mass gets the correct value, we can compute the bounds on the couplings of $\sigma$ to gauge bosons $\tilde{\xi}_G$ and fermions $\tilde{\xi}_t$ (we are explicitly assuming that all fermions couple in the same way, i.e. 
$\tilde{\xi}_t = \tilde{\xi}_b = \tilde{\xi}_l$) following the procedure in Sec.~\ref{sec:expconstr}.
The results are shown in Fig.~\ref{fig:EWPTTC}, where we plot $1\sigma$, $2\sigma$ and $3\sigma$ contours from the measured Higgs couplings.
A fairly large region around the SM limit $\tilde{\xi}_G = \tilde{\xi}_t = 1$ is still open.
In principle, there is no reason for these couplings to be close to the ones of the SM Higgs, however it is fascinating that this happens for the $\sigma$ meson in QCD~\cite{Belyaev:2013ida}. 
We also compare this allowed region with the bound from EWPTs, which is only dependent on $\tilde{\xi}_G$. The vertical grey band delimits the allowed region for $N_D=2$, corresponding to $\SU(2)_{\FC}$. A substantial overlap exists, pointing to couplings to gauge bosons that are larger than the SM values. This effects should become measurable once more precise data on the Higgs couplings are available. The intersection becomes smaller for larger gauge groups like $\Sp(4)_{\rm FC}$, which has 4 doublets, while larger $\Sp(N)_{\rm FC}$ are clearly disfavored as EWPTs push the parameters in a region excluded by the Higgs coupling measurements~\cite{Arbey:2015exa}.
These results clearly show that the TC limit is still allowed, provided that the correct value of the mass can be achieved.

\subsection{The Goldstone Higgs vacuum and beyond}

 \begin{table}[tb]
 \begin{center}
 \begin{tabular}{l|c|c|}
    & $WW$, $ZZ$ & $f\bar{f}$ \\
 & & \\
$h$ & $\cos\theta$ & $\cos\theta$ \\
$\sigma$ & $\tilde{\xi}_G \sin\theta$ & $\tilde{\xi}_f \sin\theta$ \\
$\eta$ & $0$ & $0$ \\
 & & \\
$hh$ & $\cos 2 \theta$ & $- \sin^2\theta$  \\
$\sigma h$ & $\tilde{\xi}_G \sin 2 \theta$ &  $\tilde{\xi}_f \sin\theta \cos\theta$ \\
$\sigma \sigma$ & $\tilde{\xi}^{(2)}_G \sin^2 \theta$ & $\tilde{\xi}_f^{(2)} \sin^2 \theta$ \\
$\eta \eta$ & - $\sin \theta^2$ &  $- \sin^2 \theta$ \\
\end{tabular} 
\end{center}
\caption{Couplings of one and two scalars to gauge bosons and fermions normalized to the SM values. The bilinear couplings to fermions, absent in the SM, are normalized to $m_f/v^2$.} \label{tab:couplings}
\end{table}

We will now study the general case where $\theta \neq \pi/2$, and both the pNGB and techni-Higgs contribute to the Higgs physics. In this case, the two states mix with each other, while the singlet $\eta$ (being pseudo-scalar) does not. 
For the phenomenology, what is most relevant are the modification of the couplings to the SM fermions and gauge bosons.
A summary of the couplings of the 3 scalars to the SM states normalized to the SM values can be found in Table~\ref{tab:couplings}, where we also report the couplings of two scalars to fermions, normalized to $m_f/v^2$, which are relevant for pair production of the scalars at the LHC.
The importance of such couplings for the Higgs pair production has been stressed in~\cite{Contino:2012xk}.

Here, we will keep the discussion general, so we will simply replace $h$ and $\sigma$ by the mass eigenstates $h_{1,2}$, where the lighter states $h_1$ is identified with the observed Higgs at $m_{h_1} = 125$ GeV:
\begin{equation}
\begin{pmatrix}
  h_1 \\
  h_2 \\
  \end{pmatrix}
  =
 \begin{pmatrix}
  c_{\alpha} & s_{\alpha} \\
  -s_{\alpha} & c_{\alpha} \\
  \end{pmatrix}
  \begin{pmatrix}
  h \\
  \sigma \\
  \end{pmatrix}\,.
  \label{eq:mixing h sigma}
\end{equation}
Both the mass $m_{h_2}$ and the mixing angle $\alpha$ will be considered here as independent free parameters.
It should only be reminded that $\alpha \to 0$ for $\theta \to 0$, as the EW symmetry is not broken in that limit, and also $\alpha\to \pi/2$ for $\theta \to \pi/2$ as a global U(1) subgroup will prevent mixing in the TC limit~\cite{Ryttov:2008xe} (and we need to associate the observed Higgs with $\sigma$).
The sign of $\alpha$ is not determined, however the analysis in \cite{Cacciapaglia:2014uja} shows that the mass of the light state will generically receive a negative correction from the mixing, that is reduced with respect to the prediction in Eq.~\eqref{eq:mh}.
We can therefore consider that
\begin{equation}
m_\eta > \frac{m_{h_1}}{\sin \theta}\,.
\end{equation}

We also present the trilinear couplings among scalars, which are relevant for the pair production of the discovered Higgs  
\begin{eqnarray}
g_{h^3} &=& \frac{3 m_{h}^2}{v}  \cos \theta\,, \\
g_{\sigma h^2} &=& \frac{m_{h}^2}{v} \frac{1}{\sin \theta} \left( \tilde{\xi}_m^{(1)} \cos^2\theta - 2 \tilde{\xi}_t \cos (2 \theta) \right)\,, \\
g_{\sigma^2 h} &=& \frac{m_{h}^2}{v} \frac{2 \cos \theta}{\sin \theta} \left( \tilde{\xi}_m^{(2)} - (\tilde{\xi}_t^{(2)} + \tilde{\xi}_t^2) \right)\,,
\end{eqnarray}
and of $\eta$:
\begin{eqnarray}
g_{h\eta^2} &=& \frac{m_{h}^2}{v}\cos\theta\,,\\
g_{\sigma \eta^2} &=& \frac{m_{h}^2}{v} \frac{1}{\sin \theta} \left( \tilde{\xi}_m^{(1)} \cos^2\theta + 2 \tilde{\xi}_t \sin^2 \theta \right)\,,
\end{eqnarray}
where we defined, for convenience,
\begin{equation}
\tilde{\xi}_{m}^{(1)} = \frac{\kappa_{m}^{(1)}}{4 \pi}\,, \quad \tilde{\xi}_{m}^{(2)} = \frac{\kappa_{m}^{(2)}}{16 \pi^2}\ , 
\end{equation}
and we are working in the non-diagonalized scalar basis. 
It is interesting to notice that the couplings of $\sigma$ diverge for small $\theta$: this is a sign that they are proportional to the condensation scale $f$ and thus increase for increasing condensation scale. The trilinear coupling of $\sigma$ cannot be determined as it comes directly from the strong dynamics. It should therefore be considered as an additional free parameter, also proportional to the condensation scale $f$.

\subsubsection*{Experimental constraints}

\begin{figure}[tb!]
\center
\includegraphics[scale=0.5]{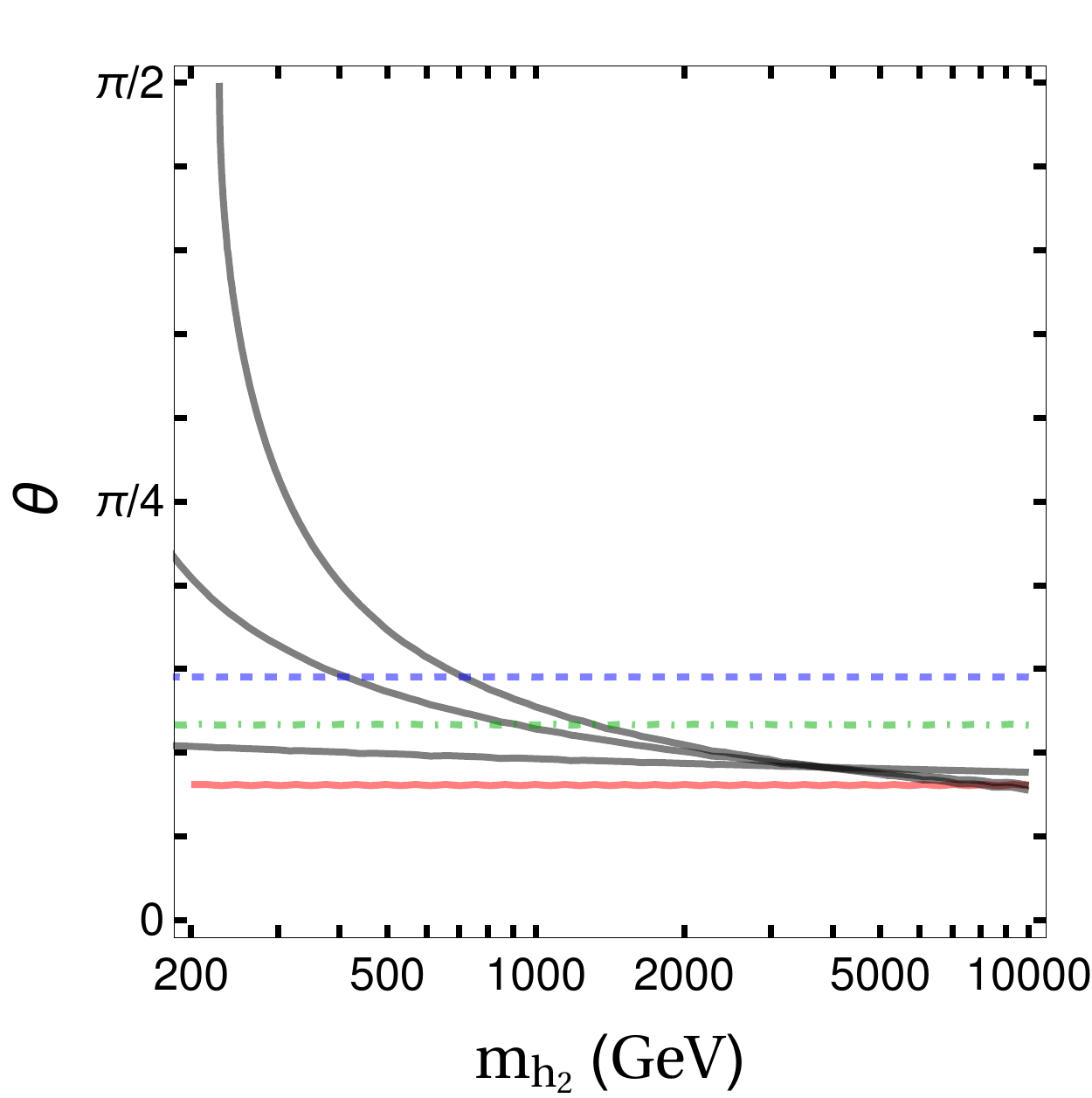}
\caption{Upper bound on $\theta$ as a function of the mass of $\sigma$. The red curve corresponds to the decoupling limit $\theta < 0.239$, while for the other lines correspond to $\tilde{\xi}_G = 0.5$, $1$ and $1.2$, while we keep $\alpha = 0$ and $N_D = 2$. Fig. from~\cite{LeCorre:2018}.}
\label{fig:EWPTpNGB}
\end{figure}

The indirect experimental constraints from the Higgs coupling measurements and EWPTs can be extracted following the procedure in Sec.~\ref{sec:expconstr}, with the only modification that both scalars $h_{1,2}$ now enter the expressions for the oblique parameters:
\begin{eqnarray}
\Delta S &=& \frac{1}{6 \pi} \left[ (1-k_{h_1}^2) \ln \frac{\Lambda}{m_{h_1}} - k_{h_2}^2 \ln \frac{\Lambda}{m_{h_2}} + N_D \sin^2 \theta \right]\,, 
\nonumber \\
\Delta T &=& - \frac{3}{8 \pi \cos^2 \theta_W} \left[ (1-k_{h_1}^2) \ln \frac{\Lambda}{m_{h_1}} - k_{h_2}^2 \ln \frac{\Lambda}{m_{h_2}} \right]\,,
\end{eqnarray}
where 
\begin{eqnarray}
k_{h_1} &=& \cos (\theta-\alpha) + (\tilde{\xi}_G -1) \sin \theta \sin \alpha\,, \nonumber \\
k_{h_2} &=& \sin (\theta-\alpha) +(\tilde{\xi}_G -1) \sin \theta \cos \alpha\,.
\end{eqnarray}
As a starter, we can consider a situation where the mixing between the two states is small, i.e. $\alpha \approx 0$, while the SM-like Higgs is identified with the pNGB state. Via a large $\tilde{\xi}_G$, however, the heavier state $h_2 \approx \sigma$ can affect the bounds. The upper bounds on $\theta$, as a function of the mass $m_{h_2}$ for various couplings are shown in Fig.\ref{fig:EWPTpNGB}, showing that the bound we found in Sec.~\ref{sec:expconstr} can be significantly reduced if the $\sigma$ remains lighter than about $1$~TeV.

\begin{figure}[tbh]
\center \begin{tabular}{ccc}
$\tilde{\xi}=0.8$ & $\tilde{\xi}=1$ & $\tilde{\xi}=1.1$ \\
\includegraphics[scale=0.38]{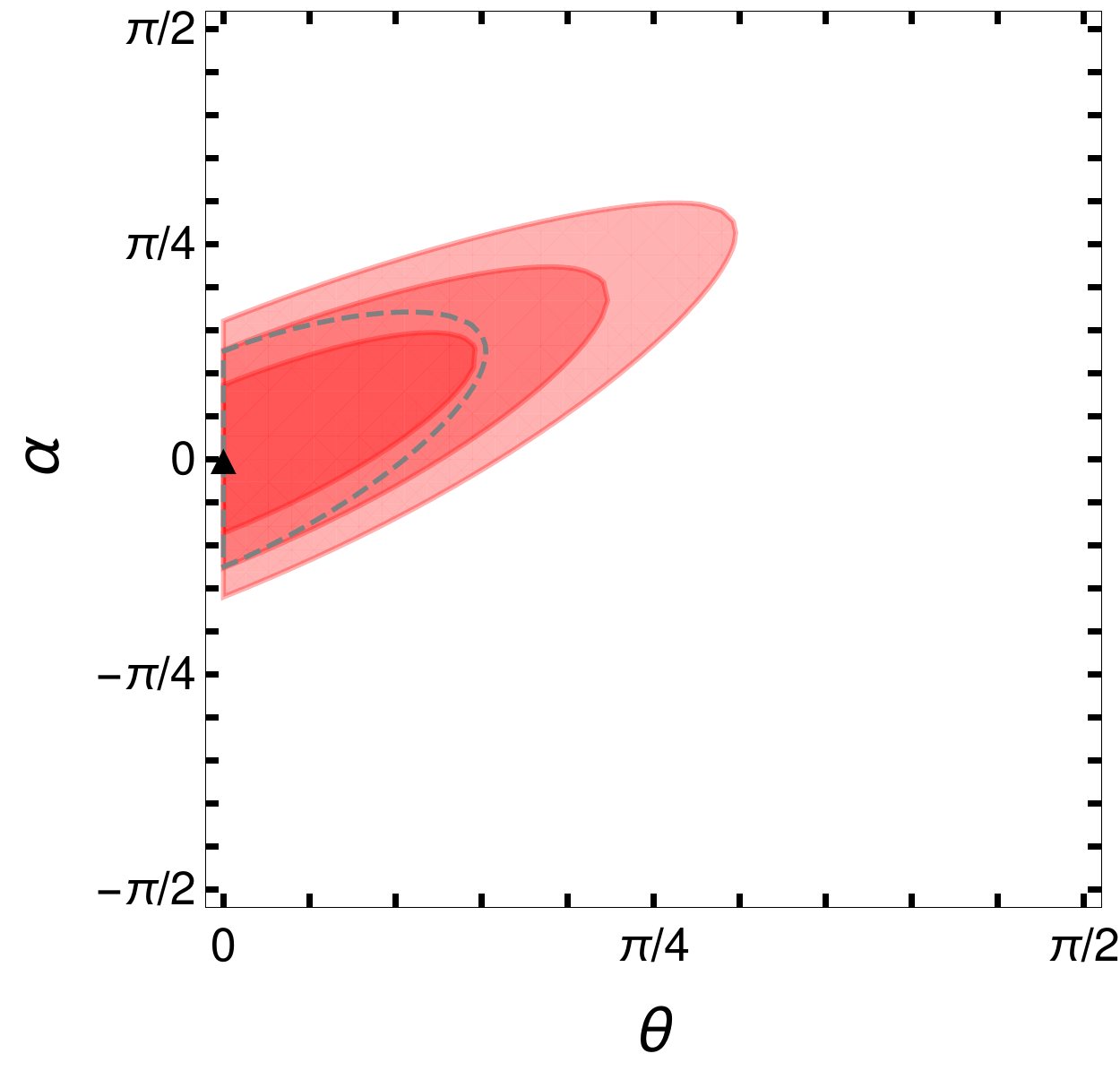} &
\includegraphics[scale=0.38]{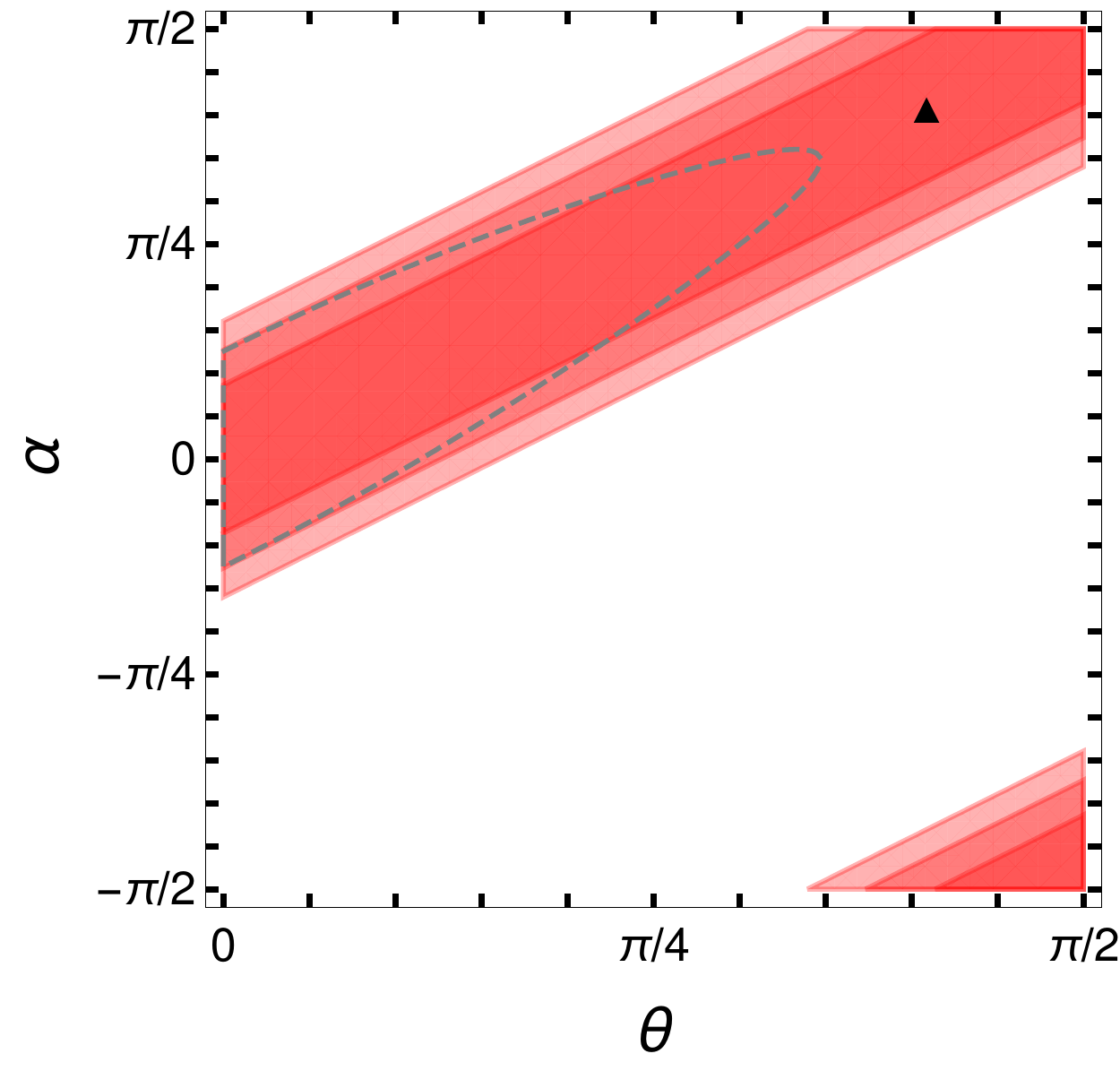} &
\includegraphics[scale=0.38]{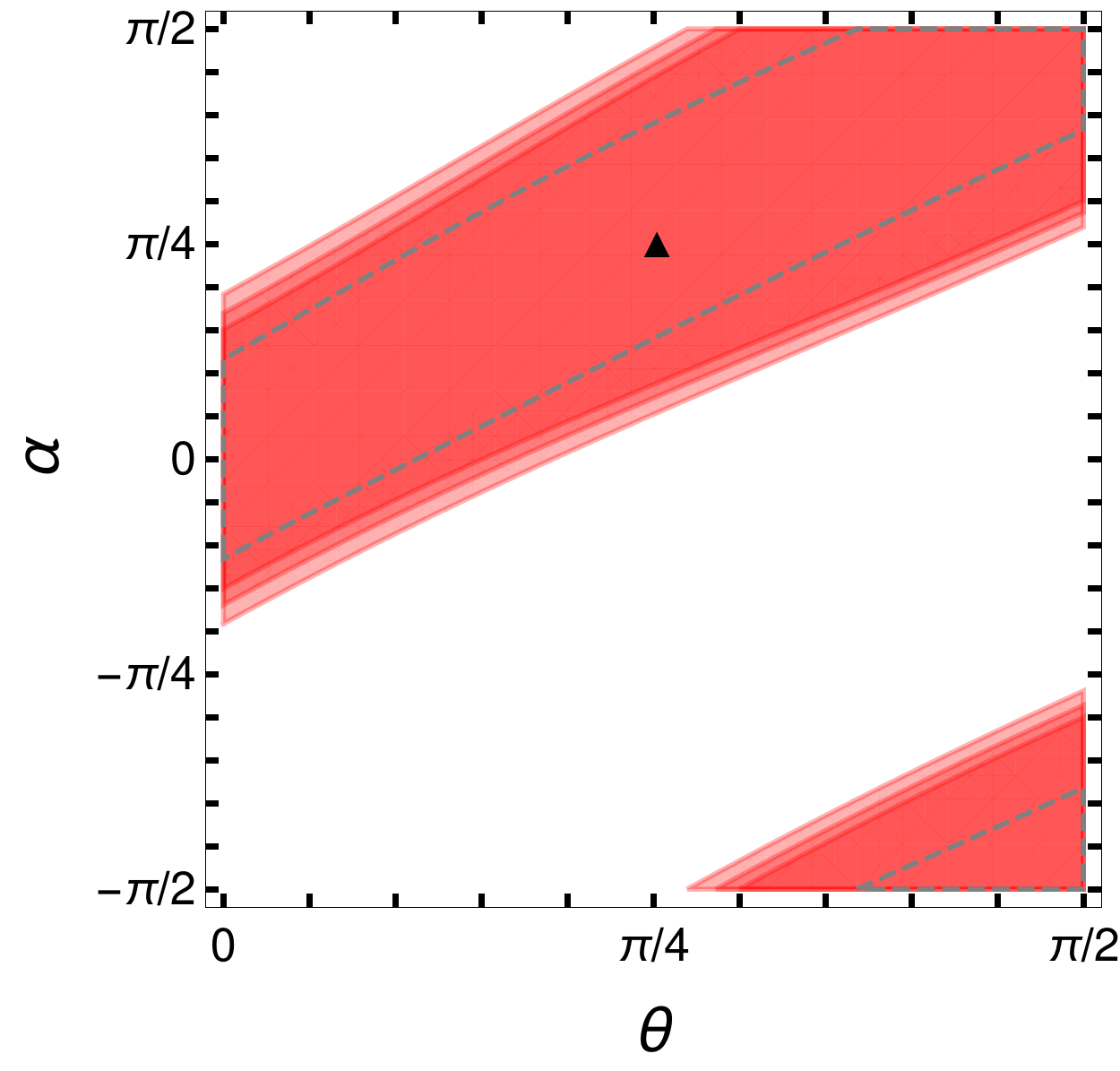} \\
\includegraphics[scale=0.38]{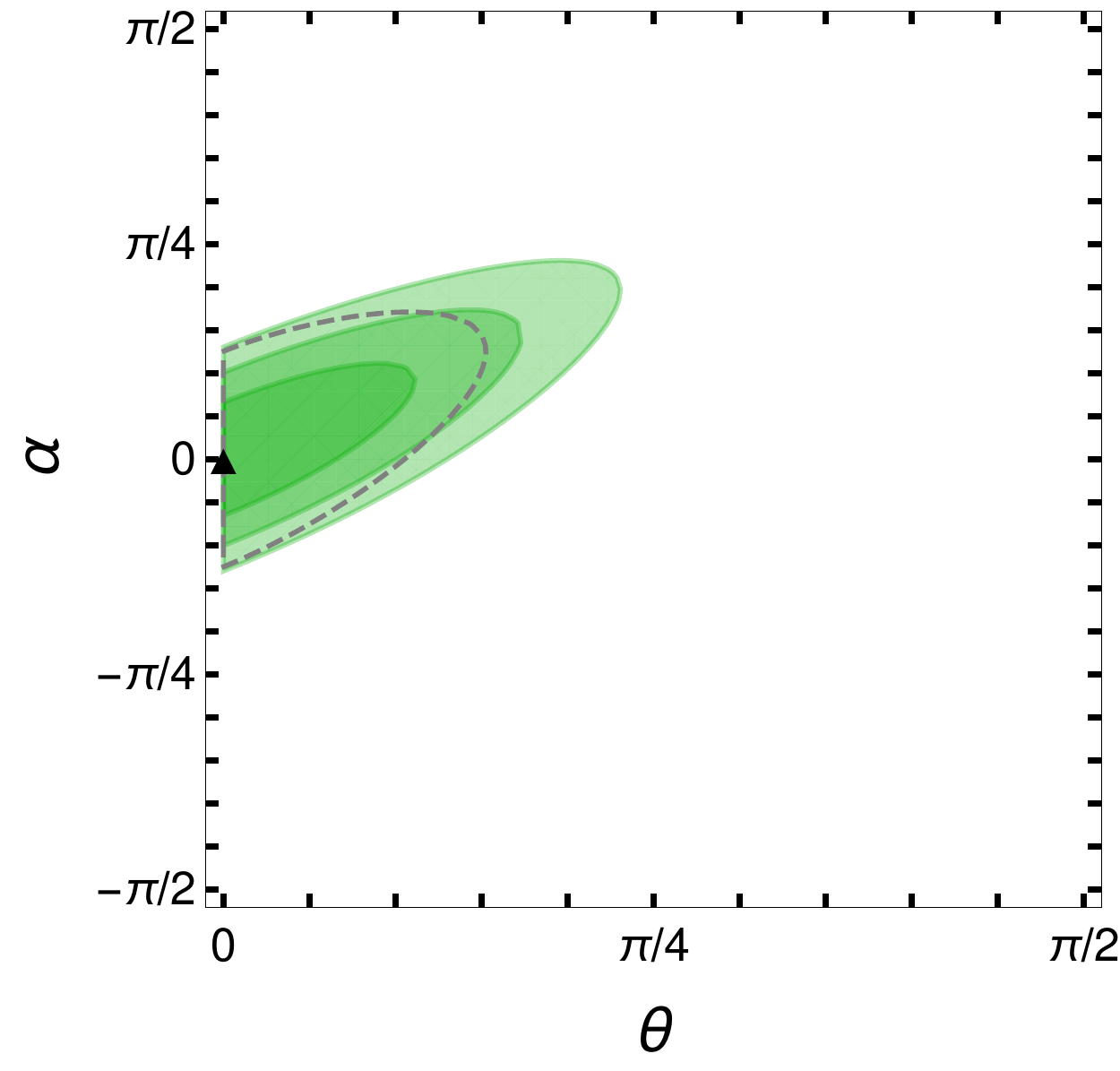} &
\includegraphics[scale=0.38]{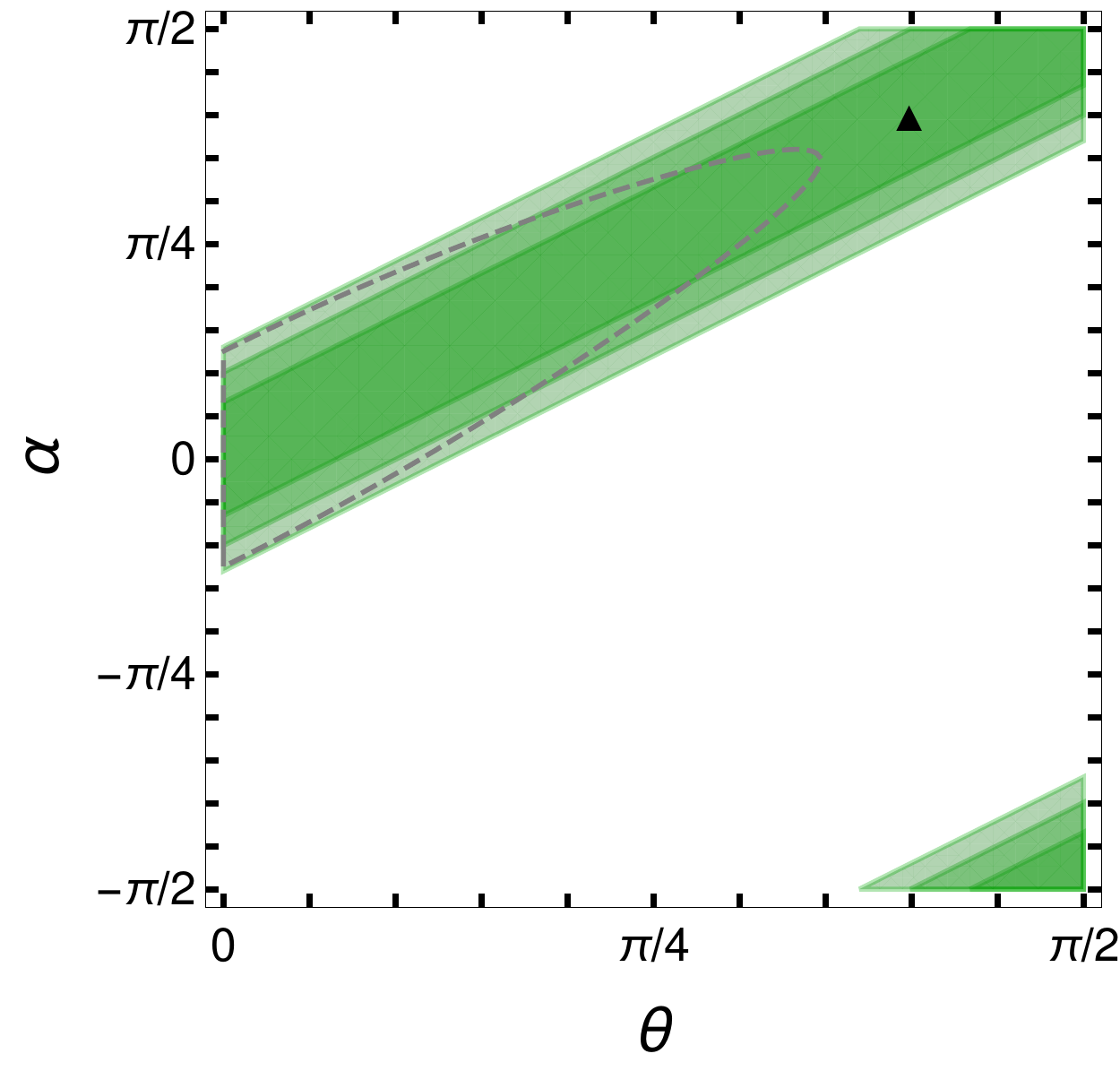} &
\includegraphics[scale=0.38]{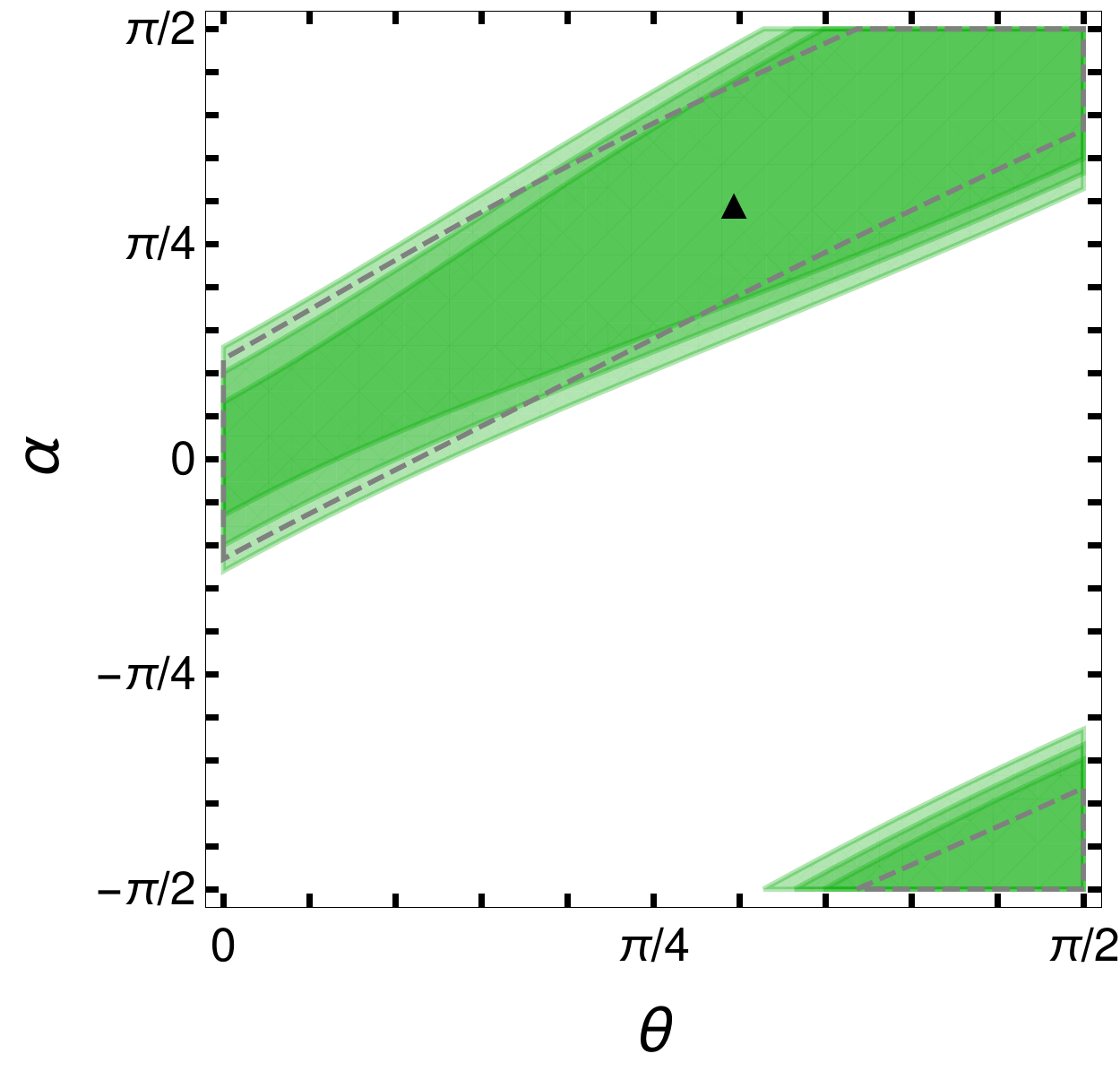} 
\end{tabular}
\caption{Regions allowed by the Higgs couplings for $m_{h_2}=1$ TeV and for Run-I (top row) and Run-II (bottom row) data, for 3 values of $\tilde{\xi}=0.8,\ 1,\ 1.1$. The grey dashed line indicates the 3$\sigma$ bound from EWPTs in the SU(2)$_{\FC}$. Fig from~\cite{LeCorre:2018}.}
\label{fig:EWPTk}
\end{figure}

We now turn our attention to the most general case. To reduce the number of unknown parameters, we fix the $\sigma$ couplings as follows: $\tilde{\xi}_G = \tilde{\xi}_t = \tilde{\xi}$. We also fix the mass of the heavier Higgs $h_2$ and plot the bounds in the plane $\theta$--$\alpha$.
In Fig.~\ref{fig:EWPTk}, we show the bounds in the cases $\tilde{\xi} = 0.8,\ 1,\ 1.1$, and $m_{h_2} = 1$ TeV.
For $\tilde{\xi}=1$ (central column) the plots show a degeneracy in the bounds from the Higgs couplings due to the fact that the couplings of both the light and heavy Higgses depend only on the difference $(\theta - \alpha)$.
On the other hand, EWPTs, in absence of any near-conformal dynamics \cite{Appelquist:1998xf} or sources of isospin breaking,  is well known do prefer small $\theta$ cutting out the TC corner.
Interestingly, however, we observe a novel way to loosen the bound on $\theta$, thanks to a positive mixing angle $\alpha$ allowing for values of $\theta$ up to $\pi/4$. This is an interesting result since it would reduce the level of fine-tuning needed to achieve either a pure pNGB or TC limit.
%
%
The situation is qualitatively different for values of $\tilde{\xi}$ different from 1, as shown in the left and right columns.
For couplings smaller than unity, the allowed regions shrink, as the contribution of the heavy Higgs, which tends to compensate for the modification of 
the Higgs couplings, becomes less important.
For larger couplings $\tilde{\xi} > 1$, the situation is very different: the EWPT allowed regions expand until the TC limit is reached, 
while the Higgs coupling bounds tend to shrink. In the right column of Fig.~\ref{fig:EWPTk}, drawn for $\tilde{\xi} = 1.1$, we see that the TC limit is allowed. 
Going to even larger values, for example $\tilde{\xi} = 1.2$ the TC limit is at odds with the measurements of the Higgs couplings but not with EWPTs. 
This is due, however, to our choice of $\tilde{\xi}_G = \tilde{\xi}_t$. We have shown in the previous section that smaller values of 
$\tilde{\xi}_t$ would reconcile the Higgs couplings with the experimental measurements.

The same interplay between the pNGB Higgs and the techni-Higgs has been discussed in the framework of top partial compositeness in~\cite{BuarqueFranzosi:2018eaj}.



\section{Minimal Composite Dark Matter}
\label{sec:compoDM}

One of the earliest models of composite Dark Matter (cDM) is a stable particle emerging from Technicolor: Nussinov pioneered the idea of the lightest techni-baryon as a cDM candidate~\cite{Nussinov:1985xr}. This fermionic bound state had a natural connection to the observed baryon density by way of the electroweak sphalerons. 
Another possibility is a techniquark-technigluon bound state in minimal walking Technicolor~\cite{Kouvaris:2007iq}. Here the cDM can be Majorana in nature and its relic abundance is determined by thermal freeze-out arguments. 
In both scenarios, the cDM would be the lightest (quasi)-stable composite state carrying a $B'$ charge of a theory of dynamical electroweak breaking featuring a spectrum of technibaryons $(B')$ and technipions.
Generically, baryonic-type cDM candidates tend to be very heavy, in the multi-TeV range. 

Another possibility is that the cDM candidate arises as a pNGB of the strong dynamics, thus having a mass that is parametrically smaller than the scale of confinement. This would allow for lighter cDM candidates, even in models where the strong dynamics is associated to the EW scale \cite{Kouvaris:2007iq}. 
In the following, we will consider this possibility in detail in the context of the template model introduced in this Section.

\subsection{TIMP} \label{sec:TIMP}

As a first example, we will consider a candidate that appears in a minimal Technicolor theory, where the Higgs boson is associated
to a light composite state. We will thus consider our template in the same setup we used to study the properties of the Higgs in Sec.~\ref{sec:technihiggs},
with the vacuum aligned to the Technicolor limit $\theta=\pi/2$.
This model has been proposed and studied in~\cite{Ryttov:2008xe,Gudnason:2006yj}, and can be tested at collider experiments due to the light cDM mass~\cite{Foadi:2008qv}.
A $\U(1)$ subgroup of $\SU(4)$ is thus left unbroken, corresponding to the generator $X^5$ of \eq{eq:Xgen}, which is now preserved by the TC-vacuum.
The two pNGBs, $h$ and $\eta$, form a complex scalar $\phi = \frac{h + i \eta}{\sqrt{2}}$, with zero electric charge, and the pNGB matrix from \eq{eq:SigmapNGB} can be written as
\beq
\Sigma = \left[ \cos \frac{x}{f}\; 1 + \frac{\sqrt{2}}{x} \sin \frac{x}{f}\; \begin{pmatrix}  0 & \phi^\dagger (i\sigma^2) \\  \phi (i\sigma^2) & 0 \end{pmatrix} \right] \cdot \Sigma_0\,
\eeq 
where $x = \sqrt{2 \phi^\ast \phi}$.
The LO chiral Lagrangian reads
\begin{align}
\cL_{(p^2)} = \frac{1}{2} \left( 1+\frac{f^2}{x^2} \sin^2 \frac{x}{f} \right) (\partial_\mu \phi)^\ast \partial^\mu \phi + \left( m_W^2 W_\mu^+ W^{-,\mu} + \frac{1}{2} m_Z^2 Z_\mu Z^\mu \right) \cos^2 \frac{x}{f} \nonumber \\
+ \frac{1}{2 x^2} \left( 1-\frac{f^2}{x^2} \sin^2 \frac{x}{f}  \right) (\phi^2 \partial_\mu \phi^\ast \partial^\mu \phi^\ast + (\phi^\ast)^2 \partial_\mu \phi \partial^\mu \phi)\,.
\end{align}
It clearly shows that it is invariant under the $\U(1)$, and that couplings of the scalar $\phi$ to a single EW gauge boson vanish.

\begin{figure*} 
\begin{center}
 \includegraphics[width=.45\textwidth]{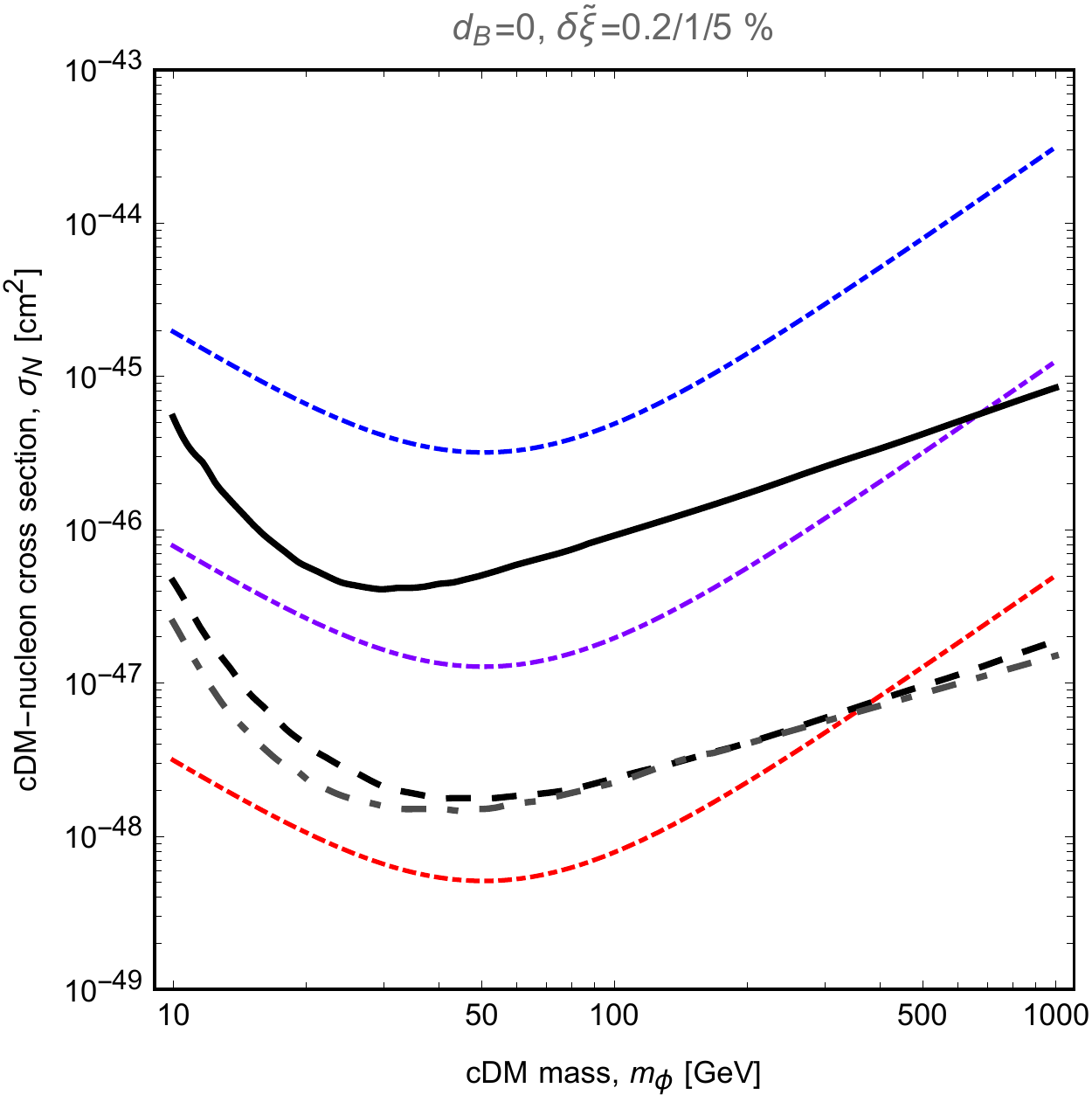} \hspace{0.2cm}
  \includegraphics[width=.45\textwidth]{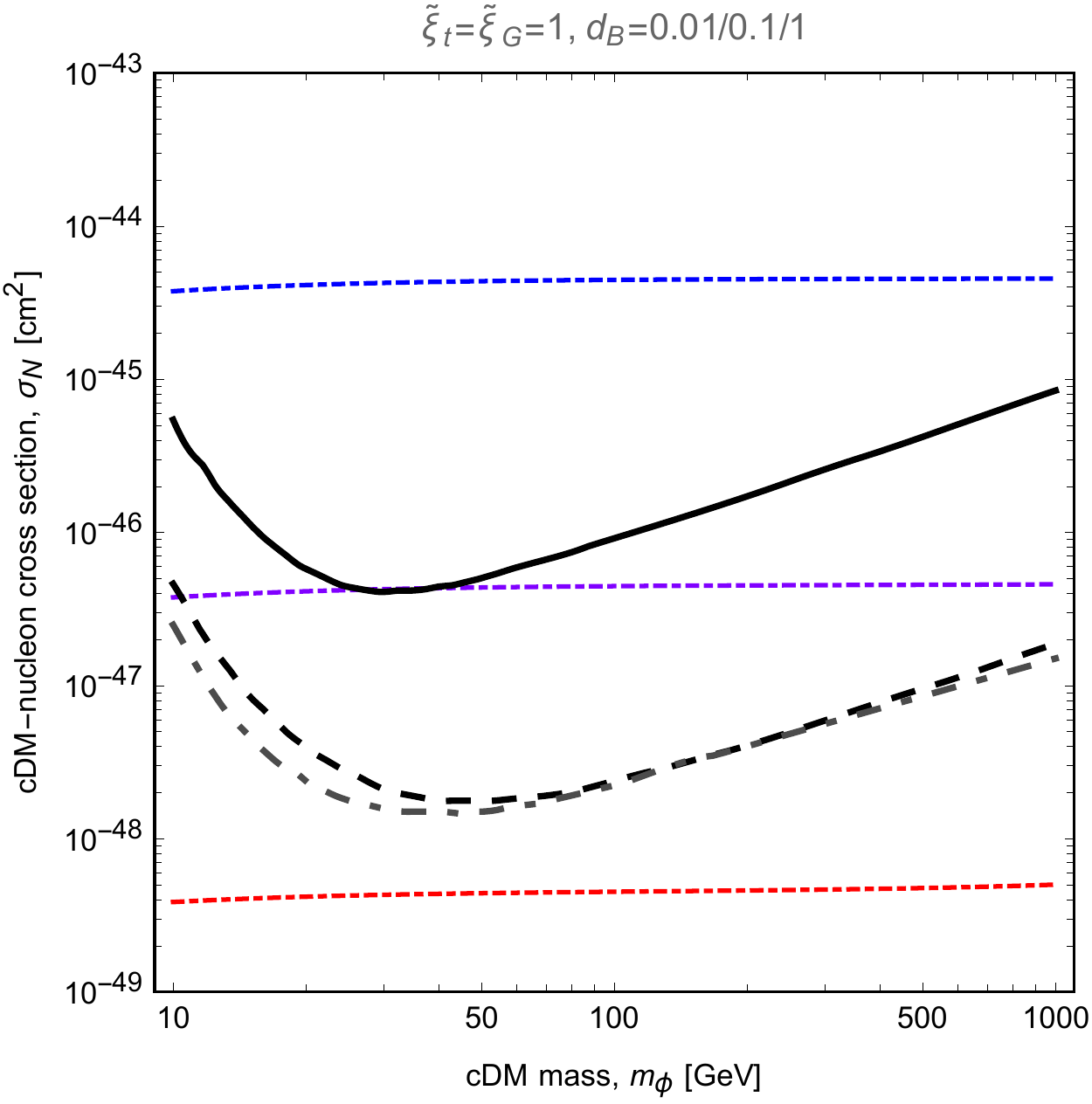}

\caption{ Resulting direct detection limits on pNGB DM in the $\SU(2)_{\rm FC}$ gauge theory with 2 Dirac flavors in the fundamental representation, with $\Lambda_{\rm FC} = 2.5$~TeV.  In the left plot, we neglect the charge radius coupling and vary $\tilde{\xi}_t,\ \tilde{\xi}_G$ within the allowed region (see Sec.~\ref{sec:techniHiggs_TCvac}), with $\delta \tilde{\xi} = |\tilde{\xi}_t - \tilde{\xi}_G| = 0.002$ (red), $0.01$ (purple) and $0.05$ (blue). In the right plot we assume SM-like couplings for the techni-Higgs, and show $d_B = 0.01$ (red), $0.1$ (purple) and $1$ (blue). The black lines show the experimental bounds from XENON1T (solid) with the 20 ton projection (dashed) \cite{Aprile:2018dbl}, while the grey dashed one shows the projections for the future LUX-ZEPLIN experiment~\cite{Akerib:2018dfk}.}
\label{fig:TIMPpNGB}
\end{center}
\end{figure*}

The mass of the cDM candidate can be computed along the same lines of the previous Sections: from Eq.~\eqref{eq:massTIMP1}, neglecting the effect of the couplings to the singlet (techni-Higgs), we find
\beq
m_\phi^2 =  2 f^2 \left( C_t {y'_t}^2 - \frac{3 g^2 + {g'}^2}{2} C_g \right) = 2 C_t m_t^2 - C_g (2 m_W^2 + m_Z^2)\,.
\eeq 
For $C_t \sim C_g \sim 1/8$, the cDM mass is $m_\phi \sim 70$~GeV. 
This example of pNGB cDM has been studied on the lattice~\cite{Lewis:2011zb,Hietanen:2013fya} where it has been found that the form factor obeys vector meson dominance~\cite{Hietanen:2013fya}. At low-energies the dominant interaction of cDM $\phi$ with the SM comes from charge radius interactions
\be  \label{eq:TIMPphoton}
\Delta \mathscr{L}_{\gamma} =  i {\rm e}~\frac{d_B}{\Lambda_\FC^2}  \phi^{*} \overleftrightarrow{\partial_{\mu}} \phi \partial_{\nu} F^{\mu \nu}\,,
\ee
and Techni-Higgs interactions
\be  \label{eq:TIMPhiggs}
\Delta \mathscr{L}_{H} =  \frac{d_{1}}{\Lambda_{\rm FC}}\ H \partial_{\mu}\phi^{*} \partial^{\mu} \phi  + \frac{d_{2}}{\Lambda_{\rm FC}} m_{\phi}^{2}\ H \phi^{*}\phi\,.
\ee
Note that $H \equiv \varphi$ is the Techni-Higgs field described in Sec.~\ref{sec:technihiggs}.
The first term in the Lagrangian~\eqref{eq:TIMPhiggs} comes from the couplings of the techni-Higgs to the pNGBs, i.e. the function $\kappa_G$ of Eq.~\eqref{eq:Lsigma}, while the second comes from the potential.
They can be computed in terms of the Lagrangian in Eq.~\eqref{eq:Lsigma}, and given by
\be
\frac{d_1}{\Lambda_{\rm FC}} = \frac{\kappa_G^{(1)}}{4 \pi f} = \frac{2 \tilde{\xi}_G}{v}\,, \qquad \frac{d_2}{\Lambda_{\rm FC}} =  -\frac{2 \tilde{\xi}_t}{v} - 2 \frac{\tilde{\xi}_t - \tilde{\xi}_G}{v} C_g \frac{2 m_W^2 + m_Z^2}{m_\phi^2}\,,
\ee
where the latter can be computed starting from the loop induced mass term, with couplings of the tecnhi-Higgs from Eq.~\eqref{eq:Vpotsigma}.
Following the results in Sec.~\ref{sec:techniHiggs_TCvac}, see Eq.~\eqref{eq:deftildexi}, we relate such couplings to the coupling modifiers of the techni-Higgs, which is thus SM-like for $\tilde{\xi}_G = \tilde{\xi}_t = 1$.\footnote{This relation was not used in Ref.~\cite{Hietanen:2013fya}, where $d_{1,2}$ were considered free couplings.}
On the other hand, the charge radius coupling $d_B$ can only be determined by means of a Lattice computation.
Therefore this model is automatically ``isospin-violating'' since the charge radius term $\Delta \mathscr{L}_{\gamma}$ couples only to the protons while the Higgs interaction $\Delta \mathscr{L}_{h}$ couples to all nucleons. 
The computation of the spin--independent cross sections can be done by following Ref.~\cite{Hietanen:2013fya}: interestingly, in the limit of vanishing momentum transfer, the contribution of the Higgs couplings is proportional to 
\beq
d_H = - \frac{m_\phi^2}{v} \frac{d_1 + d_2}{\Lambda_{\rm FC}} = 2 (\tilde{\xi}_t - \tilde{\xi}_G ) \left( \frac{m_\phi^2}{v^2} + C_g \frac{2 m_W^2 + m_Z^2}{v^2} \right)\,,
\eeq
thus the cross section vanishes for $\tilde{\xi}_G = \tilde{\xi}_t$, which is true in particular for a SM-like techni-Higgs. 
We show some results in Fig.~\ref{fig:TIMPpNGB}, where the solid black line corresponds to the current strongest bound from XENON1T~\cite{Aprile:2018dbl} (similar bounds were obtained by LUX~\cite{Akerib:2016vxi} and PandaX-II~\cite{Cui:2017nnn}). The dashed lines report the projections for future experiments: the 20 ton case of XENONnT~\cite{Aprile:2018dbl} in black and LUX-ZEPLIN~\cite{Akerib:2018dfk} in grey. For the model curves, we made two choices: in the left panel we assume $d_B = 0$ (corresponding to isospin symmetric models) while choosing allowed values for the $\tilde{\xi}$ parameters (see Sec.~\ref{sec:techniHiggs_TCvac}); in the right panel we fix SM-like couplings for the techni-Higgs and show results for $d_B = 0.01,\ 0.1,\ 1$. The left plot shows that even small deviations from the SM-like couplings of the order of few percent could give rise to exclusion by current direct detection limits, while future experiment will be able to probe sub-percent values. Furthermore, due to the mass dependence of the effective coupling, large masses above few hundred GeV are more strongly constrained.
On the left plot we show the effect of $d_B$, which gives roughly $m_\phi$--independent cross sections, with sensitivity of current limits starting at $d_B = 0.1$. The model cross sections show an ''average'' nucleon cross section on Xenon, defined as
\begin{equation}
\sigma_N = \sigma_p \frac{| Z + (A-Z) f_n/f_p |^2}{A^2}\,, 
\end{equation}
where $A=131$ and $Z=54$ for Xenon atoms, while $f_p$ and $f_n$ are effective couplings to protons and neutrons respectively (see~\cite{Hietanen:2013fya} for more details), and $\sigma_p$ is the cross section on protons.


\subsection{Goldstone Dark Matter and Higgs}

The minimal template model, $\SU(4)/\Sp(4)$, does not have enough space in the coset to feature a stable pNGB in the Goldstone-Higgs limit, $\theta \neq \pi/2$. In fact, the only potential cDM candidate, the singlet $\eta$~\cite{Frigerio:2012uc}, decays promptly via couplings generated by the WZW term of Eq.~\eqref{eq:SWZW}, as $\eta \to W^+ W^-,\ ZZ,\ Z\gamma$~\cite{Galloway:2010bp,Cacciapaglia:2014uja}. It is therefore necessary to extend the minimal template in order to obtain at the same time pNGB Higgs and Dark Matter.

This can be easily done by adding one more Dirac flavor to the matter content of Table~\ref{tab:tab1}: the two additional FC-fermions, $\tcf^{5,6}$, can thus transform either as an $\SU(2)_R$ doublet (Case A) or an $\SU(2)_L$ doublet (Case B)~\cite{Cai:2018tet}. The symmetry breaking pattern is now extended to $\SU(6)/\Sp(6)$, featuring 14 pNGBs: they always include two Higgs doublets $H_{1,2}$, two pseudo-scalar singlets $\eta_{1,2}$, one scalar singlet $\varphi_0$, and a triplet $\Delta_{R/L}$ of $\SU(2)_R$ (Case A) or of $\SU(2)_L$ (Case B). In both cases, some of the pNGBs can remain stable if the misalignment of the vacuum happens only along the direction of one of the two Higgs doublets, say $H_1$. A detailed study of the vacuum structure in Case A can be found in ~\cite{Cai:2018tet}. The pNGB properties are summarized in Table~\ref{tab:tabSU6Sp6} in the vacuum preserving a cDM candidate.
Besides the quantum numbers of the triplets $\Delta_{L/R}$, the two cases are distinguished by the type of dark symmetry:

\begin{table}[tb] \begin{center}
\begin{tabular}{c|c|c|}
   &  pNGB Higgs sector  &  DM  sector\\
\hline
Case A & $\begin{array}{c} H_1 \equiv 2_{\pm 1/2} \\ \eta_1 \equiv 1_0 \\ \eta_2 \equiv 1_0 \end{array}$ & $\begin{array}{c} H_2 \equiv 2_{\pm 1/2} \\ \varphi_0 + i \Delta_R^0 \equiv 1_0 \\ \Delta_R^\pm \equiv 1_{\pm 1} \end{array}$ \\ \hline
Case B & $\begin{array}{c} H_1 \equiv 2_{\pm 1/2} \\ \eta_1 \equiv 1_0 \\ \eta_2 \equiv 1_0 \end{array}$ & $\begin{array}{c} H_2 \equiv 2_{\pm 1/2} \\ \varphi_0 \equiv 1_0 \\ \Delta_L \equiv 3_{0} \end{array} $ \\
\hline
\end{tabular}
\caption{Quantum numbers under $\SU(2)_L$ and hypercharge for the pNGBs in the $\SU(6)/\Sp(6)$ models A and B. The DM sector features a global $\U(1)_{\rm DM}$ in Case A, and a $\mathbb{Z}_2$ discrete symmetry in Case B. The two pseudo-scalar singlets $\eta_{1,2}$ decay into a pair of SM gauge bosons via the WZW interactions.} \label{tab:tabSU6Sp6} 
\end{center} \end{table}

\begin{itemize}

\item[Case A:] A global $\U(1)_{\rm DM}$ symmetry, contained in the unbroken $\Sp(6)$, is preserved, under which the second Higgs doublet $H_2$, the triplet $\Delta_R$ and the singlet $\varphi_0$ form charged combinations. This case has been studied in detail in ~\cite{Cai:2018tet}.

\item[Case B:] A $\mathbb{Z}_2$ symmetry is left preserved, under which the second doublet $H_2$, the triplet $\Delta_L$ and the singlet $\varphi_0$ are odd. This case has been studied in ~\cite{Cai:2019cow}.

\end{itemize}

In both cases, the cDM candidate will be the lightest neutral mass eigenstate of the stable lot of pNGBs\footnote{We are implicitly assuming that the pNGBs are the lightest composite states of the theory, i.e. other dark--charged resonances are heavier.}, which naturally mix via the potential. The two cases have, however, a crucial difference: in Case A, the cDM candidate is a complex scalar field, while in Case B it's a one-component pseudo-scalar. This implies that, in Case A, via mixing with the doublet $H_2$, the cDM candidate will have a non-vanishing coupling to the $Z$ boson, proportional to $(1-c_\theta) \approx s_\theta^2$, which induces a large  spin-independent cross section on nuclei. Therefore, most of the masses for Case A are excluded by direct detection unless $s_\theta \ll 10^{-2}$~\cite{Cai:2018tet}. In Case B, being the cDM a real pseudo-scalar, no coupling to the $Z$ is allowed, thus direct detection bounds are milder.
Thanks to the couplings to the SM generated by the potential and top couplings, the pNGB Dark Matter candidate can be considered as a thermal relic: this has been considered in various cosets in ~\cite{Ma:2017vzm,Ballesteros:2017xeg,Balkin:2017aep,Balkin:2018tma,Cacciapaglia:2018avr,Cacciapaglia:2019vcb,Ramos:2019qqa}.
In ~\cite{Cai:2019cow} an alternative non-thermal mechanism has been proposed: this is based on the idea that the vacuum of the theory could be aligned to Technicolor vacuum at high temperatures, i.e. close to the phase transition leading to the condensates. In this vacuum the theory resembles the one described in the previous section, where a global $\U(1)$ remains unbroken and an asymmetry in charged pNGBs can be generate along with baryogenesis. In the $\SU(6)/\Sp(6)$ model, there are also $\mathbb{Z}_2$--odd states that carry $\U(1)$ charges, and the asymmetry stored in them can be preserved until today once the vacuum rotates away from the Technicolor alignment.

In any case of pNGB Dark Matter, couplings to the Higgs arise in a similar fashion to Eq.~\eqref{eq:TIMPhiggs}: however, being the Higgs $h$ itself a pNGB, the derivative coupling vanishes, $d_1=0$.
This fact can be easily understood as the kinetic operator in the chiral Lagrangian cannot break the shift symmetry in the pNGB fields, so that the only coupling can be generated by the explicit breaking terms in the potential. This fact, however, is only true in our vacuum parametrization, see Section~\ref{sec:vacuumchoice}. In the most common choice in the literature, where the pNGB Higgs is allowed to develop a vacuum expectation value, derivative couplings are also generated~\cite{Frigerio:2012uc}, however the two approaches must give the same physical results.
In ~\cite{Balkin:2018tma}, the fact that a pNGB Dark Matter can only have derivative couplings to the Higgs has been used to explain why direct detection rates, which rely on very low energy processes, are naturally suppressed. However, as we just explained, this is a basis-dependent statement, and the true reason behind the suppression of direct detection should be looked for in the smallness of the terms explicitly breaking the shift invariance.

\subsection{(not) the SIMPlest miracle}

In a recent paper \cite{Hochberg:2014dra} the authors revived an alternative mechanism \cite{Carlson:1992fn,deLaix:1995vi} for achieving the observed dark matter relic density. Instead of using  $2\to2$ annihilation processes, they assume that a dominant $3\to2$ number-changing process occurs in the dark sector involving strongly interacting massive particles (SIMPs). This process reduces the number of dark particles at the cost of heating up the sector. However, the presence of hot dark matter is problematic for structure formation, which means that at the time of freeze-out, the dark matter particles must be in thermal equilibrium with the SM ones. This, in turn, requires small couplings between the dark and the SM sectors. In this way the energy from the dark sector can be transferred to the standard model via scattering processes.

The coupling between the two sectors allows for direct and indirect detection, while the large self-interactions can play a role in structure formation, by solving the {\it core vs.~cusp} problem \cite{deBlok:2009sp}. Compared to the WIMP paradigm, where the dark particles are typically expected to be around the TeV scale, this model can yield masses around a few $100$~MeV. This is an interesting alternative to the WIMP paradigm given the fact that current experiments are putting  substantial constraints on this old paradigm. These constraints can, however, be alleviated or even be completely offset within the recently proposed {\it safe dark matter} paradigm \cite{Sannino:2014lxa}. 
  
A follow-up paper \cite{Hochberg:2014kqa} introduced an explicit realization of the SIMP mechanism based on an underlying strongly coupled sector described via chiral perturbation theory. In this set-up, the pNGBs constitute the cDM particles and a key role is played by the time-honoured Wess-Zumino-Witten (WZW) term \cite{Wess:1971yu,Witten:1983tw,Witten:1983tx}. The WZW term is non-vanishing in theories where the coset space of the symmetry breaking pattern has a non-trivial fifth homotopy group. This topological term introduces a 5-point pion interaction, making it an ideal candidate for the $3\to2$ annihilation process. In QCD, for example, the term describes the annihilation of two kaons into three pions. Here we shall only be concerned with the symmetry breaking pattern $\SU(2N_f) \to \Sp(2N_f)$ for $N_f = 2$. The simplest realization of this breaking pattern comes from an underlying $\Sp(2)_{\rm FC}=\SU(2)_{\rm FC}$ gauge group, thus matching our minimal template, but in general it can be realized for any $\Sp(N_c)$ gauge group. The actual pattern of chiral symmetry breaking depends on the number of flavors, colors and matter representation. 
 
The computations performed in \cite{Hochberg:2014kqa} make use of the first non-vanishing order in the chiral expansion for the $3 \to 2$ and $2\to 2$ processes. For the $3\to 2$ annihilation process, it is one order higher than the related $2\to2$ self-interaction process. It is, however, important to analyse the physical results via a consistent next-to-leading (NLO) and next-to-next-to-leading order (NNLO) chiral perturbative treatment. The leading order analysis is phenomenologically unreliable because it is outside the range of convergence and higher order corrections increase the tension with respect to the phenomenological constraints making it hard for the SIMPlest realization to be phenomenologically viable~\cite{Hansen:2015yaa}. 

\subsubsection{Consistent Setup}

Since the $3\to2$ process emerges at the four-derivative level, while the the $2\to2$ have leading $\cO(p^2)$ contribution, a potential mismatch in power counting is introduced when both are considered at their respective leading contribution. 
In the SIMP set-up, the solution of the Boltzmann equation depends on the thermally averaged cross section for the $3\to2$ process. It returns a value for the pion decay constant, as a function of the pion mass, such that the expected relic density is obtained. These values are subsequently substituted into the cross section for the $2\to2$ self-interaction, which is constrained by observations of e.g.~the bullet cluster \cite{Markevitch:2003at}. Technically, the mismatch arises because the two processes, unless computed to the same order, disagree on the definition of the physical pion mass $m_\pi$ and decay constant $f_\pi$. At NLO they can be written as:
\begin{align}
 m_\pi^2 &= m^2\left[1 + \frac{m^2}{f^2}(a_mL+b_m) + \cO\left(\frac{m^4}{f^4}\right)\right]\,, \\
 f_\pi &= f\left[1 + \frac{m^2}{f^2}(a_fL+b_f) + \cO\left(\frac{m^4}{f^4}\right)\right]\,.
\end{align}
Here $m$ and $f$ are  bare parameters from the LO chiral Lagrangian, $a_i$ and $b_i$ are combinations of NLO LECs and $L$ is a log term defined as
\begin{equation}
 L = \frac{1}{16\pi^2}  \log\left(\frac{m_\pi^2}{\mu^2}\right)\,,
\end{equation}
$\mu$ being the renormalization scale.
 The amplitude should also be expressed in terms of the physical quantities defined above, and not the bare ones. From this argumentation it is clear that a consistent calculation requires both processes to be calculated at the same order in the chiral expansion.

In practice it means that one has to go beyond leading order when computing the $2\to 2$ process. The consistent computation has been done in ~\cite{Hansen:2015yaa}, upon which most of the results in this sections are based.  We will, therefore, follow this strategy and compute full NLO corrections to the $2\to2$ cross sections. Going to NLO will also allow us to extend the convergence of the series, thus obtaining more reliable results. To check the stability of the perturbative expansion we will also estimate the NNLO corrections, by only retaining the direct contributions from the loop diagrams and neglecting the finite contributions from the NNLO operators. We will see that the NNLO results are already much closer to the NLO, than the NLO are to the LO for a much wider range of pion masses.

\subsubsection{Scattering and Relic Density}
The calculation of the relic density in the SIMP scenario follows a slightly modified
 Boltzmann equation for the $3\to2$ process, which reads 
\begin{equation}
 \label{boltzmann}
 \dot{n} + 3Hn = -(n^3-n^2n_{eq})\langle\sigma v^2\rangle_{3\to2}\,,
\end{equation}
where $n(t)$ is the pion number density and $H$ is the Hubble constant. Following the argument in \cite{Hochberg:2014dra}, we neglect contributions from the $2\to2$ annihilation into standard model particles because they are sub-dominant. To solve the differential equation we rewrite it in terms of the dimensionless quantities $Y=n/s$ and $x=m_\pi/T$, with $s$ being the entropy and $T$ the temperature. For the entropy and matter density we used the standard definitions
\begin{equation}
 \rho = \frac{\pi^2}{30}g_e(T)T^4\,,\qquad s = \frac{2\pi^2}{45}h_e(T)T^3\,,
\end{equation}
where $g_e$ and $h_e$ are the effective degrees of freedom contributing to the entropy and density, respectively. In this notation, the differential equation becomes
\begin{equation}
 \frac{dY}{dx} = -\sqrt{\frac{4\pi^5g_*}{91125G}}\frac{m_\pi^4}{x^5}(Y^3-Y^2Y_{eq})\langle\sigma v^2\rangle_{3\to2}\,,
\end{equation}
where $g_*(x)$ combines information from both $g_e(T)$ and $h_e(T)$ into a single function for the effective degrees of freedom (the function is different from the function used in the $2\to2$ case). The equilibrium function $Y_{eq}$ can be written as
\begin{equation}
 Y_{eq} = \frac{45N_\pi x^2}{4\pi^4 g_e(m_\pi/x)}K_2(x)\,,
\end{equation}
where $K_2(x)$ is the modified Bessel function of the second kind and $N_\pi$ is the number of pions. The thermally averaged scattering cross section reads \cite{Hochberg:2014kqa} 
\begin{equation}
 \langle\sigma v^2\rangle_{3\to2} = \frac{75\sqrt{5}}{512\pi^5x^2}\frac{m_\pi^5}{f_\pi^{10}}\frac{N_c^2}{N_\pi^3}\,,
\end{equation}
where $N_c$ is the number of colors ($N_c = 2$ in the minimal case) and our definition of the pion decay constant $f_\pi$ differs by a factor of one-half compared to the original paper. For each value of the pion mass, the value of $f_\pi$ is chosen in such a way that the solution to the Boltzmann equation saturates the expected relic density.  The above cross section derived from a single diagram generated by the WZW topological term we introduced in Eq.\eq{eq:SWZW}.

To determine the pion-pion scattering amplitude, the authors of ~\cite{Hansen:2015yaa} followed the formalism of the non-linearly realized chiral Lagrangian of ~\cite{Bijnens:2011fm}, which is equivalent to the one we defined in Eq.\eq{Gasser-Leutwyler-Lagrangian} by a redefinition of the LECs. We will omit the details here, and refer the reader to ~\cite{Hansen:2015yaa}.
At NLO there are three new diagrams contributing to the amplitude for pion-pion scattering (see \cite{Bijnens:2011fm} for details and diagrams) and at NNLO there are twelve additional diagrams. The scattering amplitude can be written as
\begin{align}
\begin{split}
 T(s,t,u) &= \xi^{abcd}B(s,t,u) + \xi^{acbd}B(t,u,s) \\
 &+ \xi^{adbc}B(u,s,t) + \delta^{ab}\delta^{cd}C(s,t,u) \\
 &+ \delta^{ac}\delta^{bd}C(t,u,s) + \delta^{ad}\delta^{bc}C(u,s,t)\,,
\end{split}
\end{align}
where $\xi^{abcd}$ is a group theoretical factor that depends on the breaking pattern and the number of flavors. 
At NLO there is only one diagram for the $3\to2$ process while at the NNLO there are three more diagrams. However, only two of these diagrams are non-zero in the approximation of vanishing NNLO LECs.

\subsubsection{Results}

We are now ready to discuss the physical results both at the NLO and NNLO. 
At the NLO, we retain the LECs that uniquely specify the underlying strongly coupled theory. These constants are  expected to be a number of order unity suppressed by a loop factor $1/(16\pi^2)$. To determine the LECs one can either use experiments, when available (as for QCD), or perform lattice simulations when the underlying theory is known \cite{Lewis:2011zb,Hietanen:2013fya,Hietanen:2014xca}. For QCD the NLO LECs are known \cite{Bijnens:2014lea} and most of them are of the order $\cO(10^{-4})$, interestingly this is an order of magnitude smaller than na\"ively expected. In the current model, their value is thus randomized using a Gaussian distribution with spread  $5\times10^{-4}$ and zero mean. By averaging over a large number of samples we arrive at a NLO band that reflects the size of the contributions from those LECs.

\begin{figure*}[t]
\begin{center}
 \includegraphics[width=0.49\linewidth]{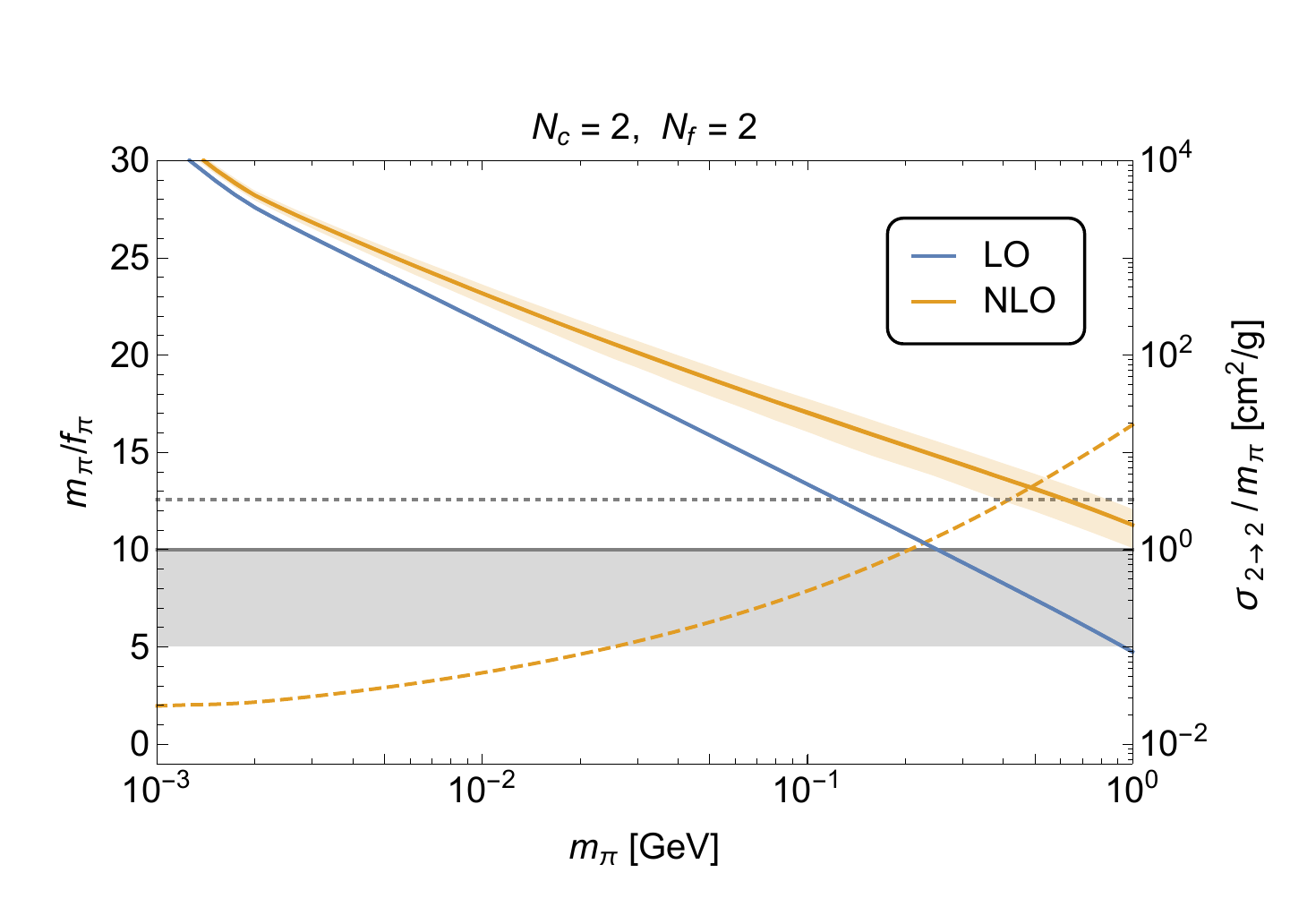}
 \includegraphics[width=0.49\linewidth]{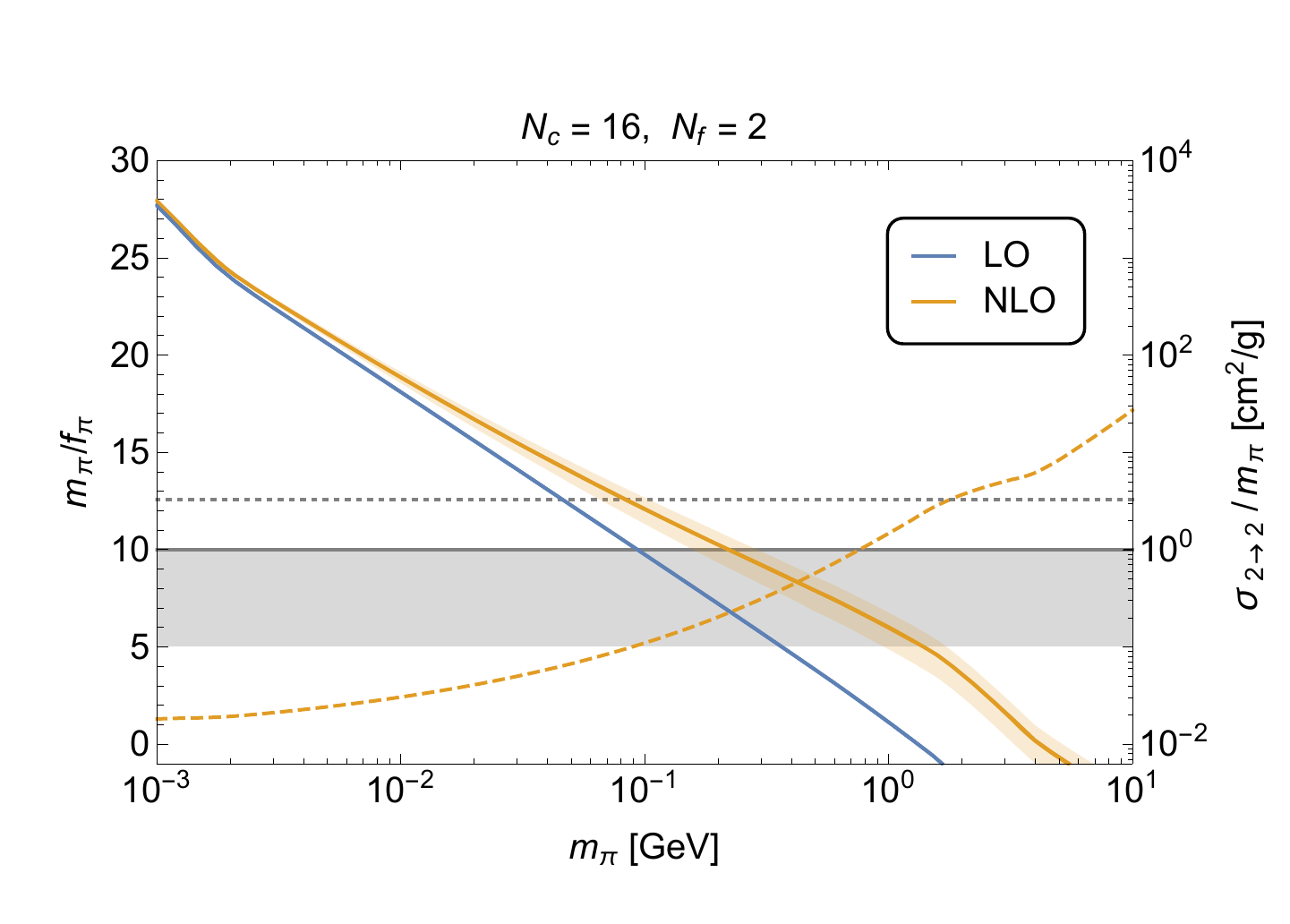}
 \caption{\textit{Dashed} lines belong to the left axis and \textit{solid} lines to the right axis. The dashed orange line is the solution $m_\pi/f_\pi$ to the Boltzmann equation and the dashed horizontal line is the upper perturbative limit $m_\pi/f_\pi=4\pi$. The two solid lines are the cross section for the $2\to2$ self-interactions at LO (blue) and NLO (orange). The band on the orange line is the uncertainty from the low-energy constants. The solid grey band is the upper limit on the self-interactions. Figs from ~\cite{Hansen:2015yaa}.}
 \label{fig:NLO}
\end{center}
\end{figure*}

In Figure~\ref{fig:NLO} we plot the solution to the Boltzmann equation (dashed orange line) and the cross section for the $2\to2$ self-interactions (solid lines) at LO (blue line) and NLO (orange line). The band on the orange line is the estimated effect from the low-energy constants. The horizontal dashed line is the upper perturbative limit $m_\pi/f_\pi=4\pi$. The cross section for the self-interactions is constrained from above by the solid grey line. The constraint from the bullet cluster \cite{Markevitch:2003at} is $\sigma/m_\pi \lesssim 1$ cm$^2$/g but simulations of halo structures \cite{Zavala:2012us,Rocha:2012jg} suggest that the limit might be closer to $\sigma/m_\pi \lesssim 0.1$ cm$^2$/g. This uncertainty is reflected in the grey band.
The model is phenomenologically relevant when the dashed orange line is below the perturbative limit and the self-interactions are below the solid grey line. 

However, for $N_c=2$ these two criteria are never met simultaneously when the NLO corrections are properly taken into account. The results demonstrate that the NLO corrections are crucial to establish the phenomenological viability of the model. The right panel of Figure~\ref{fig:NLO} shows that by substantially increasing the number of colors one may still hope to meet the phenomenological criteria although in this case a more careful use of the large-$N_c$ expansion is needed.

Because we assume small couplings to the standard model, there is also a lower limit on the dark matter mass of around 10 MeV. In fact, lighter masses will change the effective number of neutrino species \cite{Boehm:2013jpa} which is in tension with the Planck data \cite{Ade:2013zuv}.

Since, for the region of phenomenological interest, the NLO corrections are quite sizeable (especially for small values of $N_c$) we also estimate the NNLO corrections. To consistently include these higher order corrections, we re-determined the solution to the Boltzmann equation using the thermally averaged cross section
\begin{equation}
 \langle\sigma v^2\rangle_{3\to2}^{NNLO} = \langle\sigma v^2\rangle_{3\to2}^{NLO}\left(1 + \frac{m_\pi^2}{f_\pi^2}(a_wL+b_w)\right)\,,
\end{equation}
at NNLO. The term $a_wL$ comes from the loop corrections whereas the last term $b_w$ would contain the NNLO LECs if we had included them. As it turns out, the solution is relatively insensitive to the value of $b_w$ assuming it is of order $\cO(1/(16\pi^2))$. This means that setting the low-energy constants to zero in the  amplitude for the $3\to2$ process is a reasonable approximation. For the self-interactions we estimate the effects from just the NLO LECs in the same way as before.

\begin{figure*}[t]
\begin{center}
 \includegraphics[width=0.49\linewidth]{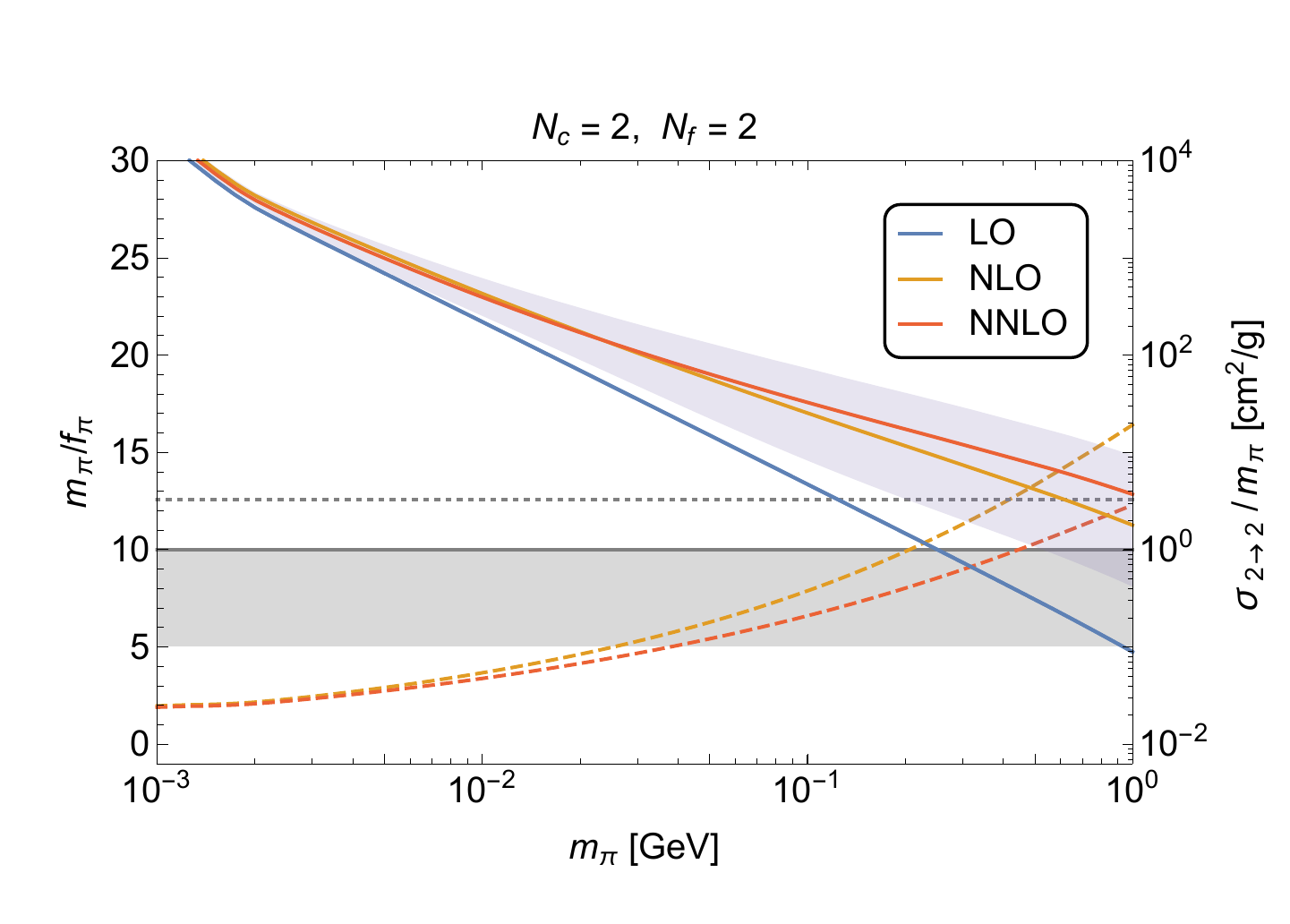}
 \includegraphics[width=0.49\linewidth]{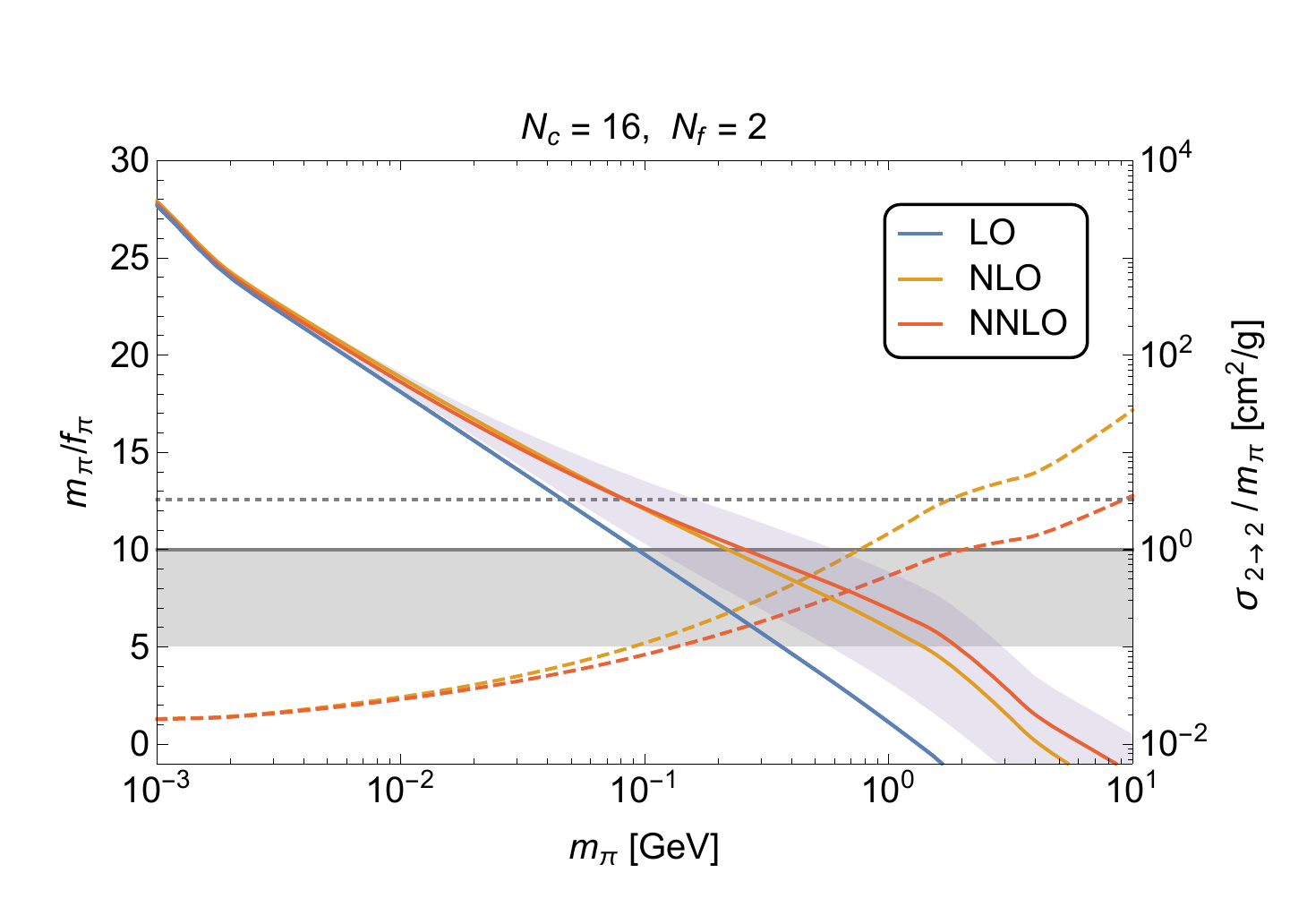}
 \caption{Same as Fig.~\ref{fig:NLO} including NNLO corrections. Figs from ~\cite{Hansen:2015yaa}.}
 \label{fig:NNLO}
\end{center}
\end{figure*}

In Figure~\ref{fig:NNLO} we show the solution to the Boltzmann equation (dashed red line) and the $2\to2$ scattering cross section (solid red line) at NNLO. There are two important points to make: (a) the solution $m_\pi/f_\pi$ decreases, making the investigation more amenable to a perturbative analysis, while (b) the cross section seems to increase further, strengthening the tension with the self-interaction constraints. If the solution did not decrease one would have seen a more dramatic increase in the NNLO cross section. This underlines the importance of a consistent calculation. Despite the small difference between the NLO and NNLO cross sections, we observe a more prominent effect from the LECs as indicated by the purple band. There are two reasons for the increased uncertainty. It is partly due to the fact that the N-LECs enter more frequently in the NNLO result, but also because the squared norm of the NLO amplitude enters in the NNLO cross section. Because this quantity is strictly positive, the average of a Gaussian distribution deviates from zero.

\begin{figure*}[t]
\begin{center}
 \includegraphics[width=0.49\linewidth]{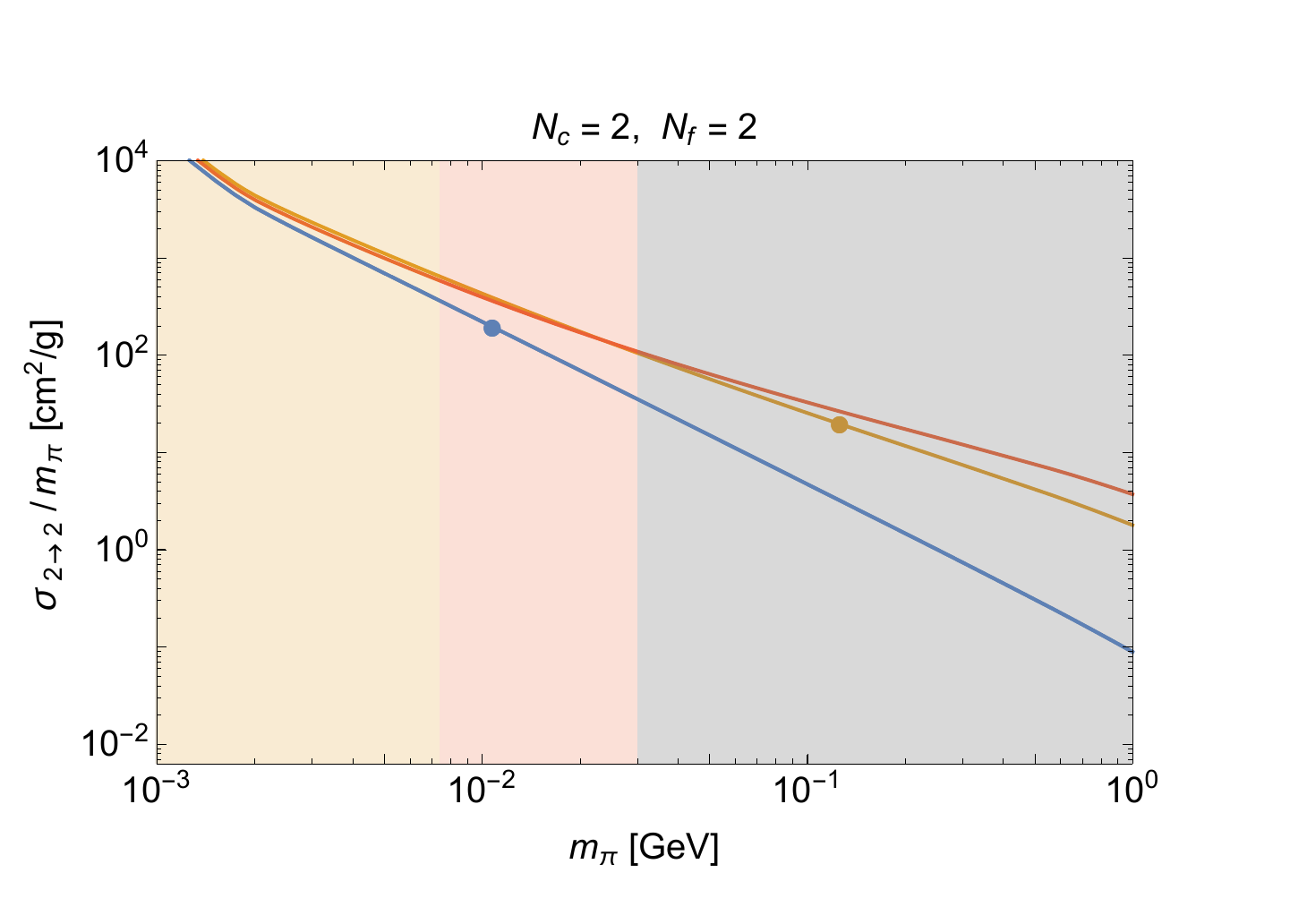}
 \includegraphics[width=0.49\linewidth]{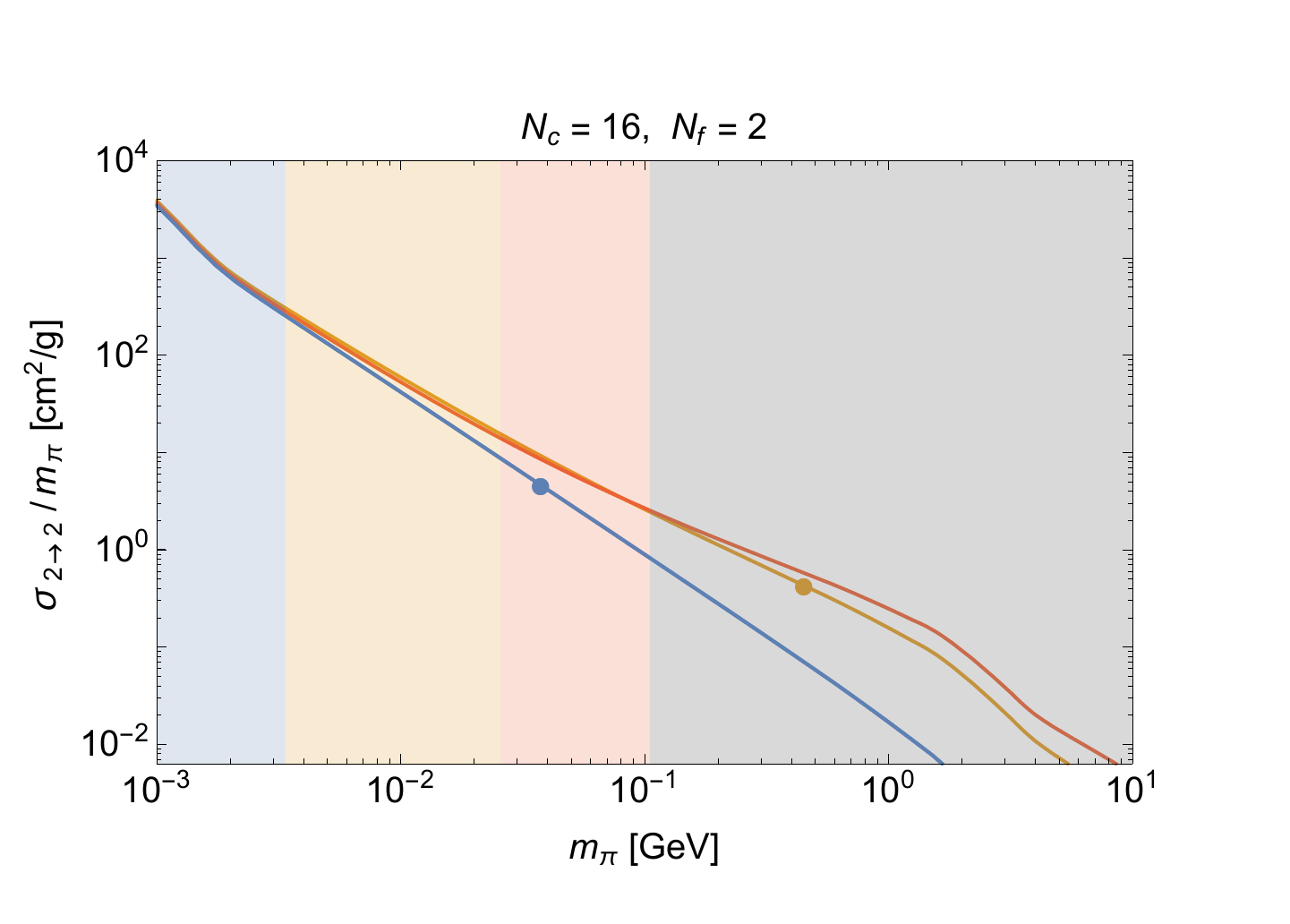}
 \caption{The shaded regions indicate where the different orders in perturbation theory can be trusted up to around 20\%. The blue line is LO, orange line is NLO and red line is NNLO. The dots show for, each order, when the next order corrections are of the same magnitude. The estimates are obtained neglecting the LECs.  Figs from ~\cite{Hansen:2015yaa}.}
 \label{fig:FLAG}
\end{center}
\end{figure*}

In Figure~\ref{fig:FLAG} we provide regions of validity where the different orders in perturbation theory can be trusted up to around 20\%. The dots show where each order breaks down because the corrections from the next order are of equal magnitude. For the NNLO case the values are rough estimates from extrapolating via the previous orders. In the grey region the theoretical uncertainty of the NNLO expansion is more than 20\%. We therefore estimate that for $N_c=2$ perturbation theory at NNLO is reasonable up to around 30 MeV. However, besides the aforementioned constraint coming from the effective number of neutrino species for masses below 10 MeV, this region is excluded by the constraint on the self-interactions. Even by allowing the NNLO treatment to be used beyond this point, the model cannot easily meet the phenomenological constraints. It is clear from the figure that the claimed phenomenologically relevant region \cite{Hochberg:2014kqa} is far beyond LO perturbation theory.

For the case of $N_c=16$ one has control up to around 110 MeV at NNLO, but here the theory is still in tension with the constraints on self-interactions. However, after this point the theory meets the constraints for plausible combinations of low-energy constants.

One can introduce additional breaking of the flavor symmetry which, in principle, can reduce the $2\to2$ scattering amplitude \cite{Hochberg:2014kqa}, but it will not change the range of convergence for the different perturbative orders.


By consistently using chiral perturbation theory we studied the phenomenological viability of an interesting class of cDM models. In these models the relic density of the dark pions is achieved via $3\to2$ number-changing processes making use of the WZW term. We determined both the $3\to2$ and the $2\to2$ processes to the NLO and NNLO order in the chiral expansion and showed that higher order corrections substantially affect the LO result of \cite{Hochberg:2014kqa} for the $2\to 2$ processes. At the NLO and NNLO we have shown that  the SIMPlest models \cite{Hochberg:2014kqa}, with a moderate number of underlying colors, are at odds with phenomenological constraints. However, for sufficiently many colors, or via additional breaking of the flavor symmetry, it is still possible to find small regions in the parameter space consistent with phenomenological constraints.

These results are also of immediate interest for different realizations of composite dark matter \cite{Ryttov:2008xe} and/or composite (Goldstone) Higgs \cite{Appelquist:1999dq,Cacciapaglia:2014uja,Arbey:2015exa} featuring the same pattern of chiral symmetry breaking.  Furthermore first principle lattice simulations of the underlying dynamics \cite{Lewis:2011zb,Hietanen:2013fya,Hietanen:2014xca} are able to provide crucial information on the low-energy constants \cite{Arthur:2014zda}.



\chapter{SM Fermion mass generation, a critical overview }
\label{SMFMgenration}

The SM Higgs allows for a direct and efficient way to give mass to the SM fermions. The masses arise via renormalizable Yukawa interactions of the from $ \overline{\psi}_{R}  H \psi_L + \mbox{h.c.}$ with $\psi$ being a generic SM fermion. A drawback of this sector is the rigid structure of the SM Yukawa interactions: the number and phases of each coupling is dictated by the quantum numbers of the fields involved. As a consequence, there is a single source of CP violation in the quark sector, and it is not allowed to include additional sources that would be needed, for example, to generate the observed baryon asymmetry. On the contrary, extensions of the SM in which the Higgs sector is replaced by a new theory inevitably lead to a  richer structure and novel theoretical and phenomenological opportunities.

 In this chapter we concentrate on the mechanisms that have been envisioned to endow SM fermions with masses in models of composite dynamics. 

\section{Time-honored approaches} \label{sec:ETC}
In any composite model in which the Higgs arises as composite bilinear of the underlying FC-fermions, one has that the SM Yukawa couplings amount to four-fermion operators. The latter can be naturally interpreted as a low energy operator induced by a new strongly coupled gauge interaction emerging at energies higher than the electroweak theory. This mechanism has been termed Extended Technicolor (ETC) \cite{Eichten:1979ah,Dimopoulos:1979es}. 
 
\begin{figure}[tp]
\begin{center}
\mbox{
\subfigure{\resizebox{!}{0.23\linewidth}{\includegraphics[clip=true]{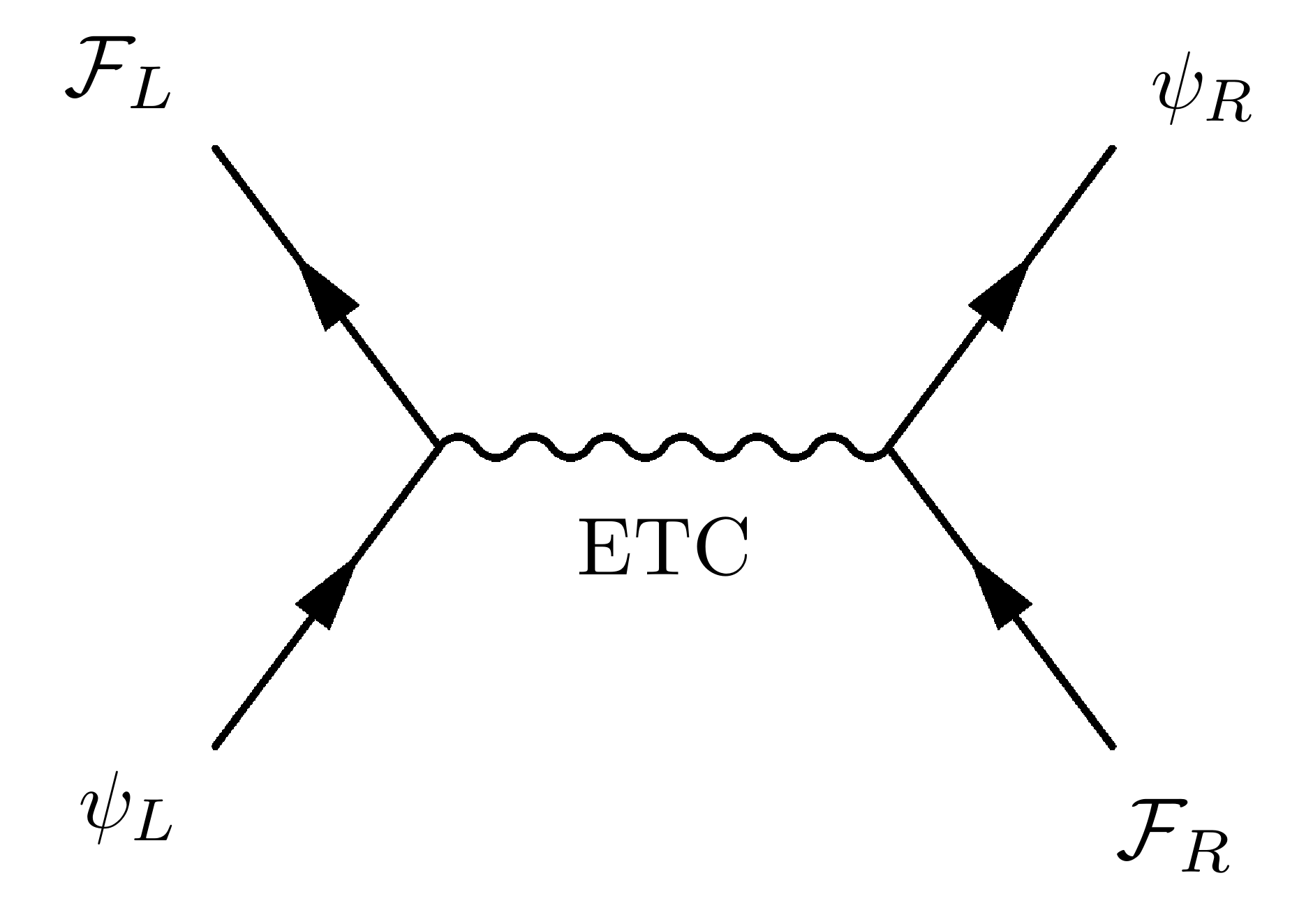}}}}\qquad \qquad
\subfigure{\resizebox{!}{0.23\linewidth}{\includegraphics[clip=true]{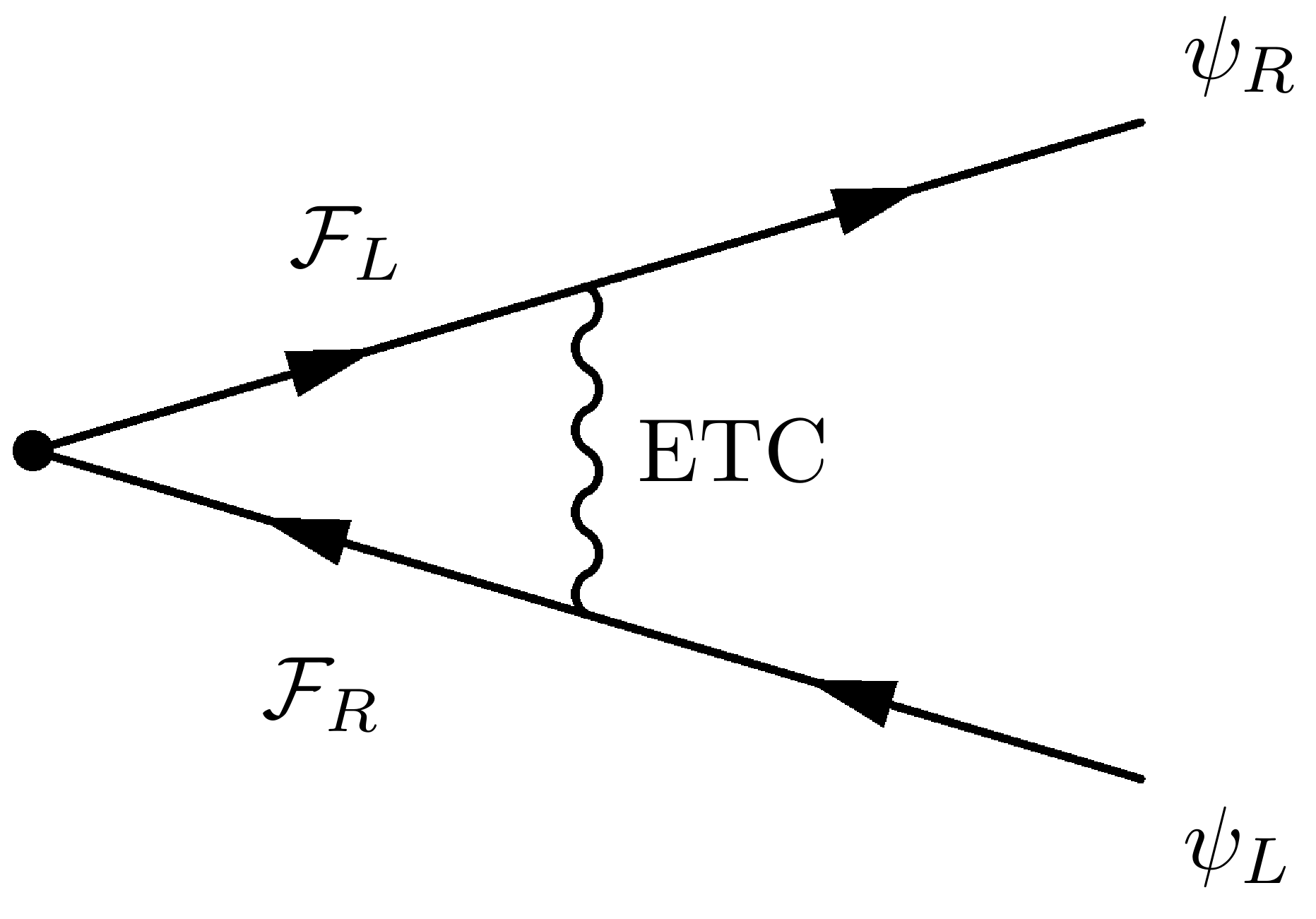}}}
\caption{Left panel: ETC  gauge boson interaction involving
FC-fermions and SM fermions. Right panel: Diagram contribution to the mass to the SM fermions, where the black dot on the left-side stands for the $\tcf$-condensate.}
\label{etcint}
\end{center}
\end{figure}

 Here we will describe the simplest ETC model in which the ETC interactions connect the chiral symmetries of the FC-fermions to those of the SM fermions (see left panel of Fig.~\ref{etcint}). To agree with earlier notation in this section we indicate with ${\cal F}$ the FC-fermions and address all SM fermions with $\psi$. 

When Fun. Color chiral symmetry breaking occurs it leads to the diagram in the right panel of
Fig.~\ref{etcint}. Let's start with the case in which the ETC dynamics is represented by an $\SU(N_\text{ETC})$ gauge group with: 
\beq N_\text{ETC} = N_\FC + N_g \ , \eeq
and $N_g=3$ is the number of SM
generations. In order to give masses to all SM fermions, in this scheme one needs a condensate for each SM fermion. This can be achieved by using as FC-fermions a complete generation of quarks and leptons (including a right-handed neutrinos), but now charged with respect to the FC interactions.  

The ETC gauge group is assumed to spontaneously break $N_g$--times down to
$\SU(N_\FC)$, thus permitting
three different mass scales, one  for each SM family. This type of TC with associated ETC is
termed the \emph{one family model} \cite{Farhi:1979zx}.
The SM heavy masses emerge by the
breaking occurring at the lowest energies and the light masses via the breaking
at highest energy scales. 
This model does not, per se, explain how the
gauge group is broken several times, neither is the breaking of weak isospin
symmetry accounted for. For example we cannot explain why neutrinos have masses much smaller than the associated electrons. See, however, \cite{Appelquist:2004ai} for progress on these issues. Schematically one assumes that the gauge group $\SU(N_\FC + 3)$ breaks to  $\SU(N_\FC + 2)$ at a scale 
$\Lambda_1$ giving mass to the first generation of fermions of the order of  $m_1 \sim {4\pi
  (F_\pi^\FC)^3}/{\Lambda_1^2}$. A second breaking occurs leaving behind $\SU(N_\FC + 1)$ at a dynamical scale $\Lambda_2 $ yielding a second generation mass of the order of $m_2 \sim{4\pi
  (F_\pi^\FC)^3}/{\Lambda_2^2}$. Finally, the last breaking
$\SU(N_\FC )$ at scale 
$\Lambda_3$ yields the last generation mass $m_3 \sim {4\pi
  (F_\pi^\FC)^3}/{\Lambda_3^2}$. 
 Starting from the low energy FC theory, the priority is therefore on discussing the interactions that generate a mass for the top (and the remainder of the third generation).
 
 For the template model of Chapter \ref{ch:su2}, a concrete example of ETC for the top has been presented in~\cite{Cacciapaglia:2015yra}. In this model, the FC group $\SU(2)_\FC$ and the QCD $\SU(3)_c$ are partially unified in a $\SU(5)$ group, where three $\tcf$ are embedded together with the quark doublet and the right-handed top in $\bf 5$'s of $\SU(5)$. The total gauge group of the model is therefore extended to
 \beq
 \SU(5) \times \SU(2)'_\FC \times \SU(3)'_c \times \SU(2)_L \times \U(1)_Y\,,
 \eeq
 where the breaking
 \beq
  \SU(5) \times \SU(2)'_\FC \times \SU(3)'_c \to \SU(2)_\FC \times \SU(3)_c
 \eeq
is assumed at the scale $\Lambda_3$. The off-diagonal gauge boson in $\SU(5)$, will therefore generate the appropriate four-fermion interactions that correspond to the top mass operators, see Eq.~\eqref{eq:EFCD}. 
 
Without specifying an ETC one can write down the most general type of four-fermion operators involving FC-fermions $\tcf$ and ordinary fermionic fields $\psi$.  Following the notation of Hill and Simmons \cite{Hill:2002ap}, we write:
\beq \mathcal{L} = \alpha_{ab}\frac{\bar {\cal F}\gamma_\mu T^a{\cal F}\ \bar\psi \gamma^\mu
  T^b\psi}{\Lambda_\text{ETC}^2} +
\beta_{ab}\frac{\bar {\cal F}\gamma_\mu T^a{\cal F}\ \bar {\cal F}\gamma^\mu
  T^b{\cal F}}{\Lambda_\text{ETC}^2} + 
\gamma_{ab}\frac{\bar\psi\gamma_\mu T^a\psi\ \bar\psi\gamma^\mu
  T^b\psi}{\Lambda_\text{ETC}^2} \ , \eeq
where the $T$'ss are unspecified ETC generators. After performing a Fierz rearrangement one has:
\beq \alpha_{ab}\frac{\bar {\cal F}T^a{\cal F}\ \bar\psi T^b\psi}{\Lambda_\text{ETC}^2} +
\beta_{ab}\frac{\bar {\cal F}T^a{\cal F}\ \bar {\cal F}T^b{\cal F}}{\Lambda_\text{ETC}^2} +
\gamma_{ab}\frac{\bar\psi T^a\psi\ \bar\psi T^b\psi}{\Lambda_\text{ETC}^2}
+ \ldots \  \label{etc} \eeq
The coefficients parametrize the ignorance on the specific ETC physics. To be more specific, the $\alpha$-terms, after the FC condensation, lead to mass terms for the SM fermions
\beq m_q \approx \frac{\alpha}{\Lambda_\text{ETC}^2}\langle \bar
{\cal F}{\cal F}\rangle_\text{ETC} \ , \label{eq:mqETC}\eeq
where $m_q$ is the mass of {\em e.g.}~a SM quark and $\langle \bar {\cal F}{\cal F}\rangle_\text{ETC}$ is the
FC condensate evaluated at the ETC scale. Note that we have not explicitly considered the different scales for the different generations of ordinary fermions but this should be taken into account for any realistic model.  Typically, one can associate the coefficient to a gauge boson mediator, leading to $\alpha/\Lambda_\text{ETC}^2 \to g_{\text{ETC}}^2/M_{\text{ETC}}^2$.

\begin{figure}[tbh]
\begin{center}
\includegraphics[width=7truecm,height=4truecm]{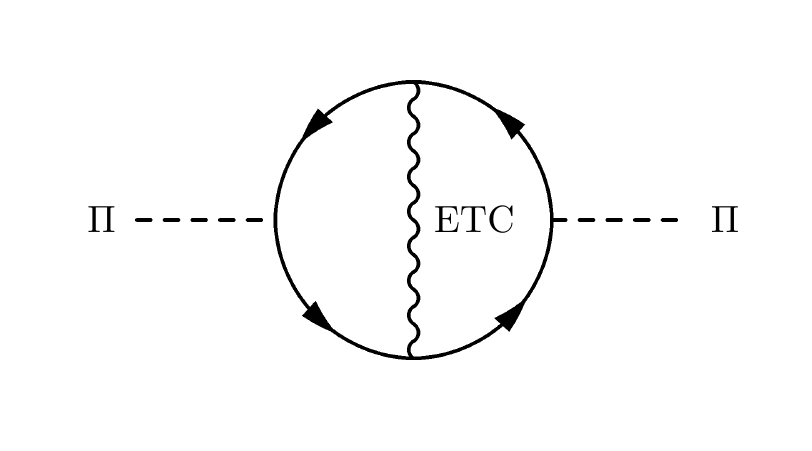}
\caption{Leading contribution to the mass of the TC pNGBs via an exchange of an ETC gauge boson.} \label{masspgb}
\end{center}
\end{figure}

The $\beta$-terms of Eq.~(\ref{etc}) provide masses for
pNGBs and for techni-axions
\cite{Hill:2002ap}, see Fig.~\ref{masspgb}. 
The last class of terms, namely the $\gamma$-terms of
Eq.~\eqref{etc}, are expected to induce flavor-chaging neutral currents (FCNCs). For example, it may generate the following terms:
\beq \frac{\gamma}{\Lambda_\text{ETC}^2}(\bar s\gamma^5d)(\bar s\gamma^5d) +
\frac{\gamma'}{\Lambda_\text{ETC}^2}(\bar \mu\gamma^5e)(\bar e\gamma^5e) + 
\ldots \ , \label{FCNC} \eeq
where $s,d,\mu,e$ denote the strange and down quarks, the muon
and the electron, respectively. The first term is a $\Delta S=2$
flavor-changing neutral current interaction affecting the
$K_L-K_S$ mass difference which is measured accurately. The experimental bounds on these type of operators, together with the  {\it naive} assumption that ETC will generate $\gamma$-terms with coefficients of order one, leads to a constraint on the ETC scale to be of the order of, or larger than, $10^3$
TeV \cite{Eichten:1979ah}. This should be the lightest ETC scale which in turn puts an upper limit on how large the ordinary fermionic masses can be. The naive estimate is that  one can account up to around 100 MeV mass for a QCD-like technicolor theory, implying that many quark mass values cannot be achieved.

The second term of Eq.~(\ref{FCNC}) induces flavor
changing processes in the leptonic sector such as $\mu\rightarrow e\bar ee,
e\gamma$ which are not observed, thus leading to even stronger bounds on the ETC scale.
It is clear that, both for the precision measurements  and the fermion masses, a better theory of flavor is needed. For the ETC dynamics interesting  developments recently appeared in the literature  \cite{Ryttov:2010kc,Ryttov:2010jt,Ryttov:2010fu,Chen:2010er}. We note that nonperturbative chiral gauge theories dynamics is expected to play a relevant role in models of ETC since it allows, at least in principle, the self breaking of the gauge symmetry. Recent progress on the phase diagrams of these theories has appeared in \cite{Sannino:2009za}.  

%

Note that the analysis presented so far is based on the assumption that the dynamics in the confined phase is similar to that of QCD.
However, important modifications may occur, thus changing the main conclusions about flavor physics. In the following sections, we will present alternative types of dynamics used in the context of SM fermion mass generation.

\subsection{Walking}

To better understand in which direction one should go to modify the QCD dynamics, we analyze the FC condensate. 
The value of the condensate used when giving mass to the ordinary fermions should be evaluated not at the FC scale but at the ETC one. Via the renormalization group evolution one can relate the condensate at the two scales via:
\beq \langle\bar {\cal F}{\cal F}\rangle_\text{ETC} =
\exp\left(\int_{\Lambda_\FC}^{\Lambda_\text{ETC}}
\text{d}(\ln\mu)\gamma(\alpha(\mu))\right)\langle\bar {\cal F}{\cal F}\rangle_\text{TC} \ ,
\label{rad-cor-tc-cond}
\eeq 
where $\gamma$ is the anomalous dimension of the FC-fermion mass operator. The limits of integration
 correspond at high energy with the ETC scale and at low energy with the FC one.
For technicolor theories with a running of the coupling constant similar to the one in QCD, i.e.
\beq \alpha(\mu) \propto \frac{1}{\ln\mu} \ , \quad {\rm for}\ \mu >
\Lambda_\text{TC} \ , \eeq
this implies that the anomalous dimension of the techniquark masses $\gamma \propto
\alpha(\mu)$. When computing the integral one gets
\beq \langle\bar {\cal F}{\cal F}\rangle_\text{ETC} \sim
\ln\left(\frac{\Lambda_\text{ETC}}{\Lambda_\FC}\right)^{\gamma}
\langle\bar {\cal F}{\cal F}\rangle_\FC \ , \label{QCD-like-enh} \eeq
which is a logarithmic enhancement of the operator. We can hence neglect this correction and use directly the value of the condensate at the FC scale when estimating the generated fermionic mass. Thus, Eq.~\eqref{eq:mqETC} evaluates to
\beq m_q \approx \frac{g_\text{ETC}^2}{M_\text{ETC}^2}\Lambda_\FC^3 \ , \qquad 
 \langle \bar {\cal F}{\cal F}\rangle_\FC \sim \Lambda_\FC^3 \ . \eeq

\begin{figure}
\centering
\begin{tabular}{cc}
\resizebox{6.0cm}{!}{\includegraphics{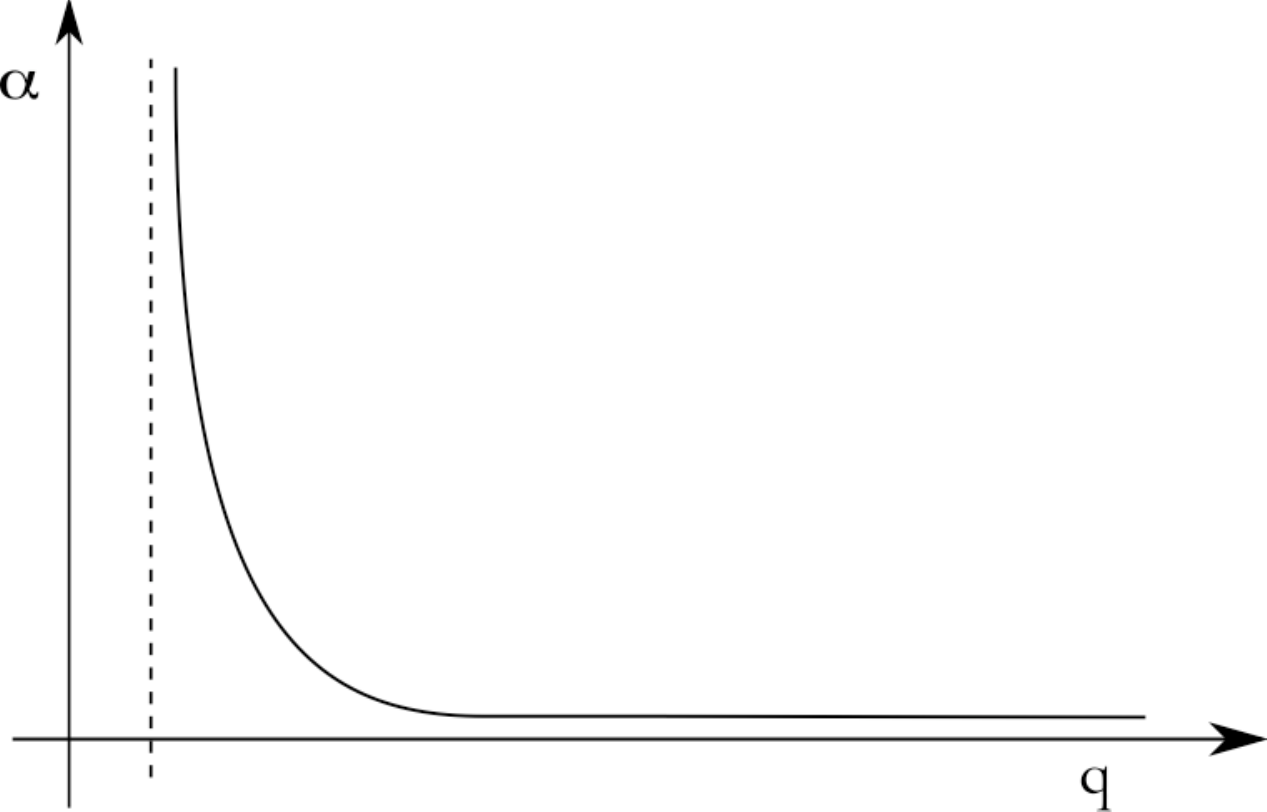}} ~~~&~~ \resizebox{6.0cm}{!}{\includegraphics{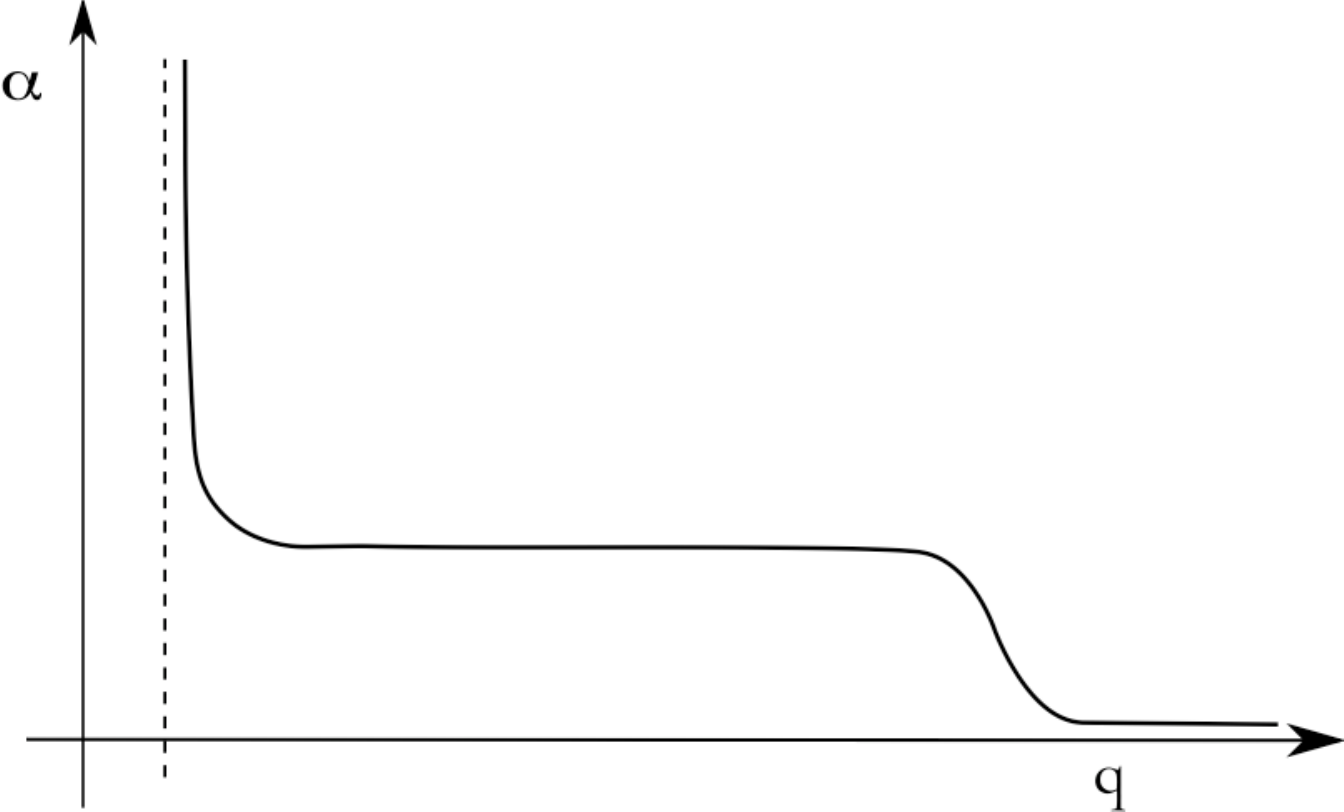}} \\&\\
&~~~~\resizebox{6.0cm}{!}{\includegraphics{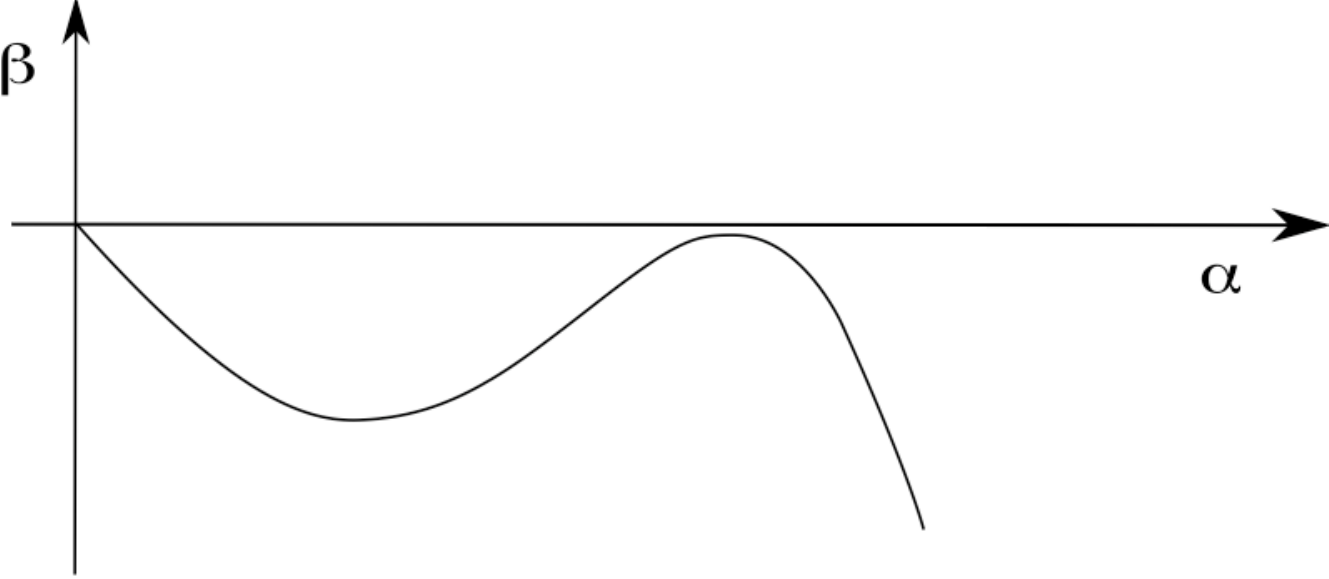}} 
\end{tabular}
\caption{Top left panel: QCD-like behavior of the coupling constant as function of the momentum (Running). Top right panel: walking-like behavior of the coupling constant as function of the momentum (Walking). Bottom right panel: cartoon of the beta function associated to a generic walking theory.}
\label{walkbeta}
\end{figure}

The tension between having to reduce the FCNCs and at the same time provide a sufficiently large mass for the heavy fermions in the SM (as well as for the pNGBs) can be reduced if the dynamics of the underlying FC theory is different from the one of QCD. 
By computing the value of the FC condensate, it has been shown that a theory where the gauge coupling varies slowly instead of \emph{running} between $\Lambda_{\text{ETC}}$ and $\Lambda_\FC$ can feature a larger condensate at the ETC scale with respect to the one at the FC scale. Thus, the estimate of the SM fermion masses given before can be significantly enhanced.
 This can be achieved if the theory has a near-conformal Infra-Red (IR) fixed point. This kind of dynamics has been denoted as of {\it walking} type. In Fig.~\ref{walkbeta}, the comparison between a running and walking behavior of the coupling is qualitatively represented. 

 In the walking regime, Eq.~\eqref{QCD-like-enh} is replaced by
\beq \langle\bar {\cal F}{\cal F}\rangle_\text{ETC} \sim
\left(\frac{\Lambda_\text{ETC}}{\Lambda_\FC}\right)^{\gamma(\alpha^*)}
\langle\bar {\cal F}{\cal F}\rangle_\FC \ , \label{walking-enh} \eeq
which shows a power-law dependence on the scales instead of the lodarithmic one of QCD dynamics \cite{Yamawaki:1985zg,Holdom:1984sk,Holdom:1981rm,Appelquist:1986an}. Here, $\gamma$ is evaluated at the would-be fixed point value of the coupling constant $\alpha^*$.  Walking can help resolving the problem of FCNCs in
composite models since, with a large enhancement of the $\langle\bar
{\cal F}{\cal F}\rangle$ condensate at the ETC scale, the four-Fermi operators of types $\alpha$ and $\beta$ are enhanced by a factor $\Lambda_\text{ETC}/\Lambda_\FC$ to the power $\gamma$, while the ones involving only SM fermions are not enhanced.

Although {\it walking} is not a must for a successful composite model, it is a fact that a near conformal theory would still be useful to reduce the contributions to the precision data and, possibly, provide a light dilaton-like composite Higgs \cite{Dietrich:2005jn}. The reason being that, for a continuous quantum phase transition, a dilaton-like mode is expected to appear in the walking regime  to enforce approximate conformal invariance~\cite{Leung:1985sn,Bardeen:1985sm,Yamawaki:1985zg,Sannino:1999qe,Hong:2004td,Dietrich:2005jn,Appelquist:2010gy}.  
This subject has recently received renewed interest~\cite{Hong:2004td,Dietrich:2005jn,Goldberger:2008zz,Appelquist:2010gy,Hashimoto:2010nw,Matsuzaki:2013eva,Golterman:2016lsd,Hansen:2016fri,Golterman:2018mfm} due to recent lattice studies ~\cite{Appelquist:2016viq,Appelquist:2018yqe,Aoki:2014oha,Aoki:2016wnc,Fodor:2012ty,Fodor:2017nlp,Fodor:2019vmw} that reported evidence of the presence of a light singlet scalar particle in the spectrum.

\subsection{Ideal walking\label{sect:idealwalking}}

There are several issues associated with the original idea of walking: 
\begin{itemize}
\item{Since the number of flavors cannot be changed continuously, it is not possible to  get arbitrarily close to the lower end of the conformal window. This applies to the technicolor theory {\it in isolation}, i.e. before coupling it to the SM and without taking into account the ETC interactions. }
\item{It is hard to achieve large anomalous dimensions of the fermion mass operator even near the lower end of the conformal window for ordinary gauge theories.}
\item{It is not always possible to neglect the interplay of the four fermion interactions on the dynamics. }
\item{There exist the logical possibility that, as function of the number of flavors, the theory jumps \cite{Sannino:2012wy} out of the conformal window rather than walking out of it. }
\end{itemize}
In \cite{Fukano:2010yv} it has been argued that it is possible to address simultaneously the problems above, if the conformal to non-conformal transition is smooth, by taking into account the effects of the four-fermion interactions on the phase diagram of strongly interacting theories for any matter representation as function of the number of colors and flavors.  A positive effect is that the anomalous dimension of the mass is expected, near the transition, to be of order unity at the lower boundary of the new conformal window.  This analysis derives from the study of the gauged Nambu-Jona-Lasinio phase diagram \cite{Kondo:1993jq}. 

This scenario, named ``Ideal Walking'', predicates that the  theory, before taking into account the four-fermion interactions, must feature an infrared fixed point. By increasing the four-fermion coupling one expects to arrive at a critical value above which conformality is lost. This is expected to happen when the four-fermion coupling, from being an IR irrelevant operator becomes an IR marginal to relevant one. This is naturally expected to occur when the Fc-fermion mass anomalous dimension is around unity. 

The Ideal Walking \cite{Fukano:2010yv} paradigm has been proven to exist via first-principle lattice simulations for the first time in \cite{Rantaharju:2019nmh,Rantaharju:2017eej}. The gauge theory investigated on the lattice is an  SU(2) gauge theory with two adjoint Dirac fermions augmented with four-fermion interactions. The theory without four-fermion interactions is known to achieve a non-perturbative IR fixed point with a mass anomalous dimension around $\gamma \sim 0.5$. Once the four-fermion interactions are added the anomalous dimension grows with the four-fermion coupling till a value slightly larger than unity. If the four-fermion coupling is further increased chiral symmetry breaks and conformality is lost. Remarkably the phase transition from the chirally broken to the conformally restored side is approached smoothly in the four-fermion coupling with a measurable critical exponent for the fermion condensate proportional $(y_c - y)^\delta$ with $\delta \sim 0.69$ and $y^2/\Lambda_{ETC}$ the four-fermion coupling \cite{Rantaharju:2019nmh,Rantaharju:2017eej}.

 \subsection{Walking Spectrum}
 \label{sec:electroweak}

Any strongly coupled dynamics, including the walking type, generates a spectrum of resonances whose natural splitting in mass is of the order of the intrinsic scale of the theory. In particular, spin-1 resonances are important as ther re known to play a primary role in generating corrections to the EWPTs. In order to extract predictions  for the composite vector spectrum and couplings in presence of a strongly interacting sector and an asymptotically free gauge theory, we make use of the time-honored Weinberg sum rules \cite{Weinberg:1967kj}, amended by the results found in \cite{Appelquist:1998xf} that allow us to treat walking and running theories in a unified way.

\subsubsection{Weinberg sum rules}
The Weinberg sum rules  are linked to the  two point vector-vector minus axial-axial vacuum polarization which is known to be sensitive to chiral symmetry breaking.  We define
\begin{equation}
i\Pi_{\mu \nu}^{a,b}(q)\equiv \int\!d^4x\, e^{-i qx}
\left[<J_{\mu,V}^a(x)J_{\nu,V}^b(0)> -
 <J_{\mu,A}^a(x)J_{\nu,A}^b(0)>\right] \ ,
\label{VA}
\end{equation}
within the underlying strongly coupled gauge theory, where
\begin{equation}
\Pi_{\mu \nu}^{a,b}(q)=\left(q_{\mu}q_{\nu} - g_{\mu\nu}q^2 \right) \,
\delta^{a b} \Pi(q^2) \ .
\end{equation}
Here $a,b=1,...,n_f^2-1$, label the flavor
currents and the SU(N$_f$) generators are normalized according to
$\rm{Tr} \left[T^a T^b\right]= (1/2) \delta^{ab} $.  The
function $\Pi(q^2)$ obeys the unsubtracted dispersion relation
\begin{equation}
\frac{1}{\pi} \int_0^{\infty}\!ds\, \frac{{\rm Im}\Pi(s)}{s + Q^2}
=\Pi(Q^2) \ ,
\label{integral}
\end{equation}
where $Q^2=-q^2 >0$, and the constraint
$\displaystyle{-Q^2 \Pi(Q^2)>0}$ holds for $0 < Q^2 < \infty$~\cite{Witten:1983ut}. The discussion above applies to the standard chiral symmetry breaking pattern $\SU(N_f)\times \SU(N_f) \rightarrow \SU(N_f)$, but it is generalizable to any breaking pattern, including the template model of Chapter~\ref{ch:su2}.

Since we are taking the underlying theory to be asymptotically
free, the behavior of $\Pi(Q^2)$ at asymptotically high momenta is
the same as in ordinary QCD, i.e. it scales like $Q^{-6}$~\cite{Bernard:1975cd}. Expanding the left-hand side of the dispersion relation
thus leads to the two conventional spectral function sum rules
\begin{equation}
\frac{1}{\pi} \int_0^{\infty}\!ds\,{\rm Im}\Pi(s) =0
\label{spectral1}
\quad {\rm and} \quad
\frac{1}{\pi} \int_0^{\infty}\!ds\,s \,{\rm Im}\Pi(s) =0 \ .
\end{equation}
Walking dynamics affects only the second sum rule \cite{Appelquist:1998xf} which is more sensitive to large but not asymptotically large momenta due to fact that the associated integrand contains an extra power of $s$.

We now saturate the absorptive part of the vacuum
polarization. We follow reference \cite{Appelquist:1998xf} and hence divide the
energy range of integration in three parts. In the light resonance part, the    
integral is saturated by the
pNGBs along with massive vector and
axial-vector states. If we assume, for example, that there is only a
single, zero-width, vector multiplet and a single, zero-width, axial
vector multiplet, then
\begin{equation}
{\rm Im}\Pi(s)=\pi F^2_V \delta \left(s -M^2_V \right) - \pi F^2_A
\delta \left(s - M^2_A \right) - \pi F^2_{\pi} \delta \left(s \right)
\ .
\label{saturation}
\end{equation}
The zero-width approximation is valid to leading order in the large-$N$ expansion for fermions in the fundamental representation of the gauge group (even narrower states are found for fermions in higher dimensional representations). Since we are working near a conformal fixed point, the large $N$ argument for the width is not directly applicable. We will nevertheless use this simple model for the spectrum
to  illustrate the effects of a near critical IR fixed point. 

The first Weinberg sum rule implies:
\begin{equation}
F^2_V - F^2_A = F^2_{\pi}\ ,
\label{1rule}
\end{equation}
where $F^2_V$ and $F^2_A$ are the vector and axial meson decay
constants.  This sum rule holds for walking and running dynamics. {A
more general representation of the resonance spectrum would, in principle, replace
the left hand side of this relation with a sum over vector and axial
states. However the heavier resonances should not be included since in the approach of \cite{Appelquist:1998xf} the walking dynamics in the intermediate energy range is already approximated by the exchange of underlying fermions. The walking is encapsulated in the dynamical mass dependence on the momentum dictated by the gauge theory. The introduction of heavier resonances is, in practice, double counting. Note that the approach is in excellent agreement with the Weinberg approximation for QCD, since in this case, the  approximation automatically returns the known results.  }

The second sum rule receives important contributions from throughout
the near conformal region and can be
expressed in the form of:
\begin{equation}
F^2_V M^2_V - F^2_A M^2_A = a\,\frac{8\pi^2}{d(R)}\,F_{\pi}^4,
\label{2rule-2}
\end{equation}
where $a$ is a coefficient expected to be positive and $\mathcal{O}(1)$, and $d(R)$ is the dimension of the representation of the underlying fermions.  
We have generalized the result of reference \cite{Appelquist:1998xf}  to the case in which the fermions belong to a generic representation of the gauge group. In the case of running dynamics the right-hand side of the previous equation vanishes.  

We stress that $a$ is a non-universal quantity depending on the details of the underlying gauge theory. A reasonable measure of how large $a$ can be is given by a function of the amount of walking which is the ratio of the scale above which the underlying coupling constant starts running divided by the scale below which chiral symmetry breaks.
The fact that $a$ is positive and of order one in walking dynamics is also supported, indirectly, via the results of Kurachi and Shrock \cite{Kurachi:2006ej}. At the onset of conformal dynamics, the axial and the vector resonances are degenerate, i.e. $M_A=M_V=M$, thus using the first sum rule one finds, via the second sum rule, 
\beq
a = \frac{d({\rm R})M^2}{8\pi^2 F^2_{\pi}}\ ,
\eeq
leading to a numerical value of about $4\div 5$ from the approximate results in \cite{Kurachi:2006ej}.  We will however use only the constraints coming from the generalized Weinberg sum rules  expecting them to be less model dependent.
The $S$ parameter is related to the absorptive part  of the
vector-vector minus axial-axial vacuum polarization as follows:
\begin{equation}
S=4\int_0^\infty \frac{ds}{s} {\rm Im}\bar{\Pi}(s)= 4\pi
\left[\frac{F^2_V}{M^2_V} - \frac{F^2_A}{M^2_A} \right] \ ,
\label{s-def}
\end{equation}
where ${\rm Im}\bar{\Pi}$ is obtained from ${\rm Im}\Pi$ by
subtracting the Goldstone boson contribution.

Other attempts to estimate the $S$ parameter for walking technicolor
theories have been made in the past \cite{Sundrum:1991rf} showing a reduction of the $S$ parameter. $S$ has also been evaluated using computations inspired by the original AdS/CFT correspondence \cite{Maldacena:1997re} in \cite{Hong:2006si,Hirn:2006nt,Piai:2006vz,Agashe:2007mc,Carone:2007md,Hirayama:2007hz}. Recent attempts to use AdS/CFT inspired methods can be found in \cite{Dietrich:2009af,Dietrich:2008up,Dietrich:2008ni,Nunez:2008wi,Fabbrichesi:2008ga}.   

Kurachi, Shrock and Yamawaki \cite{Kurachi:2007at} have further confirmed the results presented in \cite{Appelquist:1998xf} with their computations tailored at describing four dimensional gauge theories near the conformal window.
The present approach \cite{Appelquist:1998xf} is more physical since it is based on the
nature of the spectrum of states associated directly to the underlying gauge theory.  

Note that we will be assuming a rather conservative approach in which the $S$ parameter, although reduced with respect to the case of a running theory, is bounded by the naive $S$ parameter \cite{Sannino:2010ca,Sannino:2010fh,DiChiara:2010xb}. After all, other sectors of the theory such as new leptons can further reduce or completely offset a positive value of $S$ due solely to the technicolor theory.



\section{Partial compositeness paradigm with fermions} \label{sec:PCwfermions}

An alternative mechanism behind the generation of fermion masses has been proposed in~\cite{Kaplan:1991dc} and goes under the name ``partial compositeness'': instead of coupling a bilinear of SM fermions to the strong sector, as discussed in section~\ref{sec:ETC}, each SM fermion features a linear coupling to a spin-1/2 operator in the strong sector. The main motivation behind this approach was to avoid the generation of dangerous FCNCs, which appear in the traditional approaches.
We will first discuss how this mechanism can be implemented in models with a gauge-fermion microscopic description. 
A more general discussion of the partial compositeness scenario are reported in the next sections.

The first requirement imposed by fermion partial compositeness on the strong sector is to be able to construct spin-1/2 bound states. This is already impossible in the template theory introduced in Chapter~\ref{ch:su2}. In general, there are two possibilities:

\begin{itemize}

\item[1)] Dim-5 chromomagnetic-type coupling:
\begin{equation}
\frac{\alpha_{\rm med}^2}{4 \pi} \frac{\kappa}{\Lambda}\  \psi \sigma^{\mu \nu}  \lambda^a \tcf\ F^a_{\mu \nu}\,,
\end{equation}
where $F^a$ are the FC-gluon fields. The spin-1/2 bound state is thus formed by the fermions with FC-gluons. Typically, these kind of couplings are loop-induced, thus explaining the one loop factor appearing in front of the operator.

\item[2)] Dim-6 four-fermion interactions:
\begin{equation}
\frac{\kappa}{\Lambda^2}\ \psi \tcf\ \tcf \tcf\,.
\end{equation}
In this case, the bound state is made of three FC-charged fermions.

\end{itemize}

Higher dimensional couplings could also be possible, however they carry a stronger suppression for the couplings in the confined phase by higher powers of $\Lambda$. The two we listed, while of different dimension, feature similar suppression due to the loop nature of the former. The case 1) has also a key drawback: the fermion $\tcf$ involved needs to be in the adjoint irrep of the FC. Due to the required multiplicity  (at least a QCD-color triplet to couple to quarks), this scenario would prevent the theory from confining by loss of asymptotic freedom.

The latter case is, therefore, the most promising one. Remarkably, bound states of three fermions are only possible for three models: $\SU(3)_{\rm FC}$ with $\tcf$ in the fundamental ({\it \`a la} QCD), $\SU(6)_{\rm FC}$ with two-index anti-symmetric, and $G_2$ with fundamentals. An explicit model based on the $\SU(3)_{\rm FC}$ case has been proposed in~\cite{Vecchi:2015fma}, containing overall 7 flavors of light $\tcf$'s. This theory is likely to be outside of the conformal window, and to confine at low energies. A walking phase could be introduced by adding additional heavy flavors to the theory~\cite{Hasenfratz:2016gut}, however the anomalous dimensions of the FC-baryon operators may be too small to provide a realistic model for top partial compositeness~\cite{Pica:2016rmv}.

Another intriguing possibility is that the FC-baryon may be made of different species of fermions~\cite{Barnard:2013zea,Ferretti:2013kya}, i.e. fermions transforming under different irreps of the FC group. The main advantage of this approach is that QCD charges can be sequestered to the second specie of fermions, whilst it is only the first one that generates the composite Higgs and participate to the EWSB. By requiring that the theory remains asymptotically free (i.e., it confines at low energies), only a limited number of possibilities remain, as listed in ~\cite{Ferretti:2013kya}. Furthermore, asking that the theory is outside of the conformal window, i.e. it condenses and generates a mass gap, limits the choices of models to only 12~\cite{Ferretti:2014qta}. As a by-product, this class of models is only able to generate partial compositeness for the top quark (and eventually the bottom). Following the nomenclature in ~\cite{Belyaev:2016ftv}, we will name the models M1--M12. For concreteness, we will now focus on a specific model, i.e. M8, which was originally proposed in ~\cite{Barnard:2013zea}. As we will see, this model can be seen as an extension of the template model of Sec.~\ref{sec:compoHiggs}, having the same symmetry breaking pattern in the EW sector.

\begin{table}[tb] \begin{center}
\begin{tabular}{c|c|ccc|ccc|}
   &  $\SP(4)_{\rm FC}$  & $\SU(3)_c$ &  $\SU(2)_L$ & $\U(1)_Y$ & $\SU(4)_\tcf$ & $\SU(6)_\chi$ & $\U(1)$\\
 & & & & & & & \\
$\tcf^1$  & \multirow{4}{*}{${\tiny{\yng(1)}}$} & \multirow{2}{*}{${\bf 1}$} & \multirow{2}{*}{${\bf 2}$} & \multirow{2}{*}{$0$} & \multirow{4}{*}{${\tiny{\yng(1)}}$} & \multirow{4}{*}{$\bf 1$}  & \multirow{4}{*}{$\bf 1$}  \\
$\tcf^2$   &   &   &  &  & &  & \\
$\tcf^3$ &  &  \multirow{2}{*}{${\bf 1}$} &  \multirow{2}{*}{${\bf 1}$} & $-1/2$ & & &  \\
$\tcf^4$ & & & & $1/2$ & & & \\
 & & & & & & & \\
$\chi^a$ & \multirow{2}{*}{${\tiny{\yng(1,1)}}$} & ${\bf 3}$ & \multirow{2}{*}{${\bf 1}$} & $2/3$ & \multirow{2}{*}{${\bf 1}$} & \multirow{2}{*}{${\tiny{\yng(1)}}$} & \multirow{2}{*}{${\bf -1/3}$} \\
$\chi^{\bar{a}}$ & & ${\bf \overline{3}}$ & & $-2/3$ & & & \\
\end{tabular}
\caption{Quantum numbers under the FC group, the SM gauge groups and the global symmetries  of the fundamental constituent fermions of Model M8~\cite{Barnard:2013zea}. The indices $a$ and $\bar{a}$ indicate the QCD color charge. All the spinors in the table are left-handed.} \label{tab:M8} 
\end{center} \end{table}

Compared to the minimal template, the FC gauge group is extended to $\SP(4)_{\rm FC}$~\footnote{We recall that the minimal symplectic group is $\SP(2) \equiv \SU(2)$.},
where the fermions $\tcf$ are in the fundamental and have the same EW charges as in Table~\ref{tab:tab1}. The additional fermions, that we call $\chi$, transform in the two-index irrep of $\SP(4)_{\rm FC}$, ${\tiny{\yng(1,1)}} = {\bf 5}$, and carry QCD charges as shown in Table~\ref{tab:M8}: they thus form a Dirac fermion charged as a quark under QCD and carrying hypercharge $2/3$ (like the right-handed top). As $\tcf$ and $\chi$ are in pseudo-real and real irreps respectively of $\SP(4)_{\rm FC}$, the maximal global symmetry of the model is
\begin{equation}
\mathcal{G}_{\rm global} = \SU(4)_\tcf \times \SU(6)_\chi \times \U(1)\,,
\end{equation}
where the $\U(1)$ is a combination of a phase redefinition of the two species of fermions that has no anomaly with the FC gauge interactions. Note that there is also an anomalous $\U(1)$, analogous to the axial charge in QCD. 
Upon condensation of the two fermion species, $\langle \tcf \tcf \rangle \neq 0$ and $\langle \chi \chi \rangle \neq 0$, the global symmetries are broken as follows:
\begin{eqnarray}
\SU(4) \to \SP(4) & \Rightarrow & N(\pi_\tcf) = 5\,, \nonumber \\
\SU(6) \to \SO(6) & \Rightarrow & N(\pi_\chi) = 20\,, \\
\U(1) \to \varnothing & \Rightarrow & N(\pi_1) = 1\,; \nonumber
\end{eqnarray}
where the $\U(1)$ is broken by both condensates (a second singlet pseudo-scalar is present but acquires a potentially large mass due to the FC anomaly, in analogy to the $\eta'$ in QCD). A priori, we cannot know if both condensates are generated: if one of them were to vanish, the corresponding global symmetry would remain preserved and no pNGBs would be present in the spectrum. Furthermore, the global symmetry would have global anomalies that need to matched in the confined phase by massless composite states~\cite{tHooft:1980xss}. In this model, in fact, it is possible to keep the $\SU(6)_\chi$ unbroken and match the global anomalies by the presence of a massless baryon that may play the role of the operator coupling to the top quark, as shown in~\cite{Cacciapaglia:2015vrx} (see also~\cite{Gertov:2019yqo}). 
Preliminary lattice studies of the dynamics of this model seem to suggest that both condensates are non-zero~\cite{Bennett:2019cxd} (for more details on the Lattice results, see Sec.~\ref{sec:PCSp4}, while a study based on a Nambu-Jona-Lasinio description is in~\cite{Bizot:2016zyu}). In the following, we will therefore assume that both condensates are formed.

In addition to the 5 pNGBs in the EW sector, which have the same properties as in the template model in Sec.~\ref{sec:compoHiggs}, the low energy spectrum contains $20$ pNGBs from the $\chi$ condensation and one associated to the global $\U(1)$. Before discussing top partial compositeness, we will address the phenomenology of these states, which are expected to be light.

\subsection{Phenomenology of the additional pNGBs}

The $20+1$ additional pNGBs generated by the $\langle \chi \chi \rangle$ condensate have the SM quantum numbers shown in Table~\ref{tab:pNGBsM8}. The QCD charged ones, emerging from the $\chi$ condensate, in particular, couple dominantly to the top quark via partial compositeness~\cite{Cacciapaglia:2015eqa}, where the sextet only decay is $\pi_6 \to t_R t_R$. The main bound on this state (and the octet) derive from QCD-driven pair production followed by decays into tops, which leads to a 4-top final state and bounds of the order of TeV for the masses~\cite{Cacciapaglia:2015eqa,Xie:2019gya}.
Note that the masses for the colored pNGBs come dominantly from QCD corrections, and are of the order of the decay constant in the $\chi$ sector.

\begin{table}[tb] \begin{center}
\begin{tabular}{c|c|ccc|cc|}
   &   & $\SU(3)_c$ &  $\SU(2)_L$ & $\U(1)_Y$ & $\Sp(4)_\tcf$ & $\SO(6)_\chi$\\
 & & & & & & \\
\multirow{3}{*}{$\pi_\chi$} & $\pi_8$ & ${\bf 8}$ & ${\bf 1}$ & $0$ & \multirow{3}{*}{$1$} & \multirow{3}{*}{${\tiny{\yng(2)}}$}  \\
  & $\pi_6$  & $\bf 6$ & $\bf 1$ & $\bf 4/3$  &  &  \\
  & $\pi_6^\ast$  & $\bf \bar{6}$ & $\bf 1$ & $\bf -4/3$  &  &  \\
 & & & & & & \\
$\pi_1$ & $a$ & $\bf 1$ & $\bf 1$ & $\bf 0$ & $\bf 1$ & $\bf 1$ \\
\end{tabular}
\caption{Quantum numbers of the additional pNGBs in the model M8.} \label{tab:pNGBsM8} 
\end{center} \end{table}

While the sextet is a rather exotic state, unique to this template model M8, one remarkable point is that the octet $\pi_8$ and the singlet $a$ are ubiquitous to all models~\cite{Belyaev:2016ftv}. Their decays, however, are different for different models and are sensitive to the details of the underlying dynamics, namely the FC group and the fermion irreps. This is particularly true for the couplings to gauge bosons, which are generated by the topological anomalies, and whose coefficients can be predicted in terms of the properties of the underlying model.

For instance, the color octet has couplings to two gluons and to one gluon and one neutral EW gauge boson, in the form:
\begin{equation}
\mathcal{L}_{\rm WZW} \supset \frac{\alpha_s}{8 \pi f_\chi} C_{gg} \ \pi_8^a\ \epsilon^{\mu \nu \rho \lambda} G^b_{\mu \nu} G^c_{\rho \lambda} d^{abd} + \frac{\sqrt{\alpha_s \alpha}}{8 \pi \cos^2 \theta_W f_\chi} C_{gB}\ \pi_8^a \epsilon^{\mu \nu \rho \lambda} G^b_{\mu \nu} B_{\rho \lambda} \delta^{ab}\,,
\end{equation}
where $d^{abc} = \frac{1}{4} \mbox{Tr} [\lambda^a \{ \lambda^b, \lambda^c\}]$ and $B$ is the hypercharge gauge boson. The ratio of the two coefficients only depends on the hypercharge of $\chi$:
\begin{equation}
\frac{C_{gB}}{C_{gg}} = 4 Y_\chi\,.
\end{equation}
This implies that for M8, $Y_\chi=2/3$, the branching ratios appear in the ratio:
\begin{equation}
\mbox{BR} (\pi_8 \to gg) : \mbox{BR} (\pi_8 \to g \gamma) : \mbox{BR} (\pi_8 \to g Z) = 1 : 0.19 : 0.06\,.
\end{equation}
For other models with $Y_\chi = 1/3$, the branchings in $g \gamma$ and $g Z$ are reduced by a factor $1/4$.
In the case where the coupling to tops is suppressed, the decay mode into jets or jet-photon become the best channel to search for the ubiquitous octet~\cite{CidVidal:2018eel}.

\begin{figure*}[tbh]
\centering
\includegraphics[width=0.49\textwidth]{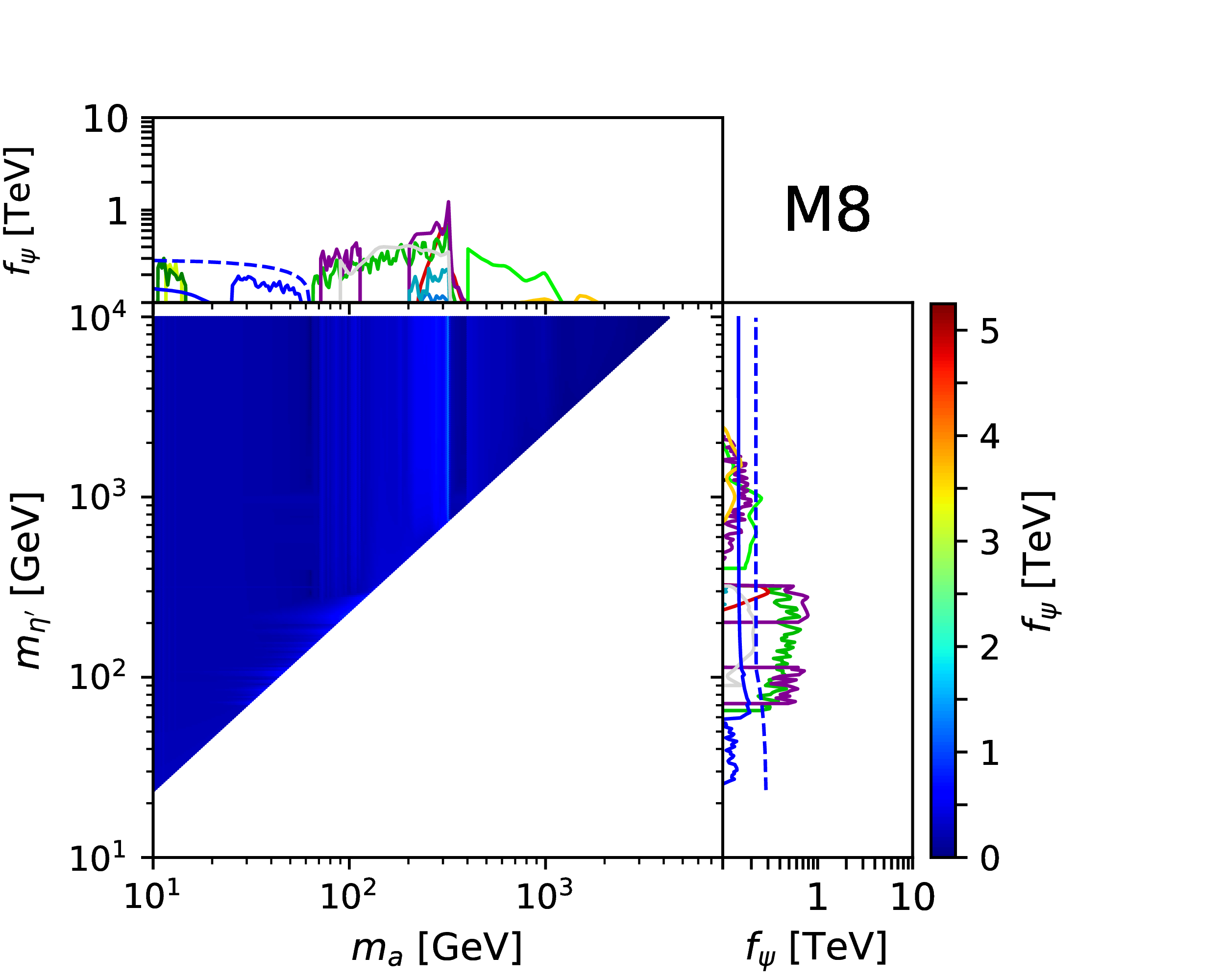}
\includegraphics[width=0.49\textwidth]{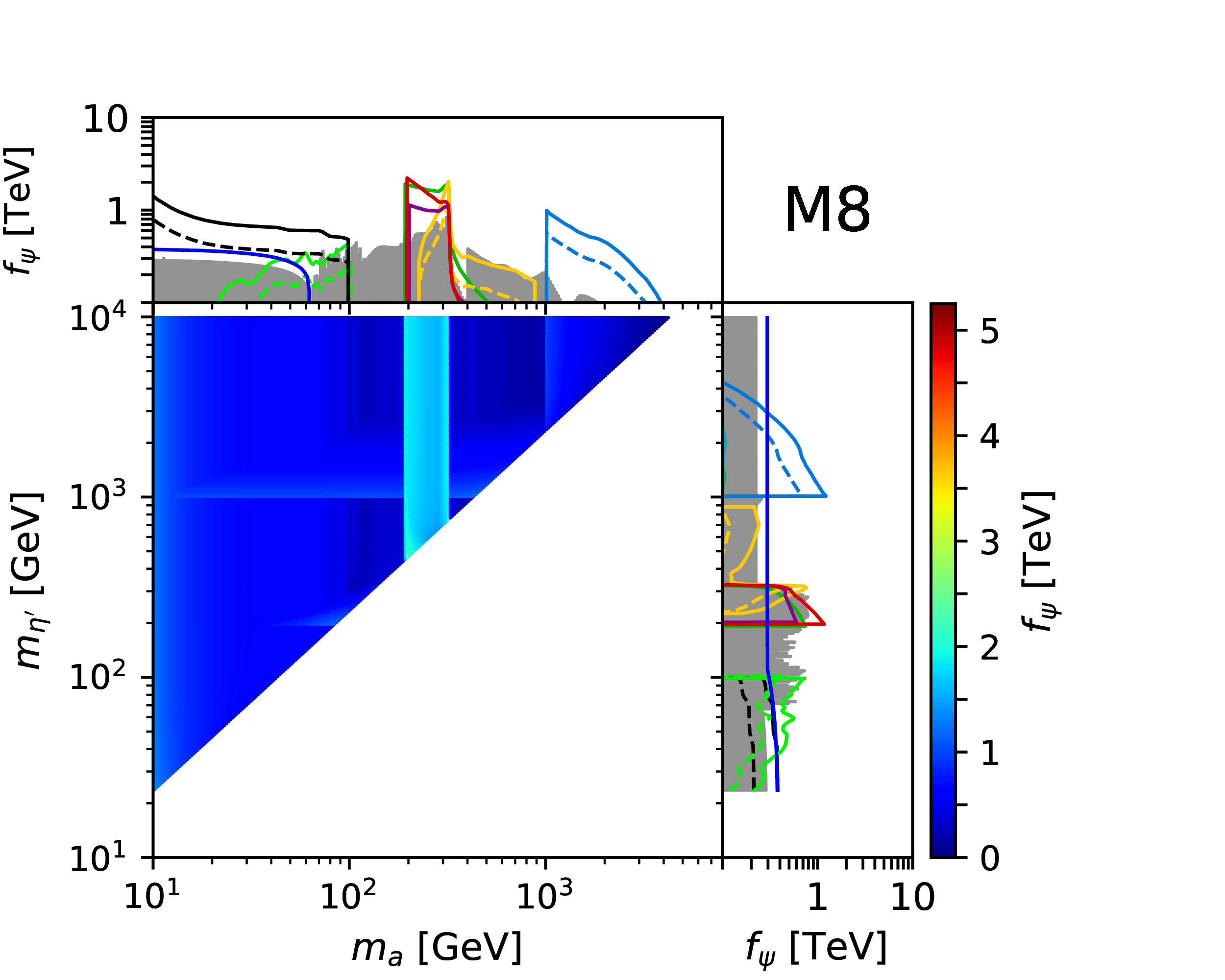}
\includegraphics[width=0.49\textwidth]{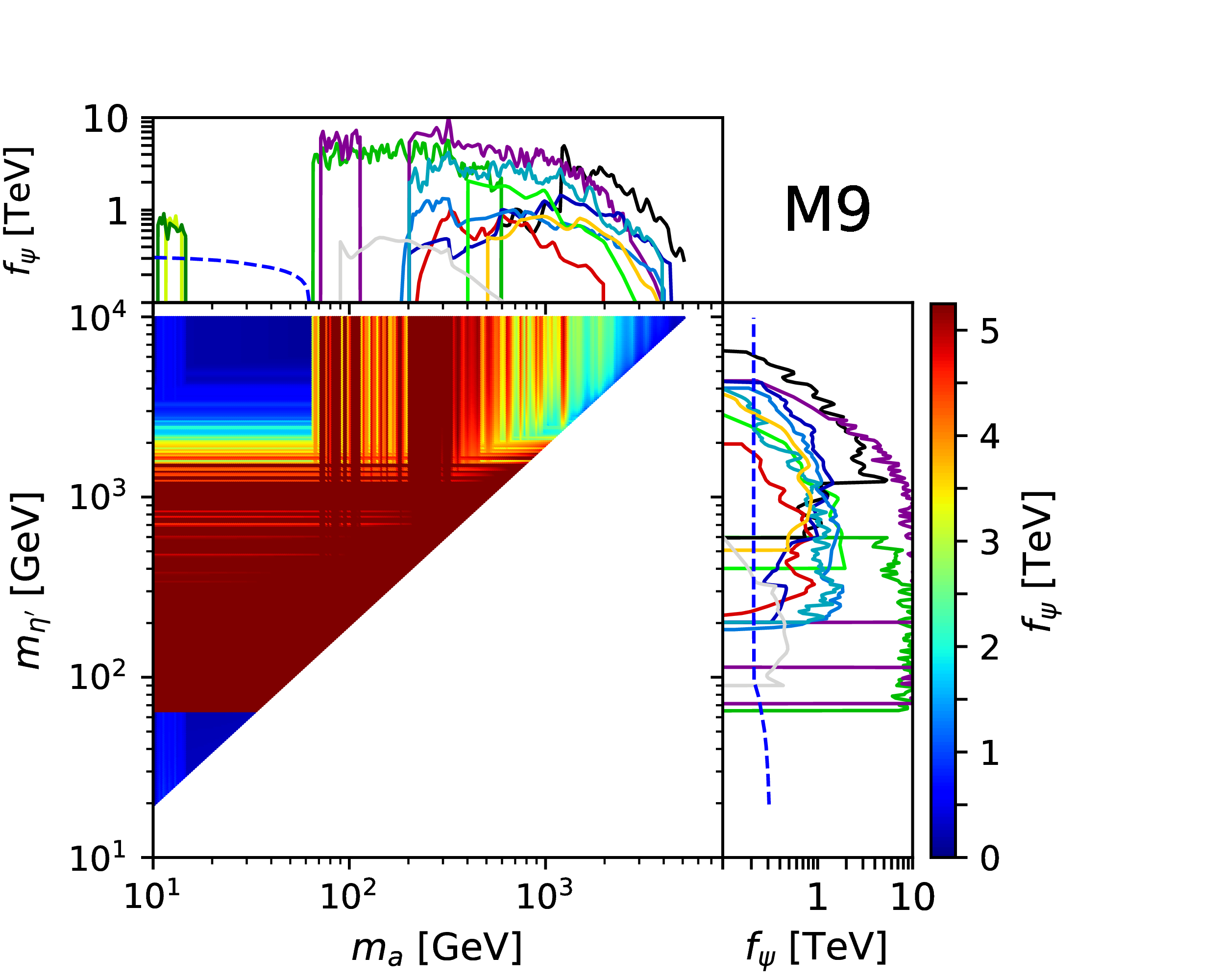}
\includegraphics[width=0.49\textwidth]{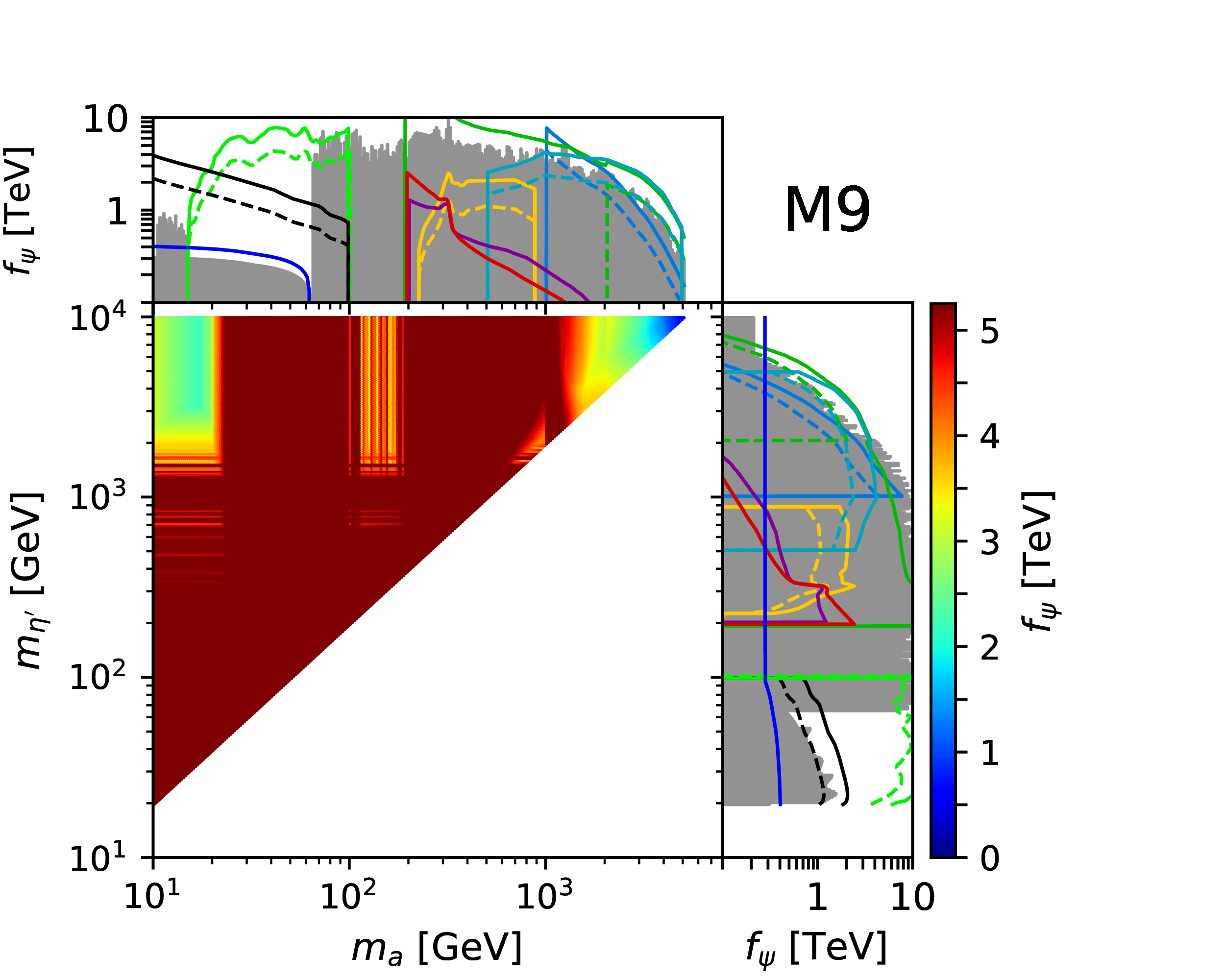}
\parbox[t]{0.49\textwidth}{ $ $ \\[-8pt] \includegraphics[width=0.49\textwidth]{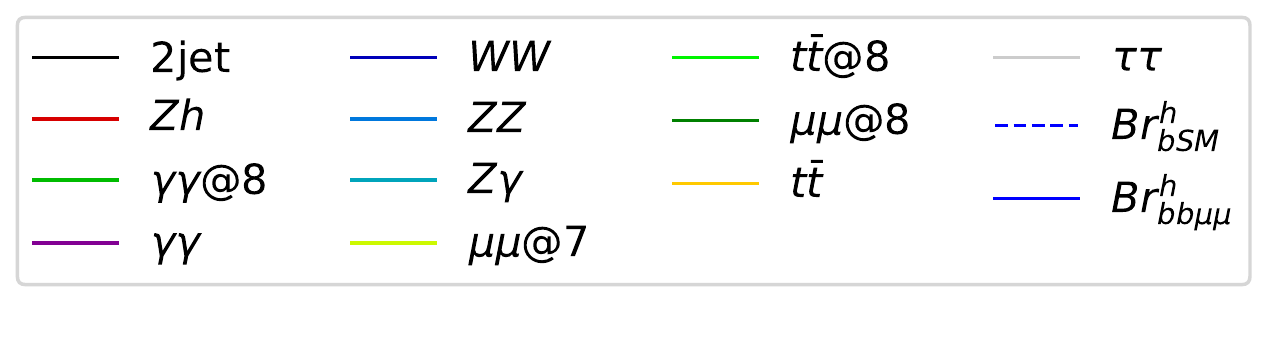}}
\parbox[t]{0.49\textwidth}{ $ $ \\[-8pt] \includegraphics[width=0.49\textwidth]{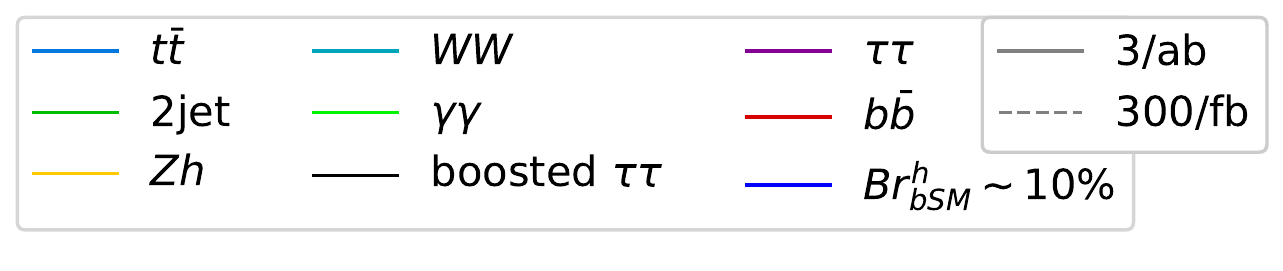}}
\caption{ Heat-plots showing the lower bounds on the Higgs decay constant $f_\psi$ in the mass plane of the two pseudo-scalars. The white triangle is not accessible by the masses in each model. The side-bands show the limits from each individual final state. On the left column, we show the current Run-I and Run-II bounds; on the right column, we show the projections at the High-Luminosity LHC run (the solid grey band summarizes the current bounds for comparison). More details in the text. Here we show model M8 (top row) and model M9 (bottom row). Figure from ~\cite{Cacciapaglia:2019bqz}.}
\label{fig:boundsM8M9}
\end{figure*}

Another ubiquitous state is the singlet $a$ associated to the non-anomalous $\U(1)$. This state can be very light as it acquires mass only via a current mass for the fermions $\chi$ and $\tcf$~\cite{Belyaev:2016ftv}. In addition, a heavier pseudo-scalar $\eta'$, associated to the anomalous $\U(1)$ charge, could also be light enough to be produced at the LHC. The couplings and mass mixing of the two states have been studied in detail in ~\cite{Belyaev:2016ftv}, to which we refer the interested reader for further details. 
What is remarkable about these two states is that the couplings to gauge bosons (generated via WZW terms) and to the top quarks (generated by partial compositeness) can be calculated in terms of the properties of the underlying dynamics. Thus, even models that have the same symmetries, like M8 and M9, and the same low energy effective description~\cite{DeGrand:2016pgq}, will have remarkably different characteristics. As an example, in Fig.~\ref{fig:boundsM8M9} we show the current bounds on the Higgs decay constant coming from the non-observation of any of the two pseudo-scalars~\cite{Cacciapaglia:2019bqz}. The plot shows comparatively M8 and the twin model M9, which share the same low energy effective theory. It is interesting to notice that the bound on the compositeness scale for M9, coming from the pseudo-scalar production, are already stronger than the indirect bounds coming from precision tests in the EW sector, see Sec.~\ref{sec:EWPT}. On the other hand, M8 has much weaker reach due to a smaller coupling to gluons that reduces the leading gluon-fusion production channel.
Another interesting feature, shared by all models, is the gap in sensitivity in the mass window $14~\mbox{GeV} < m_a < 65$~GeV: this is due to the fact that most low invariant mass resonances would not pass the trigger requirements at the LHC (this is particularly true for the low mass di-photon search). This window may possibly be closed by searches in di-tau~\cite{Cacciapaglia:2017iws}, di-photon~\cite{Mariotti:2017vtv} or di-muon.

\subsection{Top partial compositeness}

In the template model of Table~\ref{tab:M8}, the 3-fermion bound states are made of two $\tcf$'s and one $\chi$. These chimera baryons have 3 different spin-structures:
\begin{eqnarray}
\mathcal{O}_A &=& \langle \chi \tcf \tcf \rangle = [{\bf 6}, {\bf 6}]_{5/3}\,, \nonumber \\
\mathcal{O}_{\bar{A}} &=& \langle \chi \overline{\tcf} \overline{\tcf} \rangle = [{\bf \bar{6}}, {\bf 6}]_{-7/3}\,, \label{eq:opsAAbarAdj} \\
\mathcal{O}_{Adj, s} &=& \langle \overline{\chi} \overline{\tcf} \tcf \rangle = [{\bf 15} \oplus {\bf 1}, {\bf \bar{6}}]_{-1/3}\,, \nonumber
\end{eqnarray}
where we indicate their quantum numbers under the global symmetries $[\SU(4)_\tcf, \SU(6)_\chi]_{\U(1)}$.
In the confined phase, each of the three operators will match to a set of baryons, which transform under the unbroken symmetries $\{ \Sp(4)_\tcf, \SO(6)_\chi \}$:
\begin{eqnarray}
\mathcal{O}_A &\to& \{ {\bf 5}, {\bf 6} \} \oplus \{ {\bf 1}, {\bf 6} \}\,, \nonumber \\
\mathcal{O}_{\bar{A}} &\to& \{ {\bf 5}, {\bf 6} \} \oplus \{ {\bf 1}, {\bf 6} \}\,, \nonumber \\
\mathcal{O}_{Adj} &\to& \{ {\bf 5}, {\bf 6} \} \oplus \{ {\bf 10}, {\bf 6} \}\,, \\
\mathcal{O}_s &\to&  \{ {\bf 1}, {\bf 6} \}\,, \nonumber
\end{eqnarray}
Finally, therefore, the low energy theory will have 4 different baryons: $\mathcal{B}_A = \{ {\bf 5}, {\bf 6} \}$, $\mathcal{B}_S = \{ {\bf 10}, {\bf 6} \}$ and $\mathcal{B}_1 = \{ {\bf 1}, {\bf 6} \}$, which can be matched to different operators written in terms of underlying fermions.
To further understand how the top fields can be coupled to the above baryons, we need to decompose each one in terms of the SM quantum numbers $(\SU(3)_c, \SU(2)_L)_{\U(1)_Y}$ as follows:
\begin{eqnarray}
\mathcal{B}_0 &=& (3,1)_{2/3} \oplus \mbox{h.c.}\,, \nonumber \\
\mathcal{B}_A &=& (3, 2)_{1/6} \oplus (3,2)_{7/6} \oplus (3,1)_{2/3} \oplus \mbox{h.c.}\,, \\
\mathcal{B}_S &=& (3, 2)_{1/6} \oplus (3,2)_{7/6} \oplus (3,3)_{2/3}  \oplus (3,1)_{2/3}  \oplus (3,1)_{-1/3}  \oplus (3,1)_{5/3} \oplus\mbox{h.c.}\,.\nonumber 
\end{eqnarray}
The presence of the h.c. means that each baryon is a massive Dirac fermion, as expected. We also notice that the component with the same quantum numbers of the quark doublet, present in both $\mathcal{B}_A$ and $\mathcal{B}_S$, is always accompanied by a custodial doublet with which it forms a bi-doublet: this is one of the basic requirements to avoid large corrections to the $Z$ couplings of the bottom quark~\cite{Agashe:2006at}. The right-handed top, on the other hand, can couple to all three baryons. It should also be noted that $\mathcal{B}_S$ contains a right-handed bottom partner, such that this model can in principle also generate the bottom quark mass.

At this stage, the underlying dynamics provides the key ingredients to build a low energy effective theory of partial compositeness. A pedagogical introduction to the general procedure can be found in ~\cite{Marzocca:2012zn}, where the reader can find how to introduce the baryon fields in the chiral Lagrangian of Eq.~\eqref{eq:LchiLO}. Here we will mainly focus on what can be learned from the underlying dynamics. To start, looking at Eq.~\eqref{eq:opsAAbarAdj} we see that the operators couplings to the top fields always transform as two-index irreps of $\SU(4)_\tcf$~\cite{Golterman:2017vdj}. To build a concrete model, one should therefore choose one of the operators in Eq.~\eqref{eq:opsAAbarAdj}, and embed the SM fermion fields in a spurion transforming in a matching irrep. This choice may seem arbitrary, but is in fact linked to the operator that has the largest anomalous dimension in the walking regime (see next section for further discussion) and to the mechanism that generates the four-fermion coupling. At an effective level, though, one has a free choice to make. The choice will then determine the low energy properties of the partial composite top, and of the composite Higgs. 
For concreteness, we will focus here on the simplest choice of $\mathcal{B}_A$: the two needed couplings will therefore read
\begin{equation} \label{eq:LPC}
\mathcal{L}_{\rm PC} \supset y_L f \mbox{Tr} [U^\dagger Q_L U^\ast \mathcal{B}_A] + y_R f \mbox{Tr} [U^\dagger T_R U^\ast \mathcal{B}_A]\,,
\end{equation}
where $U = e^{i \sum_{j=1}^5 X^j \pi^j/f}$ is the non-linearly transforming pNGB matrix, and $Q_L$ and $T_R$ are the SM fields embedded in the conjugate anti-symmetric irrep of $\SU(4)_\tcf$. The two couplings $y_{L/R}$ parametrize the mixing between the elementary top fields and the composite partners.
There are two ways to analyze the impact of the top partners to the low energy theory:
\begin{itemize}

\item{A-} \emph{Baryon-dominance}: in this approach, one assumes that the dominant effects are due to the propagation of the lightest baryons. Thus, the top mass is generated by diagonalising the mixing matrix with the baryons in the multiplet, while contributions to the Higgs potential can be computed at one loop level, with divergences properly removed~\cite{Marzocca:2012zn}.

\item{B-} \emph{Spurion-analysis}: in this approach, one assumes that the baryons are heavier than the pNGBs, thus their effect is encoded in low energy operators built in terms of the spurions containing the top couplings to the baryon operators~\cite{Golterman:2017vdj,Alanne:2018wtp}.

\end{itemize}
The two approaches, however, are not equivalent even though they give similar results. One example was given in ~\cite{Bizot:2018tds} when computing the couplings of the singlet pNGB $a$ to the top quark. In approach B, the coupling is proportional to the top mass times the $\U(1)$ charges of the two baryons, namely
\begin{equation}
g_{a \bar{t} t} = - i \frac{m_{\rm top}}{f_a} 2 \times \frac{5}{3} \sim \mathcal{O} (y^2)\,.
\end{equation}
Expanding for small mixing, this coupling is therefore quadratic in the linear couplings (as the top mass is, $m_{\rm top} \sim y_L y_R$). On the other hand, following approach A, by diagonalising the mass matrix as well as the couplings, one obtains
\begin{equation}
g_{a \bar{t} t} = - i \frac{m_{\rm top}}{f_a} \frac{5}{3} (\sin^2 \theta_L + \sin^2 \theta_R) \sim \mathcal{O} (y^4)\,,
\end{equation}
where $\sin \theta_{L/R} \propto \frac{y_{L/R} f}{M_\mathcal{B}}$ are the mixing angles of the left- and right-handed top respectively with the composite partners.
This example clearly shows that the two approaches are not equivalent, as they lead to parametrically different results for the couplings of a pNGB.
In fact, the two approaches are complementary: A may be trusted if the lightest baryon mass is light compared to the condensation scale, so that it can be safely included in the low energy effective Lagrangian, while approach B is likely to give the correct leading effect if the baryons are all heavier that the pNGB masses.
In the literature, approach A has been most widely used, and we refer the interested reader to ~\cite{Panico:2015jxa} for more details.

Defining an underlying dynamics can also help studying the origin of the partial compositeness couplings. For instance, the two couplings in Eq.~\eqref{eq:LPC} could arise from the following four-fermion interactions:
\begin{equation}
\frac{c_L}{\Lambda_{\rm F}^2}\ q_L^{a,i} \chi^{\bar{a}} \tcf^i \tcf^4 + \frac{c_R}{\Lambda_{\rm F}^2}\  t_R^a \overline{\chi^a} \overline{\tcf}^3 \overline{\tcf}^4 + \mbox{h.c.}
\end{equation}
These couplings are generated by some unspecified new physics at a scale $\Lambda_{\rm F}$. In order to generate large enough couplings at low energy, one has two options: either $\Lambda_{\rm F} \sim \Lambda_{\rm FC}$, or the theory enters a walking regime above the condensation scale. In either case, the underlying models discussed so far need to be UV complete right above $\Lambda_{\rm FC}$, thus they can only be considered as underlying descriptions rather than UV completions of the low energy effective theories.

\subsection{Towards UV complete models}

To UV complete the models of partial compositeness described in this section, it is crucial to fulfil the following requirements:
\begin{enumerate}
\item Generate the four-fermion interactions responsible for top partial compositeness;
\item Extend the model so that it features a walking phase above the condensation scale $\Lambda_{\rm FC}$;
\item Extend the model to generate masses for all fermions, either via partial compositeness or via bi-linear couplings. 
\end{enumerate}
It is a daring endeavour to be able to successfully construct a UV completion, which is compatible with all low energy constraints. The main challenge is due to sufficiently splitting the scale where flavor violation is generated and the low energy condensation scale. Nevertheless, a few attempts are present in the literature, and we review their main approaches here.

\begin{itemize}

\item A scenario based on partial unification of the FC gauge interactions has been proposed in ~\cite{Cacciapaglia:2019dsq}, where an example for the model in Table~\ref{tab:M8} has been proposed. The main idea is to partly unify the FC gauge group with the SM ones, so that gauge mediation could explain the couplings of the top quark. In the example, the unified gauge group is $\SU(8)_{\rm PS} \times \SU(2)_L \times \SU(2)_R$, {\it {\`a} la} Pati-Salam. The SM fermions, and the $\tcf$'s are thus embedded in a bi-fundamental of $\SU(8)$ and of one of the two $\SU(2)$'s. The gauge groups are broken at high scale by scalar VEVs, so that the $\SU(8)$ contains $\SP(4)_{\rm FC}$ and a Pati-Salam quark-lepton $\SU(4)$ group. Partial compositeness for the top is achieved by introducing a fermion in the 4-index anti-symmetris irrep of $\SU(8)$, which contains $\chi$. Additional fermions charged under the FC group allow for the theory to be inside the conformal window between the high scale (where the four-fermion interactions are generated) and $\Lambda_{\rm FC}$. If sufficiently large anomalous dimensions are generated in the walking phase, the breaking scale can be pushed close to the Planck scale, so that the model is technically natural.

\item Another attempt is based on the recent idea that gauge theories with large number of fermions may feature a UV interactive fixed point~\cite{Antipin:2017ebo}, thus allowing to be extrapolated to arbitrary high scales. In ~\cite{Cacciapaglia:2018avr} this idea has been applied to the models of ~\cite{Ferretti:2013kya}: the key observation is othat, in order to give mass to all SM fermions and generations, a large number of $\chi$ fermions is needed. This allows to drive all gauge couplings to a UV fixed point. The four-fermion interactions can then be generated by high-mass scalars, as long as a walking phase is also present. 

\item A proposal based on dualities has been put forward in ~\cite{Caracciolo:2012je}. The main UV construction is based on a supersymmetric $\SO(N)$ gauge theory with matter fields in the fundamental. The low energy effective theory is thus described by a Seiberg dual, where partial compositeness couplings are generated as the low energy flow of high-scale Yukawa couplings. A complete theory is however not provided, with the main challenge given by generating masses to all SM fermions.

\item In ~\cite{Cacciapaglia:2019vce} it has been shown that the top and $\tcf$ fields could be replaced by massless baryons from a confining chiral gauge theory, which confines at a scale $\Lambda_{\rm F} > 10^7$~TeV. This model would allow to generate partial compositeness for the top only. The example shown features $\SU(3)_{\rm FC}$, as in ~\cite{Vecchi:2015fma}. However, the global symmetries of the chiral gauge theory are such that the top mass is not easily generated, as additional symmetry breaking terms are needed.

\item The idea of holography has also been used to provide extra dimensional completions for composite Higgs models~\cite{Contino:2003ve}. Holography has been historically at the core of the composite Higgs revival in the early 2000's, where a model based on warped extra dimensions was proposed~\cite{Agashe:2004rs}, building on the old idea of Gauge-Higgs Unification~\cite{Hosotani:1983xw,Antoniadis:2001cv,Hosotani:2005nz}. However, it is a matter of taste if an extra dimensional theory can be considered as a genuine UV completion. The main point here is that extra dimensional theories are intrinsically non-renormalisable, as the gauge coupling itself is dimension-full. This implies that divergences at loop level arise which need to be regulated by new operators. While warped spaces have a position dependent cut-off which can extend up to the Planck scale~\cite{Randall:1999ee}, operators suppressed by the TeV scale are also present and needed. This means that non-calculable effects at the TeV scale do arise. Furthermore, the basic idea of holography is based on a conjecture that has been tested for maximal supergravity theories in 10-dimensional spaces~\cite{Maldacena:1997re}, while the models under discussion lack both supersymmetry and the additional dimensions. We will further elaborate on this point in the next section.

\end{itemize}

\section{Extra Dimensional and Non-Lagrangian approaches}

So far we have discussed composite theories emerging from an underlying confining gauge dynamics, in analogy to the well known example of QCD. Yet, the dynamics of  a theory generating a composite Goldstone Higgs needs to be quite different from the familiar QCD one: in particular, a walking phase is needed above the condensation scale $\Lambda_{\rm FC}$ in order to address the issue of flavor. This key observation has prompted the idea that the theory behind could be simply a conformal theory that describes the conformal sector above the condensation scale. This idea is linked with the principle of holography~\cite{Contino:2003ve}, according to which a conformal 4-dimensional theory may be dual to a 5-dimensional gravity theory in warped space~\cite{Randall:1999ee}. 

Taking this scenario at face value, one can try to construct composite Higgs theories simply based on the global symmetries of the conformal sector, which thus determine the properties of the pNGBs in similar fashion to chiral symmetry breaking in QCD-like theories. In this effective field theory (EFT) based approach, the minimal model features the symmetry breaking pattern $\SO(5)/\SO(4)$~\cite{Agashe:2004rs,Agashe:2005dk}, under which the 4 pNGBs are enough to saturate the degrees of freedom of the Higgs doublet. Many other symmetry breaking patterns can be used, irrespective of their origin~\cite{Mrazek:2011iu,Bellazzini:2014yua}. In the holographic interpretation, the anomalous dimension of the operator corresponding to the Higgs is dual to the bulk mass of the 5D field associated to it. There are however more general arguments that relate to the possible values of the anomalous dimension of the Higgs, i.e. of a generic composite scalar $\phi$. In ~\cite{Rattazzi:2008pe} , by using bootstrap techniques, it has been shown that the dimension of the $\phi^2$ operator is bounded by a function of the dimension of the scalar operator:
\begin{equation}
\Delta_{\phi^2} \leq f (\Delta_\phi)\,,\quad \mbox{with}\;\; f (\Delta_\phi) \to 2 + \mathcal{O} (\sqrt{\Delta_\phi-1})\;\; \mbox{for}\;\; \Delta_\phi \to 1\,.
\end{equation}
The limit is particularly relevant for composite Higgs models, as it would imply that it may not be possible to write large enough mass terms for the SM fermions by using bi-linear interactions. The argument goes as follows: the Yukawa operator in the conformal theory would read
\begin{equation}
y_\psi \left( \frac{\mu}{\Lambda_{\rm F}} \right)^{\Delta_\phi - 1}\ \phi \bar{\psi} \psi\,,
\end{equation}
where $\mu$ is the scale of the process (fixed to $\Lambda_{\rm FC}$ for the low-energy Yukawa in the condensed phase). Thus, in order to split sufficiently the flavor scale $\Lambda_{\rm F}$ from the condensation scale, one would need $\Delta_\phi \to 1$. However, in this limit, the mass operator would acquire a dimension $\Delta_{\phi^2} \to 2$, thus becoming a relevant operator sensitive to the UV scales. In other words, the composite Higgs operator would need to have the dimension of an elementary scalar, and thus suffer from the same hierarchy problem as in the SM. The situation is however more complicated in cases with global symmetries~\cite{Rattazzi:2010yc}, which are more relevant for composite Higgs models, as there are multiple symmetry channels appearing in the bootstrap calculation. One, therefore, needs to carefully identify which channel is most relevant for the Higgs, and some counter-examples ``violating'' the bound have been found~\cite{Antipin:2014mga}.

While the argument above is not conclusive in excluding the case of bilinear couplings, it renders the idea of partial compositeness more attractive. If $\Delta_\phi > 1$, the bilinear couplings will only generate a tiny mass for all SM fermions, thus linear mixing may be the most important contribution:
\begin{equation}
\tilde{y}_L \mu\ \left( \frac{\mu}{\Lambda_{\rm F}} \right)^{\Delta_{\mathcal{B}}-5/2}\ q_L \mathcal{O}_{\mathcal{B}_L} + \tilde{y}_R \mu\ \left( \frac{\mu}{\Lambda_{\rm F}} \right)^{\Delta_{\mathcal{B}_R}-5/2}\ t_R \mathcal{O}_{\mathcal{B}_R}\,.
\end{equation}
The couplings above match the partial compositeness couplings of Eq.~\eqref{eq:LPC} as
\begin{equation}
y_{L/R} f \propto \tilde{y}_{L/R} \Lambda_{\rm FC} \left( \frac{\Lambda_{\rm FC}}{\Lambda_{\rm F}} \right)^{\Delta_{\mathcal{B}_{L/R}}-5/2}\,,
\end{equation}
so that large mixing are allowed for $\Delta_\mathcal{B} \approx 5/2$. It may seem that such value of the anomalous dimension is unrelated to that of the composite Higgs  operator $\mathcal{O}_H$, while a direct link exists in the case of bi-linear couplings. In fact, a link can be found by constructing a two-point function correlator of two baryon operators, which has the same quantum numbers as the Higgs doublet. In other words, operator product expansion would lead:
\begin{equation}
\mathcal{O}_{\mathcal{B}_L} \cdot \mathcal{O}_{\mathcal{B}_R} = \mathcal{O}_H + \dots
\end{equation} 
This simple argument proves that the dimensions of the operators that mix to the top fields, and that of the composite Higgs operator are indeed related, however it is very challenging to extract any information about such relation. Furthermore, additional two point functions corresponding, for instance, to QCD-colored scalars can be constructed:
\begin{equation}
\mathcal{O}_{\mathcal{B}_R}^\ast \cdot \mathcal{O}_{\mathcal{B}_R} = \mathcal{O}_{\pi_8} + \dots\,, \quad \mathcal{O}_{\mathcal{B}_R} \cdot \mathcal{O}_{\mathcal{B}_R} = \mathcal{O}_{\pi_6} + \dots\,, \quad \mbox{etc.}
\end{equation}
Thus the conformal theory, if consistent at all, needs to also include many more operators than the ones usually included in the effective theory, some of which could be light. The models in Table~\ref{tab:M8} introduced in the previous section, for instance, is an explicit example where an additional color octet and sextet are present.



\newcommand{\f}{\mathcal{F}}

\newcommand{\TC}{\mathrm{FPC}}
 \newcommand{\tcs}{\mathcal{S}}
\newcommand{\hc}{\; + \; \mathrm{h.c.} \;}
\newcommand{\brakets}[1]{\left\langle #1 \right\rangle}
\newcommand{\eff}{\mathrm{eff}}
 \newcommand{\transpose}{^{\mathrm{T}}}
 \newcommand{\andeq}{\quad \mathrm{and} \quad}
\newcommand{\braces}[1]{\left\lbrace #1 \right\rbrace}
\newcommand{\UU}{\mathrm{U}}
\newcommand{\LL}{\mathrm{L}}
\newcommand{\RR}{\mathrm{R}}

\newcommand{\Yf}[4]{( y_f^\ast y_f)^{a_{#1}}\phantom{}_{a_{#2}}\phantom{}^{i_{#3}i_{#4}}}
\newcommand{\SDS}[2]{(\Sigma^\dagger \overleftrightarrow{D}^\mu\Sigma)_{a_{#1}}\phantom{}^{a_{#2}}}
\newcommand{\DS}[2]{(D^\mu\Sigma)^{a_{#1}a_{#2}}}
\newcommand{\DSd}[2]{(D_\mu\Sigma^\dagger)_{a_{#1}a_{#2}}}

\newcommand{\pd}{\overset{\text{\small$\leftrightarrow$}}{\partial^\mu}}
\newcommand\DDbar{\parenbar{D}}
\newcommand{\parenbar}[2][4]{%
  \mkern#1mu
  \sbox0{$#2$}%
  \makebox[0pt][r]{\raisebox{\ht0}{$\scriptscriptstyle($}}%
  \overline{\mkern-#1mu#2\mkern-1mu}%
  \makebox[0pt][l]{\raisebox{\ht0}{$\scriptscriptstyle)$}}%
  \mkern1mu
}
\newcommand{\xbar}[2][4]{%
  \mkern#1mu
  \overline{\mkern-4mu#2\mkern-1mu}%
  \mkern1mu
}

\section{Fundamental Partial Compositeness}
\label{FPC}

Because of the several challenges one faces when providing a complete theory of fermion mass generation in models of composite Higgs, it is 
 reasonable and timely to explore composite frameworks in which, while still insisting on the composite nature of the Higgs sector, one puts aside (postpones) the naturalness argument.  This  frees us to consider wider classes of composite theories featuring, for example, also scalars charged under the FC gauge interactions. It also allows for interesting model building routes featuring unexplored dynamics that can  be investigated via first principle lattice simulations \cite{Pica:2009hc, Chivukula:2010xz, Sinclair:2014cga, Pica:2016zst}. For the scalar-phobic reader, we shall argue that these models can be viewed as an intermediate effective realization of  a more fundamental gauge-fermion dynamics~\cite{Cacciapaglia:2017cdi}.  The models that we will consider here differ from TC theories featuring a SM-like Higgs doublet that have been considered long time ago in \cite{Simmons:1988fu,Carone:1992rh}, and more recently within the context of Goldstone Higgs in~\cite{Antipin:2015jia, Galloway:2016fuo, Agugliaro:2016clv}. 

The class of models that we shall consider here ~\cite{Sannino:2016sfx,Cacciapaglia:2017cdi, Sannino:2017utc,Agugliaro:2019wtf} implement the partial compositeness framework at a microscopic level.  This new paradigm was termed \emph{Fundamental Partial Compositeness} (FPC).  According to the paradigm, we introduce fundamental FPC-scalars, with QCD color embedded in the corresponding flavor symmetry with respect to the FPC gauge theory. The FPC-scalars must be chosen in such a way to allow for Yukawa couplings involving  a SM fermion, a FPC fermion, and a FPC scalar: this is sufficient to guarantee the generation of fermion masses and Yukawa couplings at low energy. The composite baryons are built out of one FPC-fermion and one FPC-scalar, and a UV Lagrangian can be written, without the need of extra colored FPC-fermions. By construction, no large anomalous dimensions are needed and the hierarchy among the SM fermions can be achieved. Additionally, the new fundamental Yukawa structure allows for unexplored flavor dynamics \cite{Sannino:2017utc}.   Composite theories including (super) colored FPC-scalars, attempting to give masses to some of the SM fermions, appeared earlier in the literature \cite{Dobrescu:1995gz,Kagan:1994qg,Altmannshofer:2015esa}  but did not feature a Goldstone Higgs.

The simplicity of FPC relies in the fact that it enables us to construct microscopic theories that, at low-energies,  reproduce the experimentally successful predictions of the SM while remaining a valid composite alternative to the SM Higgs sector. In contrast to a purely fermionic realization of partial compositeness, no additional QCD colored sectors are required, making the model complete on its own. The Lagrangian structure can be summarized as:
\begin{align} \label{eq:TClag}
\mathcal L_\text{FPC} & =  \ - \frac 1 4 F_{\mu \nu} F^{\mu \nu}  +  i { \mF^\dagger} \bar \sigma^\mu D_\mu \mF - \left(\frac 1 2 \mF m_\mF \mF + \text{h.c.} \right) + (D_\mu S) ^\dagger D^\mu S - S^\dagger m_S ^2 S - V(S)  \nonumber \\ 
 &= \mathcal L_\text{FC} +  (D_\mu S) ^\dagger D^\mu S - S^\dagger m_S ^2 S - V(S)\,,
\end{align}
where $F_{\mu \nu}$ refers to the FPC gauge fields, $\mathcal F$ is the left-handed Weyl FPC-fermion multiplet, $S$ collectively indicates the (complex) FPC-scalars, and $V(S)$ is the scalar potential. To keep the notation light we omitted  the FPC-indices that are properly contracted to form FPC-singlet operators. Also note that up to the scalars the Lagrangian is identical to the one we used for FC, except that when gauged scalars are added    the dynamics changes. This is the reason why in this subsection we use FPC and not FC to indicate the new dynamics and its derived quantities. However, as we shall see, some of the general features stemming from the global symmetries of the FPC-fermionic sector are still shared with the original FC theory.  Up to SM interactions and super-renormalizable operators, the symmetry of the Lagrangian~\eqref{eq:TClag} takes the form of a direct product of the fermionic and scalar global symmetries, $G_F \times G_S$.  A diagonal fermion mass matrix breaks the global symmetry of the fermionic kinetic term while a diagonal scalar mass matrix keeps the scalar global symmetry unspoiled. The EW gauging, color interactions and the operators needed to generate the SM fermion masses break the FPC global symmetry. 

One can envision the underlying FPC fermion matter to transform according to either the real, pseudo-real or complex representation of the FPC gauge group. For both the pseudo-real and real FPC-color representations\footnote{Here we  always choose FPC matter to be in a  defining representation of the gauge group  which will therefore be either  $\Sp(2N)_{FPC}$ or $\SO(N)_{FPC}$ with $\SU(2)_{FPC}=\Sp(2)_{FPC}$ being the first of the symplectic groups.} we can arrange the field $S$ in a multiplet $\Phi$ as follows
\begin{align} \label{eq:grouping}
\Phi =\left( \begin{array}{cc} S \\ \overline S \end{array} \right)\,, \qquad  
\overline S =\begin{cases}
 \epsilon_\text{FPC} S^* & \SP(2N_s) \\
 \quad S^* & \SO (2N_s) 
\end{cases}
\end{align}
where on the right we indicated the corresponding global symmetries over the scalars, which are $\SP(2 N_S)$ or $\SO(2N_S)$ depending on whether the FPC-color gauge representations are pseudo-real or real. The quadratic scalar Lagrangian reads:
\begin{align}
\frac 1 2 (D_\mu \Phi )^T \left( \begin{array}{cc} &  \ \pm1 \\   1 & \end{array} \right)  (D_\mu \Phi ) - \frac 1 2 \Phi^T \left( \begin{array}{cc} & \pm {m_S^2}^T \\ m_S^2 & \end{array} \right)  \Phi\,,
\end{align}
where the plus/minus sign corresponds to $\SO(2N_S)$/$\SP(2N_S)$ and we introduce the off-diagonal matrix $\displaystyle{\omega = \left( \begin{array}{cc} &  \ \pm1 \\   1 & \end{array} \right)}$. 
Note that we  need  the  global symmetry over the scalars to be at least  $\SU(3)$ to account for QCD color, since the FPC-fermions are taken to be color singlets.  For complex FPC-color representation the maximum scalar symmetry is $\U(N_S)$. Below we will summarize the minimal FPC investigated in \cite{Cacciapaglia:2017cdi,Sannino:2017utc}, which corresponds to the template model of Chapter \ref{ch:su2}, while the generalization to real and complex representations can be found in \cite{Agugliaro:2019wtf}.

\subsection{Minimal Fundamental Partial Compositeness}
 \label{MFPC}
 The most minimal realization of FPC in terms of the number of fields is, as explained in~\cite{Cacciapaglia:2014uja,Sannino:2016sfx,Cacciapaglia:2017cdi},  the case of $\Sp(2N)_\text{FPC}$ with $N=1$ and with fields in the fundamental representation, which for this gauge group is a pseudo-real representation.  As we shall argue later it is also the most minimal realization of partial compositeness that one can construct. 
This model is the natural extension of the one introduced in Chapter~\ref{ch:su2}. Therefore, as we have already argued there, with $N_\tcf$ Weyl FPC-fermions, the maximal quantum global symmetry of the fermions before introduction any further interactions is $\SU(N_\tcf)$. The symmetries are such that $ m_\tcf$ is an antisymmetric tensor in flavor space. 

As the FPC-scalars transform according to the same representation of the FPC-fermions with respect to the new gauge group, no Yukawa interactions among the FPC-fermions and FPC-scalars can be written. As argued above this implies that, with zero mass terms, the FPC-scalars have an independent $\Sp(2N_\tcs)$ symmetry. We assume the potential $V(\tcs)$ to respect the maximum global symmetries of the FPC theory.  

 The scalar kinetic and mass term now reads: 
	\begin{equation}
	\frac{1}{2} \left(D_\mu \Phi\right) \epsilon_\TC \epsilon \left(D^{\mu} \Phi\right) -  \frac{1}{2}\Phi   \epsilon_\TC M^2_\tcs \Phi \ , \label{eq:lphi}
	\end{equation}
with 
	\begin{equation} 
	M^2_\tcs   = \begin{pmatrix} 0 & -{m^2_\tcs}^T  \\ {m^2_\tcs} & 0 \end{pmatrix} \ ,
	\end{equation}
and $ \epsilon $ is the invariant symplectic form of $ \Sp(2N_\tcs) $. 
For $ G_\TC = \Sp(2N) $ we report in Table~\ref{Table-Fields}  the elementary states of the FPC theory as well as the bilinear gauge singlets along with their global transformation properties and multiplicities.
 \begin{table}[t]
\begin{center}	\begin{tabular}{*{4}{c}}
	 States \qquad&  $ \SU(N_\tcf) $  \qquad&  $ \Sp(2N_\tcs) $ \qquad & number of states \qquad \\ \hline 
	$ \tcf $ & $\tiny\yng(1)$ & $ 1 $ & $2 N \times N_\tcf$  \\
	$ \Phi $ &  1 & $ {\tiny \yng(1)} $ & $2N \times 2N_\tcs$  \\
	\hline
	$ \Phi\Phi $ &  1 & $ 1 + {\tiny \yng(1,1)} $ & $ \displaystyle{1 + {N_\tcs(2N_\tcs-1)}}$ \\
	$ \tcf \Phi $ & $ {\tiny\yng(1)} $ & $ {\tiny\yng(1)} $ & $ 2N_\tcs N_\tcf $\\
	$ \tcf \tcf $ & $ {\tiny\yng(1,1)} $ & 1  & $ \displaystyle{{N_\tcf(N_\tcf-1)}}$ 
	\end{tabular} \end{center}
	\label{Table-Fields}
	\caption{The fundamental matter fields of the theory appear in the first two lines of the table, both transforming according to the fundamental representation of FPC.  The last three lines correspond to  the bi-linear composite FPC singlet states. The number of states counts the Weyl fermions or real scalars.}
\end{table}

When adding the EW sector, following the analysis in Chapter~\ref{ch:su2}, we  embed it within the $ \SU(N_\tcf) $ of the FPC-fermion sector. In this way the EWSB is tied to the breaking of $ \SU(N_\tcf) $ and the Higgs boson can be identified with a pNGB of the theory \cite{Kaplan:1983fs,Cacciapaglia:2014uja}.  Assuming for the scalars a positive mass squared, it is natural to expect spontaneous symmetry breaking in the fermion sector according to the pattern $\SU(N_\tcf) \rightarrow \Sp(N_\tcf)$. This pattern was established in the absence of scalars for $N_\tcf = 4$ and $G_\TC=\Sp(2)$ via first principle lattice simulations~\cite{Lewis:2011zb}. The ensuing FPC-fermion bilinear condensate is 
	\begin{equation}
	\brakets{\tcf^{a} \epsilon_\TC \tcf^{a'} } = f^2_\TC \Lambda_\TC \Sigma_0^{a a'},  
	\end{equation}
where Lorentz and FPC indices are opportunely contracted, and the $ \Sigma_0 $ matrix is an antisymmetric, two-index representation of $\SU(N_\tcf)$. We also have $\Lambda_\TC = 4\pi f_\TC$ with $\Lambda_\TC$ the composite scale of the theory and $f_\TC$ the associated pion decay constant.
 
In addition, we envision two possibilities for the FPC-scalars: the formation of a condensate $\brakets{\Phi^i \epsilon_\TC \Phi^j}$ may not happen or be proportional to the singlet of $\Sp(2 N_\tcs)$, in which case the flavor symmetry  in the scalar sector is left unbroken; or a condensate forms and breaks $\Sp(2 N_\tcs)$ generating light bosonic degrees of freedom. For simplicity  we will focus on the former case.

It is time to investigate the SM fermion mass generation. The presence of FPC-scalars in FPC models permits a new type of Yukawa interactions involving the FPC and  SM sectors. In fact each new Yukawa operator involves a FPC-fermion, a FPC-scalar and a SM fermion and the new fundamental Yukawa Lagrangian to replace the SM one reads
	\begin{equation}
	\mathcal{L}_{\mathrm{yuk}} = - \spur{i}{a} \epsilon_{ij} \Phi^{j} \epsilon_\TC \tcf^{a} \hc \ ,
	\label{eq:fund_Yukawa}
	\end{equation}  
in which we make use of the  spurion $ \psi $  transforming under the relevant global symmetries as
	\begin{equation} \label{eq:spurionpsi}
	\spur{i}{a} \equiv \left(\Psi \, y\right)^{i} \phantom{}_{a} \in {\tiny\yng(1)}_{\tcs} \otimes \overline{\tiny\yng(1)}_{\tcf} \ .
	\end{equation} 
Here $\Psi$ is a generic SM fermion and $y $ is the new Yukawa matrix. With this spurionic construction we may formally consider $ \mathcal{L}_{\mathrm{yuk}} $ an invariant of the global FPC symmetries. Additionally, the notation has the benefit that all Yukawa interactions are summarized in a single operator. Note that with the notation introduced here, the generation, color, and electroweak indices are all embedded in the global symmetries. 
At low energy, the Yukawa couplings in Eq.~\eqref{eq:fund_Yukawa} generate linear mixing of the SM fermions with spin-1/2 resonances made of one FPC-fermion and one FPC-scalar (see Table~\ref{Table-Fields}), thus implementing partial compositeness. This way of endowing masses for the SM fermions is free from long standing problems in models of composite Higgs dynamics and, as we shall review later, can be also related to previous incomplete extensions. 

Besides the SM fermions and Yukawas, the underlying theory contains two more spurions that explicitly break the flavor symmetries, that is the masses of the FPC-fermions and scalars:
	\begin{equation}
	m_\tcf \in  \overline{\tiny\yng(1,1)}_{\tcf} \otimes 1_\tcs\ , \quad M^2_\tcs \in 1_\tcf \otimes \overline{\tiny\yng(1,1)}_{\tcs}\ .
	\end{equation}
As they are dimension-full parameters, they can be inserted at the effective Lagrangian level only if an order parameter can be defined, i.e. either if the mass is small compared to the FPC scale $\Lambda_\TC$, or if they are much larger. In the latter case, one can then expand in powers of the inverse of the mass matrices.
We will start with the former case, and classify the relevant operators in terms of powers of the spurion $ \spur{i}{a} $, and then discuss  how to consistently move to the limit of large FPC-scalar masses.  
An EFT can now be constructed along the same lines as Chapter \ref{ch:su2}, where, following our assumption on the FPC-scalars, we can only include the pNGBs associated to the FPC-fermion condensation.

We are now ready to determine the effective operators emerging at the EW scale in terms of the SM fields upon consistently integrating out the heavy FPC dynamics aside from the pNGB excitations. De facto we provide the first effective field theory that matches to a concrete  and complete example of a composite theory of flavor. In turn, this allows for investigating its impact on electroweak observables and low energy flavor physics \cite{Sannino:2017utc}.

 The effective operators at the EW scale are obtained by matching the operators to the underlying composite  dynamics that can be summarized as follows  
	\begin{equation}
	\mathcal{L}_{\mathrm{EFT}} = \sum_{A} C_A \, \mathcal{O}_A + \left(\sum_{A}C'_A\, \mathcal{O}'_A \hc \right)
	\end{equation}
for the effective field theory with coefficients $ C^{(\prime)}_A $ determined by the underlying FPC dynamics. Here $ \mathcal{O}_A^{(\prime)} $ refers to the self-hermitian/complex operators respectively.

We organize  the EFT by counting the chiral dimensions of the operators \cite{Buchalla:2013eza} by generalizing the Naive Dimensional Analysis (NDA) of \cite{Georgi:1992dw}.  In any realistic FPC model the power-counting is complicated by the potential occurrence of strong Yukawa couplings. This happens because in order to achieve the correct top mass it typically requires the product $ y_{Q_3} y_t \sim 4 \pi $. Strong couplings in the chiral expansion, can potentially enhance certain operators beyond the order ascribed to them by simple counting of the chiral dimension. To alleviate this issue it was defined in \cite{Cacciapaglia:2017cdi} the effective Yukawa couplings 
	\begin{equation}
	\dfrac{y_\mathrm{fund}}{\sqrt{4\pi}} \to y,
	\end{equation} 
which are rescalings of the fundamental couplings. This allow us to treat the Yukawa couplings as perturbative, albeit with a chiral dimension lowered to 1/2 down from 1.  One should therefore keep in mind that the Yukawa parameters entering in the EFT, are different from the fundamental Yukawa couplings by a rescaling. 

For the underlying model to be fundamental, it must be possible to run a perturbative Yukawa coupling from the scale of strong gravity down to the scale of compositeness where it should become strong. The leading order beta function for the fundamental Yukawa coupling, $ y_t $, belonging to the right-handed top quark (cf. Section \ref{topandbottom}) in the presence of an $ \Sp(2N) $ FPC group is\footnote{This result differs from the analogous Eq.(32) in Ref.~\cite{Sannino:2016sfx}, however the qualitative features of the running are retained.}
	\begin{equation}
	\dfrac{\partial y_t}{\partial \ln \mu} =  \dfrac{y_t}{(4\pi)^2} \dfrac{ (4N + 10) y_t^2 - (6 N + 3) g_\TC^2 }{4}
	\end{equation}
in the absence of other Yukawa couplings. Starting the RG flow in the perturbative regime at high scales (e.g. the Planck scale) and evolving the couplings down to $ \Lambda_\TC $, one will find that $ y_t $ increases in the IR as long as $ g_\TC \geq \sqrt{(4N + 10)/(6N + 3)}\, y_t $. With $ g_\TC $ becoming strong at the scale of compositeness, we can expect that it pulls the Yukawa couplings with it. This effect, however, is outside the reach of perturbation theory and cannot be probed quantitatively, as one can only achieve $ y_t\sim 1 $, before $ g_\TC $ becomes non-perturbative. A non-perturbative study is therefore required, to quantitatively determine whether the requirement of a strong $ y_t $ is compatible with completeness of the underlying model. This tension in the top Yukawas might also be resolved by favorable strong coefficients departing from unity by $ \mathcal{O}(\mathrm{few}) $.

A comprehensive classification of the operators at LO and NLO in this minimal FPC model can be found in \cite{Cacciapaglia:2017cdi}. Below, we will give some examples, before discussing some phenomenological consequences for the fermion mass generation.
For concreteness, we will only focus here on operators containing SM fermion fields.

\subsubsection{Effective Bilinear Operators with Standard Model Fermions}
We start with operators containing only two SM fermion spurions $\psi^i_a$: such operators can be conveniently organized in terms of their chiral dimension, starting with the ope appearing at order $\mathcal{O} (p^2)$:
 	\begin{equation}
	\mathcal{O}_{\mathrm{Yuk}} = -\dfrac{f_\TC }{2} \ (\spur{i_1}{a_1} \spur{i_2}{a_2})\, \Sigma^{a_1 a_2} \epsilon_{i_1 i_2 }\ ,
	\label{eq:OH}
	\end{equation}
which generates masses for the SM fermions.
The anti-symmetric matrix $\epsilon_{i_1 i_2 }$ contracts the $\Sp(2N_\tcs)$ indices, while spinor indices are hidden with the convention that two Weyl spinors in parenthesis are contracted to a scalar.

At the next order $\mathcal{O} (p^3)$ we have the operator:
	\begin{equation} \label{eq:OPif}
	\mathcal{O}_{\Pi f} = \dfrac{i f_\TC}{2 \Lambda_\TC} \ (\spurbar{i_1}{a_1} \bar{\sigma}_\mu \spur{i_2}{a_2} )\  \Sigma _{a_1 a_3}^\dag  \overleftrightarrow{D}^\mu \Sigma ^{a_3 a_2}\  \epsilon _{i_1 i_2}\,,
	\end{equation}	
which modifies the coupling of massive gauge bosons $W$ and $Z$, contained in the covariant derivative, to the SM fermions. 

  At next order $\mathcal{O} (p^4)$ we find the dipole operators:
	\begin{align}
	\mathcal{O}_{f W} &= \frac{f_\TC }{2 \Lambda_{\TC}^2  } \ (\spur{i_1}{a_1} \sigma^{\mu \nu} \spur{i_2}{a_2} ) A_{\mu\nu}^I \left( T^{I}_{\tcf} \Sigma - \Sigma (T^{I}_\tcf)\transpose \right)^{a_1 a_2} \epsilon_{i_1 i_2}, \label{eq:OfW}\\
	\mathcal{O}_{f G} &= \frac{f_\TC }{2 \Lambda_{\TC}^2 } \  (\spur{i_1}{a_1} \sigma^{\mu \nu} \spur{i_2}{a_2} ) G^{A}_{\mu\nu} \Sigma^{a_1 a_2} \left( \epsilon T_\tcs^{A} - (T_\tcs^{A})\transpose \epsilon \right)_{i_1 i_2}, \label{eq:OfG} 
	\end{align}
where $T^k_{\tcf/\tcs}$ are the generators of $\SU(N_\tcf)$ and $\Sp(2N_\tcs)$ respectively, and $ A_{\mu \nu}^k$/$G_{\mu \nu}^k $ the field strength tensors of the relative gauge bosons (more precisely, of the gauged subgroup). We note that the gauge couplings constants have been absorbed into the generators $T^k_{\tcf/\tcs}$ to account for there being several SM gauge groups embedded into each of them.  
The two operators, \eqref{eq:OfW} and \eqref{eq:OfG}, have structures mimicking the Penguin-induced operators in the SM~\footnote{The naming of these operators are loosely inspired by the corresponding operators in the SM effective field theory~\cite{Grzadkowski:2010es}.}.

\subsubsection{Four-Fermion Operators with Standard Model Fermions} \label{sec:4fermionOps}

We now construct a consistent basis of four-fermion operators: this exercise is particularly interesting because it allows to find how many independent LECs are relevant for FCNCs. This class of operators start at order $\mathcal{O} (p^4)$. Firstly, we list the five independent operators featuring two left-handed spinors $\psi$ and two right-handed ones $\bar{\psi}$:
	\begin{align}
	\mathcal{O}_{4f}^1 &= \dfrac{1}{4 \Lambda_\TC^2} (\spur{i_1}{a_1} \spur{i_2}{a_2} ) (\spurbar{i_3}{a_3} \spurbar{i_4}{a_4} ) \Sigma^{a_1 a_2} \Sigma^\dagger_{a_3 a_4} \epsilon_{i_1 i_2} \epsilon_{i_3 i_4}\ ,\label{eq:fourfermion1} \\
	\mathcal{O}_{4f}^2 &= \dfrac{1}{4 \Lambda_\TC^2} (\spur{i_1}{a_1} \spur{i_2}{a_2} ) (\spurbar{i_3}{a_3} \spurbar{i_4}{a_4} ) \left(\delta^{a_1}_{\enspace a_3} \delta^{a_2}_{\enspace a_4} - \delta^{a_1}_{\enspace a_4} \delta^{a_2}_{\enspace a_3} \right) \epsilon_{i_1 i_2} \epsilon_{i_3 i_4}\ , \\
	\mathcal{O}_{4f}^3 &= \dfrac{1}{4 \Lambda_\TC^2} (\spur{i_1}{a_1} \spur{i_2}{a_2} ) (\spurbar{i_3}{a_3} \spurbar{i_4}{a_4} ) \Sigma^{a_1 a_2} \Sigma^\dagger_{a_3 a_4} \left(\epsilon_{ i_1 i_4} \epsilon_{ i_2 i_3} - \epsilon_{ i_1 i_3} \epsilon_{ i_2 i_4} \right)\ ,	\\
	\mathcal{O}_{4f}^4 &= \dfrac{1}{4 \Lambda_\TC^2} (\spur{i_1}{a_1} \spur{i_2}{a_2} ) (\spurbar{i_3}{a_3} \spurbar{i_4}{a_4} ) \left( \delta^{a_1}_{\enspace a_3} \delta^{a_2}_{\enspace a_4} \epsilon_{ i_1 i_3} \epsilon_{ i_2 i_4} + \delta^{a_1}_{\enspace a_4} \delta^{a_2}_{\enspace a_3} \epsilon_{ i_1 i_4} \epsilon_{ i_2 i_3}\right)\ , \\
	\mathcal{O}_{4f}^5 &= \dfrac{1}{4 \Lambda_\TC^2} (\spur{i_1}{a_1} \spur{i_2}{a_2} ) (\spurbar{i_3}{a_3} \spurbar{i_4}{a_4} ) \left( \delta^{a_1}_{\enspace a_3} \delta^{a_2}_{\enspace a_4} \epsilon_{ i_1 i_4} \epsilon_{ i_2 i_3} + \delta^{a_1}_{\enspace a_4} \delta^{a_2}_{\enspace a_3} \epsilon_{ i_1 i_3} \epsilon_{ i_2 i_4}\right)\ , \label{eq:fourfermion5}
	\end{align}
where $\bar{\psi}^{a,i} = \epsilon^{i j} \bar{\psi}^a_j$. Note also that the above operators are self-conjugate.
Similarly, one can construct five corresponding operators containing four left-handed spinors: however, we find that only three of them are truly independent. We take these three to be~\cite{Cacciapaglia:2017cdi}:
	\begin{align}
	\mathcal{O}_{4f}^6 &= \dfrac{1}{8 \Lambda_\TC^2} (\spur{i_1}{a_1} \spur{i_2}{a_2} ) (\spur{i_3}{a_3} \spur{i_4}{a_4} )  \Sigma^{a_1 a_2} \Sigma^{a_3 a_4} \epsilon_{i_1 i_2} \epsilon_{i_3 i_4}\,, \label{eq:fourfermion6} \\
	\mathcal{O}_{4f}^7 &= \dfrac{1}{8 \Lambda_\TC^2} (\spur{i_1}{a_1} \spur{i_2}{a_2} ) (\spur{i_3}{a_3} \spur{i_4}{a_4} )  \left(\Sigma^{a_1 a_4} \Sigma^{a_2 a_3} - \Sigma^{a_1 a_3} \Sigma^{a_2 a_4}\right) \epsilon_{i_1 i_2} \epsilon_{i_3 i_4}\,,   \\
	\mathcal{O}_{4f}^8 &= \dfrac{1}{8 \Lambda_\TC^2} (\spur{i_1}{a_1} \spur{i_2}{a_2} ) (\spur{i_3}{a_3} \spur{i_4}{a_4} )  \Sigma^{a_1 a_2} \Sigma^{a_3 a_4} \left(\epsilon_{i_1 i_4} \epsilon_{i_2 i_3} - \epsilon_{i_1 i_3} \epsilon_{i_2 i_4}\right)\,.
	\end{align}	
For completeness, we also show the two-dependent operators
\begin{align}
	\mathcal{O}_{4f}^9 &= \dfrac{1}{8\Lambda_\TC^2} (\spur{i_1}{a_1} \spur{i_2}{a_2} ) (\spur{i_3}{a_3} \spur{i_4}{a_4} ) \left( \Sigma^{a_1 a_3} \Sigma^{a_2 a_4} \epsilon_{i_1 i_3} \epsilon_{i_2 i_4} + \Sigma^{a_1 a_4} \Sigma^{a_2 a_3} \epsilon_{i_1 i_4} \epsilon_{i_2 i_3}\right),\\
	\mathcal{O}_{4f}^{10} &= \dfrac{1}{8\Lambda_\TC^2} (\spur{i_1}{a_1} \spur{i_2}{a_2} ) (\spur{i_3}{a_3} \spur{i_4}{a_4} ) \left( \Sigma^{a_1 a_3} \Sigma^{a_2 a_4} \epsilon_{i_1 i_4} \epsilon_{i_2 i_3} + \Sigma^{a_1 a_4} \Sigma^{a_2 a_3} \epsilon_{i_1 i_3} \epsilon_{i_2 i_4} \right), \label{eq:fourfermion10}
\end{align}
which are related to $ \mathcal{O}^{6-8}_{4f} $ via
\begin{equation}
\mathcal{O}_{4f}^6 + \mathcal{O}_{4f}^9 = 0\,, \quad \mathcal{O}_{4f}^7 +\mathcal{O}_{4f}^8 - \mathcal{O}_{4f}^{10} = 0\,. 
\end{equation}
For the case of $N_\tcf=4$ one can write another operator:
\begin{equation}
\mathcal{O}_A  = - \dfrac{1}{8 \Lambda_\TC^2} (\spur{i_1}{a_1} \spur{i_2}{a_2} ) (\spur{i_3}{a_3} \spur{i_4}{a_4} )  \epsilon^{a_1 a_2 a_3 a_4} \epsilon_{i_1 i_2} \epsilon_{i_3 i_4}\,, ~ {\rm for~}~N_\tcf=4 \ , 
\end{equation}
where $\epsilon^{a_1 a_2 a_3 a_4}$ is the fully antisymmetric 4-index matrix which is naturally linked to the ABJ anomaly of the global ${\mathrm U}(1)_{\tcf}$. However this operator is already contained in the list above because of the following operator identity:
\begin{equation}
\mathcal{O}_A =\mathcal{O}_{4f}^{10} - \mathcal{O}_{4f}^8 - \mathcal{O}_{4f}^9   = \mathcal{O}_{4f}^6 + \mathcal{O}_{4f}^7 \ , ~ {\rm for~}~N_\tcf=4 \ . 
\end{equation} 
All in all, therefore, we identify 8 independent LECs. 
 
\begin{figure}
	\includegraphics{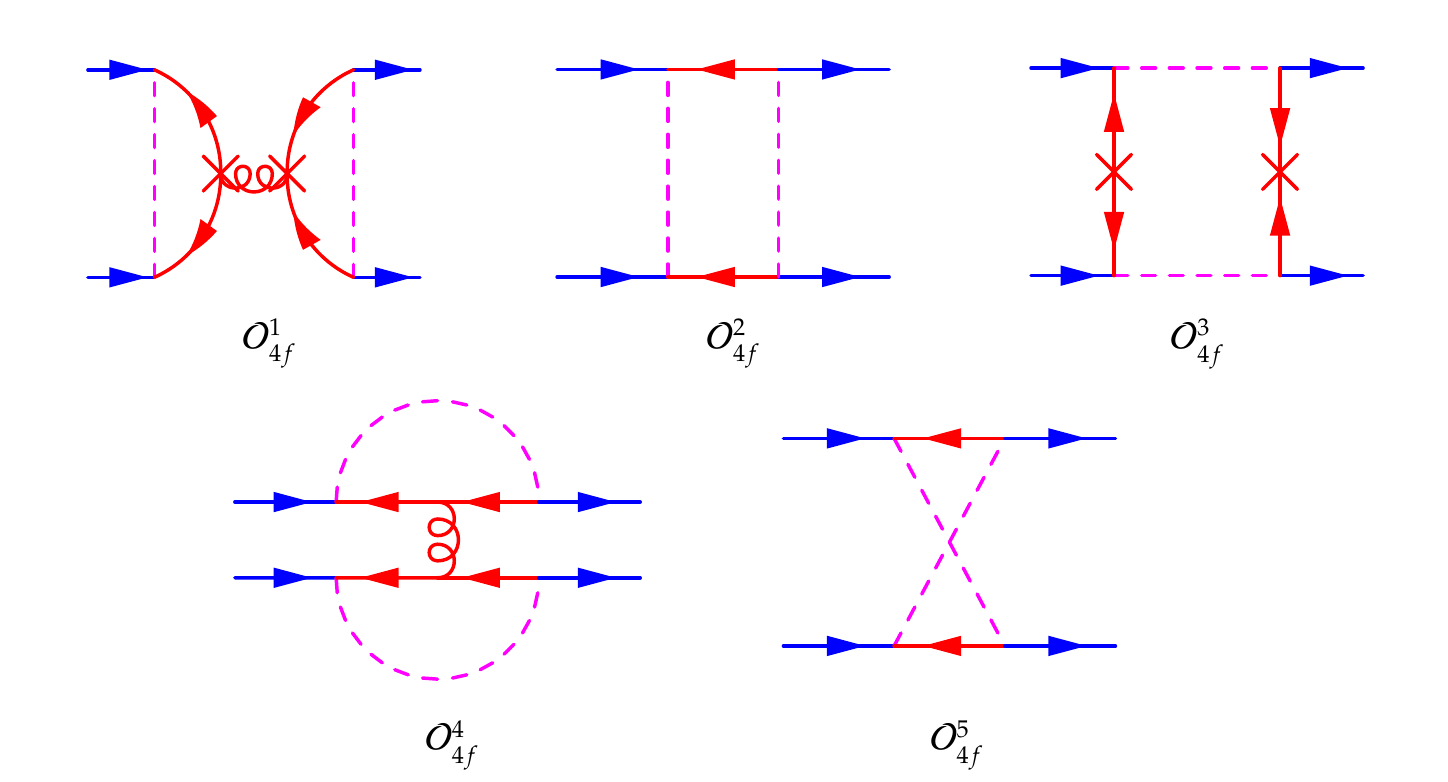}
	\caption{Representative Feynman diagrams corresponding to the operators $\mathcal{O}_{4f}^1-\mathcal{O}_{4f}^5$ in eq. (\ref{eq:fourfermion1}-\ref{eq:fourfermion5}). The blue colored lines are SM fermions, the red colored solid lines are FPC fermions, the red colored curly lines are FPC gluons, and the magenta lines are FPC scalars.} \label{fig:diagra1}
\end{figure}

\begin{figure}
	\includegraphics{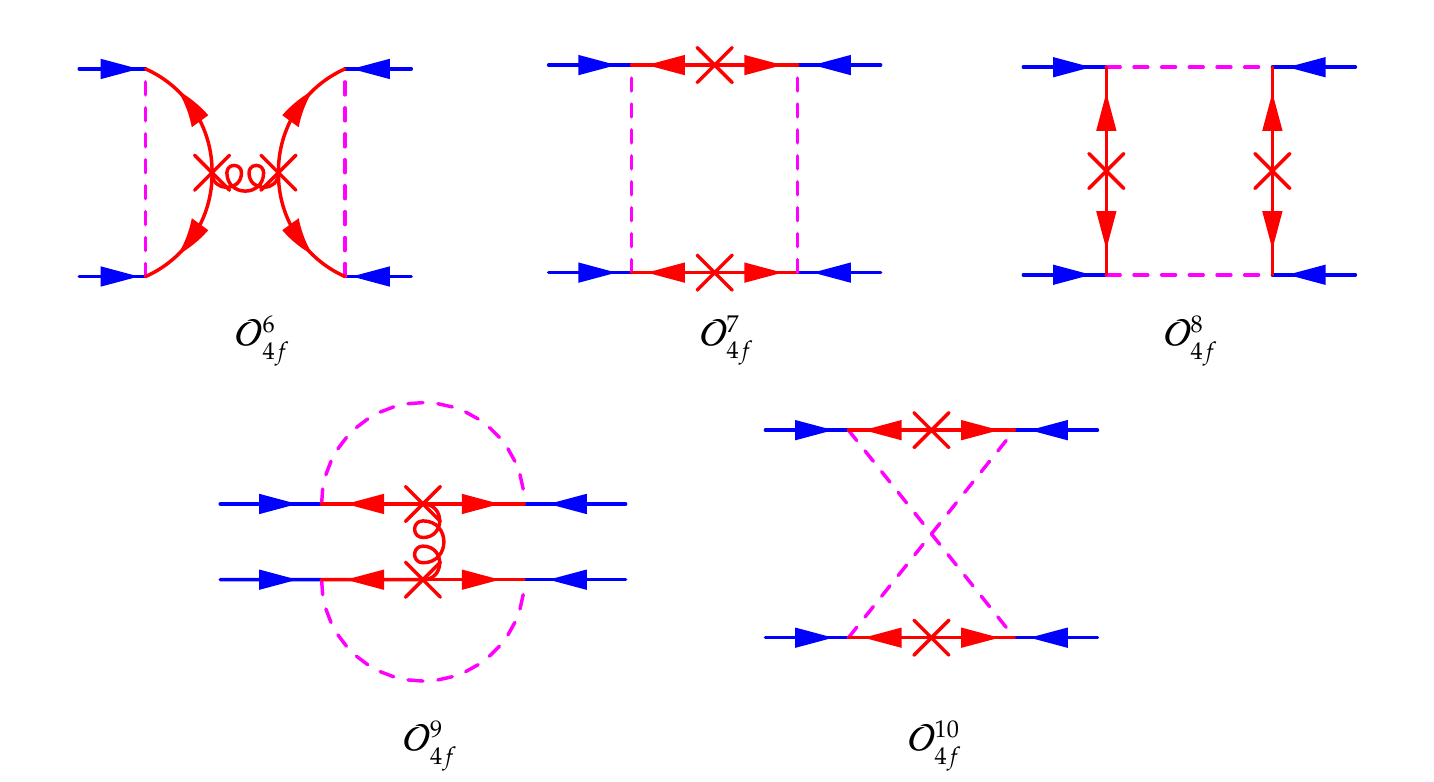}
	\caption{Representative Feynman diagrams corresponding to the operators $\mathcal{O}_{4f}^6-\mathcal{O}_{4f}^{10}$ in eq. (\ref{eq:fourfermion6}-\ref{eq:fourfermion10}). The blue colored lines are SM fermions, the red colored solid lines are FPC fermions, the red colored curly lines are FPC gluons, and the magenta lines are FPC scalars.}\label{fig:diagra2}
\end{figure}

It is useful to represent each of the ten operators $\mathcal{O}_{4f}^{1\dots10}$ in terms of representative diagrams involving  $\tcf$ and $\tcs$ loops, as  shown in Fig.~\ref{fig:diagra1} and \ref{fig:diagra2}, where the ``X'' signifies an insertion of the dynamical FPC-fermion mass, that is proportional to $\Sigma$. Thus the diagrams show how the $ \Sigma  $-dependence occurs in each operator. At a naive perturbative level (these diagrams are only mnemonics) the operators $\mathcal{O}_{4f}^{6-10}$  need mass insertion, while non-perturbatively one obtains operators such as $\mathcal{O}_A$  stemming from instanton corrections. 

The case in which the masses of the scalars are much heavier than $\Lambda_\TC$ is obtained by replacing
	\begin{equation}
	\epsilon_{i j} \to \Lambda_\TC^2 \left( \frac{1}{M_\tcs^2}\right)_{i j}
	\end{equation}
in each operator.  The large mass limit corresponds physically to integrating out the scalars, which in the naive diagrams corresponds to replacing each heavy scalar propagator with the inverse mass matrix. Of course one needs to identify diagrammatically the leading contributions in the inverse scalar mass expansion, as  shown in Fig.~\ref{fig:diagra1} and \ref{fig:diagra2}. 
  
\subsubsection{Other operators}

Other operators up to $\mathcal{O} (p^4)$ include operators deriving from integrating out the SM fermions, i.e. operators that only contains the Yukawa couplings.
Phenomenologically they are important as they include the loop-induced potential for the pNGBs, as well as corrections to the pNGB kinetic terms and contributions to the EWPTs (the S parameter). The complete list can be found in~\cite{Cacciapaglia:2017cdi}.

 \subsection{Top, bottom and EW gauge boson physics in MFPC}
\label{topandbottom}

We now specialize to the most minimal model~\cite{Sannino:2016sfx}, defined by  the choice of gauge group $G_\TC = \SU(2) \sim \Sp(2) $ and $N_\tcf = 4$ Weyl FPC-fermions in the fundamental representation. We start the analysis by studying in detail the minimal FPC-scalar sector to give mass to top and bottom alone. 
The FPC-scalar sector, therefore, only contains a single field $\tcs_t$, with quantum numbers  summarized in Table~\ref{tab:TC_states-top}: the global symmetry is  $\Sp(6)$ since $N_\tcs =3$. 

\begin{table}[t]
	\centering
	\begin{tabular}{c|ccc|ccc}
		& $\SU(3)_c$ & $\SU(2)_\LL$ & $\UU(1)_Y$ &  $ \UU(1)_B $ & $\SU(4)_\tcf$  & $\Sp (6)_\tcs$\\
		\hline
		$\f_Q$ & 1 & {\tiny\yng(1)} & 0 & & & \\
		$\f_u$ & 1 & 1 & $-\frac{1}{2} $ & 0& {\tiny\yng(1)} &  1 \\
		$\f_d$ & 1 & 1 & $\frac{1}{2}$ & &  & \\
		\hline
		$\mathcal{S}_t$
		 & $\overline{{\tiny\yng(1)}}$ & 1 & $-\frac{1}{6}$ &  $ -\tfrac{1}{3} $ & 1 &  {\tiny\yng(1)} \\
		\hline
		$Q_3$ & {\tiny\yng(1)} & {\tiny\yng(1)} & $\frac{1}{6}$ &  $\tfrac{1}{3}$ & &\\
		$u_3$ & $\overline{{\tiny\yng(1)}}$ & 1 & $-\frac{2}{3}$  &$ -\tfrac{1}{3} $ & &\\
		$d_3$ & $\overline{{\tiny\yng(1)}}$ & 1 & $\frac{1}{3}$  &$ -\tfrac{1}{3} $ & &\\
	\end{tabular}
	\caption{Fundamental technicolor states with their gauge quantum numbers and global symmetries. The table includes the 3rd generation quarks too, and the charge assignment under the baryon number $\UU(1)_B$.}
	\label{tab:TC_states-top}
\end{table}

At the fundamental Lagrangian level, the new Yukawa couplings with the SM fields read: 	
	\begin{equation}
	\mathcal{L}_{\rm {top-bottom}} = y_{Q_3} \  Q_{3,\alpha} \tcs_t \epsilon_\TC \tcf_Q^{\alpha} - y_{t} \ u_3 \tcs_t^{\ast} \tcf_d +  y_{b} \  d_3 \tcs_t^{\ast} \tcf_u  \hc \ ,
	\label{eq:Ltop_y}
	\end{equation}
where $ \alpha $ is the $ \SU(2)_\LL $ index, and $u_3$ and $d_3$ are the left-handed spinors constructed out of the charge-conjugate right-handed top and bottom singlets. The above Yukawa interactions can be written in the compact form of Eq.~\eqref{eq:fund_Yukawa} by defining a spurion
	\begin{equation}
	\spur{i}{\,a}= \left( \begin{array}{cccc}
	0 & 0 & y_{b}\  d_3 & - y_{t}\ u_3 \\
	y_{Q_3}\ q_3^{(d)} & -y_{Q_3}\ q_3^{(u)} & 0 & 0
	\end{array} \right)\,,
	\end{equation}
where each row transforms as  anti-fundamental of $ \SU(4)_\tcf $ and each column as a fundamental of   $ \Sp(6)_\tcs $~\footnote{The implicit QCD color indices of the quarks are embedded as part of $\Sp(6)$.}. Note that $ Q_{3,\alpha} = \varepsilon_{\alpha \beta} Q_3^\beta = (-q_3^{(d)}, q_3^{(u)})$ transforms as an anti-doublet of $ \SU(2)_\LL $, while $ (y_{b} d_3, -y_{t} u_3) $ as a doublet of $ \SU(2)_\RR $, consistently with the decomposition of an $ \overline{ {\tiny\yng(1)} } $ of $ \SU(4)_\tcf $.

The operator $ \mathcal{O}_{\mathrm{Yuk}} $, in Eq. \eqref{eq:OH},  is responsible for the generation of the top and bottom masses:
	\begin{equation}
	\mathcal{L}_{\rm EFT} \supset 
	 -C_{\mathrm{Yuk}}\ v \  \left( y_{Q_3} y_{b}\ q_3^{(d)} d_3 + y_{Q_3} y_{t }\ q_3^{(u)} u_3 \right)\left(1 + \dfrac{c_\theta h} {v} + \dots \right) \hc
	\label{eq:top_bot_yukawa}
	\end{equation}
where we can identify 
	\begin{equation} \label{eq:topmass}
	m_t = \abs{C_{\mathrm{Yuk}}\ y_{Q_3} y_{t}} v \andeq m_b = \abs{C_{\mathrm{Yuk}}\  y_{Q_3} y_{b}} v \ . 
	\end{equation}

\subsubsection{Couplings of the $Z$ to the bottom quark} \label{sec:Zbb}

We now turn to the operator in Eq.~\eqref{eq:OPif}, that generates corrections to the gauge couplings of the massive gauge bosons to fermions:
\begin{multline} \label{eq:Zbb}
\mathcal{O}_{\Pi f} =  \frac{g}{2 \cos \theta_W} \dfrac{f_\TC}{\Lambda_\TC} s^2_\theta\;  Z_\mu \left( |y_{Q_3}|^2\ (\bar{t}_\LL \gamma^\mu t_\LL - \bar{b}_\LL \gamma^\mu b_\LL) + |y_{b}|^2\  \bar{b}_\RR \gamma^\mu b_\RR - |y_{t}|^2\ \bar{t}_\RR \gamma^\mu t_\RR \right) \\
- \frac{g}{\sqrt{2}} \dfrac{f_\TC}{\Lambda_\TC} s^2_\theta\;  W^+_\mu \left( y_{b}^\ast y_{t}\ \bar{t}_\RR \gamma^\mu b_\RR - |y_{Q_3}|^2\ \bar{t}_\LL \gamma^\mu b_\LL \right) \hc
\end{multline}
where the SM top and bottom are in the usual Dirac spinor notation.
While the couplings of the top to the $ Z $ are poorly constrained, and $y_b$ can be taken small to reproduce the bottom mass, the coupling of the left-handed bottom to the $Z$ receives sizeable corrections proportional to $|y_{Q_3}|^2$. The well known issue is that $y_{Q_3}$ coupling cannot be too small, as it enters the formula for the top mass.
Imposing the latest constraints~\cite{Baak:2014ora,Gori:2015nqa}, we obtain the $2\sigma$ limit~\footnote{For all our numerical estimates we have used $ \Lambda_\TC = 4\pi f_\TC $.}~\footnote{ Please note that all bounds found here, are on the effective rather than the fundamental Yukawa parameters.}: 
\begin{equation}
 C_{\Pi f} |y_{Q_3}|^2 s^2_\theta < 0.043\,, \qquad \mbox{@ 95\% CL}\,. 
\end{equation}
This constraint mainly comes from the measurement of $R_b$ at LEP~\cite{ALEPH:2005ab}.
The constraint on $\theta$ from electroweak precision tests tends to ease the tension, as $ s^2_\theta \lesssim 0.1$ is generically required~\cite{Arbey:2015exa}. Furthermore, it is possible to obtain the correct top mass with a small $y_{Q_3}$ by maximising the right-handed mixing $y_t$, i.e. assuming that the right-handed top is more composite than the left-handed part.
Using Eq.~\eqref{eq:topmass}, the above bound translates into the following lower bound on the right-handed top mixing:
\begin{equation} \label{eq:boundZbb}
|y_t| \frac{\abs{C_{\rm Yuk}}}{\sqrt{C_{\Pi f}}} \gtrsim \frac{m_t}{f_\TC} \frac{1}{\sqrt{0.043}} = \frac{10~\mbox{TeV}}{\Lambda_\TC}\,,
\end{equation}
which, for reasonably low scale compositeness, $ \Lambda_\TC = 10 $ TeV, and $ C_\mathrm{Yuk} = C_{\Pi f} = 1 $, corresponds to the bound $ \abs{y_t} \gtrsim 1 $. This implies that the fundamental Yukawa coupling would have to be larger than $ 2\sqrt{\pi} \sim 3.5$. It should be mentioned that it is enough that one (or both) strong coefficient departs from unity by a factor of a few to lower the bound on the fundamental Yukawa coupling, thus allowing for perturbative values at the condensation scale.

A possible concern for the model is that the fundamental Yukawa couplings may become large enough that they cause an unwanted condensate, $ \langle f \tcf \rangle $, to form between technifermions and SM fermions (breaking both $ \SU(3) $ color and TC). An approximate Schwinger-Dyson analysis for a Yukawa model with $ \SU(2)_\LL \times \SU(2)_\RR $ symmetry would indicate that such a condensate only forms for $ y \gtrsim O(2 \pi) $~\cite{Tanabashi:1989sz,Kondo:1993ty}. If this estimate is also valid in the MFPC, it would suggest that the model  may be safe from forming a Yukawa induced condensate. However, further work is required to verify this, e.g. on the lattice. Once more, suitable $ \mathcal{O}(\mathrm{few}) $ strong coefficients would render this potential problem mute.

\subsubsection{Bounds on top effective interactions}

Potentially strong bounds also arise from the four-fermion interactions generated by the 8 operators we listed above. Expanding such operators, we find four-fermion operators in terms of the SM fields: interestingly, such terms cannot be matched to the Warsaw basis~\cite{Grzadkowski:2010es} because our theory contains non-linearities in the Higgs field. Effectively, this gives us the Wilson coefficient for each operator in terms of the fundamental Yukawa couplings, the scale of strong dynamics $ \Lambda_\TC $, and the coefficients $ C_{4f}^i $ of the strong dynamics

The phenomenologically relevant operators involve four tops, as they are directly probed at the LHC in four top final states, such as 
	\begin{equation}
	\mathcal{L}_{\mathrm{EFT}} \supset \dfrac{C_{4f}^4 + C_{4f}^5}{4 \Lambda^2_\TC } \abs{y_{t}}^4 (\bar{t}_\RR \gamma^{\mu} t_\RR) (\bar{t}_\RR \gamma_{\mu} t_\RR)  = \dfrac{C_{4f}^4 + C_{4f}^5}{4 \Lambda^2_\TC } \abs{y_{t}}^4 O_{uu}^{3333}, 
	\end{equation}
where the four $3$'s refer to the generation of each of the four fermions. ATLAS~\cite{ATLAS:2016btu} puts an upper limit on this operator at 95\% CL, yielding the constraint:
	\begin{equation}
	\dfrac{\abs{C_{4f}^4 + C_{4f}^5}}{4 \Lambda^2_\TC } \abs{y_{t}}^4 < 2.9~\mathrm{TeV}^{-2} \quad \Rightarrow \quad \abs{C_{4f}^4 + C_{4f}^5}^{1/4} |y_t| < 5.8\, \left( \frac{\Lambda_\TC}{10~\mbox{TeV}} \right)^{1/2}\,, \quad \mbox{@ 95\% CL}\,.
	\end{equation}
The above upper bound is compatible with the lower bound in Eq.~\eqref{eq:boundZbb}, and the situation improves significantly for increasing values of $\Lambda_\TC$.

In addition to the four fermion interactions, the operators $ \mathcal{O}_{fW}$ and $ \mathcal{O}_{fG} $, in Eqs \eqref{eq:OfW} and \eqref{eq:OfG}, give rise to new dipole interactions between gauge fields and the top, which are also probed at the LHC. However, currently the constraints are weaker than the one from the four-fermi interaction.

\subsection{Extension to light generations and leptons} \label{sec:light_gens}

The fundamental Lagrangian can be expanded to include all three generations of quarks and leptons. The minimal strategy~\cite{Sannino:2016sfx} is to extend the FPC-scalar sector by three extra un-colored scalars $\tcs_l$ to couple to the three generations of leptons, and two extra colored scalars $\tcs_u$ and $\tcs_c$ (corresponding to 6 complex scalars) to couple to the two light quark generations. In total, therefore, we have $N_\tcs = 12$ complex scalars, which enjoy a global $\Sp(24)_\tcs$ symmetry. The quantum numbers of both the TC and the SM fields are summarized in Table~\ref{tab:TC_states}. 

\begin{table}
	\centering
	\begin{tabular}{c|ccc|cccc}
		& $ \SU(3)_c $ & $ \SU(2)_\LL $ & $ \UU(1)_Y $ & $ \UU(1)_B $ & $ \UU(1)_\ell $ & $ \UU(3)_{g_1} $ & $ \UU(3)_{g_2} $\\
		\hline
		$\f_Q$ & 1 & {\tiny\yng(1)} & 0 & & & & \\
		$\f_u$ & 1 & 1 & $-\frac{1}{2} $ & 0 & 0 &  1 & 1\\
		$\f_d$ & 1 & 1 & $\frac{1}{2}$ & &  & & \\
		\hline
		$\mathcal{S}_q$ & $\overline{{\tiny\yng(1)}}$ & 1 & $-\frac{1}{6}$  & $ -\tfrac{1}{3} $ & 0 & {\tiny\yng(1)} & 1 \\
		$\mathcal{S}_l$ & 1 & 1 & $\frac{1}{2}$ &0 & $ -1 $ & 1 & {\tiny\yng(1)}  \\
		\hline
		$Q$ & {\tiny\yng(1)} & {\tiny\yng(1)} & $\frac{1}{6}$ & $ \tfrac{1}{3} $ & 0 &&\\
		$u$ & $\overline{{\tiny\yng(1)}}$ & 1 & $-\frac{2}{3}$ & $ -\tfrac{1}{3} $ & 0 &&\\
		$d$ & $\overline{{\tiny\yng(1)}}$ & 1 & $\frac{1}{3}$ & $ -\tfrac{1}{3} $ & 0 &&\\
		$L$ & 1 & {\tiny\yng(1)} & $-\frac{1}{2}$ & 0 & $ 1 $&& \\
		$e$ & 1 & 1 & $-1$ & 0 & $ -1 $ &&\\
		$\nu$ & 1 & 1 & $0$ & 0 & $ -1 $ && 
	\end{tabular}
	\caption{Fundamental technicolor states and SM fermions with their SM gauge quantum numbers. The table also includes the charge assignments under the baryon and lepton number $ \UU(1)_{B,\ell} $.}
	\label{tab:TC_states}
\end{table}

The complete Yukawa interactions now read~\footnote{Note that the scalars are in the conjugate representation of $ G_\SM $ as compared to the minimal model suggested in ~\cite{Sannino:2016sfx}.}:
\begin{multline}
	\mathcal{L}_{\mathrm{yuk}} =  y_Q\  Q_{\alpha} \tcs_q\epsilon_{\TC} \tcf_Q^{\alpha} -  y_u\  u \tcs_q^{\ast} \tcf_d +  y_d\  d \tcs_q^{\ast} \tcf_u + \\
	 y_\ell\  L_\alpha \tcs_l \epsilon_{\TC} \tcf_Q^\alpha -  y_\nu\  \nu \tcs_l^{\ast} \tcf_d +  y_e\  e \tcs_l^{\ast} \tcf_u -\tilde{y}_\nu\  \nu \tcs_l \epsilon_\TC \tcf_u \hc
	\label{eq:Lfund_y}
	\end{multline}
where each coupling is a $3\times 3$ matrix in flavor space, and the flavor indices are left implicit for readability. Table~\ref{tab:TC_states} also contains the symmetries $ U(3)_{g_{1,2}} $ corresponding to global approximate flavor symmetries between the 3 generations of each FPC scalar. Additionally the full model still preserves a Baryon number symmetry as does the SM. However, the lepton number symmetry is explicitly violated by the coupling $ \tilde{y}_\nu $, not surprisingly as the inclusion of such a coupling gives rise to a Majorana mass term for the right-handed neutrinos.   

Just as in the case of the top and bottom, the Yukawa interactions can be written in the more compact form from eq. \eqref{eq:fund_Yukawa} by defining the spurion field $\psi$ as (color and generation indices are, once again, left implicit)
\begin{equation}
	\spur{i}{a}=\begin{pmatrix}
	0 & 0 & y_d d & -y_u u \\
	0 & 0 & y_e e & -y_\nu \nu \\
	y_Q q^{(d)} & - y_Q q^{(u)} & 0 & 0 \\
	y_\ell l^{(e)} & - y_\ell l^{(\nu)} & \tilde{y}_\nu \nu & 0
	\end{pmatrix},
	\label{eq:spur_light_gens}
\end{equation}
where $a\in \SU(4)_\tcf$ and  $i\in \Sp (24)_\tcs$. The hierarchy of the fermion masses can be encoded either in the fundamental Yukawa couplings or in a hierarchy in the mass spectrum of the FPC-scalars.
The phenomenology of the two scenarios is different for the low energy flavor observables as well as for the spectrum of the massive composite states of the theory.
It's noteworthy that, thanks to the compact spurion form, the effect of the light generations can be expressed in terms of the same operator basis we used for the top/bottom case. Of course, at the EW scale the effect of light quarks will be negligible, as they are suppressed by the small effective Yukawas (or scalar masses). At lower energies, modifications of the couplings of the light quarks and leptons are strongly constrained by flavor physics: these effects have been studied in detail in~\cite{Sannino:2017utc}, where the potential to explain the anomalies in lepton universality in $B$--meson physics are also destablished.

%
%

\subsection{Connecting the different approaches to partial compositeness}\label{Connection}

Here we sketch the connection among the different approaches used so far for building SM extensions featuring partial compositeness.   We  start with effective approaches, based either on the construction of an EFT or extra dimensional implementations and finally, we discuss how to link FPC to purely fermionic underlying theories of PC.

The most popular approach to composite Higgs models in the literature has been to construct EFTs simply based on the symmetry breaking patterns (see ~\cite{Marzocca:2012zn,Panico:2015jxa} for a pedagogical introduction), without any reference to the underlying theory. 
As a consequence, to implement partial compositeness, the choice of the representation under which the top partners transform has been arbitrary. 
Furthermore, top partners in the EFT approach have been assumed to be the main driving force in the stabilization of the vacuum alignment along the small-$\theta$ limit: this mechanism can only work if the top partners are light~\cite{Redi:2012ha,Pomarol:2012qf} and the contribution to the pNGB potential is dominated by their loops.  Accepting the lightness of top partners with respect to the natural resonance scale, i.e. $\Lambda_\TC \sim 4 \pi f_\TC$, one is justified to include them in the EFT construction. Note however that top partners are not necessarily the only contributors to the Higgs potential~\cite{Katz:2005au,Galloway:2010bp,Cacciapaglia:2014uja}.

In the case of minimal FPC, the representation of the top partners is fixed to be the fundamental of the global symmetry $\SU (4)$. This choice has been considered problematic in the literature, as it typically leads to large corrections to the $Z$ coupling to bottoms. However, as we will see shortly, this problem only applies if the top partners are light. It is instructive to compare our general operator approach presented in Section~\ref{topandbottom} with the results one would obtain by adding the top partners to the EFT. The couplings of the top partners, that we collectively call $\mathcal{B}$, to the SM fermions can be written as:
\begin{equation}
\mathcal{L}_{\rm PC} = - \bar{y}_{Q_3}^{\rm EFT} f_\TC \ \bar{\psi}_{Q_3} \cdot \Sigma^\dagger \cdot \mathcal{B}_R - \bar{y}_t^{\rm EFT} f_\TC\ \bar{\mathcal{B}}_L \cdot \Sigma \cdot \psi_t\,, \quad \mbox{with}\;\; \psi_{Q_3} = \left( \begin{array}{c} Q_3 \\ 0 \\ 0 \end{array} \right)\;\; \mbox{and} \;\; \psi_t = \left( \begin{array}{c} 0 \\ u_3 \\ 0 \end{array} \right)\,,
\end{equation}
where the SM fermions are embedded into spurions transforming as the fundamental of $ \SU(4)_\tcf $. The symmetries associated to the scalars $\tcs$ are thus ignored. The mass of the top can be obtained by diagonalising the resulting mass matrix, yielding
\begin{equation}
m_t = 2 M_\mathcal{B} s_\theta\ 
\frac{{\bar{y}^{\rm EFT}}_{Q_3}{ f_\TC}}  {\sqrt{M_\mathcal{B}^2 
+{ \bar{y}^{{\rm EFT}^2}_{Q_3}}{ f_\TC}^2
}}
\frac{{\bar{y}^{\rm EFT}}_t f_\TC}{\sqrt{{M_\mathcal{B}}^2 + \bar{y}^{{\rm EFT}^2}_t {f_\TC}^2}} + \dots
\end{equation}
 where the dots stand for higher orders in an expansion for small $s_\theta$. This equation should be compared to Eq.~\eqref{eq:topmass}. We see that the two results coincide once we identify
\begin{equation} \label{eq:EFTmatching}
y_{Q_3/t} \dfrac{\sqrt{f_\TC}}{\sqrt{\Lambda_\TC}} \to \frac{\bar{y}^{\rm EFT}_{Q_3/t} f_\TC}{\sqrt{{M_\mathcal{B}}^2 + {\bar{y}^{{\rm EFT}^2}_{Q_3/t}} f_\TC^2}}\,, \quad C_{\rm Yuk} \Lambda_\TC \to 2 M_\mathcal{B}\,. 
\end{equation}
We see that the operator estimate matches if the mass of the top-partners is at its natural value $M_\mathcal{B} \sim \Lambda_\TC$.
The mixing between SM fermions and top partners induces corrections to the gauge couplings of the top and bottom to the massive $W$ and $Z$ too, due to the fact that the top partners are vector-like fermions~\cite{delAguila:2000rc}. In the bottom sector, we thus obtain:
\begin{equation}
\frac{g}{2 \cos \theta_W}  s^2_\theta\, Z_\mu  \frac{{\bar{y}^{{\rm EFT}^2}_{Q_3}} f^2_\TC}{M_\mathcal{B}^2 + {\bar{y}^{{\rm EFT}^2}_{Q_3}} f_\TC^2} \bar{b}_L \gamma^\mu b_L + \dots
\end{equation}
which nicely compares with Eq.~\eqref{eq:Zbb} once the identification in Eq.~\eqref{eq:EFTmatching} is taken into account.
We see, therefore, that the approach with top partners in the EFT gives the same results as the effective operators we consider, and the two actually coincide if the mass of the top partners is at the natural scale $\Lambda_\TC$. Thus, for heavy top partners, the bound from the $Z$ coupling are not problematic, as we showed in Section~\ref{sec:Zbb}.

Another effective approach to partial compositeness relies on extra dimensions: it is mainly based on adapting the conjectured correspondence of anti-de-Sitter (AdS) space-time with 4-dimensional conformal field theories~\cite{Maldacena:1997re} to non-supersymmetric scenarios. Models based on warped extra dimensions have been used to characterize composite Higgses based on a conformal underlying theory~\cite{Contino:2003ve,Agashe:2004rs}. The light Higgs is  identified with an additional polarization of gauge fields in the bulk, thus borrowing many similarities from Gauge-Higgs unification models~\cite{Hosotani:1983xw,Antoniadis:2001cv} (see also ~\cite{Hosotani:2005nz} in warped space).  
The mechanism of partial compositeness is described by fermions propagating in the bulk of the extra dimensions, as discussed in~\cite{Contino:2004vy,Cacciapaglia:2008bi}.
An extra-dimensional version of the model under study can be easily obtained by promoting the global symmetries $\SU(N_\tcf) \times \Sp (2 N_\tcs)$ to gauge symmetries in the bulk, broken by boundary conditions to the SM on the Planck brane, while on the TeV brane the breaking induced by the fermion condensate, i.e. $\SU (N_\tcf) \to \Sp (N_\tcf)$, is imposed. Composite fermions are represented by bulk fermions transforming as the bi-fundamental of the symmetries, while the mixing of the SM fermions, at the basis of partial compositeness, comes from explicit mass mixings on the Planck brane~\cite{Scrucca:2003ra}.
The theory would thus automatically describe spin-1 resonances in the form of Kaluza-Klein resonances of the gauge bosons.
The advantage of extra dimensions, which is also their limitation, is the fact that the spectrum is determined by the geometry. In the model under consideration, which is not conformal in the UV, the spectrum will hardly match the prediction of a warped extra dimension.

As argued earlier, traditional approaches hope to achieve partial compositeness via pure underlying gauge-fermion realizations. In this case the new composite fermion operators ${\cal B}$, that couple linearly to the SM fermions, must be built out of the underlying gauge-fermion dynamics. This necessarily limits its underlying composition. In addition the need to have the composite fermion operator ${\cal B}$ with a physical dimension such that the operator $ \Psi  {\cal B}$ (with $\Psi$ a generic SM fermion) is either super-renormalisable or marginal further constrains the underlying origin of ${\cal B}$.     Therefore one can schematically build ${\cal B}$  as follows:
\begin{equation}
	{\cal B} \sim {\cal F} {\cal F} {\cal F}, \quad {\cal F} {\cal F} {\cal X}, \quad {\cal F} {\cal X} {\cal X}, \quad {\cal F} {\cal X} {\cal Z}, \quad {\cal F} \sigma^{\mu \nu}{\cal G}_{\mu \nu}, 
\end{equation}
with $ {\cal X} $ and ${\cal Z}$ potentially new FPC-fermions transforming according to different representations of the gauge group and ${\cal G}_{\mu \nu} $ the technicolor field strength. Clearly which technicolor invariant composite operator can actually be built depends on the underlying dynamics. Theories in which ${\cal B}$ is made by an even larger number of fermionic degrees of freedom are strongly disfavoured because of the anomalously large anomalous dimensions that the composite fermion must have for $\Psi {\cal B}$ to be at least a marginal operator. In fact, in \cite{Pica:2016rmv} it has been argued that even realizations with three underlying fermions are challenging. 

 As noted in \cite{Sannino:2016sfx}  because any purely fermionic extension~\cite{Barnard:2013zea,Ferretti:2013kya, Vecchi:2015fma, Ferretti:2014qta} is required to have composite baryons with dimensions close to $5/2$, these baryons would presumably behave as if they were made by a fermion and a composite scalar similar to ours (see also~\cite{Caracciolo:2012je} for a supersymmetric realization).  Naively, at some intermediate energy, our description can be viewed as an effective construction of the purely gauge-fermionic one with 
 
 \begin{equation}
{\cal F}(\Phi) \sim \	{\cal B} \sim {\cal F}( { \cal F} {\cal F}), \quad {\cal F} ({\cal X} {\cal F}), \quad {\cal F} ({\cal X} {\cal X}), \quad {\cal F} ({\cal X} {\cal Z}). 
\end{equation}
 Obviously this identification is just a mnemonic and it means that the composite baryon made by $\tcf \Phi$ can describe, at an intermediate effective level, one of the composite baryons with the same quantum number and physical dimensions. A similar relation can be thought for the ${\cal F} \sigma^{\mu \nu}{\cal G}_{\mu \nu}$ operator.  
  
We can use group theory to investigate related theories.  For example, from Table I of \cite{Belyaev:2016ftv},  we learn that model M6, that features five two-index antisymmetric $\tcf$  under the technicolor gauge group $\SU(4)$ as well as three Dirac fermions in the fundamental representation ${\cal X}$~\cite{Ferretti:2014qta},  gives rise to composite baryons $\tcf {\cal X} {\cal X}$ and   $\tcf \overline{{\cal X}}  \overline{{\cal X}}$. At intermediate energies these composite baryons can be mapped into a fundamental partial composite theory featuring the same $\tcf$ fermions and six two index antisymmetric FC-scalars.



\chapter{Non perturbative Lattice results}
\label{Lattice}
To tackle composite dynamics one can use a plethora of analytic tools from effective field theories to the implementation of global symmetries to quantum constraints such as 't Hooft anomaly conditions, large $N_c$ methods \cite{tHooft:1982uvh,Sannino:2015yxa,Sannino:2007yp} and limits \cite{Sannino:2004qp,Dietrich:2005jn,Dietrich:2006cm}, holography \cite{Maldacena:1997re,Maldacena:1998im,Rey:1998bq,Gursoy:2007cb,Gursoy:2007er}, other non-perturbative methods \cite{Dobado:2019fxe,Sannino:2009za} and last but not the least  QCD-like dynamics as an analog computer \cite{Sannino:2009za}. Nevertheless to learn more about the details of a given strongly coupled theory such as its massive spectrum or to unveil its true dynamical nature we must employ first principle computations such as lattice quantum field theory. 

Therefore the lattice community is actively investigating relevant models for beyond the standard model physics of composite nature. These non-perturbative studies complement the phenomenological approach by providing valuable information on the strongly coupled dynamics.

The numerical studies of many such models present great numerical and theoretical challenges. Beyond the traditional ones \cite{Wiese:2009qsa},  the new ones crucially stem from trying to establish whether the new dynamics is QCD-like, near conformal (e.g. of walking nature), or simply conformal at low energies. Specifically: i) disentangling the (near) conformal nature of the model requires very large volumes; ii) the possible presence of light scalars renders the chiral extrapolation difficult.
Nonetheless, significant progress has been made in the last years thanks to the continuous numerical effort and the development of new techniques .

\begin{figure}[th!] 
    \center 
    \includegraphics[width=.88\textwidth]{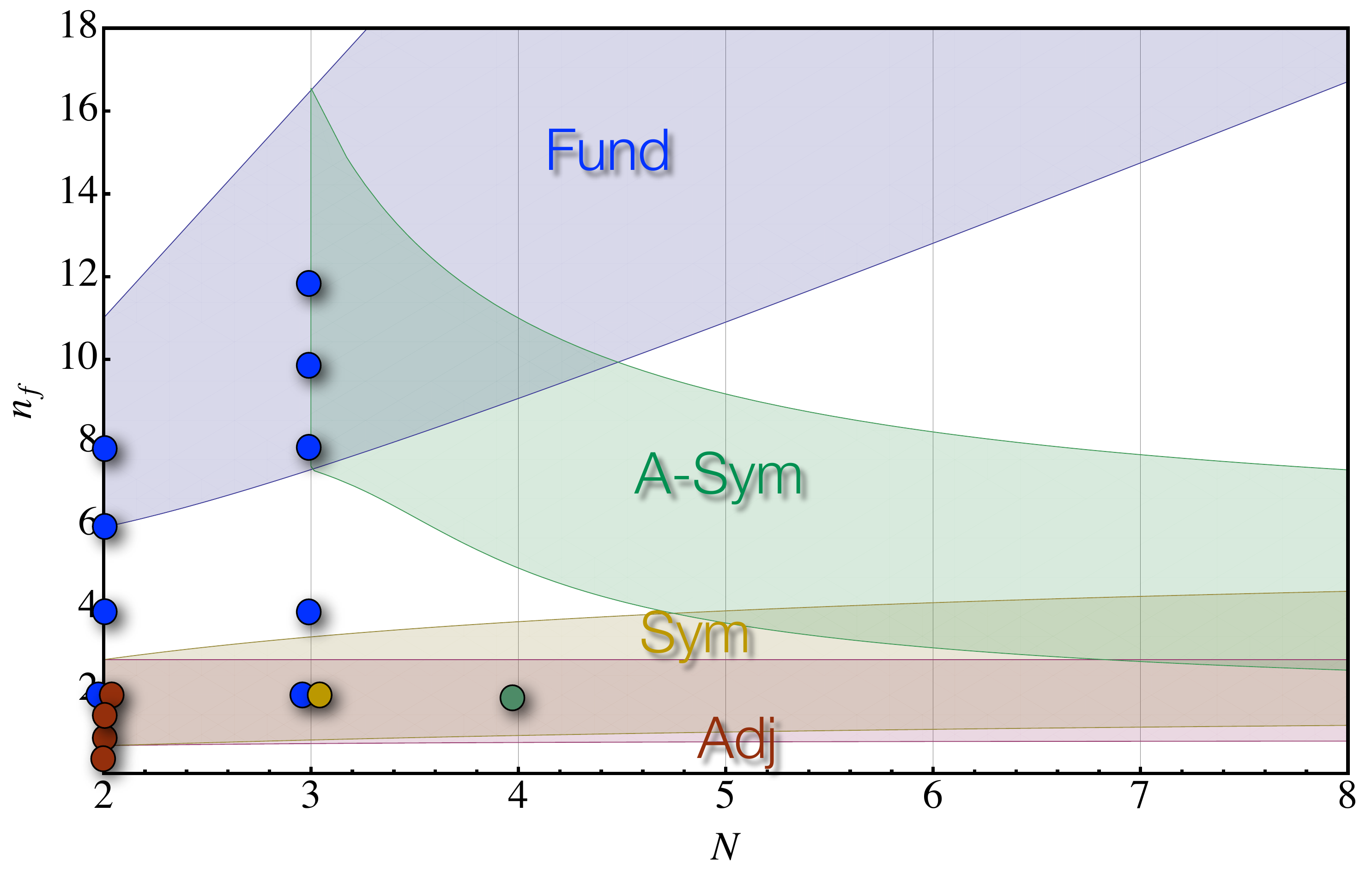} 
    \caption{Chart summarizing the majority of the models of composite dynamics studied by use of numerical lattice gauge theory simulations. The colors correspond to different representations of the condensing fermions.\label{latchart}}
\end{figure}

Many results are available for models based on the gauge groups $\SU(2)$ or $\SU(3)$ with fermions in the fundamental or higher representations (see \reffig{latchart} for a visual summary).
A great effort is undergoing to determine the precise location of the conformal window \cite{Sannino:2004qp,Dietrich:2005jn,Dietrich:2006cm},  as we shall introduce below, for two and three color models with few borderline cases remaining elusive.
Interesting models featuring light composite scalar states are being investigated in detail and many results for the spectrum of these walking TC and/or Goldstone Higgs models are already available.

In this section, we summarize some of the most relevant results from numerical lattice investigations of beyond the SM models.

\section{Conformal Window}

We start by discussing theories that feature a conformal fixed point, which allow therefore to define theories with a conformal window, or walking dynamics.

\subsection{Analytic studies of the Conformal Window}

Determining the phase structure of generic gauge theories of fundamental interactions is crucial in order to be able to select relevant extensions of the SM \cite{Sannino:2009za}. 
In particular, as explained in the previous chapters, it is interesting to find the models which exhibit near-conformal dynamics, possibly very strongly interacting and with large anomalous dimensions.
For this reason determining the phase boundaries between IR-conformal and QCD-like phases is of special interest, and many of the non-perturbative numerical studies performed via Lattice Gauge Theory (LGT) simulations for beyond the SM physics have been dedicated to this effort in the past ten years.
We will present the current status of the LGT simulations later in this chapter.
First it is important to gain a quantitative analytic understanding of the phase structure of different gauge theories by studying the results of perturbation theory (PT).

The existence of an IR conformal phase can be studied in PT by looking at the high-order perturbative expansion of the beta function and of the anomalous dimensions of some significant operators, such as fermion bilinears and baryonic operators, in a given renormalization scheme.
The existence of a physical IR conformal fixed point must not dependent on the choice renormalization scheme, but it is known that truncations of the perturbative expansion and singular choices of the renormalization scheme\footnote{A famous example is the so-called 't Hooft scheme, where the two-loops beta function is exact.} can give rise to spurious, i.e. non-physical, IR zeros of the beta function. 
It is therefore important to study the behavior and stability of the perturbative series at high order. 
Standard PT can also be complemented by other techniques, which can provide non perturbative information, such as the $\Delta$-expansion, described below, or various resummation schemes.

We consider simple models featuring an arbitrary number $n_f$ of species of fermions in a given representation $R$ of the gauge group. 
Despite the fact that models with fermions in more than one representations are the norm when building relevant extensions of the Standard Model, the limitation to only one species of fermions is needed to be able to consider high-orders in PT.\footnote{To circumvent this problem, in \cite{Ryttov:2009yw} a conjectured all-orders beta function was used to estimate the conformal window for multiple representations.}   
We focus on smallest representations up to the two-indices, as higher dimensional representations only describe models which are not asymptotically free.
Such models are also interesting on their own, due to the possible existence of non-trivial ultraviolet (UV) fixed point giving rise to a asymptotically safe scenarios \cite{Litim:2014uca,Sannino:2014lxa,Litim:2015iea,Nielsen:2015una,Molgaard:2016bqf,Bajc:2016efj,Mann:2017wzh,Abel:2017rwl,Antipin:2017ebo,Bajc:2017xwx,Sannino:2018suq}.
However, as all of the analytical and numerical work in the literature considered only fermions in lower dimensional representations, including studies of asymptotic safety, here we will report on this case.

In standard PT, one can investigate the existence of zeros of the beta function to the maximum known order in a renormalization scheme of choice, usually the $\overline{\rm MS}$ scheme.
For models with one species of massless fermions in a single representation of the gauge group, the beta function in the $\overline{\rm MS}$ scheme has been derived up to four loops in \cite{vanRitbergen:1997va,Vermaseren:1997fq,Czakon:2004bu},  with extensions to five loops in \cite{Baikov:2016tgj,Herzog:2017ohr}. The general form of these expressions is:
\begin{eqnarray}
\label{beta}
\frac{d a }{d \ln \mu^2}  =
\beta(a) = -\beta_0 a^2 - \beta_1 a^3
-\beta_2 a^4 
-\beta_3 a^5 -\beta_4 a^6 + O(a^6), \\ 
 \label{gamma}
-\frac{d\ln m }{ d \ln \mu^2}  = 
\frac{\gamma_m (a)}{2}  =  \gamma_0 a +  \gamma_1 a^2
+ \gamma_2 a^3
+ \gamma_3 a^4 + O(a^5),  \hspace{.5cm}
\end{eqnarray}
where $m=m(\mu)$ is the renormalized (running)  fermion mass, 
$\mu$ is  the renormalization scale in the $\overline{\rm MS}$ scheme and
$a=\alpha/4\pi=g^2/16\pi^2$, where $g=g(\mu^2)$, is the renormalized 
coupling constant of the theory. 
The coefficients $\beta_l$ and $\gamma_l$ depend on the number of Dirac fermions $n_f$ and their representation $R$.
The explicit values of these coefficients were extended to generic representations from the original references in \cite{Mojaza:2010cm}, while the explicit formulae for these coefficients can be found in the appendices of \cite{Pica:2010xq, Ryttov:2017dhd}.

By using the perturbative expression in Eq.\eq{beta}, one can study the existence and locations of the zeros of the beta function. Since in PT the beta function is a polynomial in $g^2$, only zeros at positive $g^2$ are considered as possibly physical and several such zeros can exist. For a given representation $R$ the existence and location of the zeros are functions of $n_f$. The general behavior of such functions for generic gauge groups was studies in \cite{Pica:2010xq}.

The most studied scenarios for composite dynamics make use of asymptotically free models close to the boundary of the conformal window.
The latter is defined as the region in parameter space where the model is asymptotically free in the UV and possesses an fixed point of the renormalization group flow in the IR.

The conformal window can be estimated from PT by using Eqs\eq{beta} and \eq{gamma}. For fixed gauge group and fermion representation, the critical number of fermions $n_f^{AF}$ above which asymptotic freedom is lost is known: $n_f^{AF}=11C_A / (4 T_F)$.
It is customary to write expressions such as this, which give rise to a fractional number of fermions, with the understanding that only integer, or half integer for Weyl fermions, are physical.

Just below $n_f^{AF}$, for $\delta\equiv n_f^{AF} - n_f >0$ and small, a ``Banks-Zaks" fixed point \cite{Banks:1981nn} arises at $a^*=\frac{4 T_F}{3C_A(7C_A+11C_F)}\delta+O(\delta^2)$. 
Using higher loop expressions for the beta function \eq{beta}, one can study the behavior of such an IR fixed point as a function of $n_f$ and its stability in terms of the perturbative order of the expansion. 
Such a PT expansion is expected to work well for small $\delta$, however the extent of the region in $n_f$ where the PT expansion can provide sensible information is not known a priori.
Once the location of the relevant fixed point $a^*$ is known from the PT beta function at some order, Eq.\eq{gamma} can be used to estimate the value of the mass anomalous dimension at the IR fixed point, $\gamma^*_m$.
This is a physical quantity, directly relevant for models of beyond the SM physics, and as such its value is independent on the renormalization scheme.
It is known \cite{Mack:1975je} that unitarity poses an upper bound $\gamma^*_m\le 2$. It is believed that $\gamma^*_m\le 1$ in the conformal window and $\gamma^*_m\approx 1$ at its lower edge.

\begin{figure}[p] 
    \begin{center}
    \includegraphics[width=.7\textwidth]{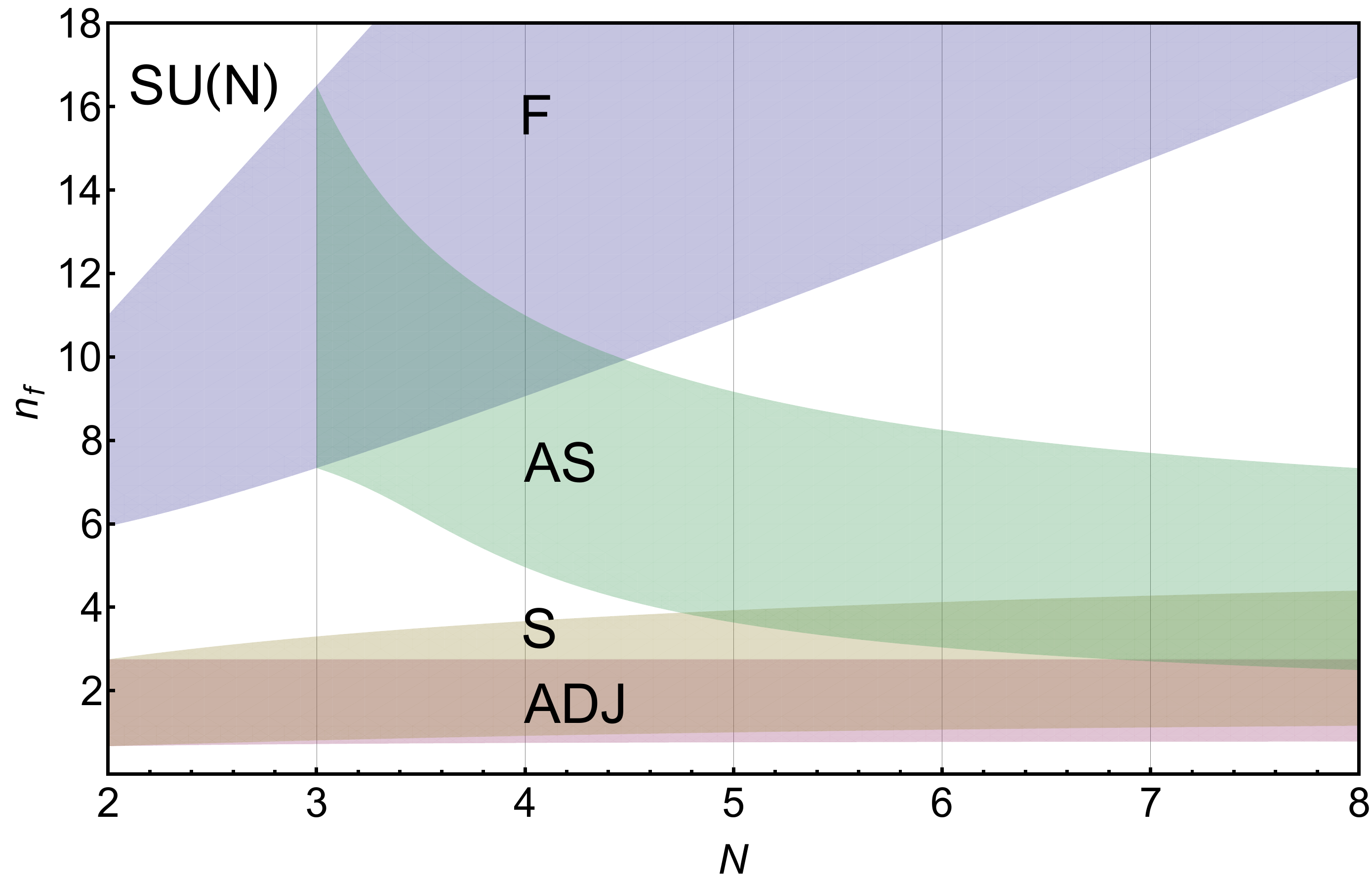}   
    \caption{Example of conformal window from PT for $\SU(N)$ groups for the fundamental representation (upper light-blue), two-index antisymmetric (next to the highest light-green), two-index symmetric (third window from the top light-brown) and finally the adjoint representation (bottom light-pink). The lower boundary corresponds to the point where the infrared fixed point disappears at four loops.}     
    \label{4loop-PD}
    \end{center}  
\end{figure} 
\begin{figure}[p]
    \begin{center}
    \includegraphics[width=.7\textwidth]{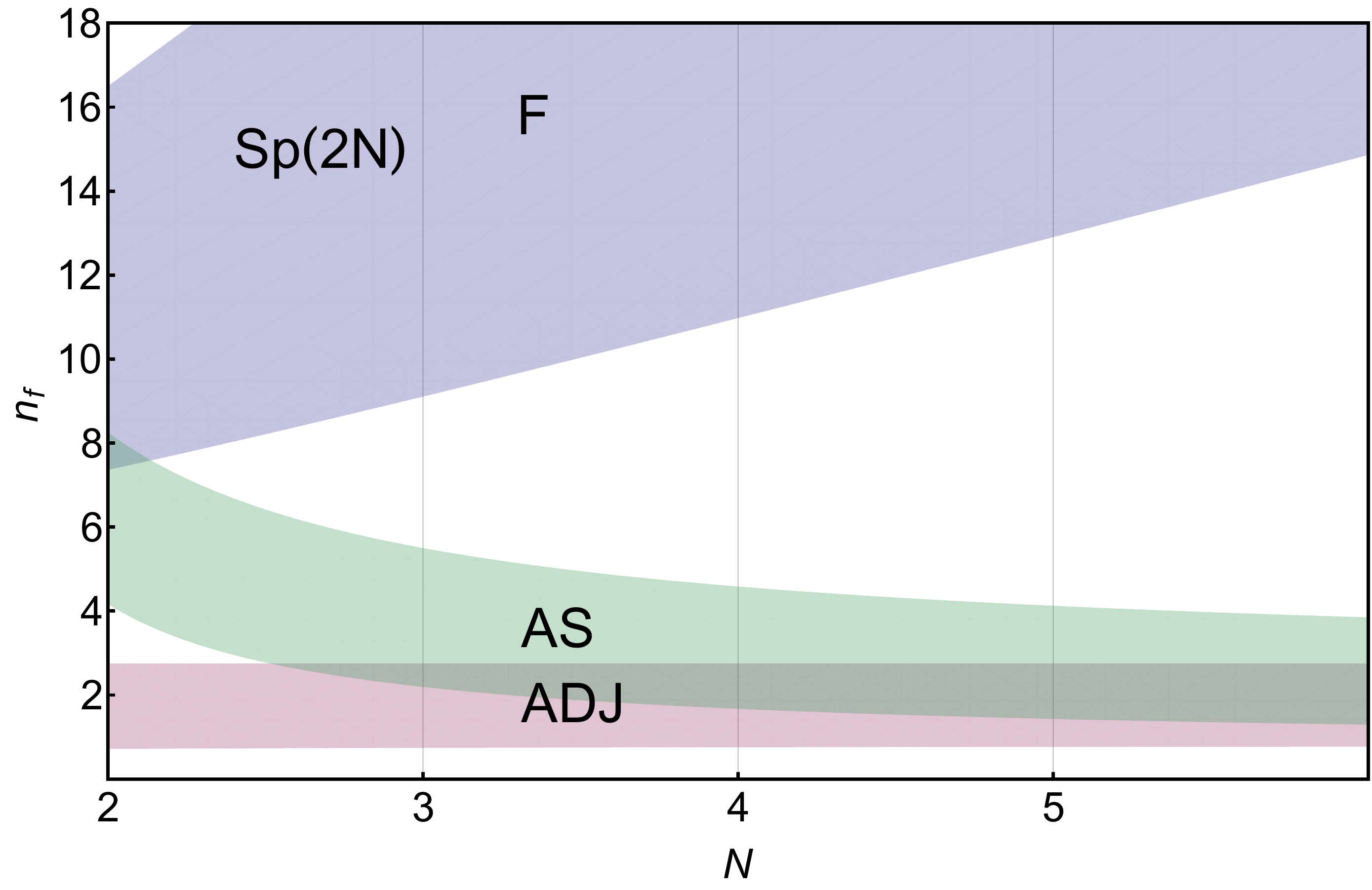}  
    \caption{Conformal window from 4-loops PT for $\Sp(2N)$ groups for the fundamental representation (upper light-blue), two-index antisymmetric (next to the highest light-green), two-index symmetric, i.e. the adjoint,  (bottom window in pink).}    
    \label{SP-4loop-PD}
    \end{center}
\end{figure}

Another physical quantity of interest is the derivative of the beta function at the IR fixed point $\beta'(a^*)$, which is, up to a numerical factor, the anomalous dimension of $\Tr(F_{\mu\nu} F^{\mu\nu})$, where $F_{\mu\nu}$ is the field-strength tensor of the confining strong gauge force.

The predictions of PT for the conformal window were studied in \cite{Ryttov:2010iz,Pica:2010xq,Mojaza:2012zd,Ryttov:2013hka,Ryttov:2013ura,Ryttov:2014nda} at four loops and more recently at five loops in \cite{Ryttov:2016ner} and subsequent works using the $\delta$ expansion discussed below \cite{Ryttov:2016asb,Ryttov:2016hal,Ryttov:2017toz,Ryttov:2017dhd,Ryttov:2017kmx}. 
A typical example of the conformal window obtained from PT at 4-loops is shown in \reffig{4loop-PD} for $\SU(N)$ gauge groups and in \reffig{SP-4loop-PD} for $\Sp(2N)$ groups (figures from \cite{Pica:2010xq}). 
  
While PT works well at the upper end of the conformal window, the problem for the determination of the conformal window is the intrinsic instability of PT in the most interesting region, the lower edge where the models become more strongly interacting.
The instabilities in PT show up as lack of convergence of the perturbative series in quantities such as $a^*$ or as non physical values of e.g. $\gamma^*_m$. 
    
To overcome this problem, different approaches have been developed, which try to capture (some of) the relevant non-perturbative physics. These include: the truncated Schwinger-Dyson equation in the so-called ladder approximation \cite{Pagels:1974se,Dietrich:2006cm}; the use a conjectured inequality between the thermal degrees of freedom in the UV and IR \cite{Appelquist:1999hr,Sannino:2005sk}; a conjectured all-orders beta function \cite{Ryttov:2007cx,Pica:2010mt}; more recently the use of the $\delta$ expansion \cite{Ryttov:2016hdp}.  
  
It is worth summarizing the properties of the most recent proposal of the $\delta$ expansion \cite{Pica:2010mt,Ryttov:2016hdp} here. 
This is a series expansion of a physical quantity, such as the anomalous dimensions at the IR fixed point, in terms of the physical parameter $n_f$ centered at the point where asymptotic freedom is lost, i.e. at $n_f=n_f^{AF}$.
The resulting power series is therefore in powers of $\delta$.
A remarkable property is that, contrary to ordinary PT, the exact n$^{th}$ order coefficient of the $\delta$ series can be computed by using ordinary PT from only the first n$^{th}$+1 loop orders of the ordinary perturbative series. 
The resulting series is therefore exact, to all loops in PT, and it is scheme independent, as it should also be obvious from its definition.
Another remarkable property is that the series, as tested in \cite{Ryttov:2016hdp}, appears to be rapidly convergent, in stark contrast to ordinary PT.
The resulting series is therefore much better behaved and can be used to study physical quantities inside the conformal window, and to estimate its lower edge.

\begin{figure}[ph!]
     \center 
     \includegraphics[width=.48\textwidth]{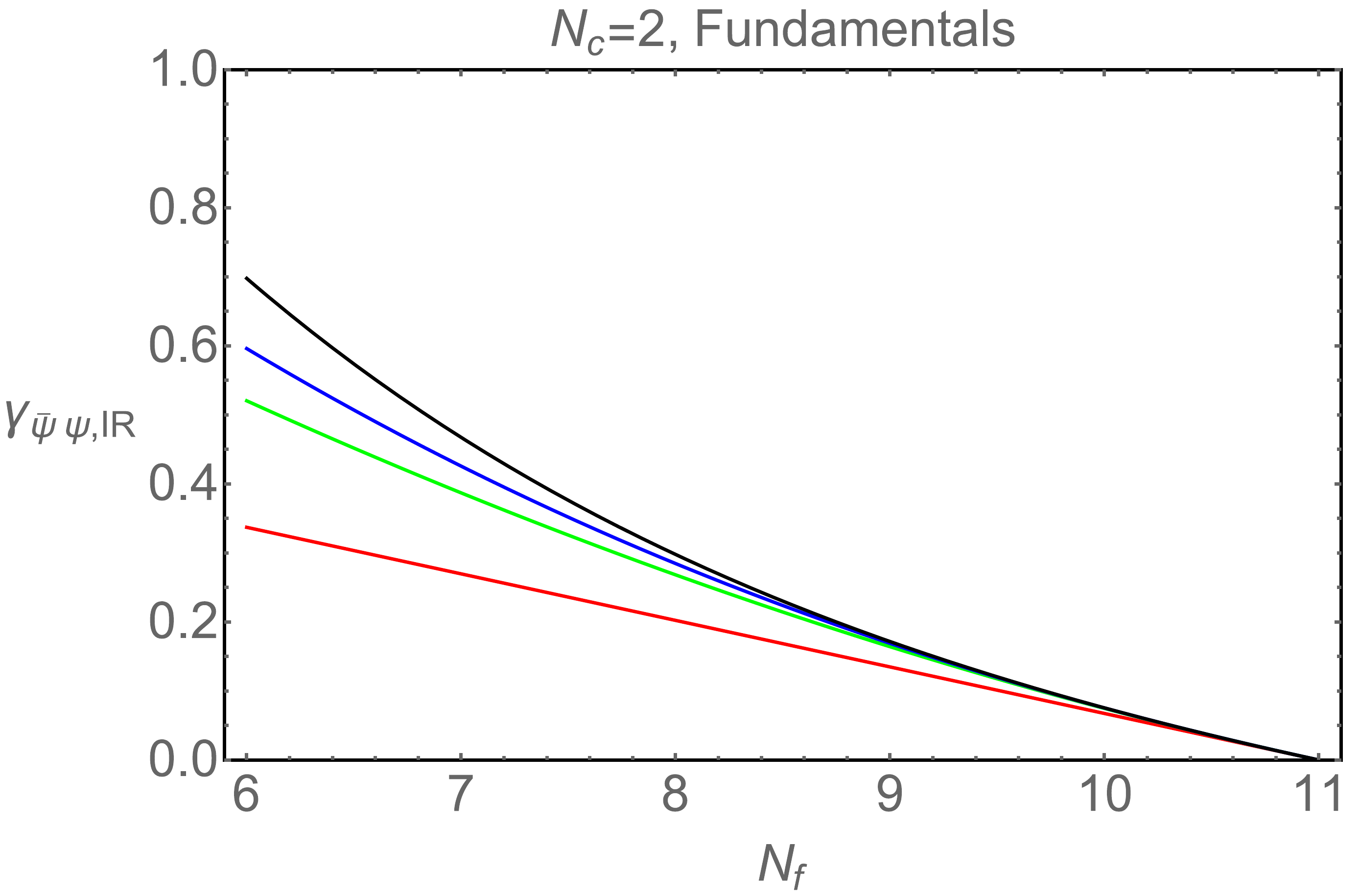} 
     \includegraphics[width=.48\textwidth]{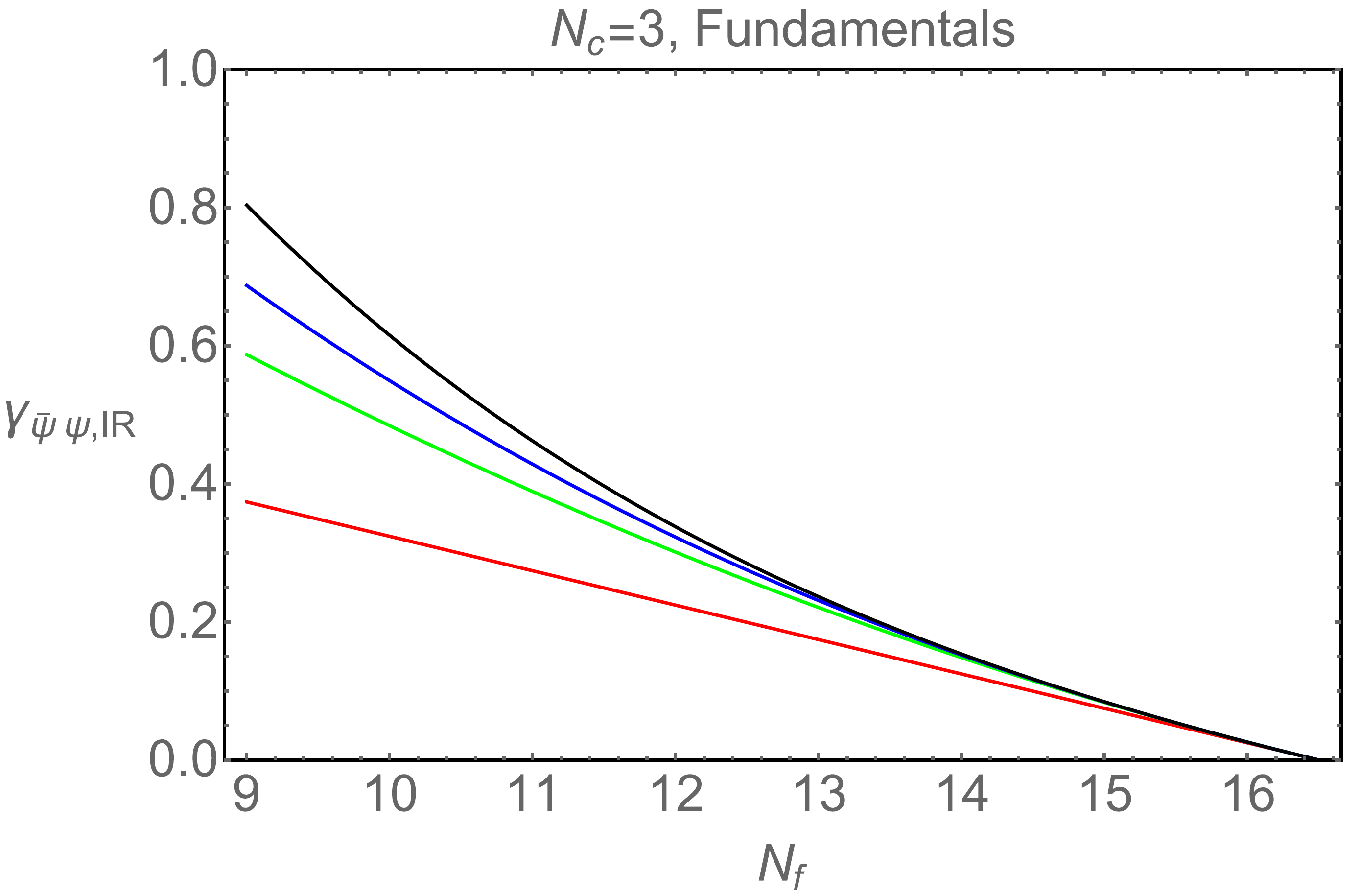}
     \includegraphics[width=.48\textwidth]{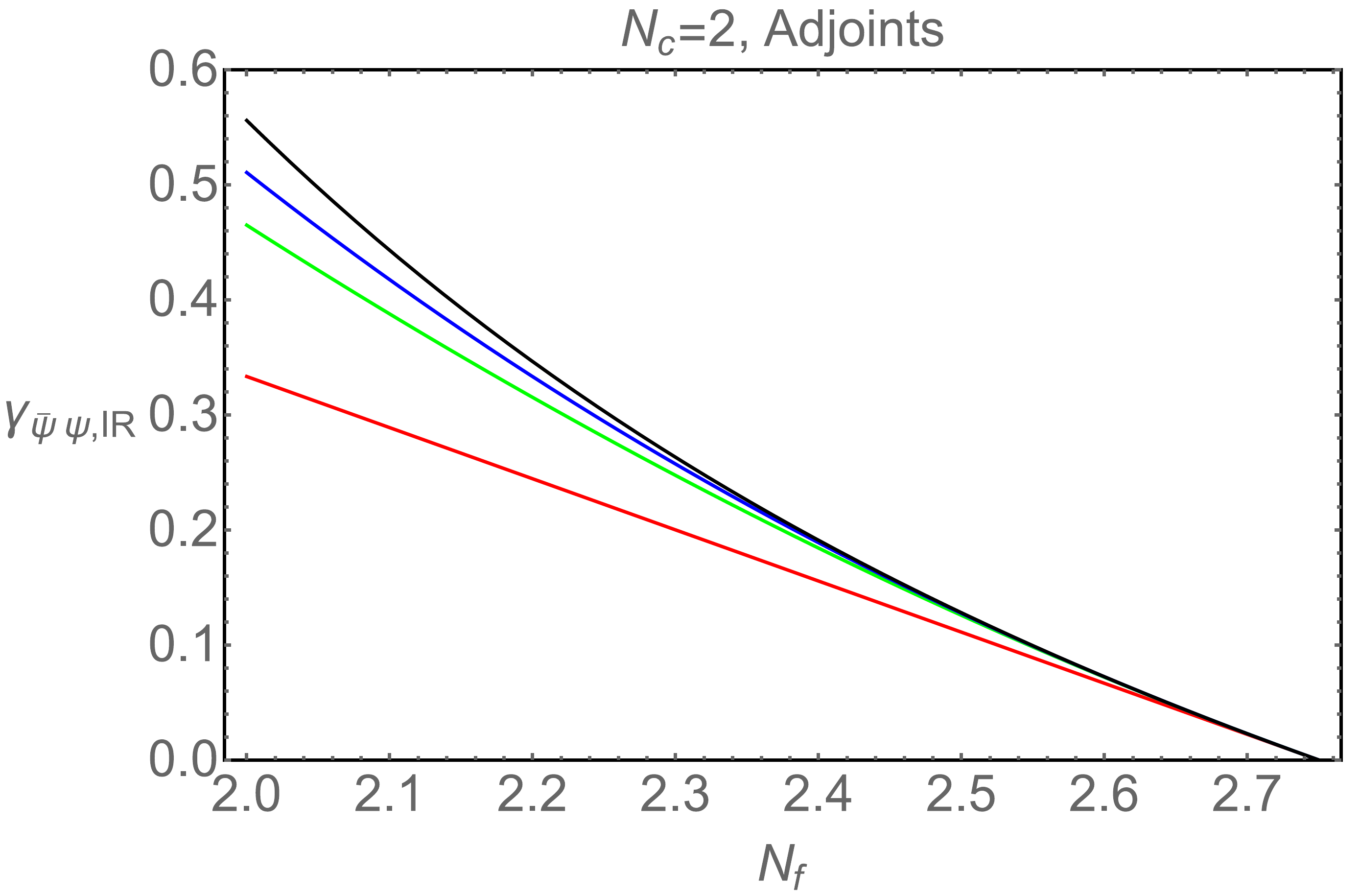}
     \includegraphics[width=.48\textwidth]{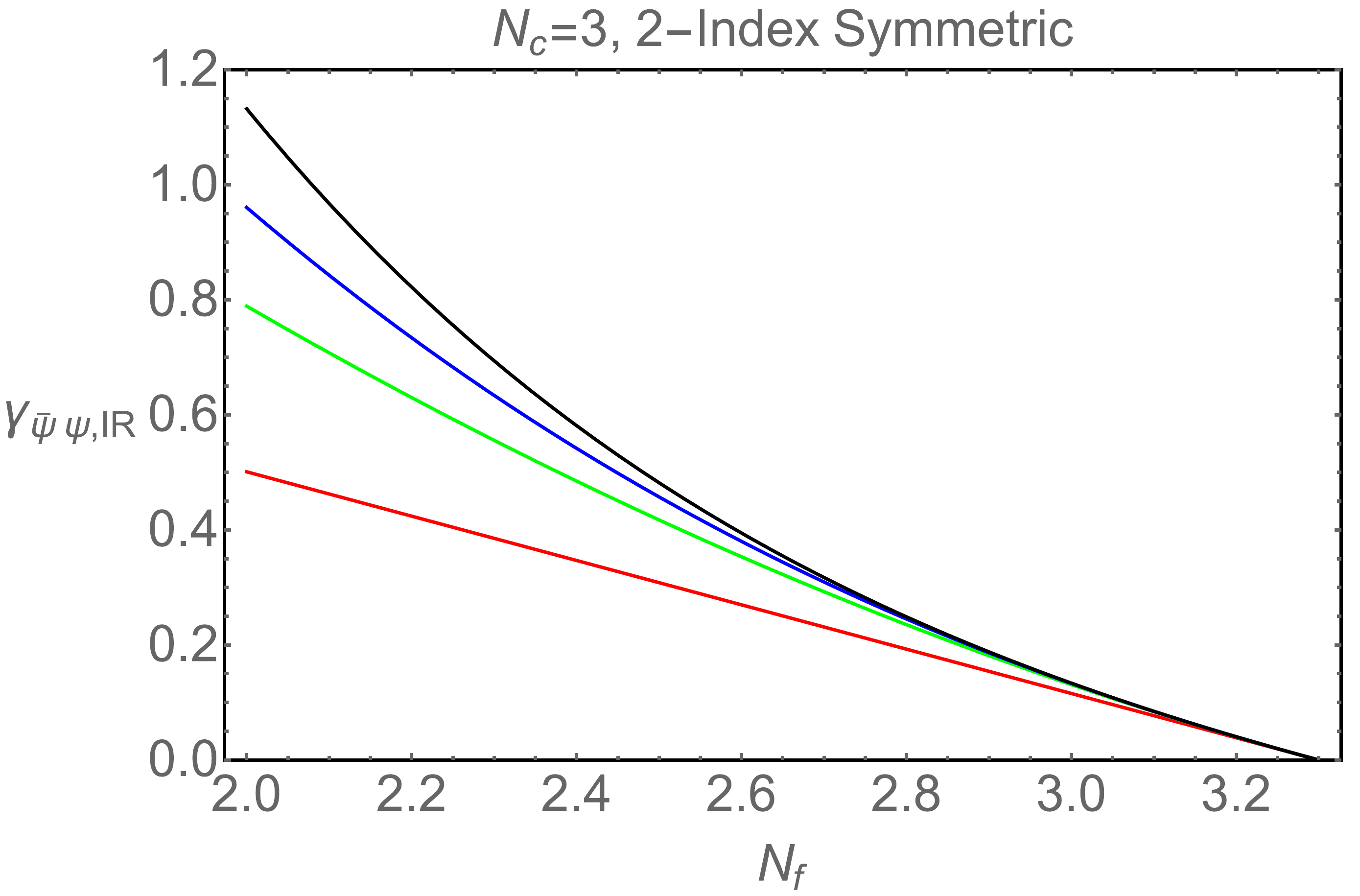}
     \caption{Estimates of the mass anomalous dimension in the $\delta$ expansion for $\SU(2)$ and $\SU(3)$ models with fermions in various representations. Different colors correspond to increasing order in the $\delta$ expansion from O($\delta$) up to O($\delta^4$). From \cite{Ryttov:2017kmx}.\label{gammadexp}} 
 \end{figure}

An example of the results from the $\delta$ expansion for the mass anomalous dimension is shown in \reffig{gammadexp} from \cite{Ryttov:2017kmx}, where various orders of the expansions are plotted from O($\delta$) up to O($\delta^4$) for different gauge groups and representations. We also note that for this quantity the series appears to have only positive coefficients, so that the resulting sum is well-behaved. 

\begin{figure}[ph!]
     \center 
     \includegraphics[width=.48\textwidth]{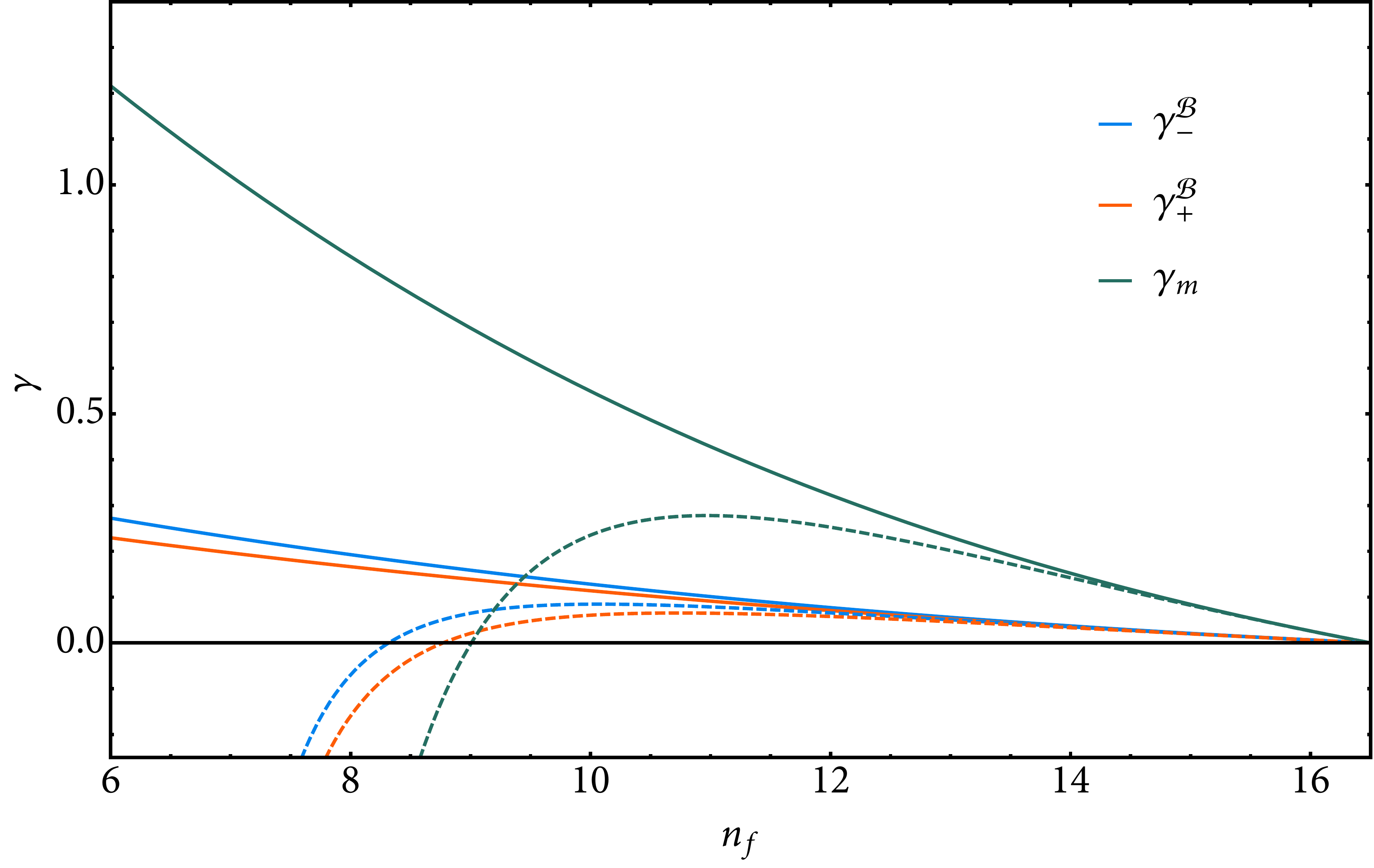} 
     \caption{Anomalous dimensions of the quark mass $\gamma_m$ and of the two 3-quark operators $\gamma_\pm^{\cal B}$ as a function of the number of fundamental Dirac fermions $n_f$ in the $\delta = n_f^{\rm{AF}}  - n_f$ expansion (solid lines). The corresponding perturbative results at three loops are represented by the dashed lines. From \cite{Pica:2016rmv}.\label{baryondexp}}
 \end{figure}

The $\delta$ expansion has also been used in \cite{Pica:2016rmv} to investigate the anomalous dimension of baryonic operators, which are relevant as top partners in  models with partial compositeness (see Sec.~\ref{sec:PCwfermions}). 
In \cite{Pivovarov:1991nk,Gracey:2012gx} the anomalous dimensions of the proton-like baryons were derived at three loop order for an $\SU(3)$ model with fermions in the fundamental, taking into account the mixing of two independent three-quarks operators.
By combining this result with the known perturbative expression of the beta function in Eq.\eq{beta}, one can derive the $\delta$ expansion for these observables.
We show in \reffig{baryondexp} the results for the anomalous dimension of the baryon operators $\gamma_\pm^{\cal B}$
as compared to a similar result for the mass anomalous dimension. As noted in \cite{Pica:2016rmv}, the anomalous dimensions of the baryonic operators are very small across all the plausible range of $n_f$ for the conformal window and, in fact, much smaller than the anomalous dimension of the mass operator.
Models of partial compositeness typically require the presence of baryonic operators with large anomalous dimensions such that 
\begin{equation}
  \frac32\le {\cal D}[{\cal B}]\le\frac52\, .
\end{equation}
This implies, for models in which the baryonic operators are composites of three fermions \cite{Ferretti:2013kya,Barnard:2013zea,Ferretti:2014qta,Vecchi:2015fma,Kaplan:1991dc}, that the anomalous dimension should be
\begin{equation}
  2\le \gamma^{\cal B}\le 3\, .
\end{equation}
The result above indicates that such large anomalous dimensions can hardly occur in the minimal template of an $\SU(3)$ model with $n_f$ fundamental fermions~\cite{Vecchi:2015fma}. 
The use of fermions in multiple representations for models of partial compositeness as advocated in \cite{Ferretti:2013kya} can somewhat increase the anomalous dimensions of some types of baryonic operators~\cite{BuarqueFranzosi:2019eee}. However it remains to be seen if the large anomalous dimensions needed for a realistic model of partial compositeness can be realized in a strong dynamics.

\subsection{Conformal Window on the Lattice}
\label{sect:cwindow}

The precise location of the conformal window is a primary objective of lattice studies beyond the SM.
The main focus has been on models based on $\SU(2)$ and $\SU(3)$ gauge groups, with many fermions in the fundamental representation, but also on the interesting case of $\SU(2)$ models with adjoint fermions.
The main new results are summarized below, showing that models with large anomalous dimensions inside the conformal window have not been found yet, despite the large number of models investigated so far. While this is not a proof, it appears unlikely that large anomalous dimensions exist in this class of models.

\subsubsection{$\SU(3)$ with fundamental fermions\label{sect:su3f}}
While this is the most familiar case for its similarities with QCD, the exact extent of the conformal window in this case is still debated. There is a general consensus that $n_f=6$ lies outside the conformal window and most groups also agree that $n_f=8$ is outside\footnote{See e.g. \cite{daSilva:2015vna} for an alternative point of view.} (studies on the spectrum of the $n_f=8$ model will be reported below in Sec.\ref{sect:scalars}).

\begin{figure}[t!] 
     \center
     \includegraphics[width=.49\textwidth]{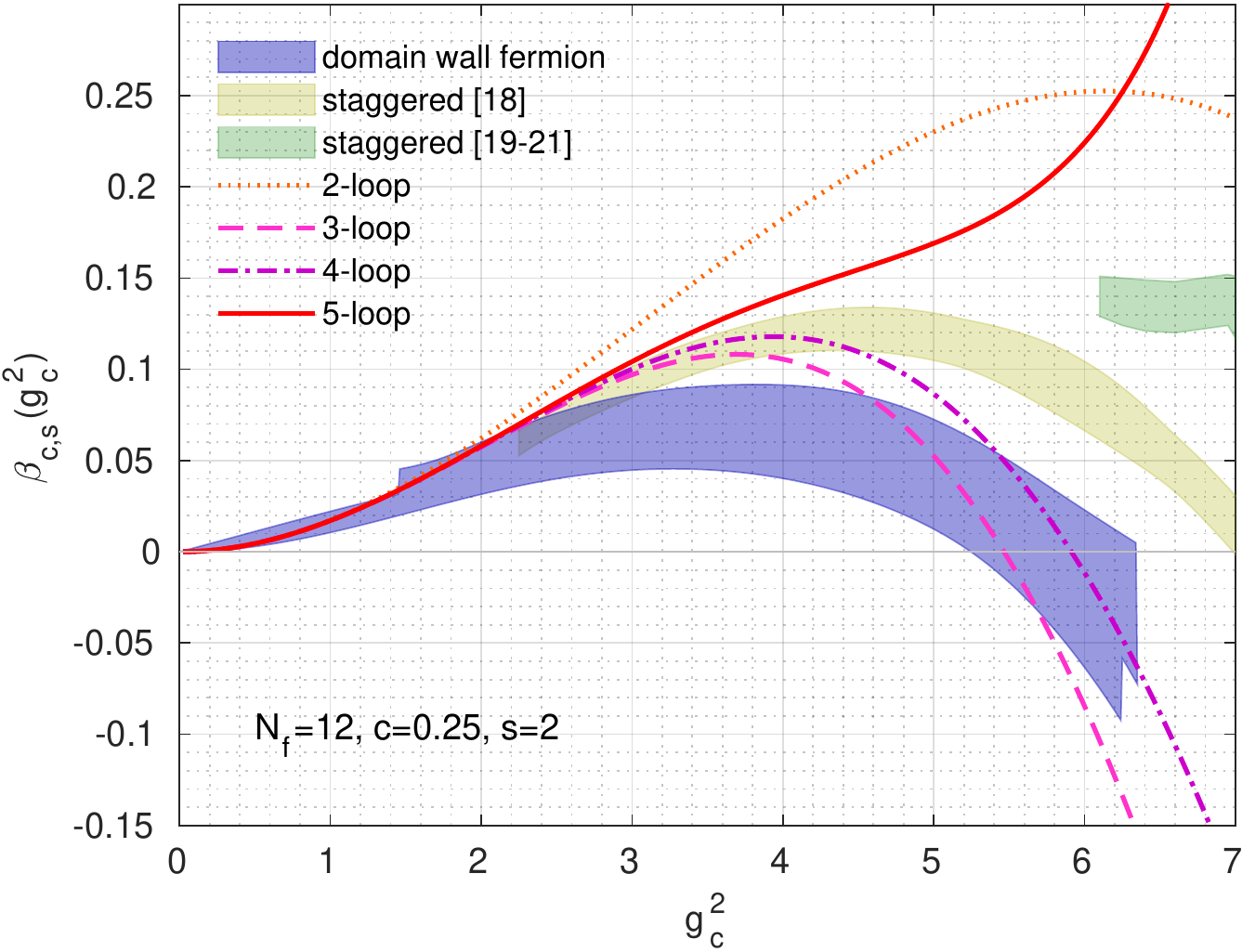}\hfill
     \includegraphics[width=.49\textwidth]{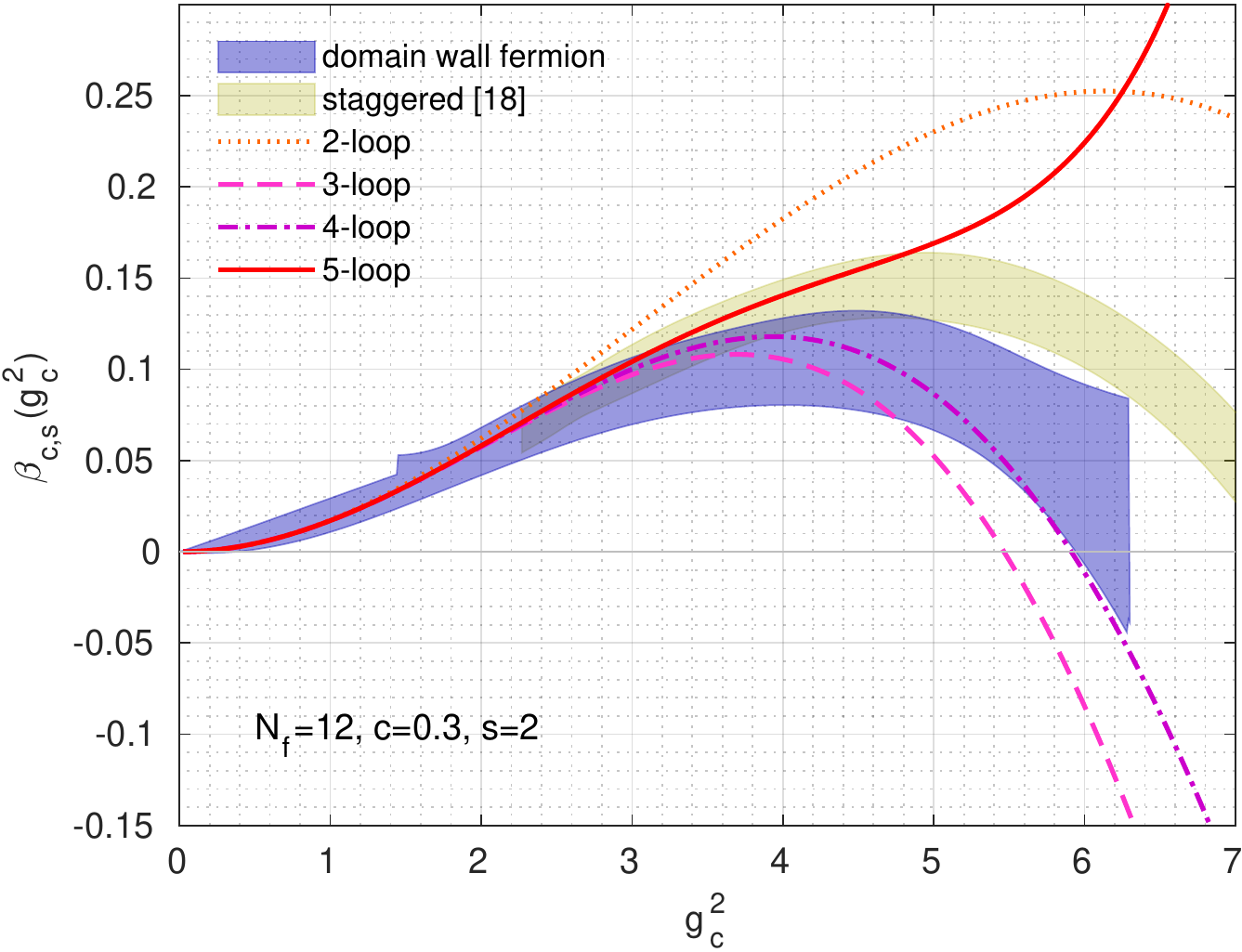}
     \caption{Comparison of the extrapolated continuum $\beta$-function for the $\SU(3)$ $n_f=12$ model with different lattice actions (staggered, domain wall) for two different renormalization schemes. From \cite{Hasenfratz:2019dpr}.}
     \label{fig:Hasenfratz}
\end{figure}

The $n_f=12$ case has been studied extensively by several groups by a variety of different methods~\cite{Appelquist:2009ty,Appelquist:2007hu, Lin:2012iw,Fodor:2011tu, Appelquist:2011dp, DeGrand:2011cu, Fodor:2012td, Aoki:2012eq, Cheng:2013eu,Cheng:2013xha, Aoki:2013zsa, Lombardo:2014pda,Hasenfratz:2011xn,Cheng:2014jba,Hasenfratz:2019dpr}. 
All these studies are consistent with the model being inside the conformal window and having a small anomalous dimension of the mass $\gamma_m^*\sim 0.2-0.3$.
The precise location of the fixed point is however subject to debate, as the measure is very sensitive to the lattice extrapolation to the continuum limits and, in particular, it was shown that very large lattice volumes are required \cite{Fodor:2016zil}, which makes a precise determination very expensive.
We report in \reffig{fig:Hasenfratz} a recent comparison among different determination of the non-perturbative $\beta$-function of the $n_f=12$ model.

The use of very large lattice volumes in the vicinity of a candidate IR fixed point is crucial for lattice determinations of the non-perturbative $\beta$-function based on the step-scaling method, in which the infinite volume limit is entangled with the continuum limit itself. While for model such as QCD where the coupling ``runs" fast and quickly become large the continuum extrapolation can be performed reliably, it is unclear if the method is reliable near a IR fixed point due to the limited range of lattice volumes available in practical lattice simulations.

The $n_f=10$ model was recently considered by two groups in \cite{Chiu:2016uui,Fodor:2018tdg,Chiu:2018edw}. The two sets of lattice simulations use different actions, staggered and domain wall, and disagree on the conclusion of the existence of an IR fixed point. While the study performed with domain wall fermions shows the presence of an IR fixed point, the staggered fermion simulations do not indicate the presence of such feature. Both studies were performed in the same renormalization scheme, so that their results, after continuum extrapolation, are directly comparable. The striking difference in behavior is 
difficult to explain in terms of systematic errors in one or both studies and more investigations are needed to pinpoint the origin of the disagreement. This difficulty led to speculations about the possible non-universal behavior of different lattice discretization in the presence of IR fixed points or near-conformal (walking) dynamics.

 \begin{figure}[b!]
     \center
     \includegraphics[width=.49\textwidth]{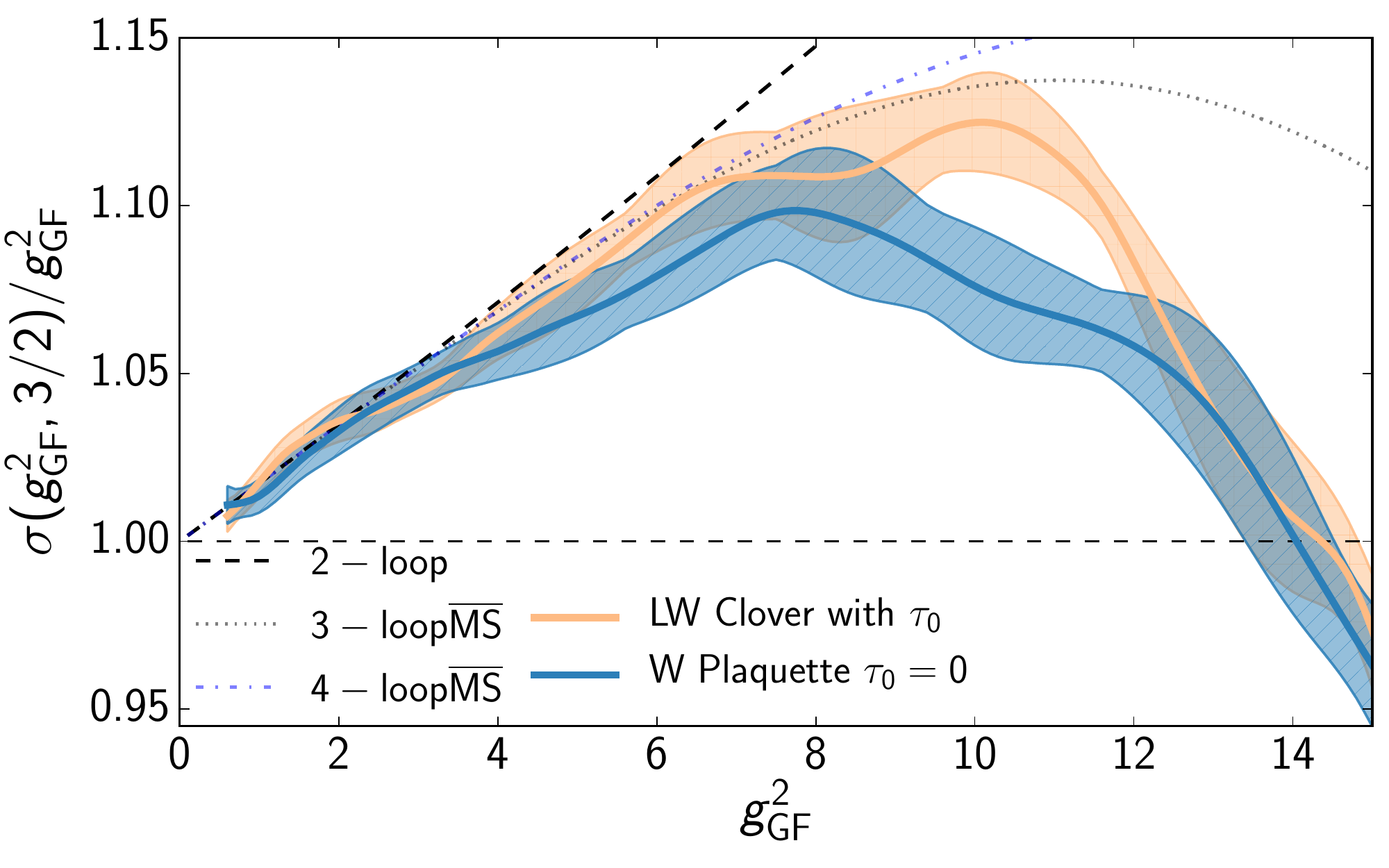}\hfill
     \includegraphics[width=.49\textwidth]{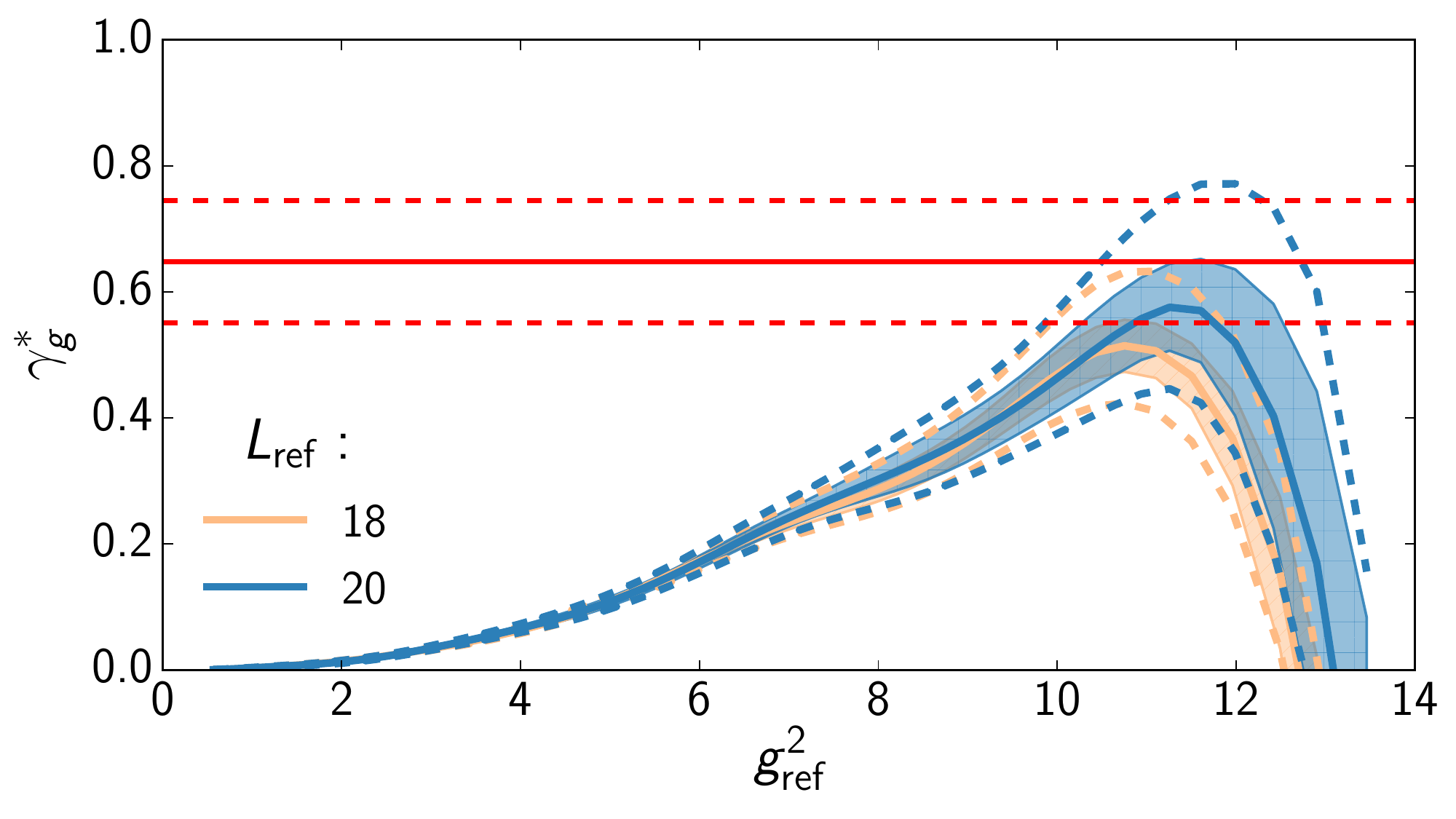}\hspace{2cm}\hfill
     \caption{Left panel: continuum step scaling function for the $\SU(2)$ $n_f=6$. Right panel: Mass anomalous dimension. From \cite{Leino:2018yfd}.}
     \label{fig:hel}
     \end{figure}
\subsubsection{$\SU(2)$ with fundamental fermions}
In the case of two colors, the location of the conformal window is more firmly established, as there is agreement among various groups that the model with $n_f\geq 8$ lies inside whereas the model with $n_f\leq 4$ lies outside \cite{Karavirta:2011zg,Ohki:2010sr,Rantaharju:2014ila,Lewis:2011zb,Hietanen:2013fya,Hietanen:2014xca,Arthur:2016dir, Leino:2018yfd}. The case $n_f=6$ is more debated and still not settled \cite{Karavirta:2011zg,Bursa:2010xn,Hayakawa:2013yfa,Appelquist:2013pqa}. In \cite{Leino:2016njf,Suorsa:2016jsf} the non-perturbative $\beta$-function in the SF gradient flow coupling scheme with an HEX-smeared Wilson-clover action was studied, which allows to reach rather strong couplings. The authors perform a careful study of the systematic errors involved and conclude in favor of the existence of an IR fixed point at a rather large value of the coupling $g^2\sim 15$, see \reffig{fig:hel} for an example of their final result. The same authors also investigated the mass anomalous dimension finding a value at the fixed point of $\gamma^\ast_m \sim 0.3$.
Given the experience with the $\SU(3)$ $n_f=12$ case, large volumes will be required to check these results in the vicinity of the observed fixed point and to confirm its existence. In fact this is the region where the raw data shows the most sensitivity to volume, as expected, and can therefore affect the continuum extrapolation in a step-scaling analysis for the extraction of the continuum $\beta$-function.

\subsubsection{$\SU(2)$ with adjoint fermions}
Most of initial effort for beyond the SM studies on the lattice was focused on the model with $n_f=2$ adjoint Dirac fermions \cite{Catterall:2007yx,Hietanen:2008mr,DelDebbio:2008zf,DelDebbio:2009fd,Catterall:2009sb,DelDebbio:2010hx,DelDebbio:2010hu,Bursa:2011ru,DelDebbio:2015byq}. The model was found to be IR conformal with an anomalous dimension of the mass $\gamma^\ast_m\sim 0.2-0.4$.\footnote{Different methods for the determination of $\gamma$ do not fully agree with each other, hinting to residual systematic errors not fully under control.} Studies for $n_f=1/2, 1, 3/2$ adjoint Dirac fermions have also been performed \cite{Athenodorou:2014eua,Bergner:2017bky}. Half integer numbers of Dirac fermions correspond to odd integer numbers of Weyl-Majorana fermions. The case of $n_f=1/2$ massless fermion corresponds to ${\cal N}=1$ supersymmetric Yang-Mills theory. In the case of $n_f=1$, the global symmetry in this case is $\SU(2)$ which is too small for a model for dynamical EWSB.
The measured observables include the masses of triplet and singlet mesons, the $0^{++}$ glueball mass, the mass of the lightest composite spin-1/2 fermion-gluon state, and the mass anomalous dimensions at a few different values of the lattice spacing. 
By studying ratios of masses for different states as a function of the bare fermion mass in the chiral limit, evidence was presented that the $n_f=1, 3/2$ models also lie inside the conformal window. The mass anomalous dimension has been determined both from the spectral quantities and from the mode number of the Dirac operator. The study confirmed the presence of very large finite size effects when approaching the chiral limit, as pointed out first in \cite{DelDebbio:2015byq}, which makes numerical simulation close to the chiral limit extremely expensive. By comparing results at two different values of the lattice spacing, a residual dependence on the lattice cutoff was observed. In particular the value of the mass anomalous dimension decreases significantly at smaller lattice spacings. For example at $n_f=1$ the (preliminary) value of $\gamma^\ast_m$ drops to $~0.75$ on the finer lattice spacing, while the value measured on coarser lattice is $~1$. Such residual dependence is observed for all models inside the conformal window, i.e. $n_f=1,3/2, 2$, and it might be due to residual finite size effects, which drive the system away from the IR conformal fixed point. More studies are needed to determine the precise value of $\gamma^\ast_m$ for these models.


\section{Models with light composite scalars\label{sect:scalars}} 

Anther class of models that may play a crucial role for models of composite dynamics includes those that feature a light composite scalar resonance. 
The phenomenological relevance of this state, analog to the $\sigma$ meson in QCD, has been highlighted in Sec.~\ref{sec:technihiggs}. Here we will focus on the recent lattice evidence of its presence.

\subsection{$\SU(3)$ with $n_f=4l+8h$ fundamental fermions}

\begin{figure}[tb]
\centering
\hfill{\includegraphics[height=0.23\textheight]{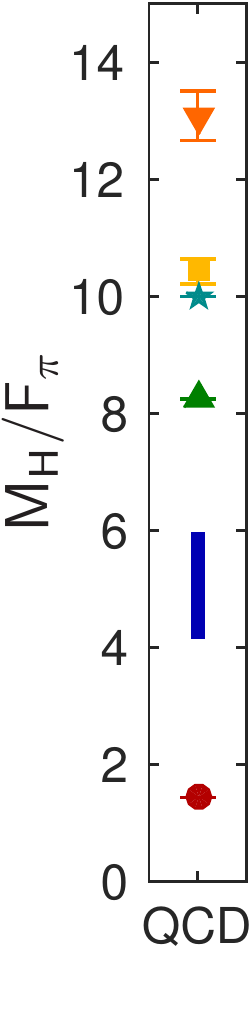}}
\hfill{\includegraphics[height=0.23\textheight]{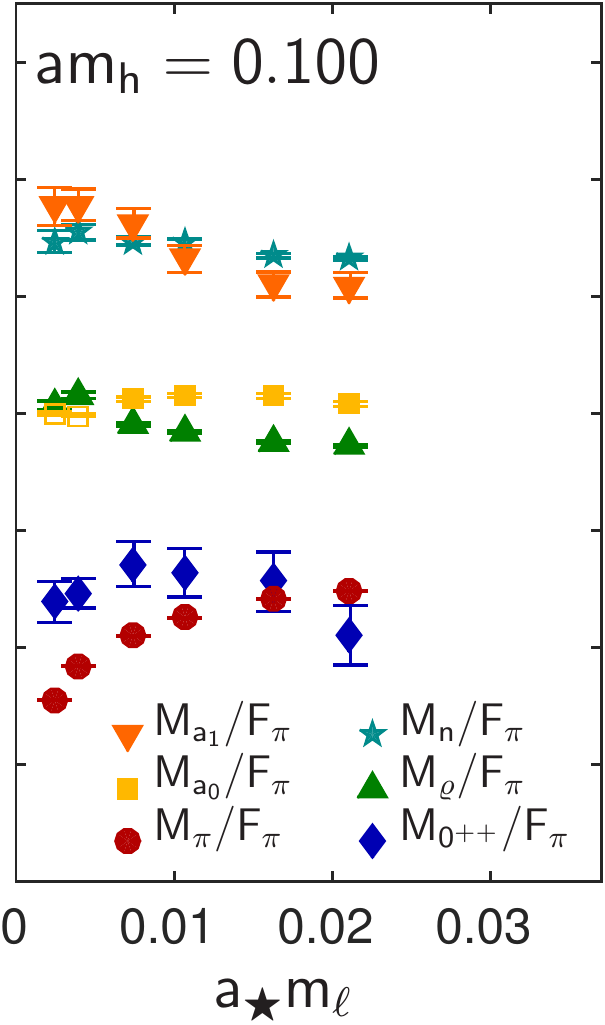}}
\hfill{\includegraphics[height=0.23\textheight]{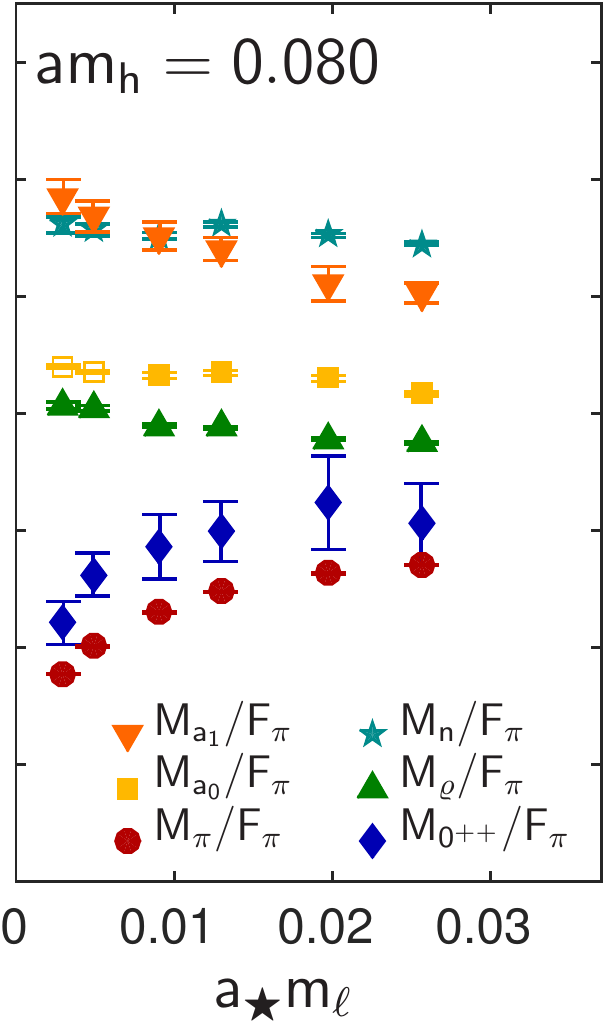}}
\hfill{\includegraphics[height=0.23\textheight]{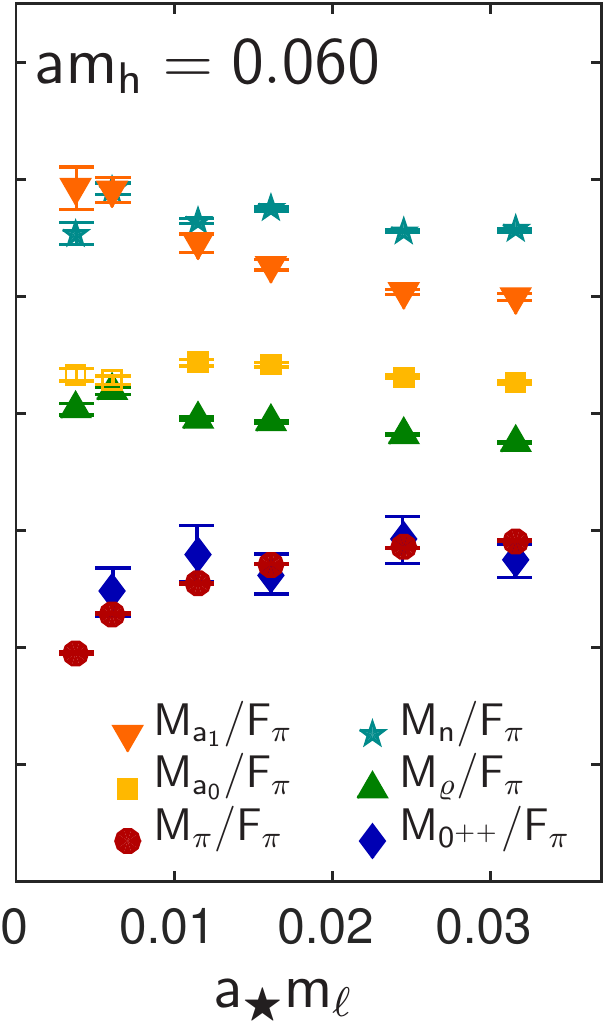}}
\hfill{\includegraphics[height=0.23\textheight]{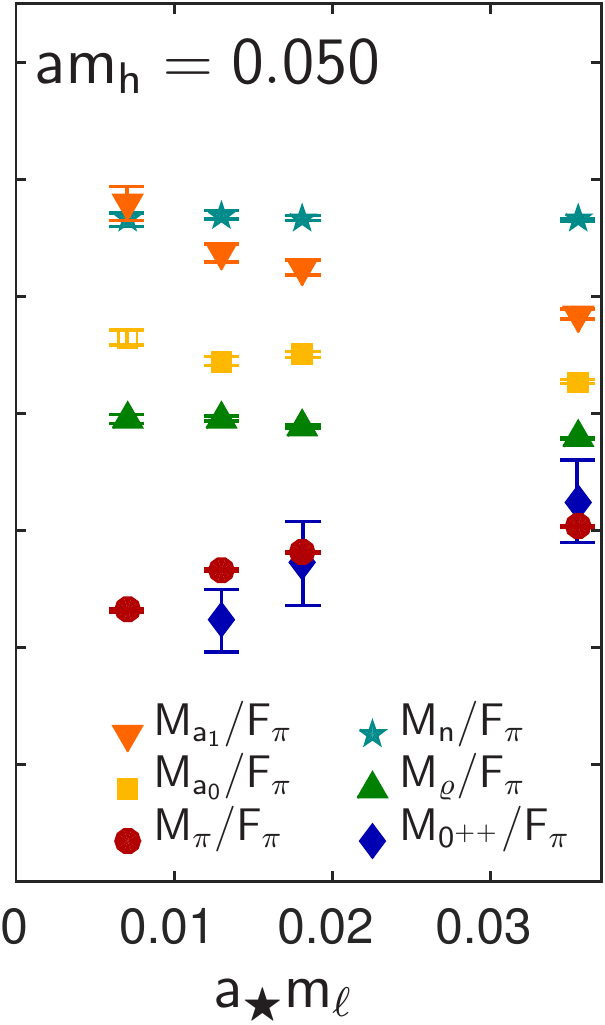}}
\hfill{\includegraphics[height=0.23\textheight]{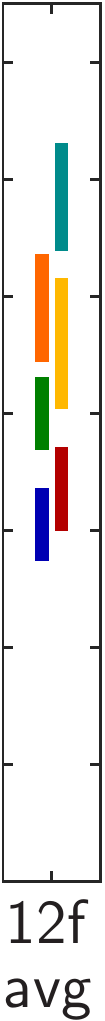}}
\caption{Spectrum of the $\SU(3)$ model with $n_f=4l+8h$ fundamental fermions. Each panel correspond to a different mass value for the heavy fermions; the leftmost panel corresponds to the physical spectrum of QCD while the rightmost panel corresponds to the $n_f=12$ model for a fixed fermion mass. All masses are normalized to the pion decay constant $F_\pi$. From \cite{Brower:2015owo}.}
\label{fig:rebbi}
\end{figure}

Based on the idea that walking can be ``generated" by continuously deforming a model with IR fixed point, one can introduce a small fermion mass to study the spectrum of the model close to an IR fixed point \cite{DelDebbio:2010hx, DelDebbio:2010hu} in the ``walking" regime. The model considered in this section is built on the assumption that the $\SU(3)$ model with $n_f=12$ fundamental fermion has an IR fixed point (see Sec~\ref{sect:su3f}). One could then introduce a mass $m_h$ for eight of the twelve fermions so that, in the limit of large $m_h$, the model is reduced to the $n_f=4$ model that is similar to ordinary QCD. For intermediate masses $m_h$, a walking regime can be generated and the spectrum of the model can be studied. In actual lattice simulations, it is also necessary to introduce a mass $m_l$ for the four light fermions, so that an extrapolation to zero $m_l$ is required for each $m_h$.

This model can be considered as a prototype for models of walking TC and for models of composite Goldstone Higgs based on the $\SU(4)\times\SU(4)\to\SU(4)$ coset~\cite{Ma:2015gra,Ma:2017vzm}. The additional heavy flavors can also be used to generate fermion partial compositeness, as in \cite{Vecchi:2015fma}.\footnote{As noted in \cite{Vecchi:2015fma}, a simple realistic model which takes into account partial compositeness effectively reduces this coset to our template $\SU(4)/\Sp(4)$.}

Results for the spectrum and scaling properties of this model were presented in \cite{Brower:2015owo,Hasenfratz:2016uar,Hasenfratz:2016gut}.
The spectrum for the light-light mesons as a function of the heavy $m_h$ and light fermion mass $m_l$ is shown in \reffig{fig:rebbi}, where meson masses are normalized by the value of the pseudoscalar decay constant $F_\pi(m_l,m_h)$. In the chiral limit $m_l\to 0$ the ratios $M_H/F_\pi$ seem to depend only weakly on the value of $m_h$, with the exception of the scalar $0^{++}$ state which seems to become much lighter as $m_h$ is reduced, i.e. when entering the ``walking regime" of the model. In fact for $a m_h\leq 0.06$ the scalar state becomes degenerate, within errors, with the pNGBs of the model. This is, in fact, a common feature observed in numerical simulations of models with light scalar states. This feature introduces a difficulty in the interpolation to the chiral limit, as the ordinary chiral theory that includes only massless GBs is not applicable. A new chiral perturbation theory that includes the light scalar mode is therefore necessary~\cite{Golterman:2016lsd}, along the lines of Eq.~\eqref{eq:sigma} in Sec.~\ref{sec:Lext}.
It is therefore still unclear precisely how light such scalar states are.
The light-heavy and heavy-heavy meson states were also investigated in \cite{Hasenfratz:2016uar}. Assuming the existence of an IR fixed point for the $n_f=12$ model, hyperscaling relations hold close to the fixed point $m_l=m_h=0$ for the masses of hadrons and decay constants, and their ratios. In \cite{Hasenfratz:2016uar} evidence is provided for this hyperscaling, which also supports the hypothesis of an IR fixed point in the $n_f=12$ system.

\begin{figure}[t!]
     \centering
     {\includegraphics[height=0.25\textheight]{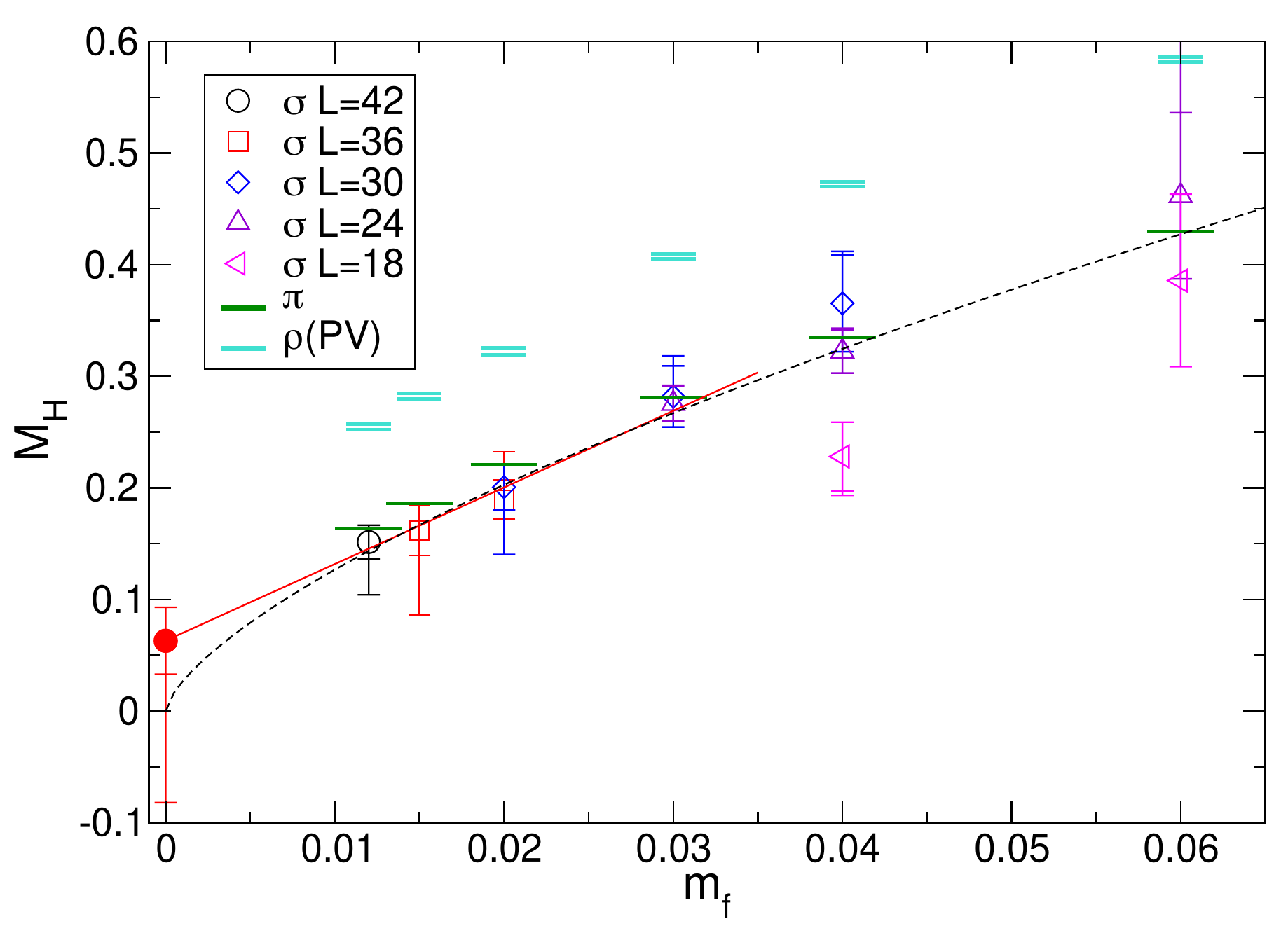}}
     \hfill{\includegraphics[height=0.25\textheight]{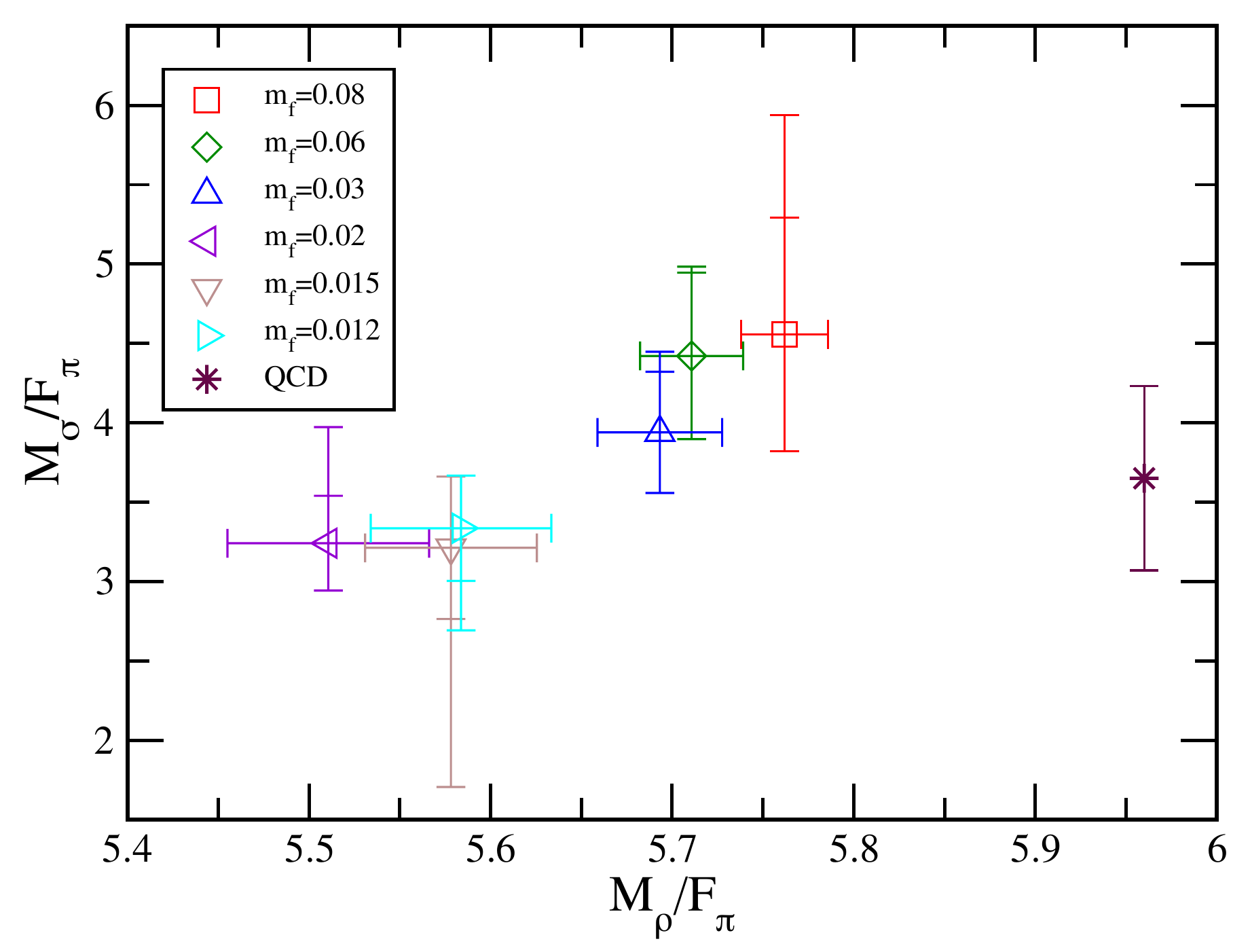}}
     \caption{Spectrum of the $\SU(3)$ model with $n_f=8$ fundamental fermions. A light scalar particle, as light as the measured pions of the model, is visible. From the LatKMI collaboration \cite{Aoki:2016wnc}.}
     \label{fig:aoki}
\end{figure}

\subsection{$\SU(3)$ with $n_f=8$ fundamental fermions}
The results from the previous section on the extent of the conformal window point to the possibility of light scalars for walking models which lie just below the onset of the conformal window. As a direct test of this hypothesis, different groups have been investigating the spectrum of the $\SU(3)$ model with $n_f=8$ fundamental fermions \cite{Aoki:2013xza, Appelquist:2014zsa, Aoki:2014oha,Appelquist:2016viq,Aoki:2016wnc,Appelquist:2018yqe}.

The LatKMI collaboration published results for spectrum of the $n_f=8$ and $n_f=4$ models in \cite{Aoki:2016wnc}, shown in \reffig{fig:aoki}, in particular for the flavor singlet meson $\sigma$. In the $n_f=8$ model, the $\sigma$ meson appears to be degenerate with the pions over the whole range of fermion masses explored, similarly to the case considered in the previous section, while for the $n_f=4$ case it remains heavier than the pions at the most chiral point investigated. This hints at the possibility of a much lighter $\sigma$ resonance than in QCD. However if one considers the ratio $m_\sigma/m_\rho$, or the ratio $m_\sigma/F_\pi$, at the most chiral point in \reffig{fig:aoki}, this is unchanged between the $n_f=4$ and $8$ models, and it is also similar to the QCD case. At face value, this would indicate that the $\sigma$ resonance does not become lighter with respect to the strong scale at the onset of the conformal window.

\begin{figure}[t!]
     \center
     \includegraphics[width=.55\textwidth]{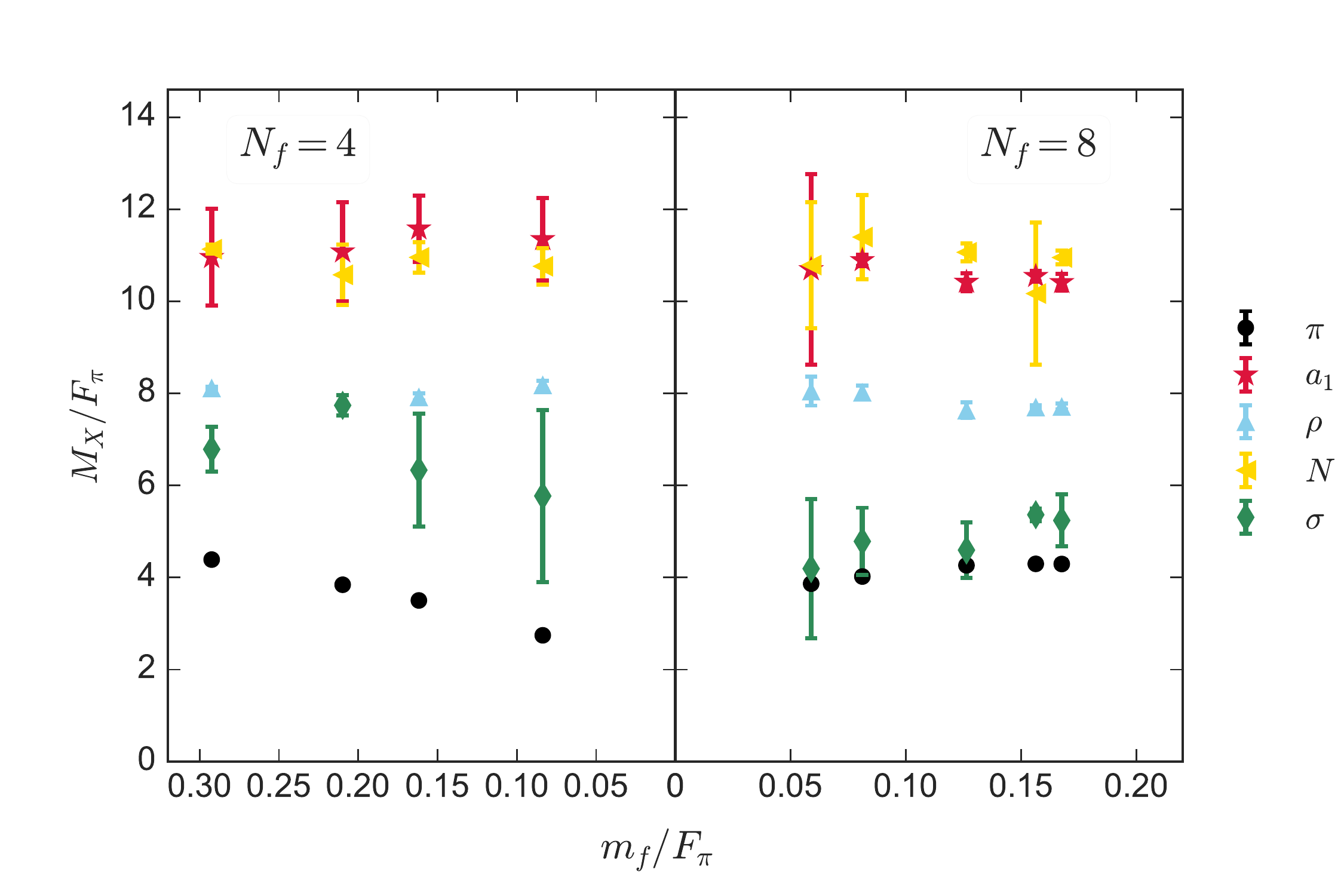}
     \caption{Comparison of the spectrum of the $\SU(3)$ models with $n_f=4,8$ fundamental fermions obtained by the LSD collaboration~\cite{Appelquist:2018yqe}. In the case of $n_f=8$ a light scalar state, comparable in mass to the pions of the model, is present, even if the ratio $m_\sigma/F_\pi$ is comparable to the model with $n_f=4$ or QCD.}
     \label{fig:lsd}
     \end{figure}

Similar results were obtained by the LSD collaboration \cite{Gasbarro:2017fmi,Appelquist:2018yqe} at a somewhat smaller value of the pion mass. The mass spectrum of the model obtained by the LSD collaboration is shown in \reffig{fig:lsd}. Very large volumes up to $64^3\times 128$ were required in order to keep systematic errors under control. In agreement with the results from the LatKMI collaboration, the mass of the $\sigma$ resonance is found to be degenerate with the pions of the model and the ratios $m_H/F_\pi$ are very similar to the QCD values, indicating only a very weak dependence of the mass ratios on the number of flavors $n_f$.
From the right panel of \reffig{fig:lsd} one should notice that both the pion and the light scalar resonance mass in units of $F_\pi(m_q)$ show only a very weak dependence on the fermion mass $m_q$ in the region explored.
If the $n_f=8$ is not inside the conformal window, then a sharp decrease of $m_\pi/F_\pi$ is expected close to the chiral limit, while the $\sigma$ resonance should remain massive.

It is therefore crucial to be able to extrapolate the current results closer to the chiral limit to establish if the scalar $\sigma$ resonance becomes much lighter for walking models.

\subsection{$\SU(3)$ with $n_f=2$ sextet fermions}

\begin{figure}[t!]
     \center
     \includegraphics[width=.8\textwidth, height=6cm]{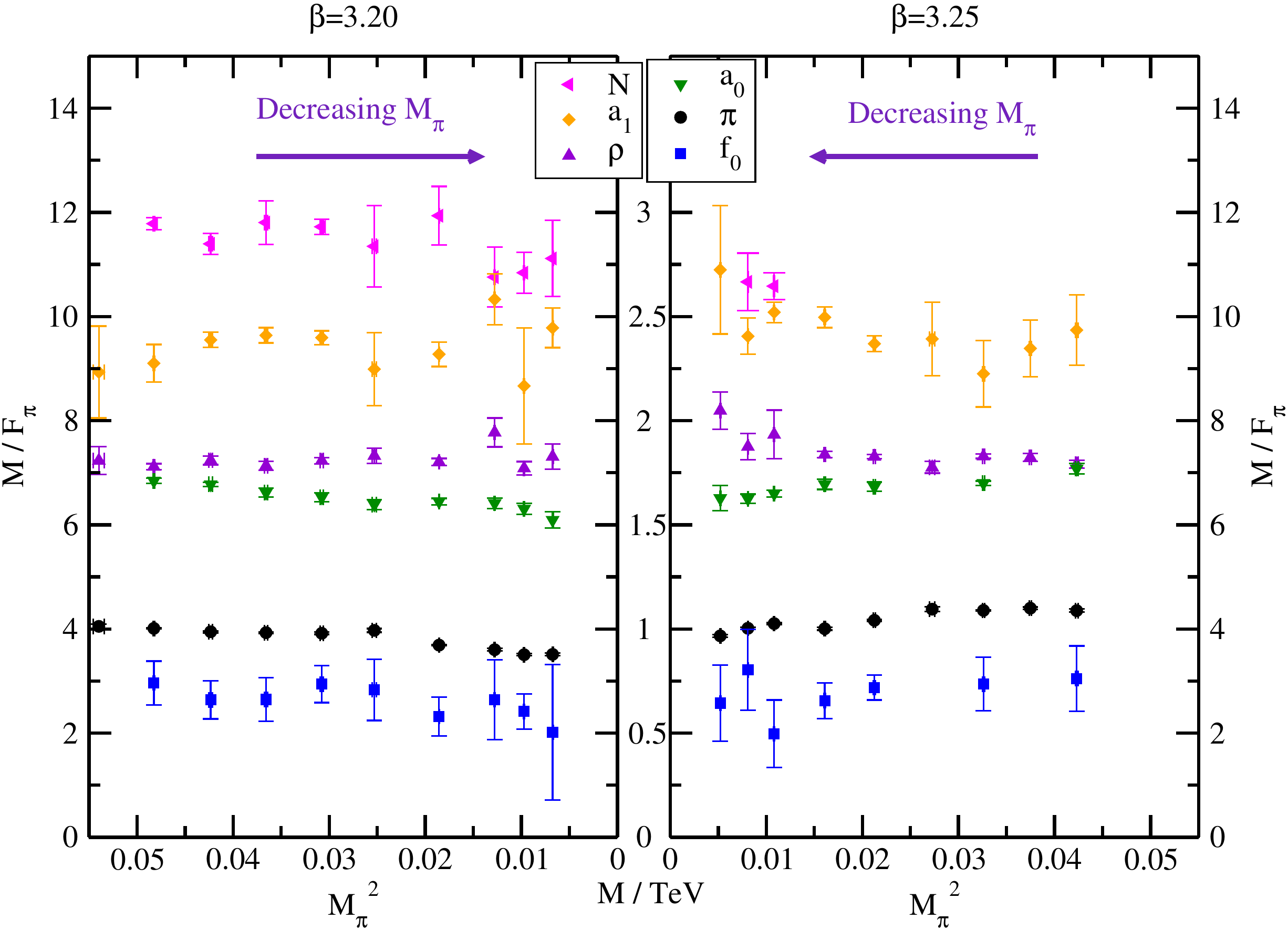}
     \caption{Spectrum of the $\SU(3)$ sextet model with $n_f=2$ two-index symmetric fermions in units of $F_\pi$.  From \cite{Fodor:2016pls}. }
     \label{fig:kuti}
     \end{figure}

A very interesting walking TC model is based on SU(3) with $n_f=2$ fermions in the two index symmetric representation (the sextet representation). The model has the three GBs, i.e. the minimal number required for a TC model, but from the higher dimensional representation of the fermions, one expects the model to be walking \cite{Dietrich:2005jn}, with, possibly, light scalar states.
The spectrum of this sextet model has been studied in detail on the lattice with staggered fermions \cite{Fodor:2012ty, Fodor:2014pqa, Fodor:2016wal, Fodor:2016pls} and with Wilson fermions \cite{Drach:2015sua,Hansen:2016sxp,Hansen:2017ejh,Hansen:2017zyo}. 

In \reffig{fig:kuti} we show the spectrum of the model as obtained by staggered fermions lattice simulations, at two different lattice spacings \cite{Fodor:2016pls}. The qualitative features of the spectrum are similar to the other walking models described above. The spectrum features a light $\sigma$ resonance over the entire range of light fermion masses explored, in fact lighter that the pions that should eventually become massless in the chiral limit if the model is chirally broken.

The issue of the model being inside or outside the conformal window has not been settled yet.
On the one hand, inspection of \reffig{fig:kuti} by eye might lead one to conclude that the model is in fact IR conformal and inside the conformal window, however a more detailed analysis has led the authors of \cite{Fodor:2016pls} to conclude that this is not the case, as the data do not follow well the expected hyperscaling behavior for the spectrum. On the other hand, the data seems to fits well with the prediction of rooted staggered chiral perturbation theory, even if the use of such effective model is not justified, given the presence of the light scalar state. As probing the model at even lighter masses in the $p$-regime would be prohibitively expensive, the authors of \cite{Fodor:2016pls} are moving to use more sophisticated analysis methods involving the cross-over regime from the $p$ to the $\epsilon$-regime, the use of random matrix theory and the use of effective models which take into account the light scalar particle in the spectrum. 
As noted above for the case of $\SU(3)$ with $n_f=10$ fundamental Dirac fermions, the use of staggered fermions implies the use of a technique called rooting, which might affect the universality class of the model and, therefore, the continuum extrapolations.

\begin{figure}[t!]
     \center
     \includegraphics[width=.49\textwidth]{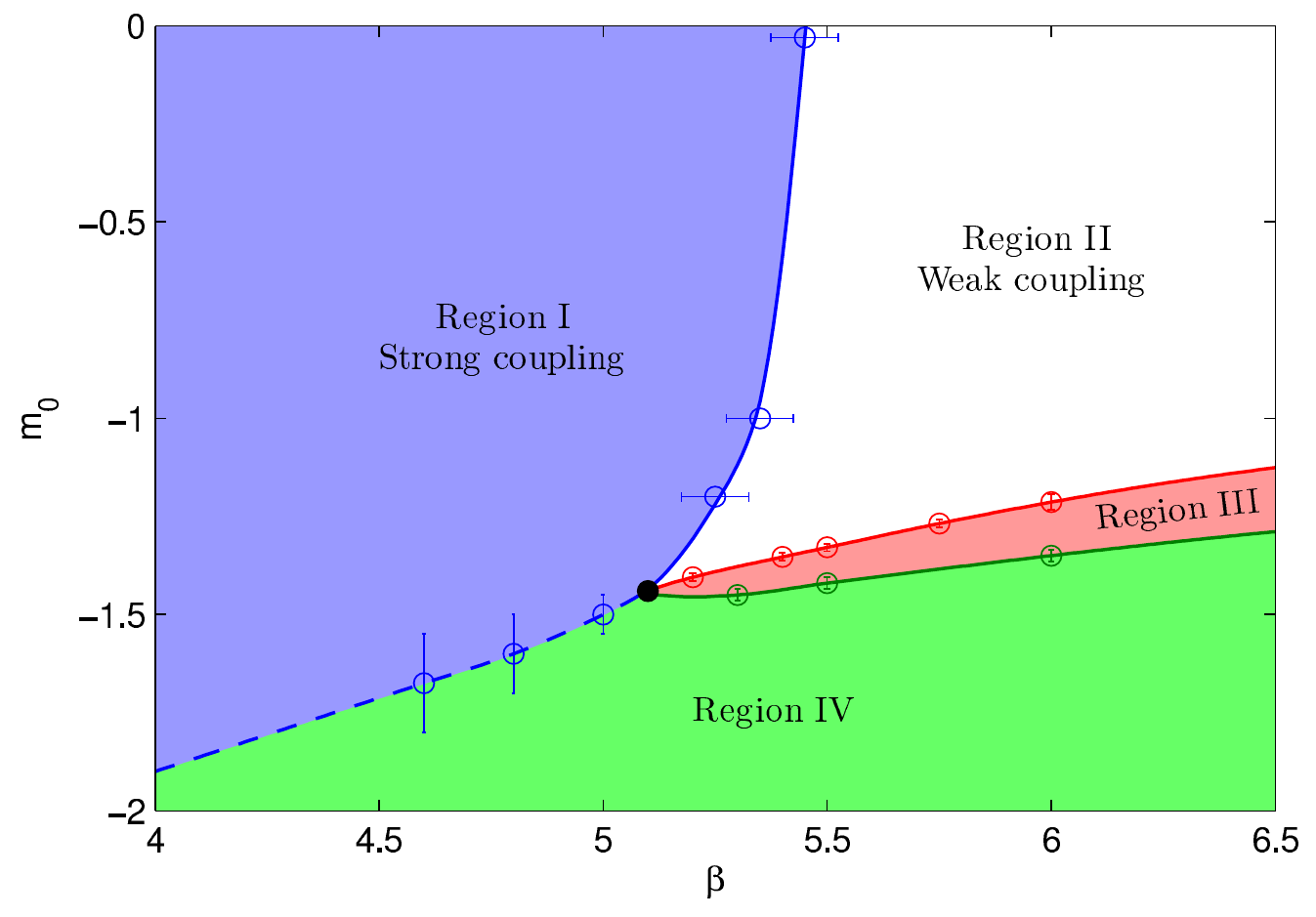}\hfill
     \includegraphics[width=.478\textwidth]{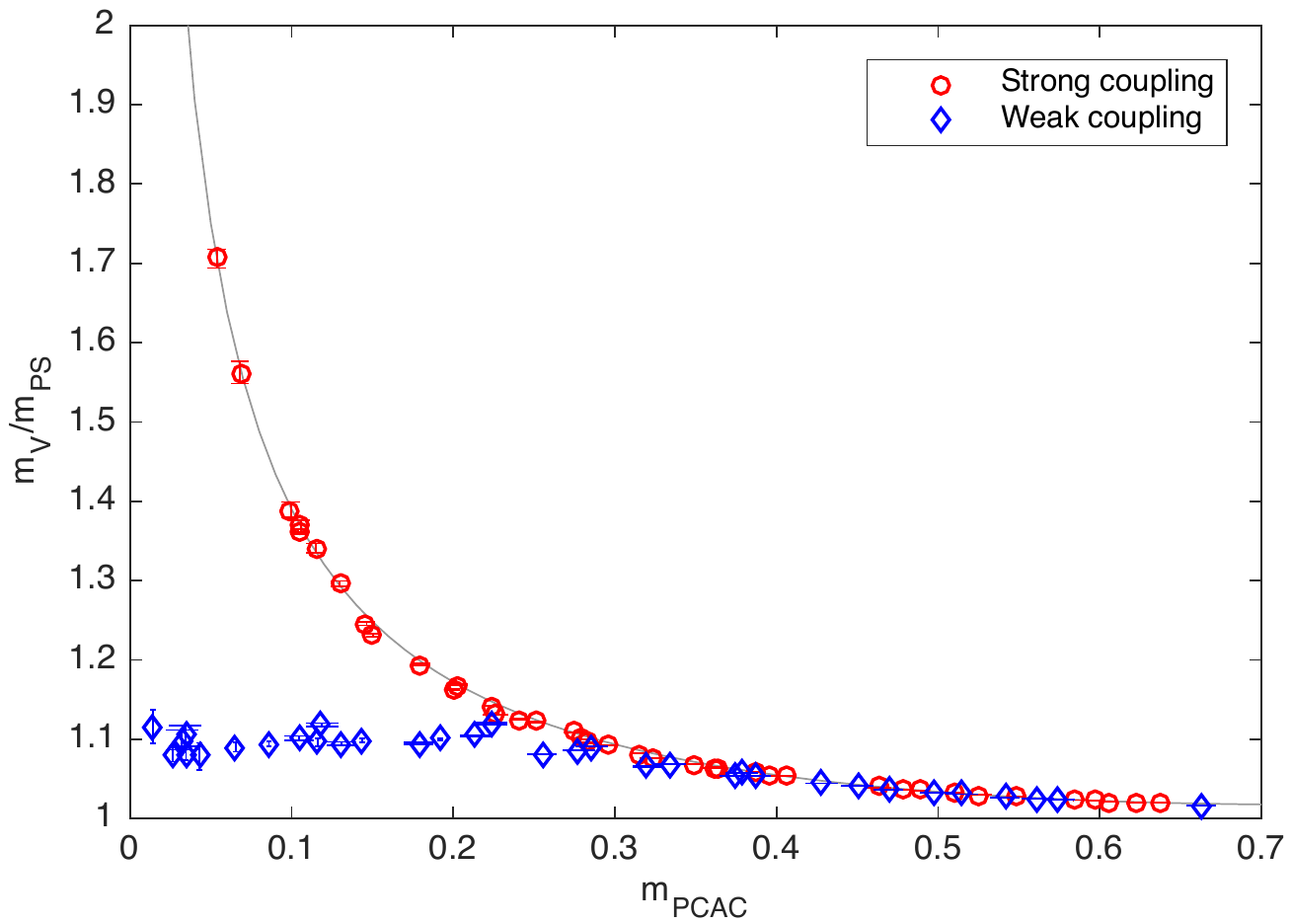}
     \caption{Left panel: Phase structure of the lattice $\SU(3)$ sextet model with Wilson fermions. Right panel: The ratio $m_\rho/m_\pi$, for different lattice spacings in the weak coupling and strong coupling phases. From \cite{Hansen:2016sxp,Hansen:2017ejh}. }
     \label{fig:hansen}
\end{figure}
The spectrum of the sextet model has also been investigated with Wilson fermions, in which the rooting procedure is not present. Similarly to the staggered fermion case, the mass spectrum can be fitted to a prediction from Wilson chiral perturbation theory, although in this case of the spectrum also fits the possibility of hyperscaling from an IR fixed point \cite{Drach:2015sua}. There is however a striking difference between the spectra obtained from staggered and Wilson fermions: in the latter case, the vector resonance never appears to become much heavier than the pion in the weak coupling phase. To better understand the behavior of the model, the full phase structure of the lattice model with Wilson fermions was studied in \cite{Hansen:2016sxp,Hansen:2017ejh}, see \reffig{fig:hansen}. A phase at strong coupling was identified, separated from the weak coupling phase by a crossover. At strong coupling there is a first order transition as the fermion mass is reduced, which becomes a continuous transition in the weak coupling phase corresponding to the line of vanishing PCAC mass. 
The behavior of several quantities, which include the mass spectrum and the scale-setting observables, such as $w_0$ and $t_0$, was studied both in the weak coupling and strong coupling phase. The results show a sharp change in the qualitative behavior of the measured quantities, see e.g. the right panel of \reffig{fig:hansen}. While at strong coupling, the observations are compatible with a chirally broken model as expected, data in the weak coupling phase do not show any clear indications of spontaneous chiral symmetry breaking.

The question of the IR conformality of this model will require the use of more data at
weak coupling on the spectrum of the model, at several lattice spacings, to show consistently the
presence or not of a critical behavior in the chiral limit.

\subsection{Ideal walking}
The ideal walking scenario, described in Sec.~\ref{sect:idealwalking}, was studied on the lattice in \cite{Rantaharju:2017eej,Rantaharju:2019nmh}. The particular model studied is based on $\SU(2)$ with 2 Dirac adjoint fermions, known to be inside the conformal window (see Sec.\ref{sect:cwindow}), plus a particular four-fermions interactions chosen so that the lattice action remains positive definite, and therefore amenable to numerical lattice simulations.

\begin{figure}[t!]
     \center
     \includegraphics[width=.25\textwidth]{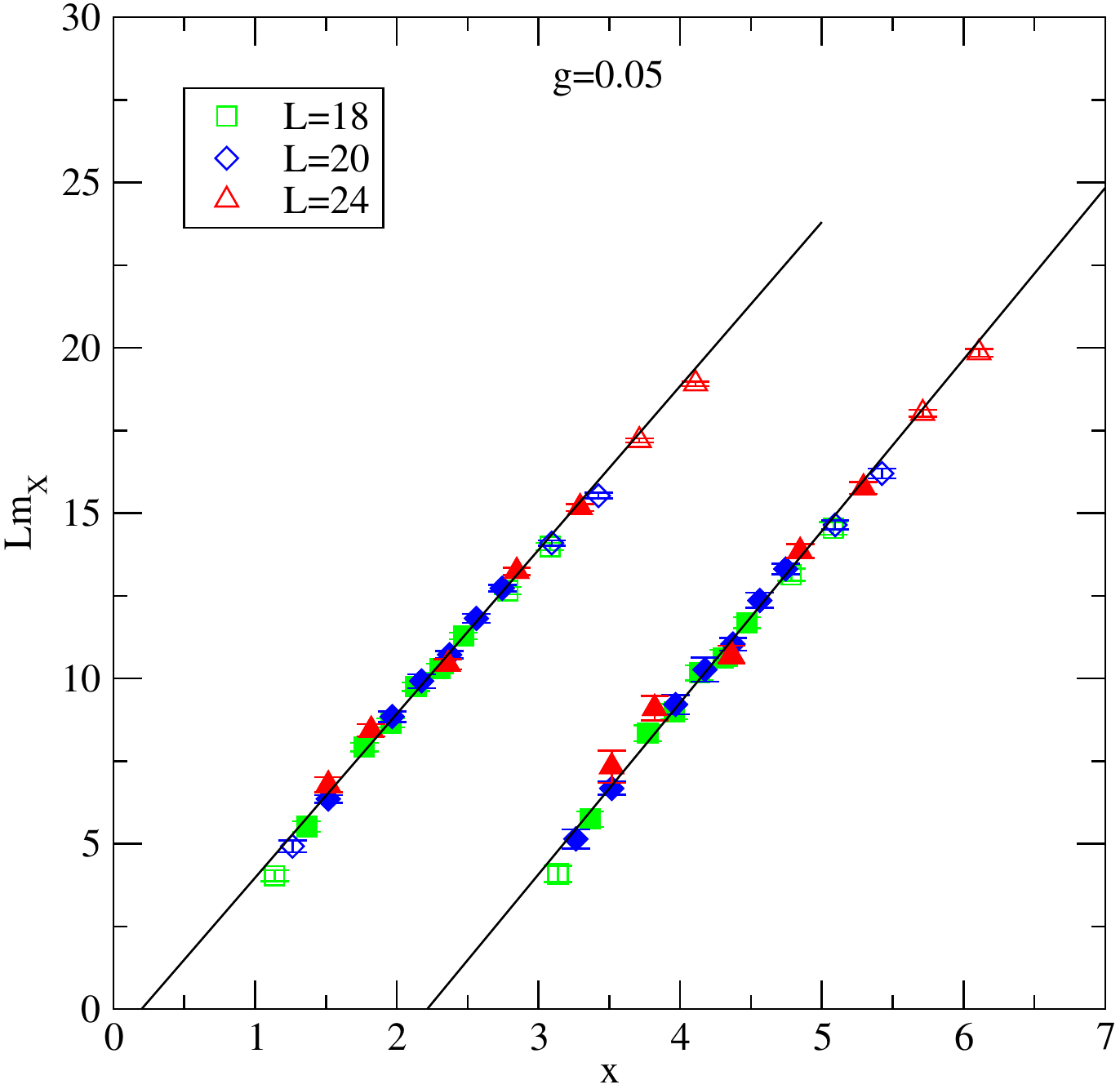}\hfill
     \includegraphics[width=.25\textwidth]{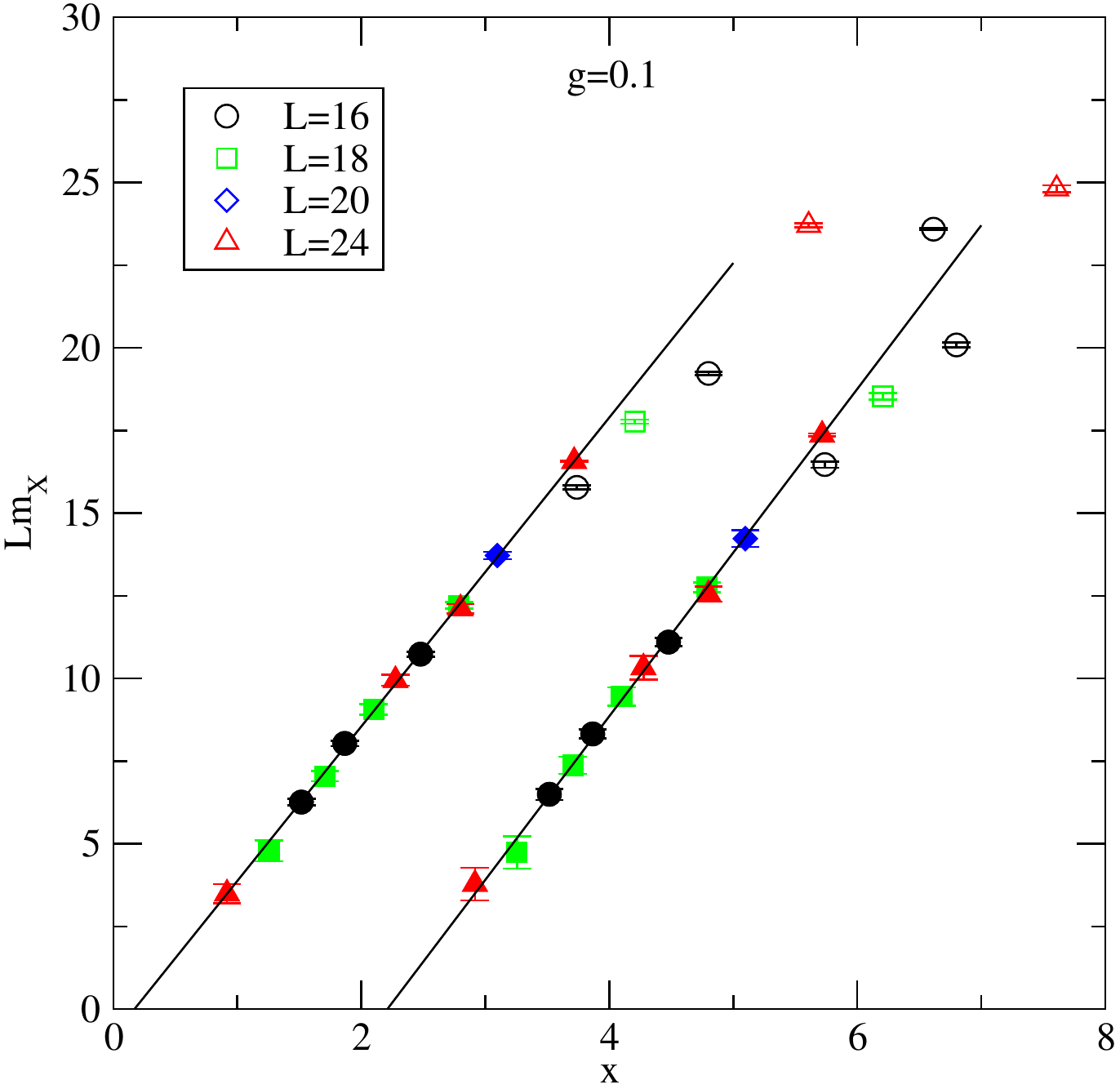}\hfill
     \includegraphics[width=.25\textwidth]{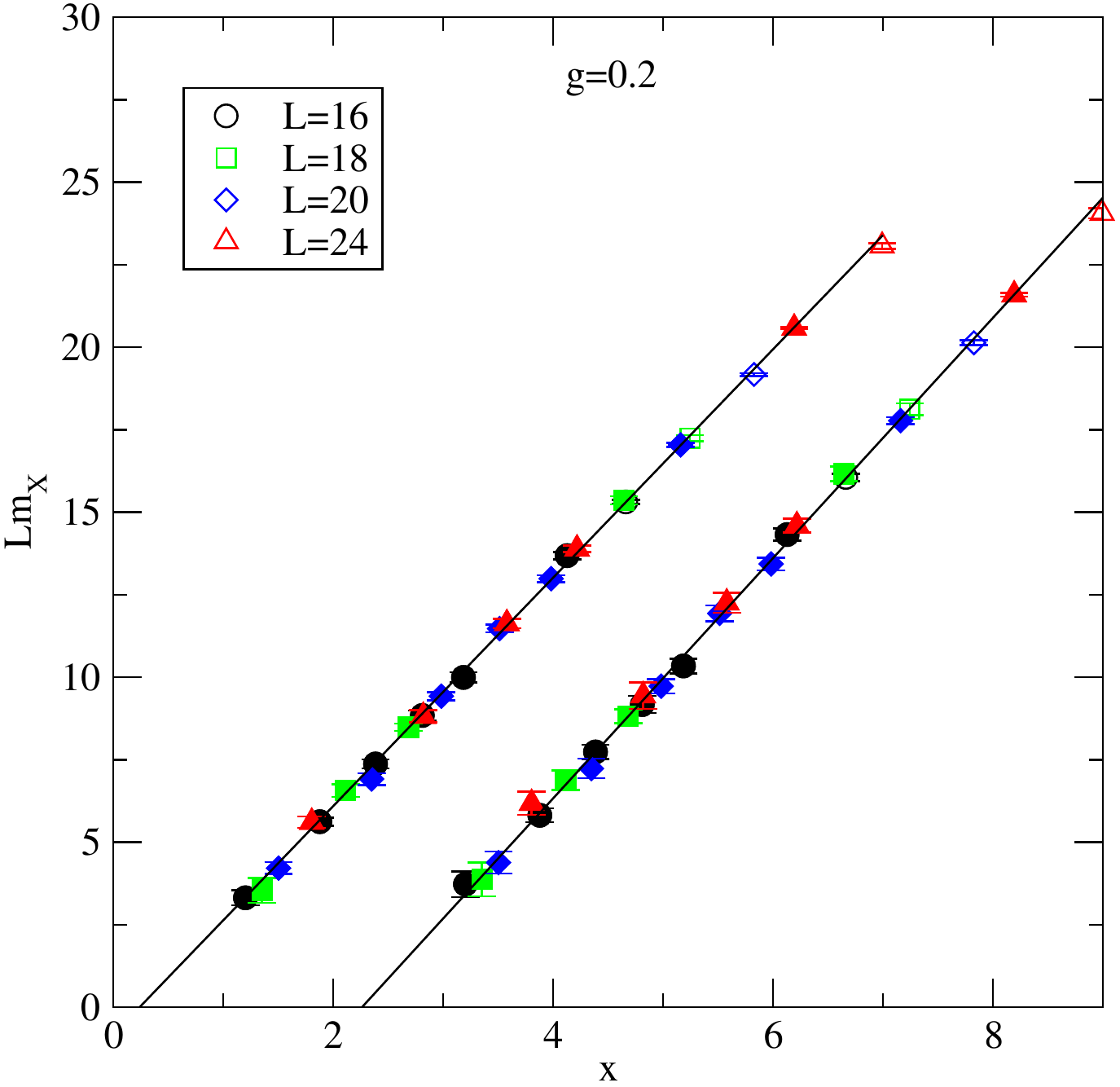}\hfill
     \includegraphics[width=.25\textwidth]{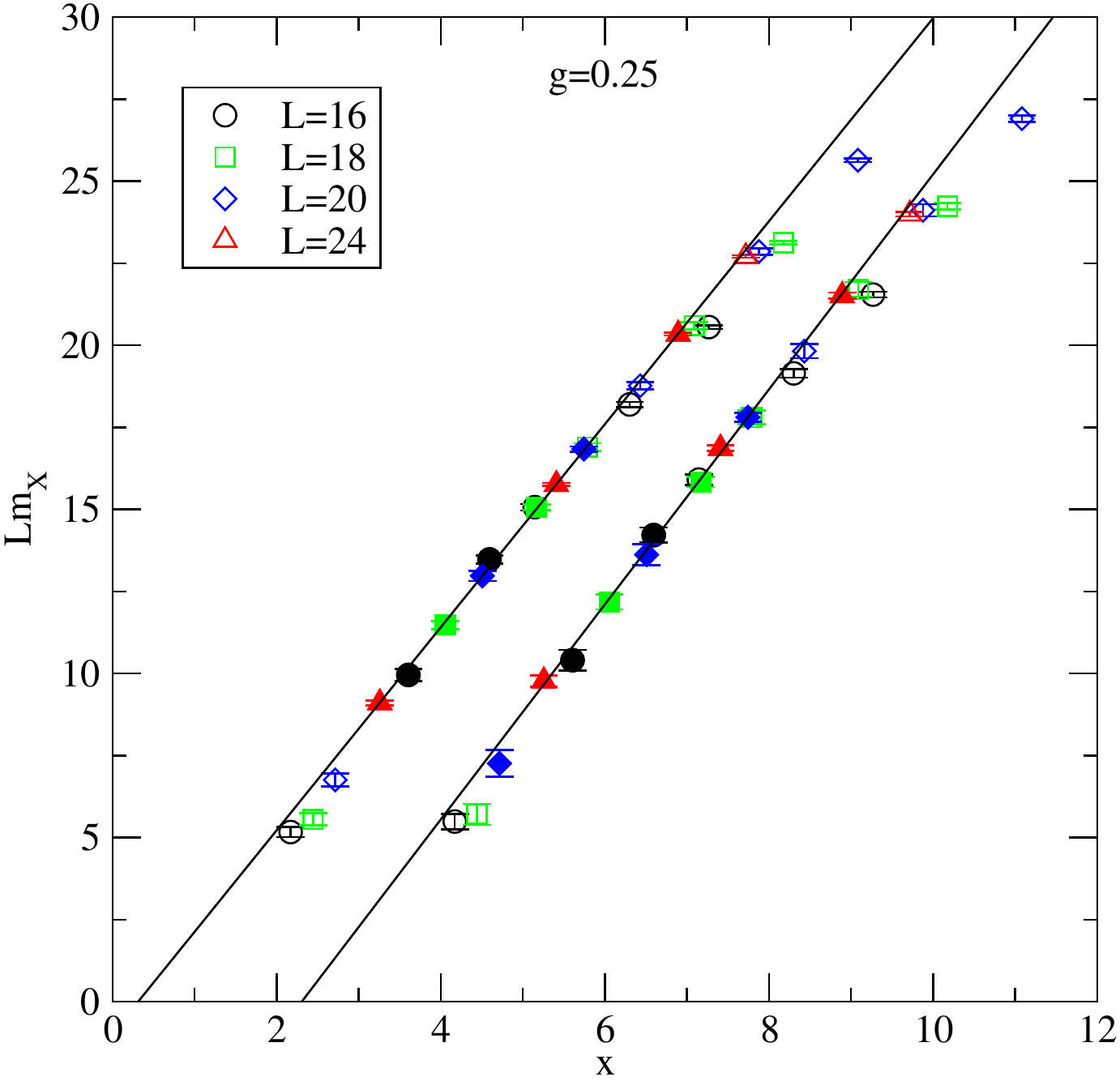}
     \caption{Finite size scaling for the spectrum of the ideal walking model studied in \cite{Rantaharju:2017eej}. Here $L$ is the lattice size, $m_X$ is a meson mass, and $x$ is the finite size scaling variable $x=L \left | m_0-m_c \right |^\frac{1}{1+\gamma_m}$. Each panel correspond to a different value of the four-fermion coupling $g$. All values are less than the critical coupling $g<g_{cr}$ for which the model loses IR conformality. From \cite{Rantaharju:2017eej}. }
     \label{fig:idealw1}
\end{figure}

In \cite{Rantaharju:2017eej}, the phase diagram of the lattice model was studied, and it was presented evidence that a small value of the four-fermion coupling $g$ does not remove the presence an the IR fixed point present at $g=0$, until a critical value of $g=g_{cr}$ is reached. 
Moreover the critical behavior for the massless theory at $g<g_{cr}$ was studied, by using finite size scaling techniques (see Fig.~\ref{fig:idealw1}). The spectrum of the model shows a scaling behavior with a critical exponent $\gamma_m$ which depends on the value of the four-fermion coupling $g$, and it seems to approach a large value $\gamma_m\approx 1$ as $g\sim g_{cr}$.  

\begin{figure}[t!]
     \center
     \includegraphics[width=.32\textwidth]{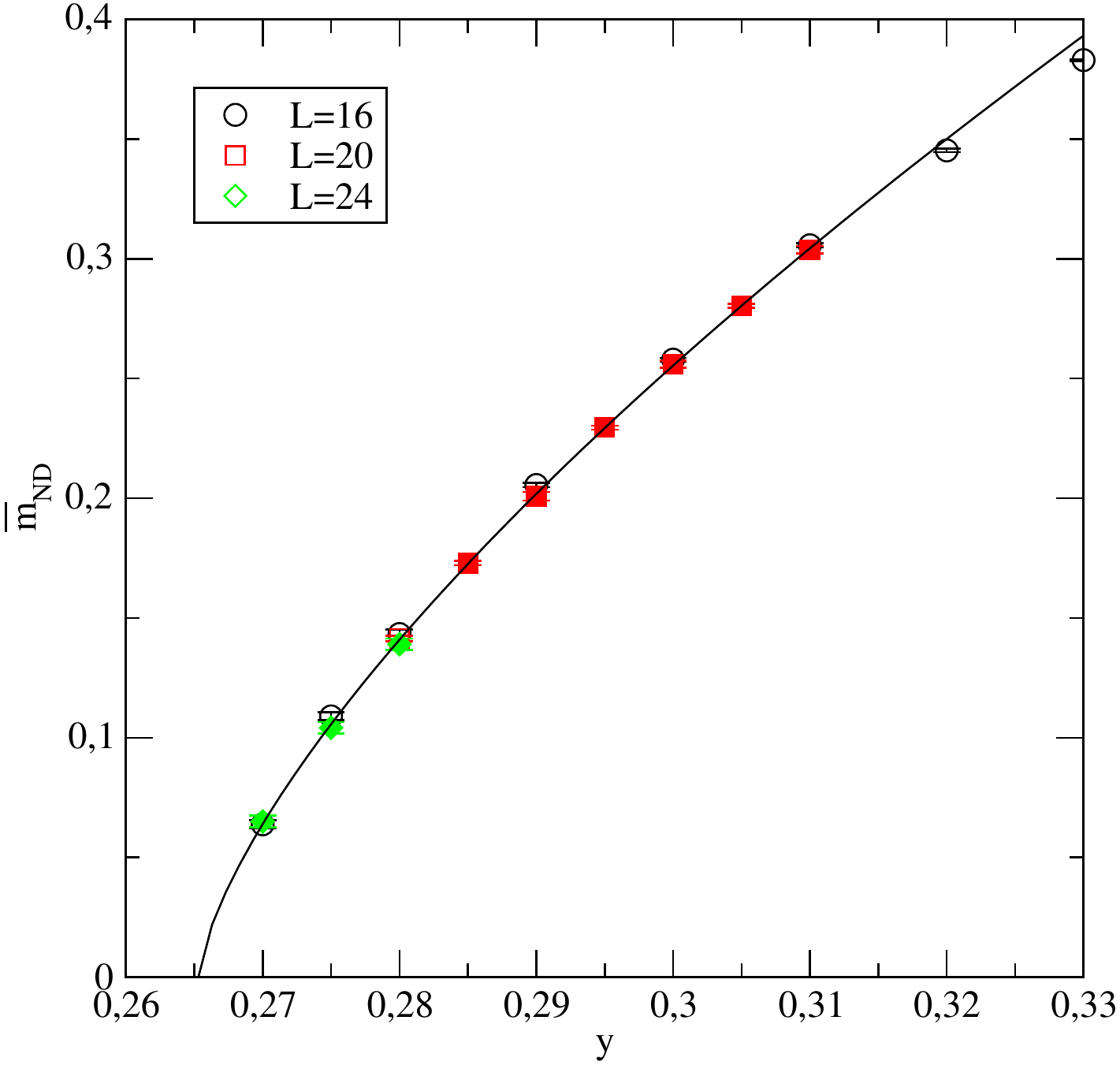}\hfill
     \includegraphics[width=.32\textwidth]{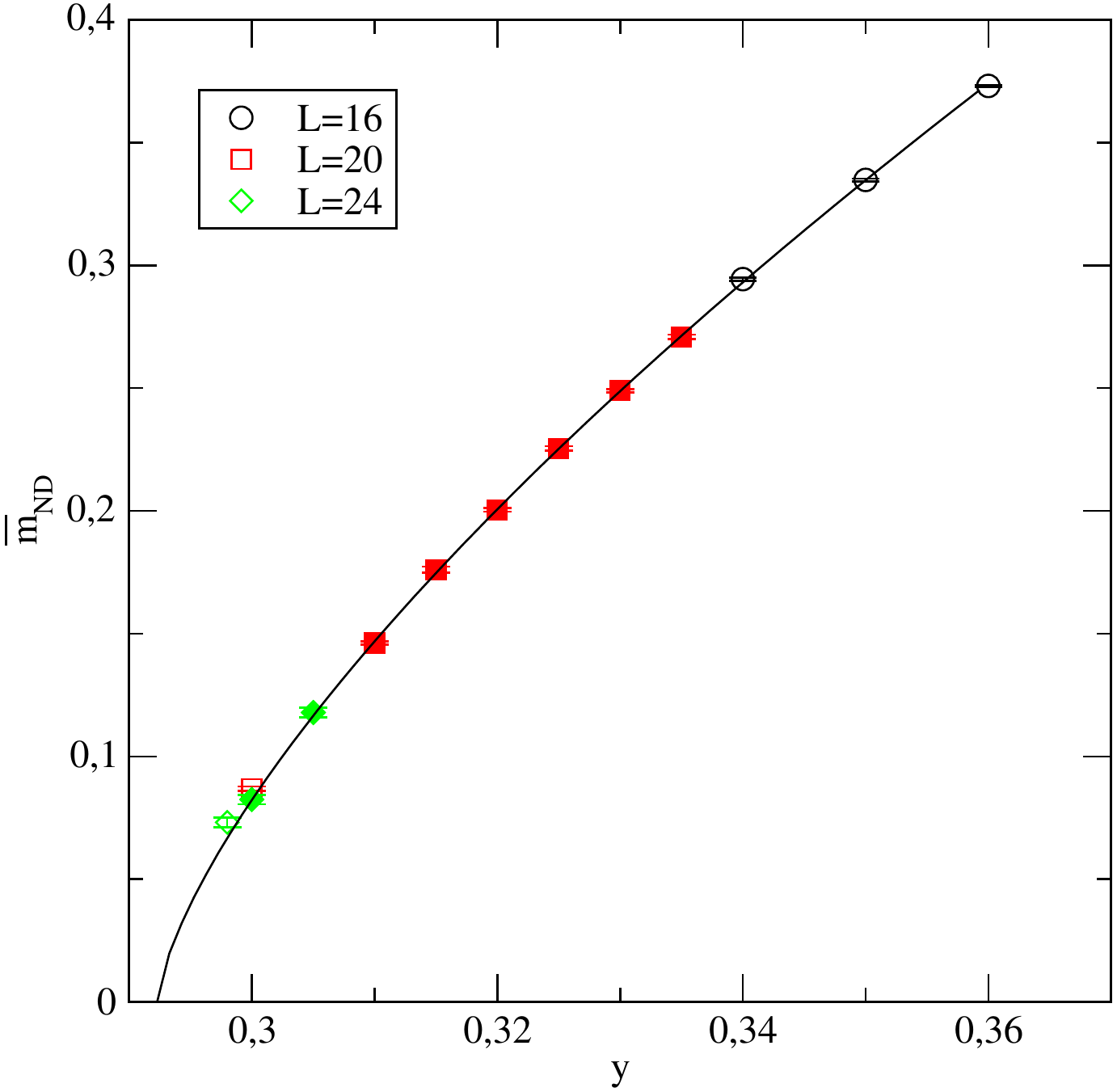}\hfill
     \includegraphics[width=.32\textwidth]{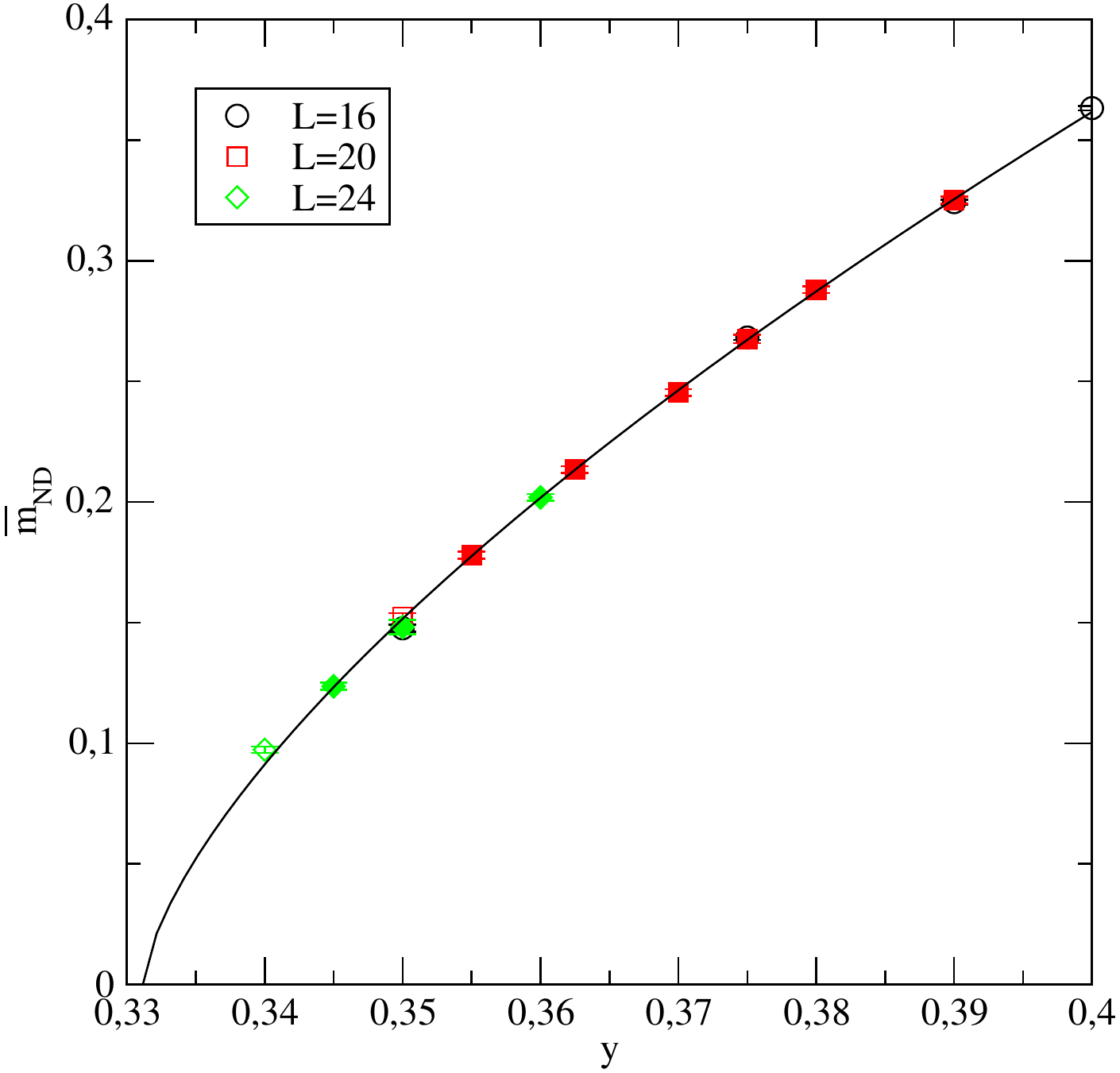}
     \caption{Scaling analysis for the approach to the conformal phase in the ideal walking model of \cite{Rantaharju:2019nmh}. Each panel correspond to a different value of the lattice spacing $\beta=2.25, 2.5, 3$. The scaling analysis shows that the transition between the two phases is continuous. From \cite{Rantaharju:2019nmh}.}
     \label{fig:idealw2}
\end{figure}

In \cite{Rantaharju:2019nmh} the region $g>g_{cr}$ was studied. In this phase, the large value of the four-fermion coupling alters the large distance behavior of the model, which results chirally broken.
The approach to the conformal phase was studied by looking at an order parameter for the conformal phase as $(g-g_{cr})\rightarrow0^+$. It was found that the approach to the conformal phase is continuous, resulting in a second order transition from the chirally broken phase to the conformal phase (see Fig.~\ref{fig:idealw2}). 
This shows that, at least in this model, the ideal walking scenario is realized. 

\section{Partial compositeness}

In this section, we review the available lattice results for models with top partial compositeness, both with only fermions as in Sec.~\ref{sec:PCwfermions} and in theories with fundamental partial compositeness with scalars of Sec.~\ref{FPC}.

\subsection{$\Sp(4)$ with fundamental and two-index antisymmetric representations} \label{sec:PCSp4}

This model has been first proposed in~\cite{Barnard:2013zea} to realize top partial compositeness with fermions, and then identified in~\cite{Ferretti:2014qta,Ferretti:2016upr} as one of the 12 feasible models.  In fact, the same dynamics can be used to realise two models of top partial compositeness: following the nomenclature of~\cite{Belyaev:2016ftv}, called M8 and M5 and distinguished by the multiplicities of the two species of fermions. This model is also a realization of the minimal model for partial compositeness discussed in Ch.~\ref{ch:su2}.

\begin{figure}[t!]
     \center
     \includegraphics[width=.32\textwidth]{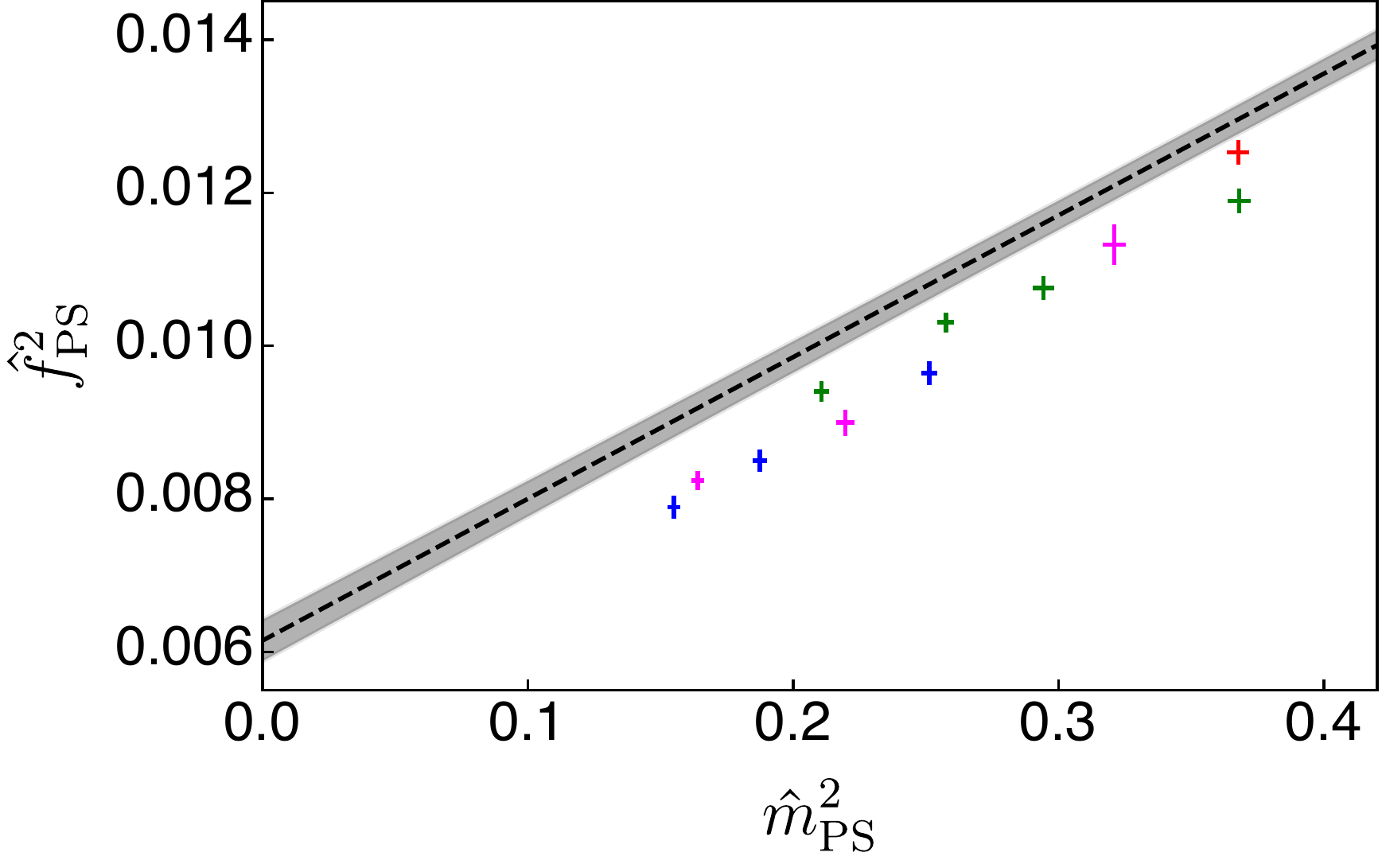}\hfill
     \includegraphics[width=.32\textwidth]{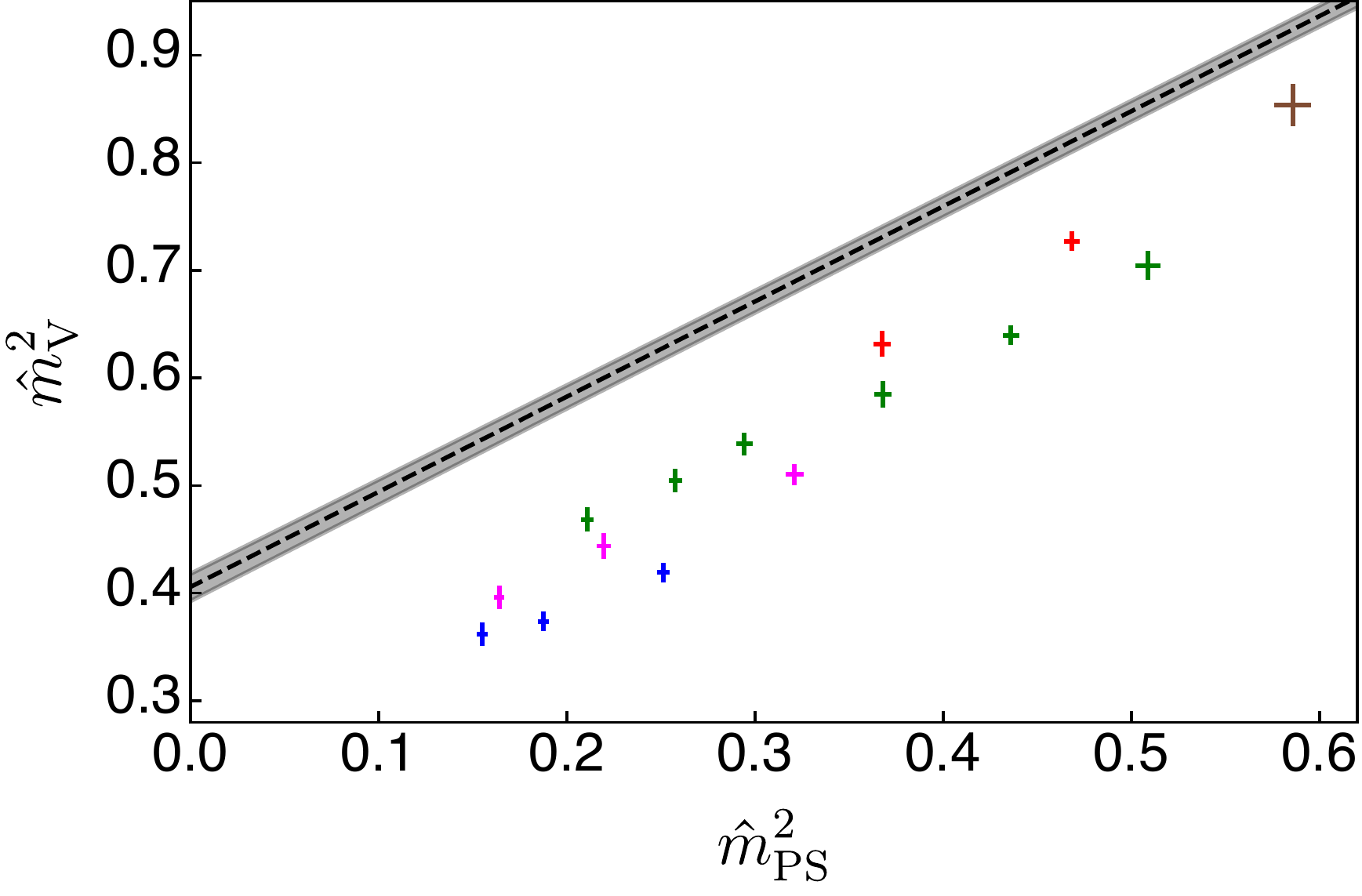}\hfill
     \includegraphics[width=.32\textwidth]{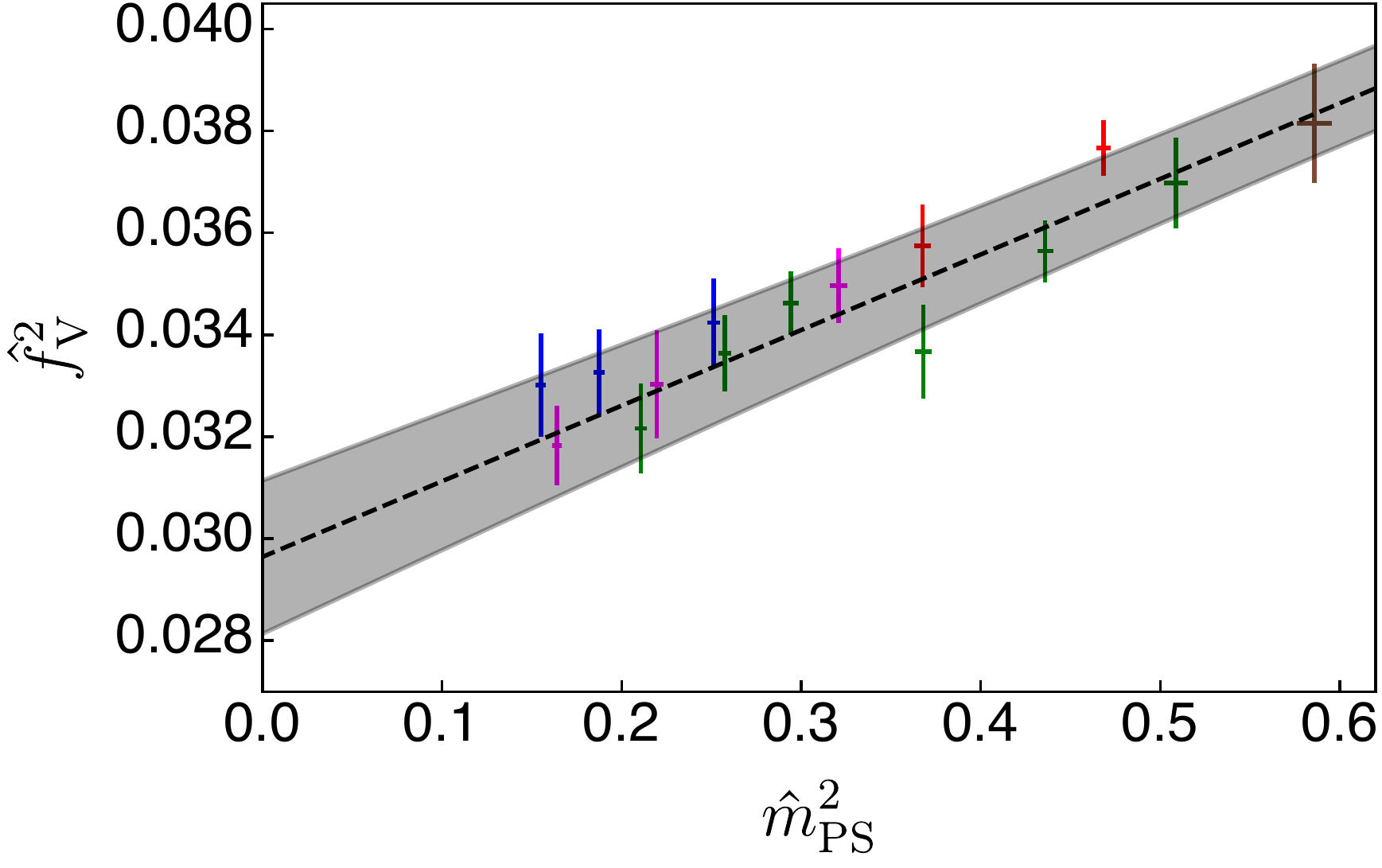} 
     \caption{Sample fo results for the continuum spectrum of $Sp(4)$ with $n_f=2$ fundamental Dirac fermions. More states are available from the original work. From \cite{Bennett:2019jzz}.}
     \label{fig:sp4}
\end{figure}

Both dynamical and quenched simulation of the $\Sp(4)$ gauge group were performed in \cite{Bennett:2017kga,Lee:2018ztv,Bennett:2019cxd,Bennett:2019jzz}.  The meson mass spectra for fermions in both the 2-index antisymmetric and fundamental representation of the gauge group were obtained in the quenched approximation \cite{Bennett:2019cxd}, while result with dynamical fermions are available only for $n_f=2$ in the fundamental representation \cite{Bennett:2019jzz}. A sample of the results are shown in Fig.~\ref{fig:sp4}.

For the dynamical spectrum, in \cite{Bennett:2019jzz} the authors perform a chiral and continuum extrapolation\footnote{The renormalization constants used are obtained from perturbation theory in this study.} of the mesonic spectrum of the model. In particular they compare their result for the ratio $M_V/f_{PS}$ with the other results available for $n_f=2$ models with gauge groups $\SU(2)$, $\SU(3)$, $\SU(4)$ and find that the ratio is approximately decreasing as the number of gauge degrees of freedom increases.
For $\Sp(4)$ they find $M_V/f_{PS}=8.08(32)$.

\subsection{$\SU(4)$ with sextet and fundamental fermions}
The case of the $\SU(5)/\SO(5)$ coset can be realized with five Majorana fermions in two-index antisymmetric (sextet) representation of SU(4). This coset has also been suggested as the base for a model of partial compositeness with three additional Dirac fundamental fermions \cite{Ferretti:2016upr}.
In the classification of~\cite{Belyaev:2016ftv}, this dynamics can be used to generate two models, namely  M6 and M11.
The odd number of fermions makes this model harder to study via lattice simulations. 

\begin{figure}[t!]
     \center
     \includegraphics[width=.51\textwidth]{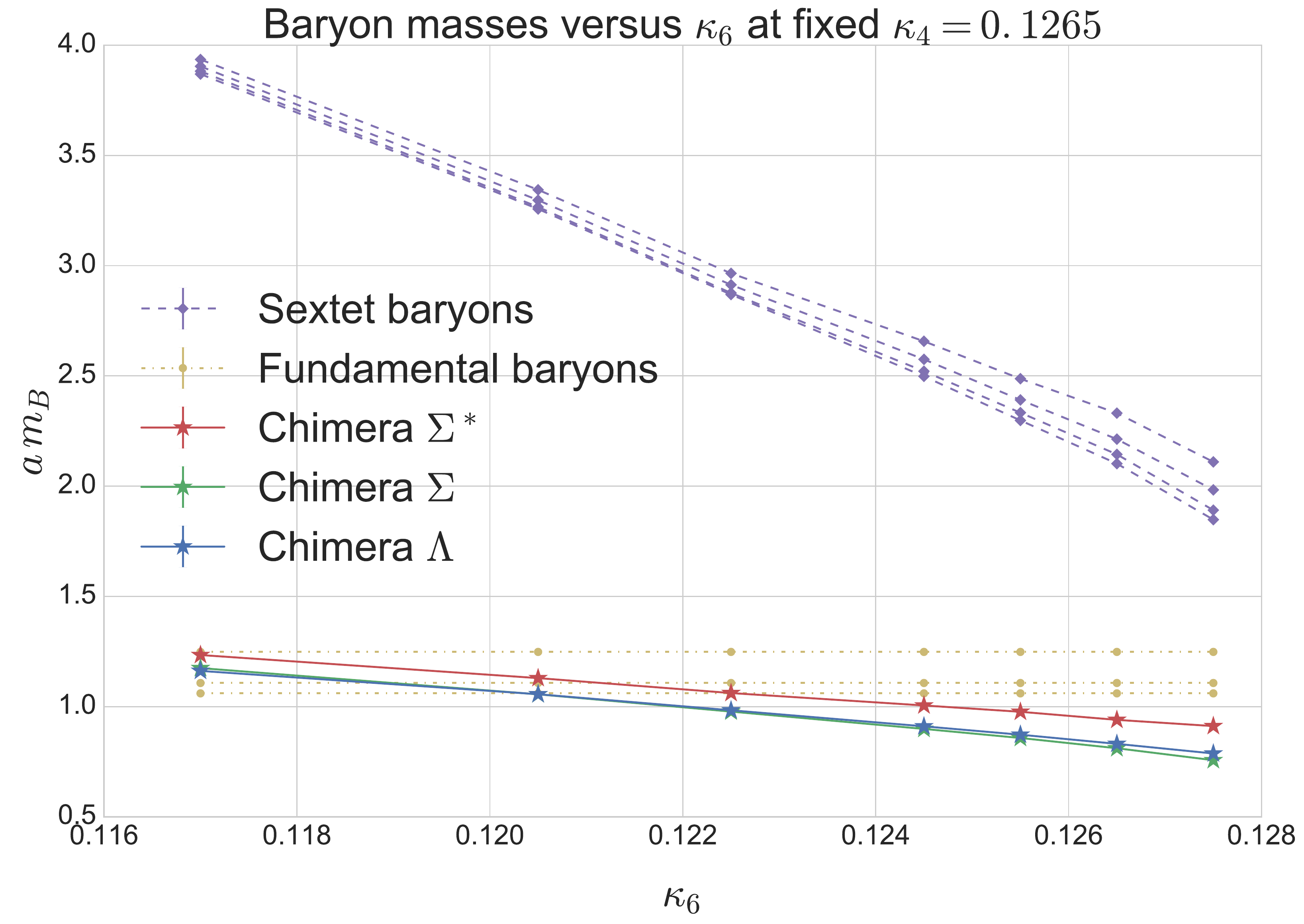}
     \caption{Spectrum of the $\SU(4)$ model with $n_f=2$ fundamental fermions and quenched sextet fermions. ``Chimera'' baryons are composite of quarks in the two representations. From \cite{DeGrand:2016mxr}. }
     \label{fig:jay}
     \end{figure}

As a first step, the case of $\SU(4)$ with four Majorana in the two-index antisymmetric representation and the case of SU(4) with two fundamental fermions plus (quenched) two-index antisymmetric fermions were studied \cite{DeGrand:2016mxr,DeGrand:2016het,Cossu:2019hse,Ayyar:2018glg,Ayyar:2019exp}.
For this simpler case, the spectrum of the model was presented in \cite{DeGrand:2016mxr}, which is shown in \reffig{fig:jay}. The model features baryons which are composite of fermions in both representations, and the quenched study shows that these can become lighter than baryon made of only one representation. This could be interesting for model building of light top-partners. However this result in the quenched approximation is not conclusive and the full spectrum of the model should be considered with dynamical fermions.

In \cite{DeGrand:2016het,Ayyar:2018glg,Ayyar:2019exp} the radiative contributions from electroweak gauge bosons to the composite Higgs potential of the $\SU(4)$ model with four Majorana fermions in the sextet representation were considered.
This calculation is similar to the electromagnetic contribution to the masses of the pions in QCD.
Electroweak gauge boson generate a potential for the composite Higgs $h$ of the form: $V(h)=C_{LR} (3g^2+g'^2)(h/F_\pi)^2+\mathcal{O}(h^4)$ where the positive constant $C_{LR}$ can be computed from the vacuum polarization function $\Pi_{LR}$. The measure of $C_{LR}$ on the lattice was presented in \cite{DeGrand:2016het, Ayyar:2019exp} and found to be roughly of the same size as its QCD counterpart. In the future this approach could be extended to the full model with fermions in two different representations.


\subsection{$\SU(2)$ with $n_f=2$ fundamental fermions and fundamental partial compositeness}
This model is the minimal realization of composite pNGB model, requiring only two fundamental fermions of $\SU(2)$, and as been discussed in detail in Ch.~\ref{ch:su2}.
The model can also be used to build a model of partial compositeness for all SM fermions \cite{Sannino:2016sfx}, which features hyper-colored scalars and it is free from Landau poles up to the Plank scale.


The spectrum of this simple theory, with only two colors and $n_f = 2$ Dirac fermions in the fundamental, has been presented in  \cite{Lewis:2011zb,Hietanen:2013fya,Hietanen:2014xca,Arthur:2014zda,Arthur:2014lma,Arthur:2016dir,Arthur:2016ozw,Pica:2016zst,Drach:2017jsh,Drach:2017btk,Janowski:2019svg}, and we already discussed the results in Sec.~\ref{sec:spectrum} (see Fig.~\ref{fig:drach}).

\begin{figure}[t!]
     \center
     \includegraphics[width=.49\textwidth]{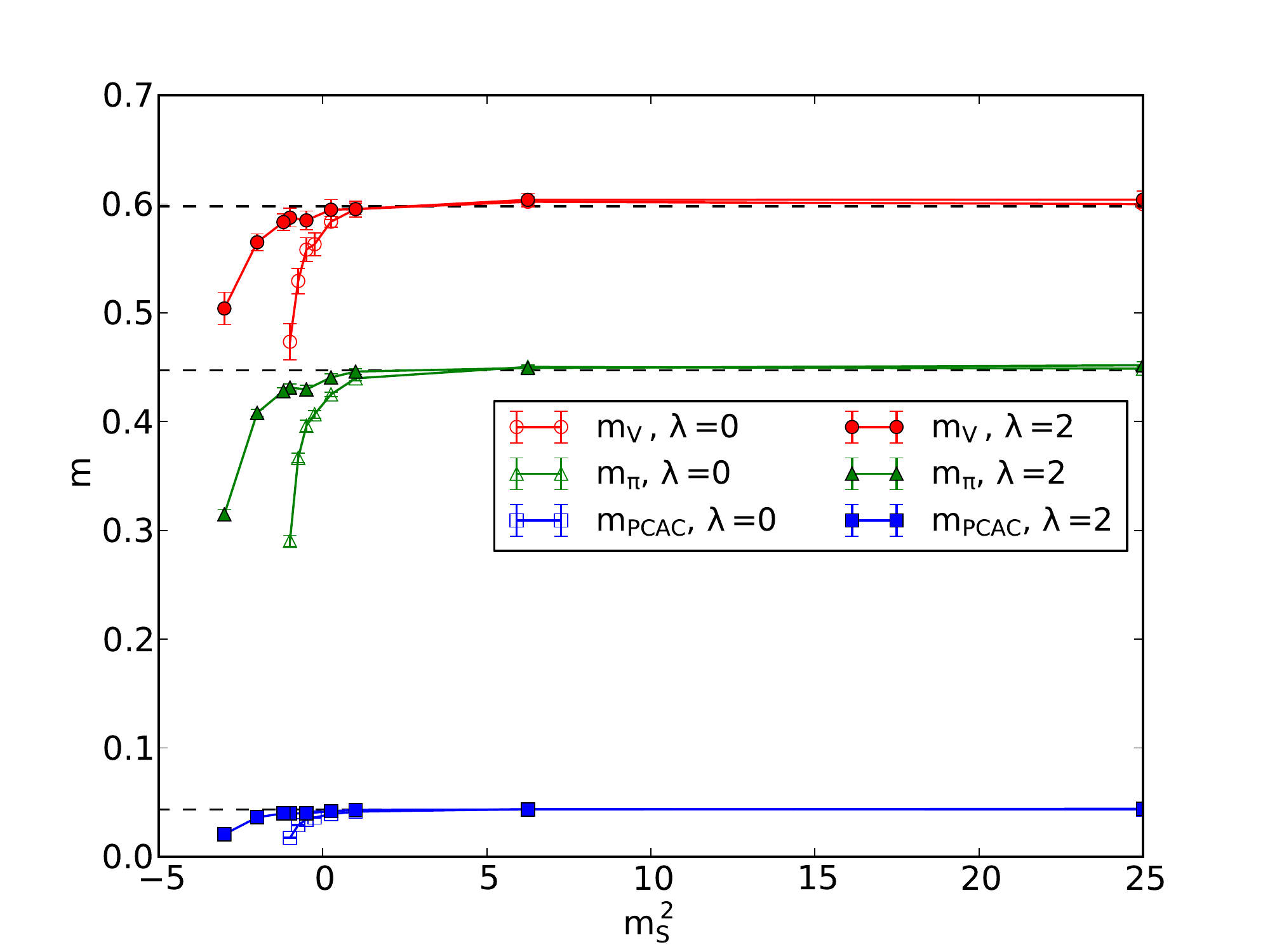}\hfill
     \includegraphics[width=.49\textwidth]{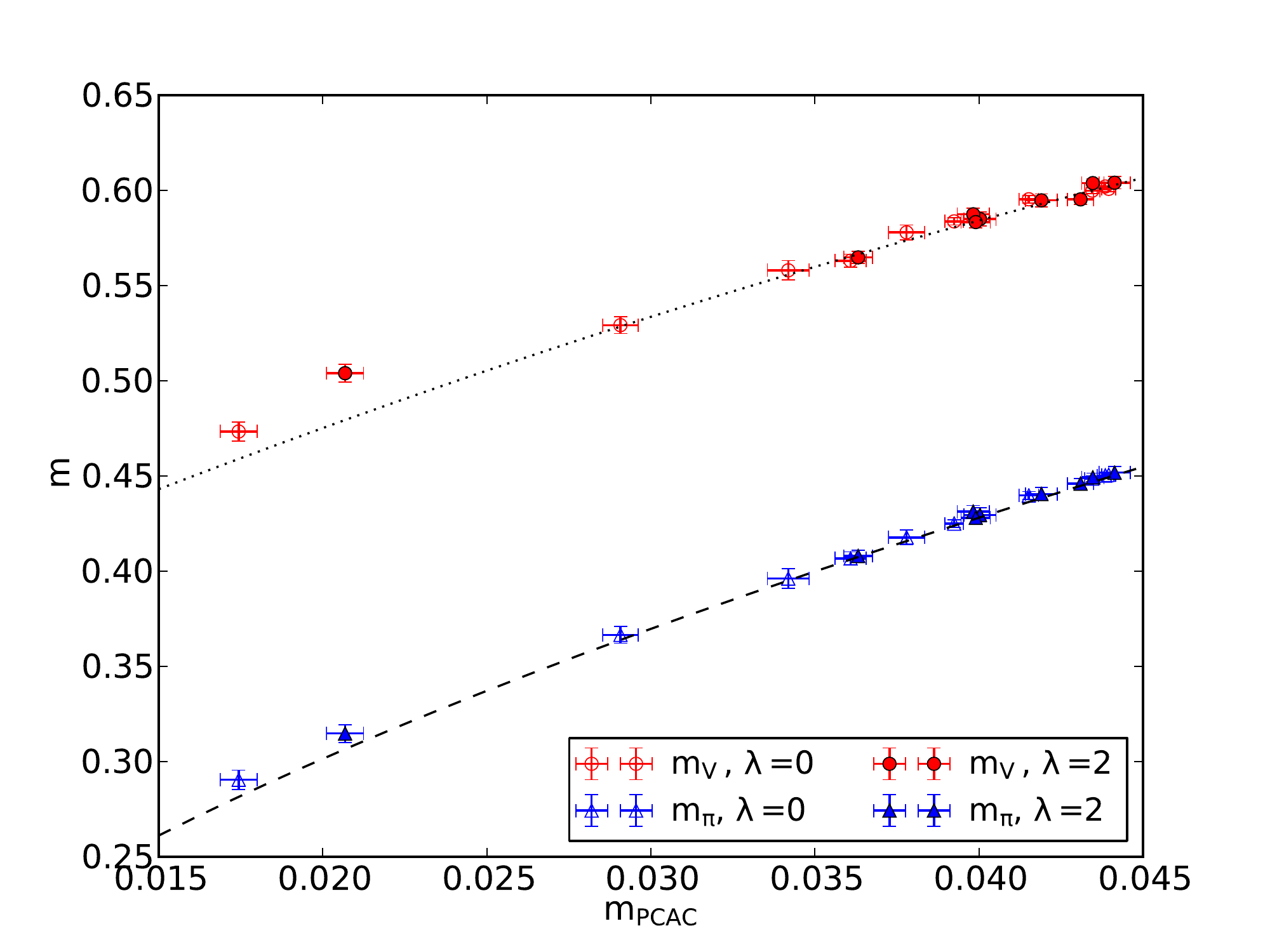}
     \caption{Spectrum of the fundamental partial compositeness model considered in~\cite{Hansen:2017mrt}. Left panel: Meson mass dependence on the scalar mass $m_s^2$ and quartic coupling $\lambda$. Right panel: The effect of the strongly coupled scalar field amount to an effective renormalization of the PCAC on the spectrum of the low-lying mesons. From \cite{Hansen:2017mrt}. }
     \label{fig:fpc}
\end{figure}

This model can be extended by $\SU(2)$--charged scalars to a model with fundamental partial compositeness, as discussed in Sec.~\ref{MFPC}.
Some preliminary results were obtained for a model based on $\SU(2)$ with $n_f=2$ fundamental fermions and one colored scalar field also in the fundamental representation~\cite{Hansen:2017mrt}.
For this particular model there is only one additional interaction operator in the Lagrangian, a quartic scalar potential with coupling $\lambda$. No Yukawa interactions are possible in the model, and the scalar and fermionic sectors interact only via gluon interactions. In~\cite{Hansen:2017mrt}, the effect of the strongly coupled scalar field on the mesonic spectrum of the model was studied. 
It was found that the presence of a relatively light scalar field changes significantly the values of the low-lying meson masses, respect to the case in which no scalar is present or, equivalently, the scalar decouples as an infinitely heavy state.
It was also found that the change in the model can be well described as an effective renormalization of the quark mass $m_{pcac}$, in the range of scalar masses $m_s^2$ and quartic coupling $\lambda$ explored (see~\reffig{fig:fpc}), with the values of the measured meson masses lying on the curves $m_X(m_{pcac})$ of the theory with no scalar fields present.  
Moreover it was found that no Higgs phase was present in the region corresponding to positive fermion masses.

These results are still preliminary and additional investigations of the model are required to have a more complete understanding of the effect of color scalar particles on the fermionic sector of the model.

\chapter{Wrapping up and our philosophy}
\label{conclusions}
The reader by now realized that, overall, composite dynamics involves  different subjects, from fundamental gauge theories to first principle numerical simulations, from other non-perturbative methods to effective field theories. Several useful reports exist delving with each of these subjects. Our philosophy has been to give the reader a sense of how composite dynamics works and can be used for model building. We hope to have achieved this by working out in detail the dynamics, and its applications, of one of the simplest models of composite dynamics that singles itself out for the plethora of possible phenomenological applications, ranging from composite Higgs physics to dark matter. The model we chose to present features only three new gauge bosons arranged in the adjoint of the $\SU(2)$ gauge group and two Dirac fermions. As argued in the previous chapters, the choice of two Dirac fermions allows for just about enough global symmetry structure to make the theory an ideal playground for several bright and dark model building ideas. To our own surprise, and probably also yours, despite the simplicity and relevance of the model, its first serious lattice simulations appeared only at the beginning of this decade. Before then little was known, except for the expected (but not tested) breaking pattern, about the overall spectrum and dynamics of the theory. Because of the low number of colors, one cannot even use the already naive large number of color limit to deduce properties about its spectrum by rescaling the QCD one. We have, therefore, also summarized the state-of-the-art of the related lattice results for this theory that includes, but is not limited to, the pattern of chiral symmetry breaking as well as the spectrum of goldstones, spin zero and spin one resonances and their fermion mass dependence. 

We have also introduced the low energy effective theory for this model and shown how to utilize it to construct an important and minimal template of composite models of the Higgs that naturally features and supports the Technicolor and the Composite Goldstone Higgs limits. In fact, we have reviewed how this model is also an outstanding example of models of composite dark matter for which the relic density can either be due to an asymmetry or  number changing operators. Within the electroweak contest we summarized, in some detail,  how to take into account the electroweak corrections once the model is used to replace the elementary Higgs sector of the Standard Model. The phenomenological constraints and predictions for collider experiments have been explained and reviewed via the effective Lagrangian approach while keeping into account the knowledge of the lattice spectrum of the theory. We spelled out the difference and similarities of the Technicolor limit of the theory and its alter ego, i.e. the composite Goldstone Higgs one. Chapter two contained all of the above as well as the aforementioned composite dark matter physics emerging from this model and its impact.  

The main crux of composite models of the Higgs is the phenomenologically consistent generation of the Standard Model fermion masses. In Chapter three we provided an up-to-date review of the challenges and reviewed different solutions and approaches ranging from the adoption of near conformal dynamics of walking nature to partial compositeness, and their possible implementations at a more fundamental level. We summarized the status and shown how to construct explicit realizations of fermion mass generation of different type, and compared them. 

Of course, underlying fundamental theories of composite dynamics can feature very distinct dynamics. This is not only because of the different quantum global symmetries that they feature and the way these can break spontaneously, but also because, depending on the number of flavors, colors and matter representation, they display a very rich phase diagram at low energies. Theories can develop infrared fixed points, display near conformal dynamics, chiral symmetry breaking and/or confinement.  This  partially constitutes what is known as the conformal window in the number of colors versus number of flavors, for given representation, of four dimensional gauge theories.  In Chapter four we provided the most adjourned summary of the lattice results, a crucial chapter that will inform our readers of what theories are most suitable for their model building needs.  
%
%
%
 
  We sincerely hope that our concise introduction to the exciting topic of composite dynamics has tickled your mind and that you decide to try to work on it yourself.  

\newpage

\appendix

\chapter{List of abbreviations}

\begin{itemize}

\item[-] BKT : Berezinskii--Kosterlitz--Thouless

\item[-] cDM : composite Dark Matter

\item[-] C.L. : Confidence Level

\item[-] ETC : Extended Technicolor

\item[-] EW : Electroweak

\item[-] EWPTs : Electroweak Precision Tests

\item[-] EWSB : Electroweak Symmetry Breaking

\item[-] FC : Fundamental Color

\item[-] FCNC : Flavor-Changing Neutral Current

\item[-] GB : Goldstone boson

\item[-] ggH : Gluon Fusion (Higgs)

\item[-] IR : Infra-Red

\item[-] LEC : Low Energy Constant

\item[-] LGT : Lattice Gauge Theory

\item[-] LO : Leading Order

\item[-] NJL : Nambu--Jona-Lasinio

\item[-] NLO : Next-to-Leading order

\item[-] pNGB : pseudo Nambu--Goldstone boson 

\item[-] PT : Perturbation Theory

\item[-] SIMP : Strongly Interacting Massive Particle

\item[-] SM : Standard Model

\item[-] TC : Technicolor

\item[-] VBF : Vector Boson Fusion

\item[-] VEV : Vacuum Expectation Value

\item[-] VH : W and Z-strahlung

\item[-] UV : Ultra-Violet

\item[-] WIMP : Weakly Interacting Massive Particle

\item[-] WZW : Wess--Zumino--Witten

\end{itemize}

\bibliography{Reviewlitt}
\bibliographystyle{unsrt}

\end{document}